\documentclass[a4paper, 11pt]{article}
\usepackage{etoolbox}
\usepackage{setspace}
\usepackage{jheppub}\makeatletter\gdef\@fpheader{}\makeatother 
\makeatletter 
\patchcmd{\maketitle}{\pagenumbering{arabic}}{}{}{} 
\patchcmd{\ps@myplain}{\pagenumbering{arabic}}{}{}{} 
\patchcmd{\maketitle}{\tableofcontents}{\begin{spacing}{.98}\tableofcontents\end{spacing}}{}{} 
\makeatother
\renewcommand\beforetochook{\pagestyle{myplain}\pagenumbering{arabic}}
\usepackage{mathtools, tabularx, graphicx, threeparttable, placeins}
\usepackage{youngtab}\Yboxdim 6.5pt\Ylinethick 0.4pt
\usepackage{dcolumn}\newcolumntype{d}[1]{D{.}{.}{#1}}
\usepackage{microtype} 
\usepackage[dvipsnames,table]{xcolor}
\usepackage[bb=dsfontserif]{mathalpha}
\usepackage[framemethod=TikZ, backgroundcolor=Aquamarine!15]{mdframed}
\mdfsetup{skipabove=0pt, skipbelow=0pt, linewidth=0pt}
\graphicspath{{Figures/}}  
\definecolor{shade1}{gray}{0.9}
\colorlet{shade}{green!40!blue!15}

\let\oldtabular\tabular
\let\endoldtabular\endtabular
\renewenvironment{tabular}{\rowcolors{2}{white}{shade}\oldtabular}{\endoldtabular}

\newcommand\mc[1]{\multicolumn{1}{c|}{#1}}

\newcommand{\R}{\mathcal{R}}
\newcommand{\U}{\textrm{U}}
\newcommand{\SU}{\textrm{SU}}
\newcommand{\USp}{\textrm{USp}}  
\def\T{\mathbb{T}}
\def\Z{\mathbb{Z}} 

\author[a]{Mudassar Sabir,}
\author[b]{Adeel Mansha,}
\author[c,d,e]{Tianjun Li,}
\author[a]{Zhi-Wei Wang}

\affiliation[a]{School of Physics, University of Electronic Science and Technology of China, \\ 2006 Xiyuan Avenue, Chengdu, China} 
\affiliation[b]{College of Physics and Optoelectronic Engineering, Shenzhen University, \\ 3688 Nanhai Avenue, Shenzhen, China}
\affiliation[c]{CAS Key Laboratory of Theoretical Physics, Institute of Theoretical Physics, \\ Chinese Academy of Sciences, \\ 55 Zhong Guan Cun East Road, Beijing, China}
\affiliation[d]{School of Physical Sciences, University of Chinese Academy of Sciences, \\ 1 Yanqihu East Road, Beijing, China}
\affiliation[e]{School of Physics, Henan Normal University, \\ 46 Jianshe East Road, Xinxiang, China}

\emailAdd{mudassar.sabir@uestc.edu.cn}
\emailAdd{tli@itp.ac.cn} 
\emailAdd{adeelmansha@alumni.itp.ac.cn}
\emailAdd{zhiwei.wang@uestc.edu.cn}

\arxivnumber{2409.09110}

\begin{document}

\title{Fermion masses and mixings in the supersymmetric Pati-Salam landscape from Intersecting D6-Branes}

\abstract{Recently, the complete landscape of three-family supersymmetric Pati-Salam models from intersecting D6-branes on a type IIA $\mathbb{T}^6/(\mathbb{Z}_2\times \mathbb{Z}_2)$ orientifold has been enumerated consisting of 33 independent models with distinct gauge coupling relations at the string scale. Here, we study the phenomenology of all such models by providing the detailed particle spectra and the analysis of the possible 3-point and the 4-point Yukawa interactions in order to accommodate all standard-model fermion masses and mixings. We find that only 17 models contain viable Yukawa textures to explain quarks masses, charged-leptons' masses, neutrino-masses, quarks' mixings and leptons' mixings. These viable models split into four classes, viz. a single model with 3 Higgs fields from the bulk and sixteen models with either 6, 9 or 12 Higgs from the $\mathcal{N}=2$ sector. The models perform successively better with the increasing number of Higgs pairs. Remarkably, the class of models with 12 Higgs naturally predicts the Dirac-type neutrino masses in normal ordering consistent with both the experimental constraints as well as the bounds from the swampland program. }

\maketitle
\flushbottom
\section{Introduction}\label{sec:Intro}

Standard Model (SM) fermions appear in chiral representations of the gauge group $\SU(3)_C \times \SU(2)_L \times \U(1)_Y$. Intersecting D6-branes in type IIA string theory provide a natural mechanism to realize chiral fermions at D6-brane intersections \cite{Aldazabal:2000cn}. D6-branes fill the four-dimensional spacetime and have three extra dimensions along the compactified directions in IIA string theory. As the latter three extra dimensions are exactly equal to half of the number of the compactified dimensions, thus two generic D6-branes intersect at one point of the extra dimensions. This intersection is where the fields arising from open strings stretched between two different D6-branes live. Family replication results from multiple intersections of D6-branes that fill four-dimensional spacetime and extend into three compact dimensions. The volumes of the cycles wrapped by D-branes determine the four-dimensional gauge couplings, while the total internal volume yields the gravitational coupling. Yukawa couplings arise from open world-sheet instantons, specifically the triangular worldsheets stretched between intersections where fields involved in the cubic coupling reside. These instanton effects are suppressed by $\exp(-A_{ijk}T)$, where $A_{ijk}$ is the area of the triangle bounded by intersections $\{i, j, k\}$ and $T$ is the string tension \cite{Cremades:2003qj}. This exponential suppression explains the fermion mass hierarchies and mixings. However, embedding the Standard Model (SM) in a Calabi-Yau compactification with three families of chiral fermions and achieving correct fermion mass hierarchies and mixings in a positively curved (de Sitter) universe with stabilized moduli has been a challenge. 

In intersecting D6-brane constructions, we usually relax the requirement of stable de Sitter and impose minimal supersymmetry. It turns out that realistic Yukawa textures with three families favors products of unitary gauge groups over the simple unitary group. And the K-theory conditions \cite{Witten:1998cd, Uranga:2000xp}, being mod~4, are more easily satisfied for $\U(2N)$ with $N \in \Z$. For instance, in trinification models, no viable three-family model meeting the stringent constraints of $\mathcal{N}=1$ supersymmetry, tadpole cancellation, and K-theory constraints has been found \cite{Mansha:2024yqz}. Consequently, the left-right symmetric Pati-Salam group, $\SU(4)_C\times \SU(2)_L\times \SU(2)_R$, emerges as the most promising choice for realistic models. The rules to construct supersymmetric Pati-Salam models on a $\T^6/(\Z_2\times \Z_2)$ orientifold from intersecting D6-branes with the requirement of $\mathcal{N}=1$ supersymmetry, tadpole cancellation and the K-theory constraints were outlined in \cite{Cvetic:2004ui, Blumenhagen:2006ci, Blumenhagen:2005mu}. Similar construction is employed in recent works \cite{Li:2019nvi, Li:2021pxo, Mansha:2022pnd, Sabir:2022hko, Mansha:2023kwq}. Recently, the complete landscape of consistent three-family $\mathcal{N}=1$ supersymmetric Pati-Salam models from intersecting D6-branes on a $\T^6/(\Z_2 \times \Z_2)$ orientifold in IIA string theory has been mapped, comprising of only 33 distinct models \cite{He:2021gug}.  

In the typical toroidal orientifold compactifications, not all of the fermions sit on the localized intersections on the same torus which results in the rank-1 problem of the Yukawa mass matrices. The viable models with rank-3 Yukawa matrices split into four classes with 3, 6, 9 or 12 Higgs fields. Note that the two light Higgs mass eigenstates arise from the linear combination of the VEVs $v^i_{u,d} = \langle H^i_{u,d}\rangle$ of the available Higgs fields present in the model \cite{Chamoun:2003pf, Higaki:2005ie}. There is a single model with 3 bulk Higgs fields while the other three classes consist of five models with 6 Higgs, eight models with 9 Higgs and three models with 12 Higgs from $\mathcal{N}=2$ sector. We systematically compute the possible three-point and four-point Yukawa couplings to accommodate the masses of up-type quarks, down-type quarks, charged leptons, and Dirac-type neutrinos\footnote{Majorana neutrino masses can be trivially generated via the type-I seesaw mechanism, as discussed in \cite{Mayes:2019isy}, where the Dirac-neutrino mass matrix is input into a seesaw mechanism to produce a Majorana mass matrix.}, as well as quarks' (CKM) mixings and leptons' (PMNS) mixings for all viable models in the landscape where Wilson fluxes are set to zero. The results of the analysis of soft terms from the supersymmetry breaking will be presented elsewhere in \cite{Sabir:2024jsx}.

The model with three Higgs fields from the bulk that can accommodate the masses of up-type quarks, down-type quarks, and charged leptons, but cannot account for quarks' (CKM) or leptons' (PMNS) mixings. Among the five 6-Higgs models, one model lacks viable four-point couplings, which prevents it from precisely accommodating the masses of down-type quarks, though it can approximately explain CKM mixings. The remaining four 6-Higgs models can precisely match the masses of up-type quarks, down-type quarks, and charged leptons, but the CKM mixings are only approximately matched. The class of eight models with 9-Higgs can precisely match the masses of up-type quarks, down-type quarks, and charged leptons, as well as the CKM mixings. However, none of these models can explain the Dirac-neutrino masses or the PMNS mixings. In the class of three 12-Higgs models, one model lacks viable four-point couplings, but using only the three-point couplings, it can exactly reproduce the correct masses of all fermions—up-type quarks, down-type quarks, charged leptons, Dirac neutrinos—and the CKM mixings, except for the leptons' mixings that remain unexplained. Remarkably, the other two models in this class can accommodate precise PMNS mixings, along with all fermion masses, by incorporating corrections from the four-point couplings. 

In refs.~\cite{Casas:2024ttx, Casas:2024clw} the tiny Yukawa couplings originating from the bulk Higgs fields to explain small neutrino masses are argued to be related to the infinite distance limit~\cite{Lee:2019wij} in the moduli space where a light of tower states, dubbed gonions~\cite{Aldazabal:2000cn}, appears signalling the decompactification of one or two compact dimensions. It was also noted in \cite{Casas:2024clw} that unlike the Yukawas from the bulk Higgs fields, the Yukawa couplings associated with $\mathcal{N}=2$ sector are not exponentially suppressed by four-dimensional dilaton which typically signals the infinite distance limit. Thus, the Yukawas arising from the $\mathcal{N}=2$ sector are insensitive to bulk moduli and as a result, the issue of decompactification of extra dimensions does not arise. 

Recent evidence from the swampland program \cite{Vafa:2005ui}, particularly from the non-SUSY AdS instability conjecture \cite{Ooguri:2016pdq} and the light fermion conjecture \cite{Gonzalo:2021fma, Gonzalo:2021zsp}, building on the earlier work of \cite{Arkani-Hamed:2007ryu, Arnold:2010qz} suggests that without additional chiral fermions with tiny masses, neutrinos must be of Dirac-type together with a bound on the lightest neutrino mass given by the cosmological constant scale as, $m_{\nu}^{\rm lightest} \lesssim \Lambda^{1/4}$. The 3D Casimir energy of the SM compactified on a circle receives a positive contribution from the lightest neutrino, which is necessary to avoid unstable non-supersymmetric AdS vacua. This constraint is only satisfied for Dirac neutrinos, which carry 4 degrees of freedom, unlike Majorana neutrinos, which only have 2 and cannot compensate for the 4 bosonic degrees of freedom from the photon and the graviton. This also avoids the inevitable lepton-number violations in the Majorana case. Hence, it is crucial in string theory to generate tiny Dirac Yukawa couplings while keeping the other Yukawa couplings and SM gauge couplings unsuppressed. Previous attempts to generate Dirac neutrino masses in intersecting D-branes primarily focused on Euclidean D2-brane instantons within local models \cite{Blumenhagen:2009qh, Cvetic:2008hi, Ibanez:2008my}. For a recent survey on this issue, see Ref.~\cite{Casas:2024clw}.

We show that the problem of obtaining tiny Dirac-neutrino Yukawa couplings while keeping the other Yukawa couplings and SM gauge couplings unsuppressed requires at least twelve Higgs fields from the $\mathcal{N}=2$ sector, which is the maximum available in the landscape, specifically in Models \hyperref[model21]{21}, \hyperref[model22]{22}, and \hyperref[model22.5]{22-dual} (corresponding to Models 19, 21, and 12 respectively in \cite{He:2021gug}). The class of models with twelve Higgs naturally predicts Dirac-type neutrino masses with the normal ordering $\sim (50.4,~10.5,~6.1)~\mathrm{meV}$ consistent with both the experimental constraints and the bounds from the swampland program based on the AdS instability conjecture \cite{Sabir:2024mfv}. 

This rest of the paper is organized as follows. We review the model building rules for constructing supersymmetric Pati-Salam models from stacks of intersecting D6-branes on a $\T^6/(\mathbb{Z}_2\times \mathbb{Z}_2)$ orientifold in section \ref{sec:orientifold}. In section \ref{sec:Yukawa} we describe the calculation of three-point and four-point Yukawa couplings in intersecting D6-brane models. We then proceed to systematically compute all possible three-point and four-point functions in all viable models in sections \ref{sec:3Higgs}, \ref{sec:6Higgs}, \ref{sec:9Higgs} and \ref{sec:12Higgs}. Finally, we conclude in section \ref{sec:conclusion}.  

\section{Pati-Salam model building from $\T^6/(\mathbb{Z}_2\times \mathbb{Z}_2)$ orientifold} \label{sec:orientifold}
In the orientifold $\T^6/(\mathbb{Z}_2\times \mathbb{Z}_2)$, $\T^6$ is a product of three 2-tori with the orbifold group $(\mathbb{Z}_2\times \mathbb{Z}_2)$ has the generators $\theta$ and $\omega$ which are respectively associated with the twist vectors $(1/2,-1/2,0)$ and $(0,1/2,-1/2)$ such that their action on complex coordinates $z_i$ is given by,
\begin{eqnarray}
	& \theta: & (z_1,z_2,z_3) \to (-z_1,-z_2,z_3), \nonumber \\
	& \omega: & (z_1,z_2,z_3) \to (z_1,-z_2,-z_3). \label{orbifold}
\end{eqnarray}
Orientifold projection is the gauged $\Omega \R$ symmetry, where $\Omega$ is world-sheet parity that interchanges the left- and right-moving sectors of a closed string and swaps the two ends of an open string as,
\begin{align}
	\textrm{Closed}: & \quad \Omega : (\sigma_1, \sigma_2) \mapsto (2\pi -\sigma_1, \sigma_2), \nonumber \\
	\textrm{Open}:   & \quad  \Omega : (\tau, \sigma) \mapsto (\tau, \pi - \sigma) ,                     
\end{align}
and $\R$ acts as complex conjugation on coordinates $z_i$. This results in four different kinds of orientifold 6-planes (O6-planes) corresponding to $\Omega \R$, $\Omega \R\theta$, $\Omega \R\omega$, and $\Omega \R\theta\omega$ respectively. These orientifold projections are only consistent with either the rectangular or the tilted complex structures of the factorized 2-tori. Denoting the wrapping numbers for the rectangular and tilted tori as $n_a^i[a_i]+m_a^i[b_i]$ and $n_a^i[a'_i]+m_a^i[b_i]$ respectively, where $[a_i']=[a_i]+\frac{1}{2}[b_i]$. Then a generic 1-cycle $(n_a^i,l_a^i)$ satisfies $l_{a}^{i}\equiv m_{a}^{i}$ for the rectangular 2-torus and $l_{a}^{i}\equiv 2\tilde{m}_{a}^{i}=2m_{a}^{i}+n_{a}^{i}$ for the tilted 2-torus such that $l_a^i-n_a^i$ is even for the tilted tori. 

The two different basis $(n^i,m^i)$ and $(n^i,l^i)$ are related as,
\begin{align}\label{basis-l-m}
	l^i & = 2^{\beta_i} (m^i + \frac{\beta_i}{2} n^i), \quad \beta_i = \begin{cases} 
	0 ~ & \mathrm{rectangular}~\T^2,                                                 \\ 
	1 ~ & \mathrm{tilted}~\T^2. \end{cases}                                          
\end{align}
We use the basis $(n^i,l^i)$ to specify the model wrapping numbers in appendix \ref{appA} while the basis $(n^i,m^i)$ is convenient to sketch the Yukawa textures in sections \ref{sec:3Higgs}, \ref{sec:6Higgs}, \ref{sec:9Higgs} and \ref{sec:12Higgs}.

The homology cycles for a stack $a$ of $N_a$ D6-branes along the cycle $(n_a^i,l_a^i)$ and their $\Omega \R$ images ${a'}$ stack of $N_a$ D6-branes with cycles $(n_a^i,-l_a^i)$ are respectively given as,
\begin{align}
	[\Pi_a ]   & =\prod_{i=1}^{3}\left(n_{a}^{i}[a_i]+2^{-\beta_i}l_{a}^{i}[b_i]\right), \nonumber \\
	[\Pi_{a'}] & =\prod_{i=1}^{3}\left(n_{a}^{i}[a_i]-2^{-\beta_i}l_{a}^{i}[b_i]\right),           
\end{align}
The homology three-cycles, which are wrapped by the four O6-planes, are given by
\begin{alignat}{2}
	\Omega \R :             & \quad & [\Pi_{\Omega \R}]             & = 2^3 [a_1]\times[a_2]\times[a_3],  \nonumber                             \\
	\Omega \R\omega :       &       & [\Pi_{\Omega \R\omega}]       & =-2^{3-\beta_2-\beta_3}[a_1]\times[b_2]\times[b_3],  \nonumber            \\
	\Omega \R\theta\omega : &       & [\Pi_{\Omega \R\theta\omega}] & =-2^{3-\beta_1-\beta_3}[b_1]\times[a_2]\times[b_3], \nonumber             \\
	\Omega \R\theta :       &       & [\Pi_{\Omega \R \theta}]      & =-2^{3-\beta_1-\beta_2}[b_1]\times[b_2]\times[a_3]. \label{orienticycles} 
\end{alignat}
The intersection numbers can be calculated in terms of wrapping numbers as,
\begin{align}
	I_{ab}  & =[\Pi_a][\Pi_b] =2^{-k}\prod_{i=1}^3(n_a^il_b^i-n_b^il_a^i),\nonumber                                               \\
	I_{ab'} & =[\Pi_a]\left[\Pi_{b'}\right] =-2^{-k}\prod_{i=1}^3(n_{a}^il_b^i+n_b^il_a^i),\nonumber                              \\
	I_{aa'} & =[\Pi_a]\left[\Pi_{a'}\right] =-2^{3-k}\prod_{i=1}^3(n_a^il_a^i),\nonumber                                          \\
	I_{aO6} & =[\Pi_a][\Pi_{O6}] =2^{3-k}(-l_a^1l_a^2l_a^3+l_a^1n_a^2n_a^3+n_a^1l_a^2n_a^3+n_a^1n_a^2l_a^3),\label{intersections} 
\end{align}
where $k=\sum_{i=1}^3\beta_i$ and $[\Pi_{O6}]=[\Pi_{\Omega \R}]+[\Pi_{\Omega \R\omega}]+[\Pi_{\Omega \R\theta\omega}]+[\Pi_{\Omega \R\theta}]$.

In order to have three families of the left chiral and right chiral standard model fields, the intersection numbers must satisfy
\begin{align}
	I_{ab} + I_{ab'} = 3 , \quad I_{ac} = -3, \quad I_{ac'} = 0. \label{eq:NoG} 
\end{align}  
 
\subsection{Constraints from tadpole cancellation and $\mathcal{N}=1$ supersymmetry}\label{subsec:constraints}
Since D6-branes and O6-orientifold planes are the sources of Ramond-Ramond charges they are constrained by the Gauss's law in compact space implying the sum of D-brane and cross-cap RR-charges must vanishes \cite{Gimon:1996rq}
\begin{table}[ht]
	\begin{center}
		\begin{tabular}{|c|c|c|}
			\hline
			Orientifold action      & O6-plane & $(n^1,l^1)\times (n^2,l^2)\times (n^3,l^3)$                    \\
			\hline
			$\Omega \R$             & 1        & $(2^{\beta_1},0)\times (2^{\beta_2},0)\times (2^{\beta_3},0)$  \\
			\hline
			$\Omega \R\omega$       & 2        & $(2^{\beta_1},0)\times (0,-2^{\beta_2})\times (0,2^{\beta_3})$ \\
			\hline
			$\Omega \R\theta\omega$ & 3        & $(0,-2^{\beta_1})\times (2^{\beta_2},0)\times (0,2^{\beta_3})$ \\
			\hline
			$\Omega \R\theta$       & 4        & $(0,-2^{\beta_1})\times (0,2^{\beta_2})\times (2^{\beta_3},0)$ \\
			\hline
		\end{tabular}
	\end{center}
	\caption{The wrapping numbers for four O6-planes.}
	\label{orientifold}
\end{table}
\begin{align}\label{RRtadpole}
	\sum_a N_a [\Pi_a]+\sum_a N_a \left[\Pi_{a'}\right]-4[\Pi_{O6}] & = 0, 
\end{align}
where the last terms arise from the O6-planes, which have $-4$ RR charges in D6-brane charge units. RR tadpole constraint is sufficient to cancel the ${\rm SU}(N_a)^3$ cubic non-Abelian anomaly while ${\rm U}(1)$ mixed gauge and gravitational anomaly or $[{\rm SU}(N_a)]^2 {\rm U}(1)$ gauge anomaly can be cancelled by the Green-Schwarz mechanism, mediated by untwisted RR fields \cite{Green:1984sg}.

Let us define the following products of wrapping numbers,
\begin{alignat}{4}
	A_a         & \equiv -n_a^1n_a^2n_a^3, & \quad B_a   & \equiv n_a^1l_a^2l_a^3, & \quad     C_a & \equiv l_a^1n_a^2l_a^3, & \quad   D_a & \equiv l_a^1l_a^2n_a^3, \nonumber          \\
	\tilde{A}_a & \equiv -l_a^1l_a^2l_a^3, & \tilde{B}_a & \equiv l_a^1n_a^2n_a^3, & \tilde{C}_a   & \equiv n_a^1l_a^2n_a^3, & \tilde{D}_a & \equiv n_a^1n_a^2l_a^3.\,\label{variables} 
\end{alignat}
Cancellation of RR tadpoles requires introducing a number of orientifold planes also called ``filler branes'' that trivially satisfy the four-dimensional ${\cal N}=1$ supersymmetry conditions. The no-tadpole condition is given as,
\begin{align}
	-2^k N^{(1)}+\sum_a N_a A_a & =-2^k N^{(2)}+\sum_a N_a B_a= \nonumber \\
	-2^k N^{(3)}+\sum_a N_a C_a & =-2^k N^{(4)}+\sum_a N_a D_a=-16,\,     
\end{align}
where $2 N^{(i)}$ is the number of filler branes wrapping along the $i^{\rm th}$ O6-plane. The filler branes belong to the hidden sector USp group and carry the same wrapping numbers as one of the O6-planes as shown in table~\ref{orientifold}. USp group is hence referred with respect to the non-zero $A$, $B$, $C$ or $D$-type.

Preserving ${\cal N}=1$ supersymmetry in four dimensions after compactification from ten-dimensions restricts the rotation angle of any D6-brane with respect to the orientifold plane to be an element of ${\rm SU}(3)$, i.e.
\begin{equation}
	\theta^a_1 + \theta^a_2 + \theta^a_3 = 0 \mod 2\pi ,
\end{equation}
with $\theta^a_j = \arctan (2^{- \beta_j} \chi_j l^a_j/n^a_j)$. $\theta_i$ is the angle between the $D6$-brane and orientifold-plane in the $i^{\rm th}$ 2-torus and $\chi_i=R^2_i/R^1_i$ are the complex structure moduli for the $i^{\rm th}$ 2-torus. ${\cal N}=1$ supersymmetry conditions are given as,
\begin{align}
	x_A\tilde{A}_a+x_B\tilde{B}_a+x_C\tilde{C}_a+x_D\tilde{D}_a     & = 0,\nonumber               \\
	\frac{A_a}{x_A}+\frac{B_a}{x_B}+\frac{C_a}{x_C}+\frac{D_a}{x_D} & < 0, \label{susyconditions} 
\end{align}
where $x_A=\lambda,\; x_B=2^{\beta_2+\beta_3}\cdot\lambda /\chi_2\chi_3,\; x_C=2^{\beta_1+\beta_3}\cdot\lambda /\chi_1\chi_3,\; x_D=2^{\beta_1+\beta_2}\cdot\lambda /\chi_1\chi_2$.

Orientifolds also have discrete D-brane RR charges classified by the $\mathbb{Z}_2$ K-theory groups, which are subtle and invisible by the ordinary homology \cite{Witten:1998cd, Cascales:2003zp, Marchesano:2004yq, Marchesano:2004xz}, which should also be taken into account~\cite{Uranga:2000xp}. The K-theory conditions are,
\begin{align}
	\sum_a \tilde{A}_a  = \sum_a  N_a  \tilde{B}_a = \sum_a  N_a  \tilde{C}_a = \sum_a  N_a \tilde{D}_a & = 0 \textrm{ mod }4 \label{K-charges}~.~\, 
\end{align}
In our case, we avoid the nonvanishing torsion charges by taking an even number of D-branes, {\it i.e.}, $N_a \in 2 \mathbb{Z}$.

\subsection{Particle spectrum}
To have three families of the SM fermions, we need one torus to be tilted, which is chosen to be the third torus. So we have  $\beta_1=\beta_2=0$ and $\beta_3=1$.
Placing the $a'$, $b$ and $c$ stacks of D6-branes on the top of each other on the third 2-torus results in additional vector-like particles from $\mathcal{N} = 2$ subsectors \cite{Cvetic:2004ui}. The anomalies from three global ${\rm U}(1)$s of ${\rm U}(4)_C$, ${\rm U}(2)_L$ and ${\rm U}(2)_R$ are cancelled by the Green-Schwarz mechanism, and the gauge fields of these ${\rm U}(1)$s obtain masses via the linear $B\wedge F$ couplings. Thus, the effective gauge symmetry is ${\rm SU}(4)_C\times {\rm SU}(2)_L\times {\rm SU}(2)_R$.
\begin{table}[t]
	\renewcommand{\arraystretch}{1.3}
	\centering
	\begin{tabular}{|c|c|}
		\hline {\bf Sector} & {\bf Representation}                                                                             \\
		\hline\hline
		$aa$                & ${\rm U}(N_a/2)$ vector multiplet                                                                \\
		                    & 3 adjoint chiral multiplets                                                                      \\
		\hline $ab+ba$      & $ {\cal M}(\frac{N_a}{2}, \frac{\overline{N_b}}{2})= I_{ab}(\yng(1)_{a},\overline{\yng(1)}_{b})$ \\
		\hline $ab'+b'a$    & $ {\cal M}(\frac{N_a}{2}, \frac{N_b}{2})=I_{ab'}(\yng(1)_{a},\yng(1)_{b})$                       \\
		\hline $aa'+a'a$    & ${\cal M} (a_S)= \frac 12 (I_{aa'} - \frac 12 I_{aO6})$ $\yng(2)$                                \\
		                    & ${\cal M} (a_A)= \frac 12 (I_{aa'} + \frac 12 I_{aO6}) $ $\yng(1,1)$                             \\
		\hline
	\end{tabular}
	\caption{General spectrum for intersecting D6-branes at generic angles, where ${\cal M}$ is the multiplicity, and $a_S$ and $a_A$ denote respectively the symmetric and antisymmetric representations of ${\rm U}(N_a/2)$. Positive intersection numbers in our convention refer to the left-handed chiral supermultiplets. }
	\label{tab:spectrum}
\end{table} 
\begin{figure*}[t]
	\centering
	\includegraphics[width=\textwidth]{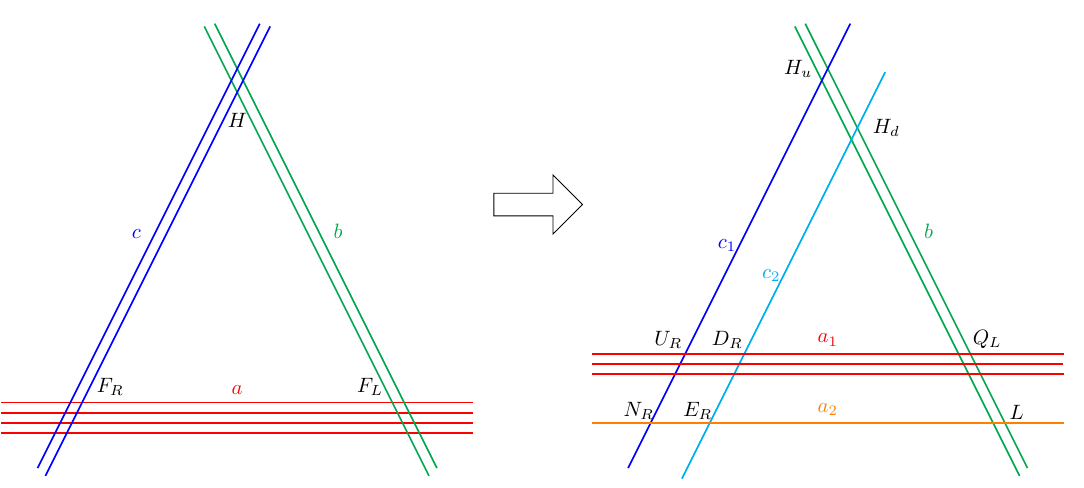}
	\caption{Pati-Salam gauge group ${\rm SU}(4)\times {\rm SU}(2)_L\times {\rm SU}(2)_R$ is broken down to the standard model gauge group ${\rm SU}(3)_C\times {\rm U}(2)_L\times {\rm U}(1)_{I3R}\times {\rm U}(1)_{B-L}$ via the process of brane splitting that corresponds to assigning VEVs to the adjoint scalars, which arise as open-string moduli associated with the positions of stacks $a$ and $c$ in the internal space.} \label{brnsplit}
\end{figure*}

Pati-Salam gauge group ${\rm SU}(4)\times {\rm SU}(2)_L\times {\rm SU}(2)_R$ is higgsed down to the standard model gauge group ${\rm SU}(3)_C\times {\rm U}(2)_L\times {\rm U}(1)_{I3R}\times {\rm U}(1)_{B-L}$ by assigning vacuum expectation values to the adjoint scalars which arise as open-string moduli associated to the stacks $a$ and $c$, see figure~\ref{brnsplit},
\begin{eqnarray}
	\textcolor{red}{a} &\rightarrow & \textcolor{red}{a_1} + \textcolor{orange}{a_2} , \nonumber \\
	\textcolor{blue}{c} &\rightarrow & \textcolor{blue}{c_1} + \textcolor{cyan}{c_2} .
\end{eqnarray}
Moreover, the ${\rm U}(1)_{I_{3R}}\times {\rm U}(1)_{B-L}$ gauge symmetry may be broken to ${\rm U}(1)_Y$ by giving vacuum expectation values (VEVs) to the vector-like particles with the quantum numbers $({\bf { 1}, 1, 1/2, -1})$ and $({\bf { 1}, 1, -1/2, 1})$ under the ${\rm SU}(3)_C\times {\rm SU}(2)_L\times {\rm U}(1)_{I_{3R}} \times {\rm U}(1)_{B-L} $ gauge symmetry from $a_2 c_1'$ intersections~\cite{Cvetic:2004ui,Chen:2006gd}.

This brane-splitting results in standard model quarks and leptons as \cite{Cvetic:2004nk},
\begin{eqnarray}
	F_L(Q_L, L_L)  &\rightarrow &  Q_L + L , \nonumber \\
	F_R(Q_R, L_R)  &\rightarrow &  U_R + D_R + E_R + N_R ~.
\end{eqnarray}
The additional exotic particles must be made superheavy to ensure gauge coupling unification at the string-scale. Similar to Refs.~\cite{Cvetic:2007ku, Chen:2007zu} we can decouple the additional exotic particles except the charged chiral multiplets under ${\rm SU}(4)_C$ anti-symmetric representation. These charged chiral multiplets can be decoupled via instanton effects in principle \cite{Blumenhagen:2006xt, Haack:2006cy, Florea:2006si}.

Three-point Yukawa couplings for the quarks and the charged leptons can be read from the following superpotential,
\begin{align}\label{eq:WY3} 
	\mathcal{W}_3 \sim  Y^u_{ijk} Q_i  U^c_j H^u_k + Y^d_{ijk} Q_i D^c_j H^d_k + Y^\nu_{ijk} L_i N^c_j H^u_k + Y^e_{ijk} L_i  E^{c}_j H^d_k  , 
\end{align}
where $Y^u_{ijk}$, $Y^d_{ijk}$, $Y^\nu_{ijk}$ and $Y^e_{ijk}$ are Yukawa couplings, and $Q_i$, $U^c_i$, $D^c_i$, $L_i$, $N^c_i$, and $E^c_i$ are the left-handed quark doublet, right-handed up-type quarks, right-handed down-type quarks, left-handed lepton doublet, right-handed neutrinos, and right-handed leptons, respectively. The superpotential including the four-point interactions is
\begin{align}\label{eq:WY4}
	\mathcal{W}_4 \sim {1\over {M_{\rm S}}} \left( Y^{\prime u}_{ijkl} Q_i U^c_j H^{\prime u}_k S^L_l + Y^{\prime d}_{ijkl} Q_i D^{c}_j H^{\prime d}_k S^L_l 
	+ Y^{\prime \nu}_{ijkl} L_i N^c_j H^{\prime u}_k S^L_l + Y^{\prime e}_{ijkl} L_i E^{c}_j H^{\prime d}_k S^L_l \right),                                   
\end{align}
where $Y^{\prime u}_{ijkl}$, $Y^{\prime d}_{ijkl}$, $Y^{\prime \nu}_{ijkl}$, and $Y^{\prime e}_{ijkl}$ are Yukawa couplings of the four-point functions, and $M_{\rm S}$ is the string scale.

\section{Yukawa couplings}\label{sec:Yukawa}

Yukawa couplings arise from open string world-sheet instantons that connect three D-brane intersections \cite{Aldazabal:2000cn}. Intersecting D6-branes at angles wrap 3-cycles on the compact space $\T^6= \T^2 \times \T^2 \times \T^2$. For instance in the case of three stacks of D-branes wrapping on a $\T^2$ the 3-cycles can be represented by the wrapping numbers in a vector form as:
\begin{align}\label{eq:branes}
	[\Pi_a]=(n_a, l_a) & \rightarrow  z_a = R(n_a + \tau l_a) \cdot x_a, \nonumber \\
	[\Pi_b]=(n_b, l_b) & \rightarrow  z_b = R(n_b + \tau l_b) \cdot x_b, \nonumber \\
	[\Pi_c]=(n_c, l_c) & \rightarrow  z_c = R(n_c + \tau l_c) \cdot x_c,           
\end{align}
where $\tau$ is the complex structure parameter, $(n_a, l_a)\in \mathbb{Z}^2$ is the wrapped 1-cycle, $x\in \mathbb{R}$ and $z_a\in \mathbb{C}$ respective to the brane $a$.
The triangles bounded by the triplet of D-branes $(z_a, z_b, z_c)$ will contribute to the Yukawa couplings \cite{Cremades:2003qj}. A  \textit{closer} condition,
\begin{align}
	z_a+z_b+z_c & = 0, \label{close} 
\end{align}
ensures that triangles are actually formed by the three branes. The Diophantine equation \eqref{eq:branes} together with the closer condition can be solved to get the following solution:
\begin{align}
	x_a & = \frac{I_{bc}}{d} x, \nonumber                 \\
	x_b & = \frac{I_{ca}}{d} x, \quad x=x_0+l,  \nonumber \\
	x_c & = \frac{I_{ab}}{d} x,                           
\end{align}
where $I_{ab}$ is the intersection number, $d=g.c.d.(I_{ab}, I_{bc}, I_{ca})$ is the greatest common divisor of the intersection numbers, $l\in\mathbb{Z}$ arises from triangles connecting different points in the covering space $\T^6$ but the same points under the lattice $\T^2$ of the triangles and $x_0\in \mathbb{R}$ depends on the relative positions of the branes and the particular triplet $(i, j, k)$ of intersection points,
\begin{align}\label{eq:ijk}
	i & =0,1,\cdots, |I_{ab}|-1, \nonumber \\
	j & =0,1,\cdots, |I_{bc}|-1, \nonumber \\
	k & =0,1,\cdots, |I_{ca}|-1,           
\end{align}
such that $x_0$ can be written as
\begin{equation}
	x_0(i,j,k) = \frac{i}{I_{ab}}+ \frac{j}{I_{bc}} + \frac{k}{I_{ca}}.
\end{equation}
Relaxing the condition that all branes intersect at the origin, we can introduce brane shifts $\epsilon_{\alpha}$, $\alpha=a,b,c$ to write a general expressions for $x_0$ as
\begin{equation}
	x_0(i,j,k) = \frac{i}{I_{ab}}+ \frac{j}{I_{bc}} + \frac{k}{I_{ca}} + \frac{d ( I_{ab}\epsilon_c + I_{ca}\epsilon_b + I_{ab}\epsilon_a)}{I_{ab} I_{bc} I_{ca}} ,\label{x0shift}
\end{equation}
where we can absorb these three parameters into only one as,
\begin{equation}
	\epsilon=\frac{I_{ab} \epsilon_c + I_{ca} \epsilon_b + I_{bc}\epsilon_a}{I_{ab} I_{bc} I_{ca}}.
\end{equation}
This is obvious due to the reparametrization invariance in $\T^2$ since we can always choose two branes to intersect at the origin and the only remaining freedom left is the shift of third brane. The formula of the areas of the triangles can then be expressed using \eqref{x0shift} as,
\begin{align}
	A(z_a, z_b)                 & = \frac{1}{2} \sqrt{|z_a|^2|z_b|^2-(\mathrm{Re} z_a \bar{z}_b)^2} \nonumber                                                \\
	\longrightarrow  A_{ijk}(l) & = \frac{1}{2} (2\pi)^2 A |I_{ab} I_{bc} I_{ca}| (\frac{i}{I_{ab}} + \frac{j}{I_{bc}} + \frac{k}{I_{ca}} + \epsilon + l)^2, 
\end{align}
where $A$ is the K\"ahler structure of the torus. Finally, the Yukawa coupling for the three states localized at the intersections indexed by $(i,j,k)$ is given as,
\begin{equation}
	Y_{ijk}= h_{qu} \sigma_{abc} \sum_{l \in Z} \exp(-\frac{A_{ijk}(l)}{2 \pi \alpha'}),
\end{equation}
where the real phase $\sigma_{abc} = {\rm sign}(I_{ab} I_{bc} I_{ca})$ comes from the full instanton contribution  \cite{Cremades:2003qj} and $h_{qu}$ is due to quantum correction as discussed in \cite{Cvetic:2003ch}. For the ease of numerical computation real modular theta function is used to re-express the summation as
\begin{equation}
	\vartheta \left[\begin{array}{c} \delta \\ \phi \end{array} \right] (t) = \sum_{l\in\Z} e^{-\pi t (\delta+l)^2} e^{2l\pi i(\delta + l) \phi}, \label{Rtheta}
\end{equation}
where the corresponding parameters are related as,
\begin{align}
	\delta & = \frac{i}{I_{ab}} + \frac{j}{I_{bc}} + \frac{k}{I_{ca}} + \epsilon, \nonumber \\
	\phi   & =0, \nonumber                                                                  \\
	t      & =\frac{A}{\alpha'} |I_{ab} I_{bc} I_{ca}|.                                     
\end{align}
Notice that the theta function $\vartheta$ is real, however $t$ can be complex while $\phi$ is an overall phase.

\subsection{Adding a $B$-field and Wilson lines}
Strings being one dimensional naturally couple to a 2-form $B$-field in addition to the metric. To incorporate the turning \emph{on} of this $B$-field leads to a \emph{complex} K\"ahler structure of the compact space $\T^2$ such that,
\begin{equation}
	J=B+iA,
\end{equation}
and the otherwise real parameter $t$ is changed to a complex parameter $\kappa$ as,
\begin{equation}
	\kappa = \frac{J}{\alpha'} |I_{ab} I_{bc} I_{ca}|.
\end{equation}
Secondly, we can also add Wilson lines around the compact directions wrapped by the D-branes. However, to avoid breaking any gauge symmetry Wilson lines must be chosen corresponding to group elements in the centre of the gauge group, i.e., a phase \cite{Cremades:2003qj}. For a triangle formed by three D-branes $a$, $b$ and $c$ each wrapping a different 1-cycle inside of $\T^2$, the Wilson lines can be given by the corresponding phases $\exp (2\pi i \theta_a)$, $\exp (2\pi i \theta_b)$, and $\exp (2\pi i \theta_c)$ respectively. The total phase picked up by an open string sweeping such triangle will depend upon the relative longitude of each segment, determined by the intersection points:
\begin{equation}
	e^{2\pi i x_a \theta_a}  e^{2\pi i x_b \theta_b} e^{2\pi i x_c \theta_c} = e^{2\pi i x(I_{bc} \theta_a + I_{ca} \theta_b + I_{ab}
		\theta_c)}.
\end{equation}
In general, considering both a $B$-field as well as Wilson lines we get a complex theta function as
\begin{equation}
	\vartheta \left[\begin{array}{c} \delta \\ 
	\phi \end{array} \right] (\kappa) = \sum_{l\in\Z} e^{\pi i \kappa (\delta+l)^2 } e^{2 \pi i(\delta + l) \phi}, \label{Ctheta}
\end{equation}
where
\begin{align}
	\delta & = \frac{i}{I_{ab}} + \frac{j}{I_{bc}} + \frac{k}{I_{ca}} + \epsilon, \nonumber \\
	\phi   & =I_{ab} \theta_c + I_{bc} \theta_a + I_{ca} \theta_b, \nonumber                \\
	\kappa & =\frac{J}{\alpha'} |I_{ab} I_{bc} I_{ca}|.                                     
\end{align}

\subsection{O-planes and non-prime intersection numbers}
To cancel the RR-tadpoles we need to introduce the orientifold O-planes that are objects of negative tension. In addition for each D-brane $a$, we must include its mirror image $a'$ under $\Omega \R$. Such mirror branes will in general wrap a different cycle $\Pi_{a*}$, related to $\Pi_{a}$ by the action of $R$ on the homology of the torus. Consequently we also need to include the triangles formed by either of the branes or their images. As an example the Yukawa coupling from the branes $a$, $b'$, and $c$ will depend on the parameters $I_{ab'}$, $I_{b'c}$, and $I_{ca}$, where the primed indexes are independent of the unprimed ones.

Furthermore, the three intersection numbers may not be coprime in general. Therefore, to avoid overcounting we need to involve the $g.c.d.$ of the intersection numbers as $d = g.c.d.(I_{ab}, I_{bc}, I_{ca})$.

Finally, to ensure that triangles are bounded by D-branes, the intersection indices must satisfy the following condition \cite{Cremades:2003qj}
\begin{equation}
	i+j+k = 0 \ {\rm~mod} \ d.  \label{selection-rule}
\end{equation}

\subsection{The general formula of Yukawa couplings}
Therefore, the most general formula for Yukawa couplings for D6-branes wrapping a compact $\T^2 \times \T^2 \times \T^2$ space can be written as,
compact space  as
\begin{equation}
	Y_{ijk}=h_{qu} \sigma_{abc} \prod_{r=1}^3 \vartheta \left[\begin{array}{c} \delta^{(r)}\\ \phi^{(r)} \end{array} \right] (\kappa^{(r)}),
\end{equation}
where
\begin{equation}
	\vartheta \left[\begin{array}{c} \delta^{(r)}\\ \phi^{(r)} \end{array} \right] (\kappa^{(r)})=\sum_{l_r\in\Z} e^{\pi i(\delta^{(r)}+l_r)^2 \kappa^{(r)}} e^{2\pi i(\delta^{(r)}+l_r) \phi^{(r)}},   \label{Dtheta}
\end{equation}
with $r=1,2,3$ denoting the three 2-tori. And the input parameters are defined by
\begin{align}
	\delta^{(r)} & = \frac{i^{(r)}}{I_{ab}^{(r)}} + \frac{j^{(r)}}{I_{ca}^{(r)}} + \frac{k^{(r)}}{I_{bc}^{(r)}} + \frac{d^{(r)} ( I_{ab}^{(r)} \epsilon_c^{(r)} + I_{ca}^{(r)} 
	\epsilon_b^{(r)} + I_{bc}^{(r)} \epsilon_a^{(r)})}{I_{ab} I_{bc} I_{ca}} + \frac{s^{(r)}}{d^{(r)}}, \nonumber \\
	\phi^{(r)}   & = \frac{I_{bc}^{(r)} \theta_a^{(r)} + I_{ca}^{(r)} \theta_b^{(r)} + I_{ab}^{(r)} \theta_c^{(r)}}{d^{(r)}}, \nonumber                                        \\
	\kappa^{(r)} & = \frac{J^{(r)}}{\alpha'} \frac{|I_{ab}^{(r)} I_{bc}^{(r)} I_{ca}^{(r)}|}{(d^{(r)})^2}. \label{eqn:Yinput}                                                  
\end{align}

The theta function defined in \eqref{Ctheta} is in general complicated to evaluate numerically. However, for the special case without $B$-field, defining $J'= -iJ = A$ and $\kappa'= -i\kappa$ the $\vartheta$ function takes a more manageable form,
\begin{align}
	\vartheta \left[\begin{array}{c} \delta                                                     \\ \phi \end{array} \right] (\kappa') &= \sum_{l\in\Z} e^{-\pi \kappa' (\delta+l)^2 } e^{2 \pi i(\delta + l) \phi}, \nonumber\\
	\stackrel{\textrm{redefine}}{\longrightarrow} \quad \vartheta \left[\begin{array}{c} \delta \\ \phi \end{array} \right] (\kappa) &= e^{-\pi \kappa \delta^2} e^{2\pi i \delta \phi} \vartheta_3 (\pi (\phi+ i \kappa \delta), e^{-\pi \kappa}), \label{Ntheta}
\end{align}
in terms of $\vartheta_3$, the Jacobi theta function of the third kind.

Pati-Salam gauge symmetry is broken down to the standard model by the process of brane-splitting as schematically shown in figure~\ref{brnsplit}, where the standard model particles are localized at their respective brane intersections. The mass hierarchies of the standard model are then easily explained by the relative shifting of the brane stacks. For instance, the left-handed quarks are localized at the intersections between the stacks $\{\textcolor{red}{a_1},~\textcolor{Green}{b}\}$ while the right-handed up-type and down-type quarks are respectively localized between stacks $\{\textcolor{red}{a_1},~\textcolor{blue}{c_1}\}$ and $\{\textcolor{red}{a_1},~\textcolor{cyan}{c_2}\}$. Thus, if we shift stack $\textcolor{cyan}{c_2}$ in the orientifold by an amount $\epsilon_{\textcolor{cyan}{c2}}$ while the stack $\textcolor{blue}{c_1}$ is unshifted ($\epsilon_{\textcolor{blue}{c_1}} = 0$), then the down-type quark masses are naturally suppressed relative to the up-type quarks. Similarly, because the left-handed and the right-handed charged leptons are respectively localized at the intersection between stacks $\{\textcolor{orange}{a_2},~\textcolor{Green}{b}\}$ and stacks $\{\textcolor{orange}{a_2},~\textcolor{cyan}{c_2}\}$, the shifting of stack $\textcolor{orange}{a_2}$ by some amount $\epsilon_{\textcolor{orange}{a_2}}$ will result in the suppression of the charged lepton masses relative to the down-type quarks. Hence, the following observed mass hierarchy is a consequence of pure geometry of the internal space,
\begin{equation}
	m_{u} > m_{d} > m_{e}.
\end{equation}

\subsection{Fitting the fermion masses and mixings} 
By running the RGE's up to unification scale, considering $\tan\beta\equiv v_u/v_d = 50$ and the ratio $m_\tau/m_b=1.58$ from the previous study of soft terms \cite{Chen:2007zu}, the diagonal mass matrices for up-type, down-type quarks and charged-leptons, denoted as $D_u$, $D_d$ and $D_e$ at the unification scale $\mu=M_X$ have been determined as \cite{Fusaoka:1998vc,Ross:2007az}, 
\begin{align}
	D_u &= m_t \left(\begin{array}{ccc}
	0.0000139 & 0       & 0   \\
	0         & 0.00404 & 0   \\
	0         & 0       & 1.  
	\end{array} \right),\label{eq:mass-upquarks}\\
	D_d &= m_b \left(\begin{array}{ccc}
	0.00141   & 0       & 0   \\
	0         & 0.0280  & 0   \\
	0         & 0       & 1.  
	\end{array} \right) \label{eq:mass-downquarks},\\
	D_e &= m_{\tau} \left(\begin{array}{ccc}
	0.000217  & 0       & 0   \\
	0         & 0.0458  & 0   \\
	0         & 0       & 1.  
	\end{array} \right). \label{eq:mass-chargedleptons}\\
	D_\nu &= m_{\nu} \left(\begin{array}{ccc}
	m_3       & 0       & 0   \\
	0         & m_2     & 0   \\
	0         & 0       & m_1 
	\end{array} \right),  \label{eq:mass-neutrinos}
\end{align}
where we have parameterized the neutrino-masses as $(m_3,m_2,m_1)$ upto an overall constant $m_\nu$. 
Experimentally, two of the mass eigenstates $m_1,~m_2$ are found to be close to each other while the third eigenvalue $m_3$ is separated from the former pair where $m_2 > m_1$ by definition. Normal ordering (NO) refers to $m_3\gg m_2>m_1$ while inverted ordering (IO) refers to $(m_2>m_1\gg m_3)$ with constraints \href{http://www.nu-fit.org/?q=node/278}{NuFIT 5.3 (2024)} \cite{Esteban:2020cvm},
\begin{mdframed}\setlength\abovedisplayskip{0pt} 
	\begin{align}
		\Delta m_{21}^2                                  & = 74.1 \pm 2.1 ~\mathrm{meV}^2, \quad \begin{cases} 
		\Delta m_{31}^2 = +2505 \pm 25 ~\mathrm{meV}^2 ~ & \mathrm{(NO)},                                      \\ 
		\Delta m_{32}^2 = -2487 \pm 27 ~\mathrm{meV}^2 ~ & \mathrm{(IO)}. \end{cases} \label{eq:constraints}   
	\end{align}
\end{mdframed} 
In the standard model, the quark matrices and the leptons matrices can always be made Hermitian by suitable transformation of the right-handed fields \cite{Mayes:2019isy, Sabir:2022hko}. For quarks, we consider the case that $M_d$ is very close to the diagonal matrix for down-type quark, which effectively means that $U^d_L$ and $U^d_R$ are very close to the unit matrix with very small off-diagonal terms, then
\begin{equation}
	V_{\rm CKM}\simeq U^u U^{d\dag}\simeq U^u,
\end{equation}
where we have transformed away the right-handed effects and made them the same as the left-handed ones.  Thus, the mass matrix of the up-type quarks becomes,
\begin{equation}
	M_u \sim V_{\rm CKM}^{\dag} D_u V_{\rm CKM}.
\end{equation} 
Employing the quarks-mixing matrix, $V_{\rm CKM}$, from \href{http://www.utfit.org/UTfit/ResultsSummer2023SM}{UTfit (2023)} \cite{UTfit:2022hsi},
\begin{align}\label{eq:VCKM}
	V_{\rm CKM} &= \left(\!\!\!\!\!\begin{array}{d{17}d{19}d{19}}
	0.97431(19)                              & 0.22517(81)                              & 0.003715(93)\, e^{-i\, 65.1(1.3)^\circ} \\
	-0.22503(83) \, e^{+i\, 0.0351(1)^\circ} & 0.97345(20)\, e^{-i\, 0.00187(5)^\circ}  & 0.0420(5)                               \\
	0.00859 (11)\, e^{-i\, 22.4(7)^\circ}    & -0.04128 (46)  \, e^{+i\, 1.05(3)^\circ} & 0.999111(20)                            
	\end{array}\! \right),  
\end{align} 
we can express the up-quark mass matrix in the mixed form as, 
\begin{align}
	\because M_u  &= V_{\rm CKM}^{\dag} D_u V_{\rm CKM} \nonumber\\
	\Rightarrow |M_u| & =  m_t \left(
	\begin{array}{ccc}
	0.000291336 & 0.00122042 & 0.00860765 \\
	0.00122042  & 0.00552729 & 0.0413384  \\
	0.00860765  & 0.0413384  & 0.998234   \\
	\end{array}
	\right).\label{eq:mixing-quarks} 
\end{align}
Similarly, using the leptons-mixing matrix, $U_{\rm PMNS}$ from \href{http://www.nu-fit.org/?q=node/278}{NuFIT}, we can express the charged-leptons matrix in the mixed form as,
\begin{align}
	\because M_e  &= U_{\rm PMNS}^{\dag} D_e U_{\rm PMNS} \nonumber\\
	\Rightarrow|M_e|  &= m_\tau \left(
	\begin{array}{ccc}
	0.0361688 & 0.108125 & 0.116698 \\
	0.108125  & 0.458457 & 0.500675 \\
	0.116698  & 0.500675 & 0.547527 \\
	\end{array}
	\right), \label{eq:mixing-chargedleptons}
\end{align}

Henceforth, we need to fit \eqref{eq:mass-downquarks}, \eqref{eq:mixing-quarks}, \eqref{eq:mixing-chargedleptons} and \eqref{eq:mass-neutrinos} to explain the SM fermions' masses and mixings by fine-tuning the Higgs VEVs against the coupling parameters.

\subsection{4-point Yukawa corrections}\label{sec:4point}

We now turn our attention to the discussion of four-point functions that affect more greatly to the masses of the lighter fermions. We are looking for four-point interactions such as
\begin{equation} \label{4pointcases}
	\phi^i_{ab} \phi^j_{ca} \phi^k_{b'c} \phi^l_{bb'} ~~\mathrm{or}~~
	\phi^i_{ab} \phi^j_{ca} \phi^k_{cc'} \phi^l_{bc'}~,~\,
\end{equation}
where $\phi^i_{x y}$ are the chiral superfields at the intersections between stack $x$ and $y$ D6-branes. The formula for the area of a quadrilateral in terms of its angles and two sides and the solutions of diophantine equations for estimating the multiple areas of the quadrilaterals from non-unit intersection numbers are given in \cite{Abel:2003yx, Abel:2003vv}. In addition to these formulae, there is a more intuitive way to calculate the area for these four-sided polygons. A quadrilateral can be always taken as the difference between two similar triangles. Therefore, since we know the classical part is
\begin{equation}
	Z_{4cl} \sim e^{-A_{quad}},
\end{equation}
it is equivalent to write \cite{Chen:2008rx},
\begin{equation}
	Z_{4cl} \sim e^{-|A_{tri}-A'_{tri}|}.
\end{equation}

\begin{figure}[t]
	\centering
	\includegraphics[width=.5\textwidth]{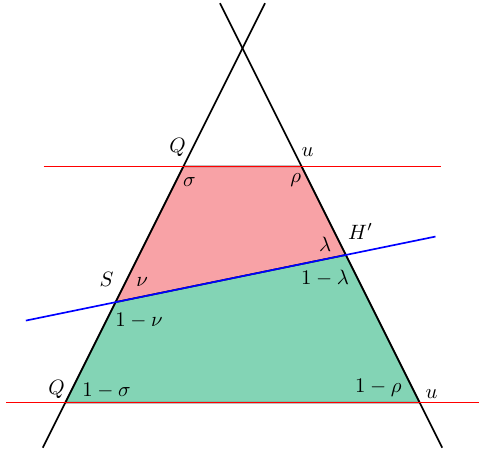}
	\caption{A picture of two quadrilaterals with different field orders. The red brane repeats in a next cycle and can still form a similar quadrilateral with the blue brane. This coupling also contributes to the four-point function.} \label{2-cases}
\end{figure}

Taking the absolute value of the difference reveals that there are two cases: $A_{tri}>A'_{tri}$ and $A_{tri}<A'_{tri}$, as shown in figure~\ref{2-cases}. From the figure we can see the two quadrilaterals are similar with different sizes, but the orders of the fields corresponding to the angles are different, which is under an interchange of $\theta \leftrightarrow 1-\theta$, $\theta=\nu, \lambda, \rho, \sigma$. These different field orders may cause different values for their quantum contributions. Here, we shall only consider the classical contribution from the 4-point interaction and ignore the quantum part which was shown to be further suppressed, consult \cite{Chen:2008rx} and references therein for details. Therefore, we are able to employ the same techniques which have developed for calculating the trilinear Yukawa couplings.

\begin{figure}[t]
	\centering
	\includegraphics[width=.5\textwidth]{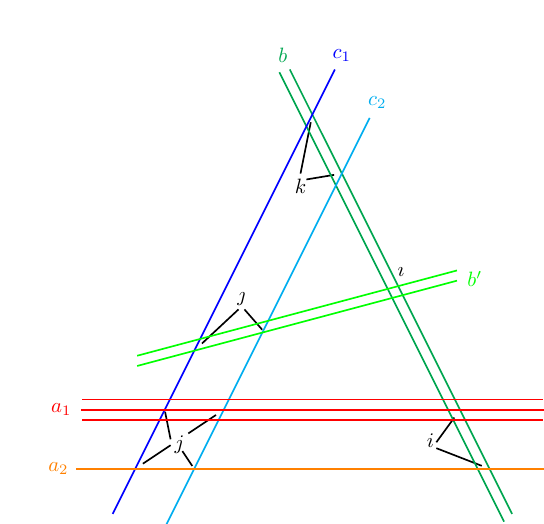}
	\caption{Figure shows the areas bounded by stacks of D-branes which give rise to the Yukawa couplings for quarks and leptons via world-sheet instantons. The Yukawa couplings of the up-type quarks are from the areas by stack $a_1$, $b$, $c_1$, the down-type quarks by stack $a_1$, $b$, $c_2$, and the leptons by $a_2$, $b$, $c_2$. The four-point function corrections to the Yukawa couplings of the up-type quarks are from the areas by stack $a_1$, $b$, $b'$, $c_1$, the down-type quarks by stack $a_1$, $b$, $b'$, $c_2$, and the leptons by $a_2$, $b$, $b'$, $c_2$. }
	\label{brane area_bb}
\end{figure} 

\begin{figure}[t]
	\centering
	\includegraphics[width=.5\textwidth]{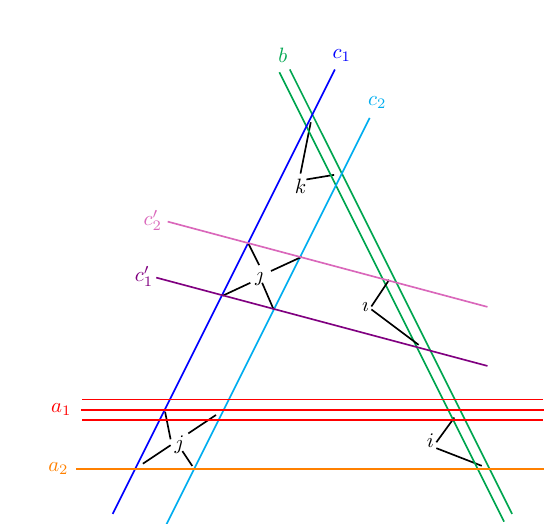}
	\caption{The figure showing the areas bounded by stacks of D-branes which give rise to the Yukawa couplings for quarks and leptons via world-sheet instantons. The Yukawa couplings of the up-type quarks are from the areas by stack $a_1$, $b$, $c_1$, the down-type quarks by stack $a_1$, $b$, $c_2$, and the leptons by $a_2$, $b$, $c_2$. The four-point function corrections to the Yukawa couplings of the up-type quarks are from the areas by stack $a_1$, $b$, $c'$, $c_1$, the down-type quarks by stack $a_1$, $b$, $c'$, $c_2$, and the leptons by $a_2$, $b$, $c'$, $c_2$. }
	\label{brane area_cc}
\end{figure}

For a quadrilateral formed by the stacks $a$, $b$, $b'$, $c$, we can calculate it as the difference between two triangles formed by stacks $a$, $b$, $c$ and $b'$, $b$, $c$. In other words, they share the same intersection $I_{bc}$.  Therefore, if we use this method to calculate the quadrilateral area, we should keep in mind that the intersection index $k$ for $I_{bc}$ remains the same for a certain class of quadrilaterals when varying other intersecting indices. Here we set indices $i$ for $I_{ab}$, $j$ for $I_{ca}$, $\imath$ for $I_{bb'}$, and $\jmath$ for $I_{cb'}$, as shown in figure~\ref{brane area_bb}. We may calculate the areas of the triangles as we did in the trilinear Yukawa couplings above \cite{Cremades:2003qj}
\begin{align}
	A_{ijk}            & =\frac{1}{2}(2\pi)^2 A_{\T^2} |I_{ab}I_{bc}I_{ca}|~ \big(\frac{i}{I_{ab}} +\frac{j}{I_{ca}}+\frac{k}{I_{bc}}+\epsilon +l \big)^2, \nonumber                              \\
	A_{\imath\jmath k} & =\frac{1}{2}(2\pi)^2 A_{\T^2} |I_{b'b}I_{bc}I_{cb'}|~ \big( \frac{\imath}{I_{b'b}} +\frac{\jmath}{I_{cb'}} +\frac{k}{I_{bc}} +\varepsilon +\ell \big)^2,  \label{Tareas} 
\end{align}
where $i$, $j$, $k$ and $\imath$, $\jmath$, $k$ are using the same selection rules as Eq. (\ref{selection-rule}). Thus, the classical contribution of the four-point functions is given by
\begin{equation}
	Z_{4cl} = \sum_{l,\ell} e^{-\frac{1}{2\pi}|A_{ijk}-A_{\imath\jmath k}|}.
\end{equation}
Note that this formula will diverge when $A_{ijk}=A_{\imath\jmath k}$, which is due to over-counting the zero area when the corresponding parameters in Eq. (\ref{Tareas}) are the same. In such a case, $Z_{4cl} = 1+\sum_{l\neq\ell} e^{-\frac{1}{2\pi}|A_{ijk}-A_{\imath\jmath k}|}$. We will not meet this special situation in our following discussion. We will consider both types of possible interactions \eqref{4pointcases} coming from considering $b'$ or from considering $c'$ as shown in figure~\ref{brane area_cc} independently.

\FloatBarrier
 
The complete list of 33 models is listed in appendix \ref{appA}. The complete perturbative particle spectra of the 33 models are tabulated in appendix \ref{appB}. The first 16 models viz. \hyperref[model1]{1}, \hyperref[model1.5]{1-dual}, \hyperref[model2]{2}, \hyperref[model3]{3}, \hyperref[model3.5]{3-dual}, \hyperref[model4]{4}, \hyperref[model5]{5}, \hyperref[model6]{6}, \hyperref[model7]{7}, \hyperref[model8]{8}, \hyperref[model9]{9}, \hyperref[model9.5]{9-dual}, \hyperref[model10]{10}, \hyperref[model11]{11}, \hyperref[model11.5]{11-dual}, and \hyperref[model12]{12} do not possess the correct form of Yukawa textures to generate the fermion masses on a single two-torus. Therefore, we will only focus on the remaining models where viable 3-point Yukawa interactions are possible.

\section{Model with 3 Higgs from the bulk}\label{sec:3Higgs}
\subsection{Model 13}\label{sec:model-13}
In Model~\hyperref[model13]{13} the three-point Yukawa couplings arise from the triplet intersections from the branes ${a, b, c}$ on the first two-torus ($r=1$) with 3 pairs of Higgs from $\mathcal{N}=2$ subsector.

\begin{figure}[t]
	\centering
	\includegraphics[width=\textwidth]{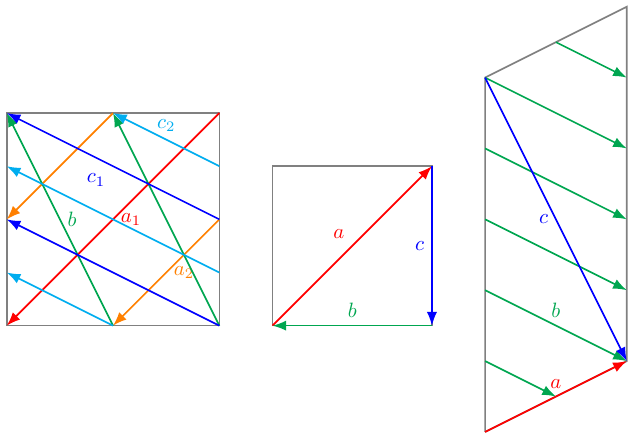}
	\caption{Brane configuration for the three two-tori in Model~\hyperref[model13]{13} where the third two-torus is tilted. Fermion mass hierarchies result from the intersections on the first two-torus.}  \label{Fig.13}
\end{figure}

Yukawa matrices for the Model~\hyperref[model13]{13} are of rank 3 and the three intersections required to form the disk diagrams for the Yukawa couplings all occur on the first torus as shown in figure~\ref{Fig.13}. The other two-tori only contribute an overall constant that has no effect in computing the fermion mass ratios. Thus, it is sufficient for our purpose to only focus on the first torus to explain the masses and the mixing in the standard model fermions.

\subsubsection{3-point Yukawa mass-matrices for Model~\hyperref[model13]{13}}
From the wrapping numbers listed in table~\ref{model13}, the relevant intersection numbers are calculated as,
\begin{align}\label{eq:Intersections.13} 
	I_{ab}^{(1)} & =-3, &   
	I_{ab}^{(2)} & =1,  &   
	I_{ab}^{(3)}&=-2, \nonumber\\
	I_{bc}^{(1)} & =3,  &   
	I_{bc}^{(2)} & =1,  &   
	I_{bc}^{(3)}&=-3, \nonumber\\
	I_{ca}^{(1)} & =3,  &   
	I_{ca}^{(2)} & =1,  &   
	I_{ca}^{(3)}&=1, 
\end{align}
As the intersection numbers are not coprime, we define the greatest common divisor,
$d^{(1)}=g.c.d.(I_{ab}^{(1)},I_{bc}^{(1)},I_{ca}^{(1)})=3$. Thus, the arguments of the modular theta function as defined in \eqref{eqn:Yinput} can be written as,
\begin{align}
	\delta^{(1)} & = \frac{i^{(1)}}{-3} + \frac{j^{(1)}}{3} + \frac{k^{(1)}}{3} +\frac{1}{9} \left(-3 \epsilon _a^{\text{(1)}}-3 \epsilon _b^{\text{(1)}}+3 \epsilon _c^{\text{(1)}}\right)+ \frac{s^{(1)}}{3},  \label{eq:delta.13} \\
	\phi^{(1)}   & = \frac{1}{3} \left(3 \theta _a^{\text{(1)}}+3 \theta _b^{\text{(1)}}-3 \theta _c^{\text{(1)}}\right) =0,                                                                                                         \\
	\kappa^{(1)} & = \frac{3 J^{\text{(1)}}}{\alpha '}, \label{eq:kappa.13}                                                                                                                                                          
\end{align}
and recalling \eqref{eq:ijk}, we have $i=\{0,\dots ,2\}$, $j=\{0,\dots ,2\}$ and $k=\{0,\dots ,2\}$ which respectively index the left-handed fermions, the right-handed fermions and the Higgs fields. Clearly, there arise 3 Higgs fields from the $bc$ sector.

The second-last term in the right side of \eqref{eq:delta.13} can be used to redefine the shift on the torus as
\begin{align}
	\epsilon^{(1)} & \equiv \frac{1}{9} \left(-3 \epsilon _a^{\text{(1)}}-3 \epsilon _b^{\text{(1)}}+3 \epsilon _c^{\text{(1)}}\right). \label{shifts13} 
\end{align}  
The selection rule for the occurrence of a trilinear Yukawa coupling for a given set of indices is given as,
\begin{equation}\label{selection-rule.13}
	i^{(1)} + j^{(1)} + k^{(1)} = 0\; \mathrm{ mod }\; 3.
\end{equation}
 
Then the 3-point Yukawa matrices take the following form
\begin{align}
	Y^{(1)}_{k=0} \sim  \left(
	\begin{array}{ccc}
	T_0 & 0   & 0   \\
	0   & 0   & T_1 \\
	0   & T_2 & 0   \\
	\end{array}
	\right),   \;\;
	Y^{(1)}_{k=1} \sim  \left(
	\begin{array}{ccc}
	0   & 0   & T_1 \\
	0   & T_2 & 0   \\
	T_0 & 0   & 0   \\
	\end{array}
	\right),   \;\;
	Y^{(1)}_{k=2} \sim  \left(
	\begin{array}{ccc}
	0   & T_2 & 0   \\
	T_0 & 0   & 0   \\
	0   & 0   & T_1 \\
	\end{array}
	\right),
\end{align}
where
\begin{align}\label{eq:3couplings13}
	T_k & \equiv  \vartheta \left[\begin{array}{c} 
	\epsilon^{(1)}+\frac{k}{3}\\  \phi^{(1)} \end{array} \right]
	(\frac{3 J^{\text{(1)}}}{\alpha '}),  \quad   k={0,\cdots,2}.
\end{align}
where we take $s^{(1)}=i$ in \eqref{eq:delta.13},
\begin{align}
	\delta^{(1)} & =  \frac{j}{3}+\frac{k}{3} \label{eq:s1.13}, 
\end{align}
and we ignore the other equivalent cases of solutions or cases with rank-1 problem. 

Furthermore, there is also a contribution from the third torus where some of the intersection numbers are greater than 1. 
Choosing the specific value of $s^{(3)}=\frac{i}{2}$. 
\begin{align}
	\delta^{(3)} & =  j-\frac{k}{3} \label{eq:s33.13} 
\end{align} 
\begin{align}\label{eq:33couplings13}
	Y^{(3)}_k \sim t_k & \equiv  \vartheta \left[\begin{array}{c} 
	\epsilon^{(3)}+\frac{k}{3}\\  \phi^{(3)} \end{array} \right]
	(\frac{6 J^{\text{(3)}}}{\alpha '}) ,  \quad   k={0,\cdots,2}.
\end{align}
Therefore the classical part of this three-point couplings is given by
\begin{align}\label{eq:Zcls13}
	Z_{3cl} &= v_1 t_2\left(
	\begin{array}{ccc}
	T_0 & 0   & 0   \\
	0   & 0   & T_1 \\
	0   & T_2 & 0   \\
	\end{array}
	\right) + v_2 t_2\left(
	\begin{array}{ccc}
	0   & 0   & T_1 \\
	0   & T_2 & 0   \\
	T_0 & 0   & 0   \\
	\end{array}
	\right) + v_3 t_2\left(
	\begin{array}{ccc}
	0   & T_2 & 0   \\
	T_0 & 0   & 0   \\
	0   & 0   & T_1 \\
	\end{array}
	\right) \nonumber \\
	& + v_4 t_0\left(
	\begin{array}{ccc}
	T_0 & 0   & 0   \\
	0   & 0   & T_1 \\
	0   & T_2 & 0   \\
	\end{array}
	\right) + v_5 t_0\left(
	\begin{array}{ccc}
	0   & 0   & T_1 \\
	0   & T_2 & 0   \\
	T_0 & 0   & 0   \\
	\end{array}
	\right) + v_6 t_0\left(
	\begin{array}{ccc}
	0   & T_2 & 0   \\
	T_0 & 0   & 0   \\
	0   & 0   & T_1 \\
	\end{array}
	\right)  \nonumber \\
	& + v_7 t_1\left(
	\begin{array}{ccc}
	T_0 & 0   & 0   \\
	0   & 0   & T_1 \\
	0   & T_2 & 0   \\
	\end{array}
	\right) + v_8 t_1\left(
	\begin{array}{ccc}
	0   & 0   & T_1 \\
	0   & T_2 & 0   \\
	T_0 & 0   & 0   \\
	\end{array}
	\right) + v_9 t_1\left(
	\begin{array}{ccc}
	0   & T_2 & 0   \\
	T_0 & 0   & 0   \\
	0   & 0   & T_1 \\
	\end{array}
	\right).
\end{align}
Then the mass matrices for up quarks, down quarks and charged leptons have the following general form:
\begin{equation}
	Z_{3}=Z_{3q}\left(
	\begin{array}{ccc}
		T_0 \left(t_1 v_1+t_2 v_4+t_0 v_7\right) & T_1 \left(t_1 v_3+t_2 v_6+t_0 v_9\right) & T_2 \left(t_1 v_2+t_2 v_5+t_0 v_8\right) \\
		T_1 \left(t_1 v_3+t_2 v_6+t_0 v_9\right) & T_2 \left(t_1 v_2+t_2 v_5+t_0 v_8\right) & T_0 \left(t_1 v_1+t_2 v_4+t_0 v_7\right) \\
		T_2 \left(t_1 v_2+t_2 v_5+t_0 v_8\right) & T_0 \left(t_1 v_1+t_2 v_4+t_0 v_7\right) & T_1 \left(t_1 v_3+t_2 v_6+t_0 v_9\right) \\
	\end{array}
	\right),\label{eq:s-i13}
\end{equation} 
It is clear that only the three linear combinations of the nine Higgs states, $t_1 v_1+t_2 v_4+t_0 v_7$, $t_1 v_3+t_2 v_6+t_0 v_9 $ and $t_1 v_2+t_2 v_5+t_0 v_8 $ contribute to the Yukawa couplings up to the normalizations factors.
  
\subsubsection{Fermion masses and mixings from 3-point functions in Model~\hyperref[model13]{13}}
In order to accommodate the fermion masses and the quark mixings, we need to fit the up-type quarks mixing matrix \eqref{eq:mixing-quarks}, the down-type quarks matrix \eqref{eq:mass-downquarks} and the masses of the charged-leptons \eqref{eq:mass-chargedleptons}.
 
We set the K\"{a}hler modulus on the first two-torus defined in \eqref{eq:kappa.13} as $\kappa^{(1)} = 30 $ which also fixes $\kappa^{(3)} = 60 $ and evaluate the couplings functions \eqref{eq:3couplings13} by setting geometric brane position parameters as $\epsilon^{(1)}_{u} = 0$, $\epsilon^{(1)}_{d} = 0.00720233$ and $\epsilon^{(1)}_{e} = 0$ which yields a nearest fit for the following VEVs,
\begin{equation}\label{eq:VEVs_13} 
	\begin{array}{ll}
		v^u{}_1 = 211479.               , & v^d{}_1 = 4977.6                \\
		v^u{}_2 = 1.4168\times 10^{11}  , & v^d{}_2 = 3.70977\times 10^{10} \\
		v^u{}_3 = 2.55874\times 10^{13} , & v^d{}_3 = 8.09994\times 10^{11} \\
	\end{array}
\end{equation}
\begin{align} 
	|M_{3u}|&=m_t \left(
	\begin{array}{ccc}
	0.000291 & 0.998234 & 0.005527 \\
	0.998234 & 0.005527 & 0.000291 \\
	0.005527 & 0.000291 & 0.998234 \\
	\end{array}
	\right)\sim M_u ~, \label{eq:Up3_13} \\
	|M_{3e}|&=m_\tau \left(
	\begin{array}{ccc}
	0.000217 & 1.       & 0.0458   \\
	1.       & 0.0458   & 0.000217 \\
	0.0458   & 0.000217 & 1.       \\
	\end{array}
	\right)\sim D_e~,\label{eq:Leptons3_13} \\ 
	|M_{3d}|&=m_b \left(
	\begin{array}{ccc}
	0.000341 & 1.       & 0.113223 \\
	1.       & 0.113223 & 0.000341 \\
	0.113223 & 0.000341 & 1.       \\
	\end{array}
	\right)~.\label{eq:Down3_13}
\end{align}
Only the masses of up-type quarks, the masses of charged leptons and the bottom quark mass are fitted with the three-point couplings. The quark mixings and the masses of the charm and the down quarks are not matched. Notice, that these results are only at the tree-level and there could indeed be other corrections, such as those coming from higher-dimensional operators, which may contribute most greatly to the charm and the down quarks' masses since they are lighter.

\subsubsection{4-point corrections in Model~\hyperref[model13]{13}}
 
The four-point couplings in Model~\hyperref[model13]{13} in table~\ref{model13} can come from considering interactions of ${a, b, c}$ with $b'$ or $c'$ on the first two-torus as can be seen from the following intersection numbers,
\begin{align}
	I_{bb'}^{(1)} & = -4, & I_{bb'}^{(2)} & = 0,  & I_{bb'}^{(3)} & = 5 ,\nonumber                   \\
	I_{cc'}^{(1)} & = -4, & I_{cc'}^{(2)} & = 0,  & I_{cc'}^{(3)} & = -1 ,\nonumber                  \\
	I_{bc'}^{(1)} & = 5,  & I_{bc'}^{(2)} & = -1, & I_{bc'}^{(3)} & = 2 . \label{eq:4Intersection13} 
\end{align}
There are 4 SM singlet fields $S_L^i$ and 5
Higgs-like state $H^{\prime}_{u, d}$. 

Let us consider four-point interactions with $c'$ with the following parameters with shifts $l=-k $ and $\ell=-\frac{k}{3} $ taken along the index $k$,
\begin{align}\label{deltas13}
	\delta & =\frac{i}{I_{ab}^{(1)}} +\frac{j}{I_{ca}^{(1)}} +\frac{k}{I_{bc}^{(1)}} +l , \nonumber               \\
	       & = \frac{j}{3}-\frac{i}{3} ,                                                                          \\
	d      & =\frac{\imath}{I_{c c'}^{(1)}} +\frac{\jmath}{I_{bc'}^{(1)}}+\frac{k}{I_{bc}^{(1)}} +\ell ,\nonumber \\
	       & = \frac{\jmath}{5}-\frac{\imath}{4} ,                                                                
\end{align}
the matrix elements $a_{i,j,\imath}$ on the first torus from the four-point functions results in the following classical 4-point contribution to the mass matrix,
\begin{equation}\label{eq:4point13} 
	Z_{4cl} =  
	\resizebox{.95\textwidth}{!}{
		$\begin{aligned}
			  & \left(                 
			\begin{array}{c}
			F_0 u_1 w_1+F_5 u_4 w_1+F_{14} u_3 w_2+F_3 u_2 w_3+F_{12} u_1 w_4+F_{17} u_4 w_4+F_6 u_3 w_5\\
			F_{10} u_3 w_1+F_{19} u_2 w_2+F_8 u_1 w_3+F_{13} u_4 w_3+F_2 u_3 w_4+F_{11} u_2 w_5\\
			F_{15} u_2 w_1+F_4 u_1 w_2+F_9 u_4 w_2+F_{18} u_3 w_3+F_7 u_2 w_4+F_{16} u_1 w_5+F_1 u_4 w_5\\
			\end{array}\right.\\
			  & \begin{array}{c}       
			F_{10} u_3 w_1+F_{19} u_2 w_2+F_8 u_1 w_3+F_{13} u_4 w_3+F_2 u_3 w_4+F_{11} u_2 w_5\\
			F_{15} u_2 w_1+F_4 u_1 w_2+F_9 u_4 w_2+F_{18} u_3 w_3+F_7 u_2 w_4+F_{16} u_1 w_5+F_1 u_4 w_5\\
			F_0 u_1 w_1+F_5 u_4 w_1+F_{14} u_3 w_2+F_3 u_2 w_3+F_{12} u_1 w_4+F_{17} u_4 w_4+F_6 u_3 w_5\\
			\end{array}\\
			  & \left.\begin{array}{c} 
			F_{15} u_2 w_1+F_4 u_1 w_2+F_9 u_4 w_2+F_{18} u_3 w_3+F_7 u_2 w_4+F_{16} u_1 w_5+F_1 u_4 w_5\\
			F_0 u_1 w_1+F_5 u_4 w_1+F_{14} u_3 w_2+F_3 u_2 w_3+F_{12} u_1 w_4+F_{17} u_4 w_4+F_6 u_3 w_5\\
			F_{10} u_3 w_1+F_{19} u_2 w_2+F_8 u_1 w_3+F_{13} u_4 w_3+F_2 u_3 w_4+F_{11} u_2 w_5\\
			\end{array}\right),
		\end{aligned}$
	} 
\end{equation} 
where $u_i, w_j$ are the VEVs and the couplings are defined as,
\begin{align}\label{eq:4couplings13}
	F_{4i} & \equiv  \vartheta \left[\begin{array}{c} 
	\epsilon^{(1)}+\frac{i}{5}\\  \phi^{(1)} \end{array} \right]
	(\frac{3 J^{\text{(1)}}}{\alpha '}),\qquad i={0,\dots,|I_{c c'}^{(1)}|-1}.
\end{align}

Since, we have already fitted the up-type quark matrix $|M_{3u}|$ exactly, so its 4-point correction should be zero,
\begin{equation}
	|M_{4u}| = 0 ,
\end{equation}
which is true by setting all up-type VEVs $u_u^i$ and $w_u^i$ to be zero. 
Therefore, we are essentially concerned with fitting down-type quarks in such a way that corresponding corrections for the charged-leptons remain negligible. The desired solution can be readily obtained by setting $\epsilon^{(1)}_{4d}=0$ with the following values of the VEVs,
\begin{equation}\label{eq:UdWd13} 
	\begin{array}{l}
		u^d{}_1 = 1 \\
		u^d{}_2 = 1 \\
		u^d{}_3 = 0 \\
		u^d{}_4 = 1 \\
	\end{array}
	\quad,\quad
	\begin{array}{l}
		w^d{}_1 = 0.00168696 \\
		w^d{}_2 = 0          \\
		w^d{}_3 = 0          \\
		w^d{}_4 = -0.0739327 \\
		w^d{}_5 = 0          \\
	\end{array}
\end{equation}
The 4-point correction to the down-type quarks' masses is given by,
\begin{equation}\label{M4d_13}
	|M_{4d}|=m_b \left(
	\begin{array}{ccc}
		0.001069  & 0.        & -0.085223 \\
		0.        & -0.085223 & 0.001069  \\
		-0.085223 & 0.001069  & 0.        \\
	\end{array}
	\right)
\end{equation}
which can be added to the matrix obtained from 3-point functions \eqref{eq:Down3_13} as,
\begin{equation}\label{M34d_13}
	|M_{3d}|+|M_{4d}|=m_b \left(
	\begin{array}{ccc}
		0.00141 & 1.      & 0.028   \\
		1.      & 0.028   & 0.00141 \\
		0.028   & 0.00141 & 1.      \\
	\end{array}
	\right) \sim D_d ~,
\end{equation} 
However, we also need to keep the corrections to charged-leptons' masses to be negligible by setting $\epsilon^{(1)}_{4e}=0$,
\begin{equation}\label{M4e_13}
	|M_{4e}|\sim \left(
	\begin{array}{ccc}
		0.000676  & 0.        & -0.053938 \\
		0.        & -0.053938 & 0.000676  \\
		-0.053938 & 0.000676  & 0.        \\
	\end{array}
	\right)\sim 0.
\end{equation} 
However, the four-point corrections to the leptons and the down quarks turn out to be of similar order such that the exact-fitting achieved for either one from the 3-point functions is spoiled. Therefore, only approximate matching for the quarks and charged-leptons can be achieved and we are not able to explain the quarks' mixings. 
\section{Models with 6 Higgs from $\mathcal{N}=2$ sector}\label{sec:6Higgs}
\subsection{Model 14}\label{sec:model-14}
In Model~\hyperref[model14]{14} the three-point Yukawa couplings arise from the triplet intersections from the branes ${a, b, c}$ on the second two-torus ($r=2$) with 6 pairs of Higgs from $\mathcal{N}=2$ subsector.

\begin{figure}[t]
	\centering
	\includegraphics[width=\textwidth]{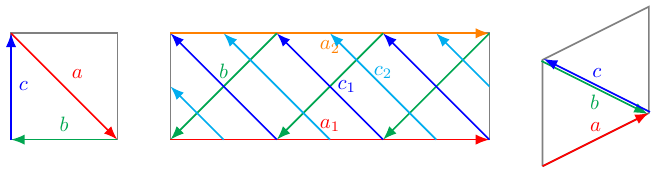}
	\caption{Brane configuration for the three two-tori in Model~\hyperref[model14]{14} where the third two-torus is tilted. Fermion mass hierarchies result from the intersections on the second two-torus.}  \label{Fig.14}
\end{figure}

Yukawa matrices for the Model~\hyperref[model14]{14} are of rank 3 and the three intersections required to form the disk diagrams for the Yukawa couplings all occur on the second torus as shown in figure~\ref{Fig.14}. The other two-tori only contribute an overall constant that has no effect in computing the fermion mass ratios. Thus, it is sufficient for our purpose to only focus on the second torus to explain the masses and the mixing in the standard model fermions.

\subsubsection{3-point Yukawa mass-matrices for Model~\hyperref[model14]{14}}
From the wrapping numbers listed in table~\ref{model14}, the relevant intersection numbers are calculated as,
\begin{align}\label{eq:Intersections.14} 
	I_{ab}^{(1)} & =-1, &   
	I_{ab}^{(2)} & =-3, &   
	I_{ab}^{(3)}&=-1, \nonumber\\
	I_{bc}^{(1)} & =-1, &   
	I_{bc}^{(2)} & =-6, &   
	I_{bc}^{(3)}&=0, \nonumber\\
	I_{ca}^{(1)} & =-1, &   
	I_{ca}^{(2)} & =-3, &   
	I_{ca}^{(3)}&=-1, 
\end{align}
As the intersection numbers are not coprime, we define the greatest common divisor,
$d^{(1)}=g.c.d.(I_{ab}^{(2)},I_{bc}^{(2)},I_{ca}^{(2)})=3$. Thus, the arguments of the modular theta function as defined in \eqref{eqn:Yinput} can be written as,
\begin{align}
	\delta^{(2)} & = \frac{i^{(2)}}{-3} + \frac{j^{(2)}}{-3} + \frac{k^{(2)}}{-6} +\frac{1}{18} \left(6 \epsilon _a^{\text{(2)}}+3 \epsilon _b^{\text{(2)}}+3 \epsilon _c^{\text{(2)}}\right)+ \frac{s^{(2)}}{3},  \label{eq:delta.14} \\
	\phi^{(2)}   & = \frac{1}{3} \left(-6 \theta _a^{\text{(2)}}-3 \theta _b^{\text{(2)}}-3 \theta _c^{\text{(2)}}\right) =0,                                                                                                          \\
	\kappa^{(2)} & = \frac{6 J^{\text{(2)}}}{\alpha '}, \label{eq:kappa.14}                                                                                                                                                            
\end{align}
and recalling \eqref{eq:ijk}, we have $i=\{0,\dots ,2\}$, $j=\{0,\dots ,2\}$ and $k=\{0,\dots ,5\}$ which respectively index the left-handed fermions, the right-handed fermions and the Higgs fields. Clearly, there arise 6 Higgs fields from the $bc$ sector.

The second-last term in the right side of \eqref{eq:delta.14} can be used to redefine the shift on the torus as
\begin{align}
	\epsilon^{(2)} & \equiv \frac{1}{18} \left(6 \epsilon _a^{\text{(2)}}+3 \epsilon _b^{\text{(2)}}+3 \epsilon _c^{\text{(2)}}\right). \label{shifts14} 
\end{align}  
The selection rule for the occurrence of a trilinear Yukawa coupling for a given set of indices is given as,
\begin{equation}\label{selection-rule.14}
	i^{(2)} + j^{(2)} + k^{(2)} = 0\; \mathrm{ mod }\; 3.
\end{equation}
Then the suitable rank-3 mass-matrix can be determined by choosing the specific value of $s^{(2)}=i$. 
\begin{align}
	\delta^{(2)} & =  -\frac{j}{3}-\frac{k}{6} \label{eq:s1.14} 
\end{align}
Here, we will ignore other equivalent cases of solutions or cases with rank-1 problem. The mass matrices for up quarks, down quarks and charged leptons have the following general form:
\begin{align}
	Z_{3} &= Z_{3q} \left(
	\begin{array}{ccc}
	T_0 v_1+T_3 v_4 & T_4 v_3+T_1 v_6 & T_5 v_2+T_2 v_5 \\
	T_2 v_3+T_5 v_6 & T_3 v_2+T_0 v_5 & T_4 v_1+T_1 v_4 \\
	T_1 v_2+T_4 v_5 & T_2 v_1+T_5 v_4 & T_0 v_3+T_3 v_6 \\
	\end{array}
	\right),\label{eq:s-i14}
\end{align} 
where $v_i = \left\langle H_{i} \right\rangle$ and the three-point coupling functions are given in terms of Jacobi theta function of the third kind as, 

\begin{align}\label{eq:3couplings14}
	T_k & \equiv  \vartheta \left[\begin{array}{c} 
	\epsilon^{(2)}+\frac{k}{6}\\  \phi^{(2)} \end{array} \right]
	(\frac{6 J^{\text{(2)}}}{\alpha '}),  \quad   k={0,\cdots,5}.
\end{align}
  
\subsubsection{Fermion masses and mixings from 3-point functions in Model~\hyperref[model14]{14}}
In order to accommodate the fermion masses and the quark mixings, we need to fit the up-type quarks mixing matrix \eqref{eq:mixing-quarks}, the down-type quarks matrix \eqref{eq:mass-downquarks} and the masses of the charged-leptons \eqref{eq:mass-chargedleptons}.
 
We set the K\"{a}hler modulus on the second two-torus defined in \eqref{eq:kappa.14} as $\kappa^{(2)} = 30 $  and evaluate the couplings functions \eqref{eq:3couplings14} by setting geometric brane position parameters as $\epsilon^{(2)}_{u} = 0$, $\epsilon^{(2)}_{d} = 0.0696666$ and $\epsilon^{(2)}_{e} = 0$ which yields a nearest fit for the following VEVs,
\begin{equation}\label{eq:VEVs_14} 
	\begin{array}{ll}
		v^u{}_1 = 0.000291336 , & v^d{}_1 = 6.8572\times 10^{-6}   \\
		v^u{}_2 = 0.117993    , & v^d{}_2 = -5.61839\times 10^{-7} \\
		v^u{}_3 = 0.998234    , & v^d{}_3 = 0.0316                 \\
		v^u{}_4 = 0.566674    , & v^d{}_4 = -2.66199\times 10^{-9} \\
		v^u{}_5 = 0.00552729  , & v^d{}_5 = 0.00144728             \\
		v^u{}_6 = 0.0163423   , & v^d{}_6 = -0.0000122672          \\
	\end{array}
\end{equation}
\begin{align} 
	|M_{3u}|&=m_t \left(
	\begin{array}{ccc}
	0.000291  & 0.00122   & 0.008608 \\
	0.00122   & 0.005527  & 0.041338 \\
	0.008608  & 0.041338  & 0.998234 \\
	\end{array}
	\right)\sim M_u ~, \label{eq:Up3_14} \\
	|M_{3e}|&=m_\tau \left(
	\begin{array}{ccc}
	0.000217  & 0.        & 0.       \\
	0.        & 0.0458    & 0.       \\
	0.        & 0.        & 1.       \\
	\end{array}
	\right)\sim D_e~,\label{eq:Leptons3_14} \\ 
	|M_{3d}|&=m_b \left(
	\begin{array}{ccc}
	0.000217  & -0.000252 & 0.000103 \\
	0.002252  & 0.0458    & 0.       \\
	-0.000012 & 0.        & 1.       \\
	\end{array}
	\right)~.\label{eq:Down3_14}
\end{align}
While the masses of up-type quarks and the charged leptons are fitted, in the down-type quarks matrix, only the mass of the bottom quark can be fitted with three-point couplings only. The masses of the charm and the down quarks are not fitted. Since the intersection number $I_{bc'}^{(2)}=0$, there are no further corrections from the four-point functions.
\subsection{Model 15}\label{sec:model-15}
In Model~\hyperref[model15]{15} the three-point Yukawa couplings arise from the triplet intersections from the branes ${a, b, c}$ on the first two-torus ($r=1$) with 6 pairs of Higgs from $\mathcal{N}=2$ subsector.

\begin{figure}[t]
	\centering
	\includegraphics[width=\textwidth]{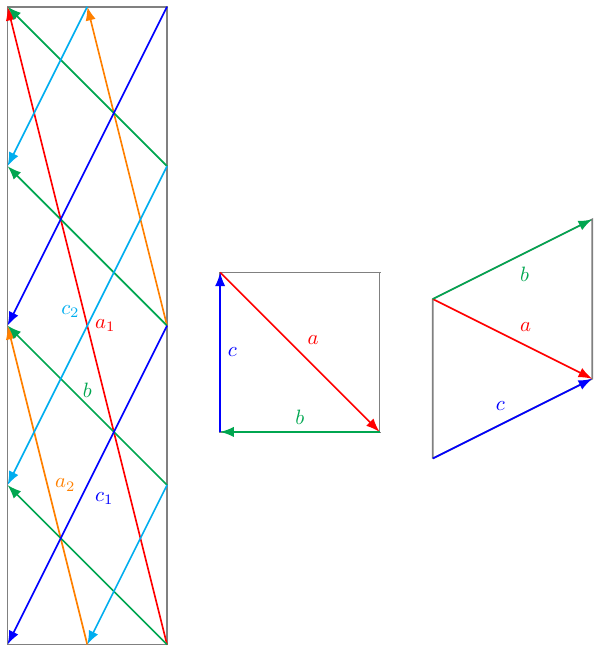}
	\caption{Brane configuration for the three two-tori in Model~\hyperref[model15]{15} where the third two-torus is tilted. Fermion mass hierarchies result from the intersections on the first two-torus.}  \label{Fig.15}
\end{figure}

Yukawa matrices for the Model~\hyperref[model15]{15} are of rank 3 and the three intersections required to form the disk diagrams for the Yukawa couplings all occur on the first torus as shown in figure~\ref{Fig.15}. The other two-tori only contribute an overall constant that has no effect in computing the fermion mass ratios. Thus, it is sufficient for our purpose to only focus on the first torus to explain the masses and the mixing in the standard model fermions.

\subsubsection{3-point Yukawa mass-matrices for Model~\hyperref[model15]{15}}
From the wrapping numbers listed in table~\ref{model15}, the relevant intersection numbers are calculated as,
\begin{align}\label{eq:Intersections.15} 
	I_{ab}^{(1)} & =3,  &   
	I_{ab}^{(2)} & =-1, &   
	I_{ab}^{(3)}&=1, \nonumber\\
	I_{bc}^{(1)} & =6,  &   
	I_{bc}^{(2)} & =-1, &   
	I_{bc}^{(3)}&=0, \nonumber\\
	I_{ca}^{(1)} & =-3, &   
	I_{ca}^{(2)} & =-1, &   
	I_{ca}^{(3)}&=-1, 
\end{align}
As the intersection numbers are not coprime, we define the greatest common divisor,
$d^{(1)}=g.c.d.(I_{ab}^{(1)},I_{bc}^{(1)},I_{ca}^{(1)})=3$. Thus, the arguments of the modular theta function as defined in \eqref{eqn:Yinput} can be written as,
\begin{align}
	\delta^{(1)} & = \frac{i^{(1)}}{3} + \frac{j^{(1)}}{-3} + \frac{k^{(1)}}{6} +\frac{1}{18} \left(-6 \epsilon _a^{\text{(1)}}+3 \epsilon _b^{\text{(1)}}-3 \epsilon _c^{\text{(1)}}\right)+ \frac{s^{(1)}}{3},  \label{eq:delta.15} \\
	\phi^{(1)}   & = \frac{1}{3} \left(6 \theta _a^{\text{(1)}}-3 \theta _b^{\text{(1)}}+3 \theta _c^{\text{(1)}}\right) =0,                                                                                                          \\
	\kappa^{(1)} & = \frac{6 J^{\text{(1)}}}{\alpha '}, \label{eq:kappa.15}                                                                                                                                                           
\end{align}
and recalling \eqref{eq:ijk}, we have $i=\{0,\dots ,2\}$, $j=\{0,\dots ,2\}$ and $k=\{0,\dots ,5\}$ which respectively index the left-handed fermions, the right-handed fermions and the Higgs fields. Clearly, there arise 6 Higgs fields from the $bc$ sector.

The second-last term in the right side of \eqref{eq:delta.15} can be used to redefine the shift on the torus as
\begin{align}
	\epsilon^{(1)} & \equiv \frac{1}{18} \left(-6 \epsilon _a^{\text{(1)}}+3 \epsilon _b^{\text{(1)}}-3 \epsilon _c^{\text{(1)}}\right). \label{shifts15} 
\end{align}  
The selection rule for the occurrence of a trilinear Yukawa coupling for a given set of indices is given as,
\begin{equation}\label{selection-rule.15}
	i^{(1)} + j^{(1)} + k^{(1)} = 0\; \mathrm{ mod }\; 3.
\end{equation}
Then the suitable rank-3 mass-matrix can be determined by choosing the specific value of $s^{(1)}=j$. 
\begin{align}
	\delta^{(1)} & =  \frac{i}{3}+\frac{k}{6} \label{eq:s1.15} 
\end{align}
Here, we will ignore other equivalent cases of solutions or cases with rank-1 problem. The mass matrices for up quarks, down quarks and charged leptons have the following general form:
\begin{align}
	Z_{3} &= Z_{3q} \left(
	\begin{array}{ccc}
	T_0 v_1+T_3 v_4 & T_2 v_3+T_5 v_6 & T_1 v_2+T_4 v_5 \\
	T_4 v_3+T_1 v_6 & T_3 v_2+T_0 v_5 & T_2 v_1+T_5 v_4 \\
	T_5 v_2+T_2 v_5 & T_4 v_1+T_1 v_4 & T_0 v_3+T_3 v_6 \\
	\end{array}
	\right),\label{eq:s-i15}
\end{align} 
where $v_i = \left\langle H_{i} \right\rangle$ and the three-point coupling functions are given in terms of Jacobi theta function of the third kind as, 

\begin{align}\label{eq:3couplings15}
	T_k & \equiv  \vartheta \left[\begin{array}{c} 
	\epsilon^{(1)}+\frac{k}{6}\\  \phi^{(1)} \end{array} \right]
	(\frac{6 J^{\text{(1)}}}{\alpha '}),  \quad   k={0,\cdots,5}.
\end{align}
  
\subsubsection{Fermion masses and mixings from 3-point functions in Model~\hyperref[model15]{15}}
In order to accommodate the fermion masses and the quark mixings, we need to fit the up-type quarks mixing matrix \eqref{eq:mixing-quarks}, the down-type quarks matrix \eqref{eq:mass-downquarks} and the masses of the charged-leptons \eqref{eq:mass-chargedleptons}.
 
We set the K\"{a}hler modulus on the first two-torus defined in \eqref{eq:kappa.15} as $\kappa^{(1)} = 30 $  and evaluate the couplings functions \eqref{eq:3couplings15} by setting geometric brane position parameters as $\epsilon^{(1)}_{u} = 0$, $\epsilon^{(1)}_{d} = 0.0696666$ and $\epsilon^{(1)}_{e} = 0$ which yields a nearest fit for the following VEVs,
\begin{equation}\label{eq:VEVs_15} 
	\begin{array}{ll}
		v^u{}_1 = 0.000291336 , & v^d{}_1 = 6.8572\times 10^{-6}   \\
		v^u{}_2 = 0.117993    , & v^d{}_2 = -5.61839\times 10^{-7} \\
		v^u{}_3 = 0.998234    , & v^d{}_3 = 0.0316                 \\
		v^u{}_4 = 0.566674    , & v^d{}_4 = -2.66199\times 10^{-9} \\
		v^u{}_5 = 0.00552729  , & v^d{}_5 = 0.00144728             \\
		v^u{}_6 = 0.0163423   , & v^d{}_6 = -0.0000122672          \\
	\end{array}
\end{equation}
\begin{align} 
	|M_{3u}|&=m_t \left(
	\begin{array}{ccc}
	0.000291  & 0.00122  & 0.008608  \\
	0.00122   & 0.005527 & 0.041338  \\
	0.008608  & 0.041338 & 0.998234  \\
	\end{array}
	\right)\sim M_u ~, \label{eq:Up3_15} \\
	|M_{3e}|&=m_\tau \left(
	\begin{array}{ccc}
	0.000217  & 0.       & 0.        \\
	0.        & 0.0458   & 0.        \\
	0.        & 0.       & 1.        \\
	\end{array}
	\right)\sim D_e~,\label{eq:Leptons3_15} \\ 
	|M_{3d}|&=m_b \left(
	\begin{array}{ccc}
	0.000217  & 0.002252 & -0.000012 \\
	-0.000252 & 0.0458   & 0.        \\
	0.000103  & 0.       & 1.        \\
	\end{array}
	\right)~.\label{eq:Down3_15}
\end{align}
While the masses of up-type quarks and the charged leptons are fitted, in the down-type quarks matrix, only the mass of the bottom quark can be fitted with three-point couplings only. The masses of the charm and the down quarks are not fitted. Notice, that these results are only at the tree-level and there could indeed be other corrections, such as those coming from higher-dimensional operators, which may contribute most greatly to the charm and the down quarks' masses since they are lighter.

\subsubsection{4-point corrections in Model~\hyperref[model15]{15}}
 
The four-point couplings in Model~\hyperref[model15]{15} in table~\ref{model15} can come from considering interactions of ${a, b, c}$ with $b'$ or $c'$ on the first two-torus as can be seen from the following intersection numbers,
\begin{align}
	I_{bb'}^{(1)} & = -8, & I_{bb'}^{(2)} & = 0, & I_{bb'}^{(3)} & = 1 ,\nonumber                    \\
	I_{cc'}^{(1)} & = 4,  & I_{cc'}^{(2)} & = 0, & I_{cc'}^{(3)} & = 1 ,\nonumber                    \\
	I_{bc'}^{(1)} & = -2, & I_{bc'}^{(2)} & = 1, & I_{bc'}^{(3)} & = -1 . \label{eq:4Intersection15} 
\end{align}
There are 4 SM singlet fields $S_L^i$ and 2
Higgs-like state $H^{\prime}_{u, d}$. 

Let us consider four-point interactions with $c'$ with the following parameters with shifts $l=-\frac{k}{2} $ and $\ell=-\frac{k}{3} $ taken along the index $k$,
\begin{align}\label{deltas15}
	\delta & =\frac{i}{I_{ab}^{(1)}} +\frac{j}{I_{ca}^{(1)}} +\frac{k}{I_{bc}^{(1)}} +l , \nonumber               \\
	       & = \frac{i}{3}-\frac{j}{3} ,                                                                          \\
	d      & =\frac{\imath}{I_{c c'}^{(1)}} +\frac{\jmath}{I_{bc'}^{(1)}}+\frac{k}{I_{bc}^{(1)}} +\ell ,\nonumber \\
	       & = \frac{\imath}{4}-\frac{\jmath}{2} ,                                                                
\end{align}
the matrix elements $a_{i,j,\imath}$ on the first torus from the four-point functions results in the following classical 4-point contribution to the mass matrix,
\begin{equation}\label{eq:4point15} 
	Z_{4cl} =  \left(
	\begin{array}{ccc}
		F_0 u_1 w_1 & 0                       & F_1 u_4 w_2             \\
		0           & F_1 u_4 w_2             & F_3 u_4 w_1+F_0 u_3 w_2 \\
		F_1 u_4 w_2 & F_3 u_4 w_1+F_0 u_3 w_2 & F_2 u_3 w_1+F_3 u_2 w_2 \\
	\end{array}
	\right)
\end{equation} 
where $u_i, w_j$ are the VEVs and the couplings are defined as,
\begin{align}\label{eq:4couplings15}
	F_{i} & \equiv  \vartheta \left[\begin{array}{c} 
	\epsilon^{(1)}+\frac{i}{4}\\  \phi^{(1)} \end{array} \right]
	(\frac{6 J^{\text{(1)}}}{\alpha '}),\qquad i={0,\dots,|I_{c c'}^{(1)}|-1}.
\end{align}

Since, we have already fitted the up-type quark matrix $|M_{3u}|$ exactly, so its 4-point correction should be zero,
\begin{equation}
	|M_{4u}| = 0 ,
\end{equation}
which is true by setting all up-type VEVs $u_u^i$ and $w_u^i$ to be zero. 
Therefore, we are essentially concerned with fitting down-type quarks in such a way that corresponding corrections for the charged-leptons remain negligible. The desired solution can be readily obtained by setting $\epsilon^{(1)}_{4d}=0$ with the following values of the VEVs,
\begin{equation}\label{eq:UdWd15} 
	\begin{array}{l}
		u^d{}_1 = 0.00002386 \\
		u^d{}_2 = 0          \\
		u^d{}_3 = 0          \\
		u^d{}_4 = -0.128723  \\
	\end{array}
	\quad,\quad
	\begin{array}{l}
		w^d{}_1 = 1 \\
		w^d{}_2 = 1 \\
	\end{array}
\end{equation}
The 4-point correction to the down-type quarks' masses is given by,
\begin{equation}\label{M4d_15}
	|M_{4d}|=m_b \left(
	\begin{array}{ccc}
		0.001193 & 0.      & -0.0178 \\
		0.       & -0.0178 & -0.0178 \\
		-0.0178  & -0.0178 & 0.      \\
	\end{array}
	\right)
\end{equation}
which can be added to the matrix obtained from 3-point functions \eqref{eq:Down3_15} as,
\begin{equation}\label{M34d_15}
	|M_{3d}|+|M_{4d}|=m_b \left(
	\begin{array}{ccc}
		0.00141   & 0.002252 & -0.017812 \\
		-0.000252 & 0.028    & -0.0178   \\
		-0.017697 & -0.0178  & 1.        \\
	\end{array}
	\right) \sim D_d ~,
\end{equation} 
However, we also need to keep the corrections to charged-leptons' masses to be negligible by setting $\epsilon^{(1)}_{4e}=0$,
\begin{equation}\label{M4e_15}
	|M_{4e}|\sim \left(
	\begin{array}{ccc}
		0.000755  & 0.        & -0.011266 \\
		0.        & -0.011266 & -0.011266 \\
		-0.011266 & -0.011266 & 0.        \\
	\end{array}
	\right)\sim 0.
\end{equation} 
Although the results appears to be an near-exact it should be noted that we have assumed a strictly symmetric CKM matrix \eqref{eq:VCKM} and therefore, the matching is only approximate for a general asymmetric matrix.
\subsection{Model 15-dual}\label{sec:model-15.5}
In Model~\hyperref[model15.5]{15-dual} the three-point Yukawa couplings arise from the triplet intersections from the branes ${a, b, c}$ on the second two-torus ($r=2$) with 6 pairs of Higgs from $\mathcal{N}=2$ subsector.

\begin{figure}[t]
	\centering
	\includegraphics[width=\textwidth]{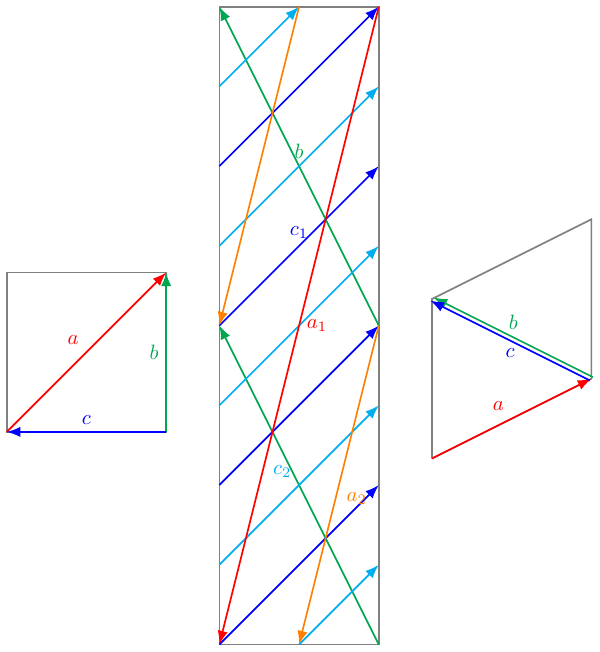}
	\caption{Brane configuration for the three two-tori in Model~\hyperref[model15.5]{15-dual} where the third two-torus is tilted. Fermion mass hierarchies result from the intersections on the second two-torus.}  \label{Fig.15.5}
\end{figure}

Yukawa matrices for the Model~\hyperref[model15.5]{15-dual} are of rank 3 and the three intersections required to form the disk diagrams for the Yukawa couplings all occur on the second torus as shown in figure~\ref{Fig.15.5}. The other two-tori only contribute an overall constant that has no effect in computing the fermion mass ratios. Thus, it is sufficient for our purpose to only focus on the second torus to explain the masses and the mixing in the standard model fermions.

\subsubsection{3-point Yukawa mass-matrices for Model~\hyperref[model15.5]{15-dual}}
From the wrapping numbers listed in table~\ref{model15.5}, the relevant intersection numbers are calculated as,
\begin{align}\label{eq:Intersections.15.5} 
	I_{ab}^{(1)} & =1,  &   
	I_{ab}^{(2)} & =-3, &   
	I_{ab}^{(3)}&=1, \nonumber\\
	I_{bc}^{(1)} & =1,  &   
	I_{bc}^{(2)} & =-6, &   
	I_{bc}^{(3)}&=0, \nonumber\\
	I_{ca}^{(1)} & =-1, &   
	I_{ca}^{(2)} & =-3, &   
	I_{ca}^{(3)}&=-1, 
\end{align}
As the intersection numbers are not coprime, we define the greatest common divisor,
$d^{(1)}=g.c.d.(I_{ab}^{(2)},I_{bc}^{(2)},I_{ca}^{(2)})=3$. Thus, the arguments of the modular theta function as defined in \eqref{eqn:Yinput} can be written as,
\begin{align}
	\delta^{(2)} & = \frac{i^{(2)}}{-3} + \frac{j^{(2)}}{-3} + \frac{k^{(2)}}{-6} +\frac{1}{18} \left(6 \epsilon _a^{\text{(2)}}+3 \epsilon _b^{\text{(2)}}+3 \epsilon _c^{\text{(2)}}\right)+ \frac{s^{(2)}}{3},  \label{eq:delta.15.5} \\
	\phi^{(2)}   & = \frac{1}{3} \left(-6 \theta _a^{\text{(2)}}-3 \theta _b^{\text{(2)}}-3 \theta _c^{\text{(2)}}\right) =0,                                                                                                            \\
	\kappa^{(2)} & = \frac{6 J^{\text{(2)}}}{\alpha '}, \label{eq:kappa.15.5}                                                                                                                                                            
\end{align}
and recalling \eqref{eq:ijk}, we have $i=\{0,\dots ,2\}$, $j=\{0,\dots ,2\}$ and $k=\{0,\dots ,5\}$ which respectively index the left-handed fermions, the right-handed fermions and the Higgs fields. Clearly, there arise 6 Higgs fields from the $bc$ sector.

The second-last term in the right side of \eqref{eq:delta.15.5} can be used to redefine the shift on the torus as
\begin{align}
	\epsilon^{(2)} & \equiv \frac{1}{18} \left(6 \epsilon _a^{\text{(2)}}+3 \epsilon _b^{\text{(2)}}+3 \epsilon _c^{\text{(2)}}\right). \label{shifts15.5} 
\end{align}  
The selection rule for the occurrence of a trilinear Yukawa coupling for a given set of indices is given as,
\begin{equation}\label{selection-rule.15.5}
	i^{(2)} + j^{(2)} + k^{(2)} = 0\; \mathrm{ mod }\; 3.
\end{equation}
Then the suitable rank-3 mass-matrix can be determined by choosing the specific value of $s^{(2)}=i$. 
\begin{align}
	\delta^{(2)} & =  -\frac{j}{3}-\frac{k}{6} \label{eq:s1.15.5} 
\end{align}
Here, we will ignore other equivalent cases of solutions or cases with rank-1 problem. The mass matrices for up quarks, down quarks and charged leptons have the following general form:
\begin{align}
	Z_{3} &= Z_{3q} \left(
	\begin{array}{ccc}
	T_0 v_1+T_3 v_4 & T_4 v_3+T_1 v_6 & T_5 v_2+T_2 v_5 \\
	T_2 v_3+T_5 v_6 & T_3 v_2+T_0 v_5 & T_4 v_1+T_1 v_4 \\
	T_1 v_2+T_4 v_5 & T_2 v_1+T_5 v_4 & T_0 v_3+T_3 v_6 \\
	\end{array}
	\right),\label{eq:s-i15.5}
\end{align} 
where $v_i = \left\langle H_{i} \right\rangle$ and the three-point coupling functions are given in terms of Jacobi theta function of the third kind as, 

\begin{align}\label{eq:3couplings15.5}
	T_k & \equiv  \vartheta \left[\begin{array}{c} 
	\epsilon^{(2)}+\frac{k}{6}\\  \phi^{(2)} \end{array} \right]
	(\frac{6 J^{\text{(2)}}}{\alpha '}),  \quad   k={0,\cdots,5}.
\end{align}
  
\subsubsection{Fermion masses and mixings from 3-point functions in Model~\hyperref[model15.5]{15-dual}}
In order to accommodate the fermion masses and the quark mixings, we need to fit the up-type quarks mixing matrix \eqref{eq:mixing-quarks}, the down-type quarks matrix \eqref{eq:mass-downquarks} and the masses of the charged-leptons \eqref{eq:mass-chargedleptons}.
 
We set the K\"{a}hler modulus on the second two-torus defined in \eqref{eq:kappa.15.5} as $\kappa^{(2)} = 30 $  and evaluate the couplings functions \eqref{eq:3couplings15.5} by setting geometric brane position parameters as $\epsilon^{(2)}_{u} = 0$, $\epsilon^{(2)}_{d} = 0.0696666$ and $\epsilon^{(2)}_{e} = 0$ which yields a nearest fit for the following VEVs,
\begin{equation}\label{eq:VEVs_15.5} 
	\begin{array}{ll}
		v^u{}_1 = 0.000291336 , & v^d{}_1 = 6.8572\times 10^{-6}   \\
		v^u{}_2 = 0.117993    , & v^d{}_2 = -5.61839\times 10^{-7} \\
		v^u{}_3 = 0.998234    , & v^d{}_3 = 0.0316                 \\
		v^u{}_4 = 0.566674    , & v^d{}_4 = -2.66199\times 10^{-9} \\
		v^u{}_5 = 0.00552729  , & v^d{}_5 = 0.00144728             \\
		v^u{}_6 = 0.0163423   , & v^d{}_6 = -0.0000122672          \\
	\end{array}
\end{equation}
\begin{align} 
	|M_{3u}|&=m_t \left(
	\begin{array}{ccc}
	0.000291  & 0.00122   & 0.008608 \\
	0.00122   & 0.005527  & 0.041338 \\
	0.008608  & 0.041338  & 0.998234 \\
	\end{array}
	\right)\sim M_u ~, \label{eq:Up3_15.5} \\
	|M_{3e}|&=m_\tau \left(
	\begin{array}{ccc}
	0.000217  & 0.        & 0.       \\
	0.        & 0.0458    & 0.       \\
	0.        & 0.        & 1.       \\
	\end{array}
	\right)\sim D_e~,\label{eq:Leptons3_15.5} \\ 
	|M_{3d}|&=m_b \left(
	\begin{array}{ccc}
	0.000217  & -0.000252 & 0.000103 \\
	0.002252  & 0.0458    & 0.       \\
	-0.000012 & 0.        & 1.       \\
	\end{array}
	\right)~.\label{eq:Down3_15.5}
\end{align}
While the masses of up-type quarks and the charged leptons are fitted, in the down-type quarks matrix, only the mass of the bottom quark can be fitted with three-point couplings only. The masses of the charm and the down quarks are not fitted. Notice, that these results are only at the tree-level and there could indeed be other corrections, such as those coming from higher-dimensional operators, which may contribute most greatly to the charm and the down quarks' masses since they are lighter.

\subsubsection{4-point corrections in Model~\hyperref[model15.5]{15-dual}}
 
The four-point couplings in Model~\hyperref[model15.5]{15-dual} in table~\ref{model15.5} can come from considering interactions of ${a, b, c}$ with $b'$ or $c'$ on the second two-torus as can be seen from the following intersection numbers,
\begin{align}
	I_{bb'}^{(1)} & = 0, & I_{bb'}^{(2)} & = -4, & I_{bb'}^{(3)} & = -1 ,\nonumber                    \\
	I_{cc'}^{(1)} & = 0, & I_{cc'}^{(2)} & = 8,  & I_{cc'}^{(3)} & = -1 ,\nonumber                    \\
	I_{bc'}^{(1)} & = 1, & I_{bc'}^{(2)} & = -2, & I_{bc'}^{(3)} & = 1 . \label{eq:4Intersection15.5} 
\end{align}
There are 4 SM singlet fields $S_L^i$ and 2
Higgs-like state $H^{\prime}_{u, d}$. 

Let us consider four-point interactions with $b'$ with the following parameters with shifts $l=\frac{k}{2} $ and $\ell=\frac{k}{3} $ taken along the index $k$,
\begin{align}\label{deltas15.5}
	\delta & =\frac{i}{I_{ab}^{(2)}} +\frac{j}{I_{ca}^{(2)}} +\frac{k}{I_{bc}^{(2)}} +l , \nonumber               \\
	       & = -\frac{i}{3}-\frac{j}{3} ,                                                                         \\
	d      & =\frac{\imath}{I_{b b'}^{(2)}} +\frac{\jmath}{I_{bc'}^{(2)}}+\frac{k}{I_{bc}^{(2)}} +\ell ,\nonumber \\
	       & = -\frac{\imath}{4}-\frac{\jmath}{2} ,                                                               
\end{align}
the matrix elements $a_{i,j,\imath}$ on the second torus from the four-point functions results in the following classical 4-point contribution to the mass matrix,
\begin{equation}\label{eq:4point15.5} 
	Z_{4cl} =  \left(
	\begin{array}{ccc}
		F_0 u_1 w_1 & 0                       & F_3 u_4 w_2             \\
		0           & F_3 u_4 w_2             & F_1 u_4 w_1+F_0 u_3 w_2 \\
		F_3 u_4 w_2 & F_1 u_4 w_1+F_0 u_3 w_2 & F_2 u_3 w_1+F_1 u_2 w_2 \\
	\end{array}
	\right)
\end{equation} 
where $u_i, w_j$ are the VEVs and the couplings are defined as,
\begin{align}\label{eq:4couplings15.5}
	F_{i} & \equiv  \vartheta \left[\begin{array}{c} 
	\epsilon^{(2)}+\frac{i}{4}\\  \phi^{(2)} \end{array} \right]
	(\frac{6 J^{\text{(2)}}}{\alpha '}),\qquad i={0,\dots,|I_{b b'}^{(2)}|-1}.
\end{align}

Since, we have already fitted the up-type quark matrix $|M_{3u}|$ exactly, so its 4-point correction should be zero,
\begin{equation}
	|M_{4u}| = 0 ,
\end{equation}
which is true by setting all up-type VEVs $u_u^i$ and $w_u^i$ to be zero. 
Therefore, we are essentially concerned with fitting down-type quarks in such a way that corresponding corrections for the charged-leptons remain negligible. The desired solution can be readily obtained by setting $\epsilon^{(2)}_{4d}=0$ with the following values of the VEVs,
\begin{equation}\label{eq:UdWd15.5} 
	\begin{array}{l}
		u^d{}_1 = 0.00002386 \\
		u^d{}_2 = 0          \\
		u^d{}_3 = 0          \\
		u^d{}_4 = -0.128723  \\
	\end{array}
	\quad,\quad
	\begin{array}{l}
		w^d{}_1 = 1 \\
		w^d{}_2 = 1 \\
	\end{array}
\end{equation}
The 4-point correction to the down-type quarks' masses is given by,
\begin{equation}\label{M4d_15.5}
	|M_{4d}|=m_b \left(
	\begin{array}{ccc}
		0.001193 & 0.      & -0.0178 \\
		0.       & -0.0178 & -0.0178 \\
		-0.0178  & -0.0178 & 0.      \\
	\end{array}
	\right)
\end{equation}
which can be added to the matrix obtained from 3-point functions \eqref{eq:Down3_15.5} as,
\begin{equation}\label{M34d_15.5}
	|M_{3d}|+|M_{4d}|=m_b \left(
	\begin{array}{ccc}
		0.00141   & -0.000252 & -0.017697 \\
		0.002252  & 0.028     & -0.0178   \\
		-0.017812 & -0.0178   & 1.        \\
	\end{array}
	\right) \sim D_d ~,
\end{equation} 
However, we also need to keep the corrections to charged-leptons' masses to be negligible by setting $\epsilon^{(1)}_{4e}=0$,
\begin{equation}\label{M4e_15.5}
	|M_{4e}|\sim \left(
	\begin{array}{ccc}
		0.000755  & 0.        & -0.011266 \\
		0.        & -0.011266 & -0.011266 \\
		-0.011266 & -0.011266 & 0.        \\
	\end{array}
	\right)\sim 0.
\end{equation} 
Although the results appears to be an near-exact it should be noted that we have assumed a strictly symmetric CKM matrix \eqref{eq:VCKM} and therefore, the matching is only approximate for a general asymmetric matrix.
\subsection{Model 16}\label{sec:model-16}
In Model~\hyperref[model16]{16} the three-point Yukawa couplings arise from the triplet intersections from the branes ${a, b, c}$ on the first two-torus ($r=1$) with 6 pairs of Higgs from $\mathcal{N}=2$ subsector.

\begin{figure}[t]
	\centering
	\includegraphics[width=\textwidth]{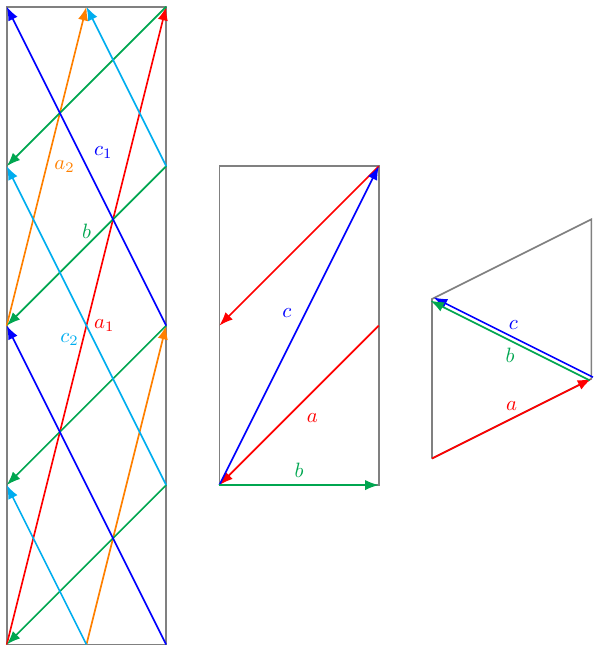}
	\caption{Brane configuration for the three two-tori in Model~\hyperref[model16]{16} where the third two-torus is tilted. Fermion mass hierarchies result from the intersections on the first two-torus.}  \label{Fig.16}
\end{figure}

Yukawa matrices for the Model~\hyperref[model16]{16} are of rank 3 and the three intersections required to form the disk diagrams for the Yukawa couplings all occur on the first torus as shown in figure~\ref{Fig.16}. The other two-tori only contribute an overall constant that has no effect in computing the fermion mass ratios. Thus, it is sufficient for our purpose to only focus on the first torus to explain the masses and the mixing in the standard model fermions.

\subsubsection{3-point Yukawa mass-matrices for Model~\hyperref[model16]{16}}
From the wrapping numbers listed in table~\ref{model16}, the relevant intersection numbers are calculated as,
\begin{align}\label{eq:Intersections.16} 
	I_{ab}^{(1)} & =3,  &   
	I_{ab}^{(2)} & =1,  &   
	I_{ab}^{(3)}&=1, \nonumber\\
	I_{bc}^{(1)} & =-6, &   
	I_{bc}^{(2)} & =1,  &   
	I_{bc}^{(3)}&=0, \nonumber\\
	I_{ca}^{(1)} & =-3, &   
	I_{ca}^{(2)} & =1,  &   
	I_{ca}^{(3)}&=-1, 
\end{align}
As the intersection numbers are not coprime, we define the greatest common divisor,
$d^{(1)}=g.c.d.(I_{ab}^{(1)},I_{bc}^{(1)},I_{ca}^{(1)})=3$. Thus, the arguments of the modular theta function as defined in \eqref{eqn:Yinput} can be written as,
\begin{align}
	\delta^{(1)} & = \frac{i^{(1)}}{3} + \frac{j^{(1)}}{-3} + \frac{k^{(1)}}{-6} +\frac{1}{18} \left(-6 \epsilon _a^{\text{(1)}}-3 \epsilon _b^{\text{(1)}}+3 \epsilon _c^{\text{(1)}}\right)+ \frac{s^{(1)}}{3},  \label{eq:delta.16} \\
	\phi^{(1)}   & = \frac{1}{3} \left(-6 \theta _a^{\text{(1)}}-3 \theta _b^{\text{(1)}}+3 \theta _c^{\text{(1)}}\right) =0,                                                                                                          \\
	\kappa^{(1)} & = \frac{6 J^{\text{(1)}}}{\alpha '}, \label{eq:kappa.16}                                                                                                                                                            
\end{align}
and recalling \eqref{eq:ijk}, we have $i=\{0,\dots ,2\}$, $j=\{0,\dots ,2\}$ and $k=\{0,\dots ,5\}$ which respectively index the left-handed fermions, the right-handed fermions and the Higgs fields. Clearly, there arise 6 Higgs fields from the $bc$ sector.

The second-last term in the right side of \eqref{eq:delta.16} can be used to redefine the shift on the torus as
\begin{align}
	\epsilon^{(1)} & \equiv \frac{1}{18} \left(-6 \epsilon _a^{\text{(1)}}-3 \epsilon _b^{\text{(1)}}+3 \epsilon _c^{\text{(1)}}\right). \label{shifts16} 
\end{align}  
The selection rule for the occurrence of a trilinear Yukawa coupling for a given set of indices is given as,
\begin{equation}\label{selection-rule.16}
	i^{(1)} + j^{(1)} + k^{(1)} = 0\; \mathrm{ mod }\; 3.
\end{equation}
Then the suitable rank-3 mass-matrix can be determined by choosing the specific value of $s^{(1)}=-i$. 
\begin{align}
	\delta^{(1)} & =  -\frac{j}{3}-\frac{k}{6} \label{eq:s1.16} 
\end{align}
Here, we will ignore other equivalent cases of solutions or cases with rank-1 problem. The mass matrices for up quarks, down quarks and charged leptons have the following general form:
\begin{align}
	Z_{3} &= Z_{3q} \left(
	\begin{array}{ccc}
	T_0 v_1+T_3 v_4 & T_4 v_3+T_1 v_6 & T_5 v_2+T_2 v_5 \\
	T_2 v_3+T_5 v_6 & T_3 v_2+T_0 v_5 & T_4 v_1+T_1 v_4 \\
	T_1 v_2+T_4 v_5 & T_2 v_1+T_5 v_4 & T_0 v_3+T_3 v_6 \\
	\end{array}
	\right),\label{eq:s-i16}
\end{align} 
where $v_i = \left\langle H_{i} \right\rangle$ and the three-point coupling functions are given in terms of Jacobi theta function of the third kind as, 

\begin{align}\label{eq:3couplings16}
	T_k & \equiv  \vartheta \left[\begin{array}{c} 
	\epsilon^{(1)}+\frac{k}{6}\\  \phi^{(1)} \end{array} \right]
	(\frac{6 J^{\text{(1)}}}{\alpha '}),  \quad   k={0,\cdots,5}.
\end{align}
  
\subsubsection{Fermion masses and mixings from 3-point functions in Model~\hyperref[model16]{16}}
In order to accommodate the fermion masses and the quark mixings, we need to fit the up-type quarks mixing matrix \eqref{eq:mixing-quarks}, the down-type quarks matrix \eqref{eq:mass-downquarks} and the masses of the charged-leptons \eqref{eq:mass-chargedleptons}.
 
We set the K\"{a}hler modulus on the first two-torus defined in \eqref{eq:kappa.16} as $\kappa^{(1)} = 30 $  and evaluate the couplings functions \eqref{eq:3couplings16} by setting geometric brane position parameters as $\epsilon^{(1)}_{u} = 0$, $\epsilon^{(1)}_{d} = 0.0696666$ and $\epsilon^{(1)}_{e} = 0$ which yields a nearest fit for the following VEVs,
\begin{equation}\label{eq:VEVs_16} 
	\begin{array}{ll}
		v^u{}_1 = 0.000291336 , & v^d{}_1 = 6.8572\times 10^{-6}   \\
		v^u{}_2 = 0.117993    , & v^d{}_2 = -5.61839\times 10^{-7} \\
		v^u{}_3 = 0.998234    , & v^d{}_3 = 0.0316                 \\
		v^u{}_4 = 0.566674    , & v^d{}_4 = -2.66199\times 10^{-9} \\
		v^u{}_5 = 0.00552729  , & v^d{}_5 = 0.00144728             \\
		v^u{}_6 = 0.0163423   , & v^d{}_6 = -0.0000122672          \\
	\end{array}
\end{equation}
\begin{align} 
	|M_{3u}|&=m_t \left(
	\begin{array}{ccc}
	0.000291  & 0.00122   & 0.008608 \\
	0.00122   & 0.005527  & 0.041338 \\
	0.008608  & 0.041338  & 0.998234 \\
	\end{array}
	\right)\sim M_u ~, \label{eq:Up3_16} \\
	|M_{3e}|&=m_\tau \left(
	\begin{array}{ccc}
	0.000217  & 0.        & 0.       \\
	0.        & 0.0458    & 0.       \\
	0.        & 0.        & 1.       \\
	\end{array}
	\right)\sim D_e~,\label{eq:Leptons3_16} \\ 
	|M_{3d}|&=m_b \left(
	\begin{array}{ccc}
	0.000217  & -0.000252 & 0.000103 \\
	0.002252  & 0.0458    & 0.       \\
	-0.000012 & 0.        & 1.       \\
	\end{array}
	\right)~.\label{eq:Down3_16}
\end{align}
While the masses of up-type quarks and the charged leptons are fitted, in the down-type quarks matrix, only the mass of the bottom quark can be fitted with three-point couplings only. The masses of the charm and the down quarks are not fitted. Notice, that these results are only at the tree-level and there could indeed be other corrections, such as those coming from higher-dimensional operators, which may contribute most greatly to the charm and the down quarks' masses since they are lighter.

\subsubsection{4-point corrections in Model~\hyperref[model16]{16}}
 
The four-point couplings in Model~\hyperref[model16]{16} in table~\ref{model16} can come from considering interactions of ${a, b, c}$ with $b'$ or $c'$ on the first two-torus as can be seen from the following intersection numbers,
\begin{align}
	I_{bb'}^{(1)} & = 8,  & I_{bb'}^{(2)} & = 0,  & I_{bb'}^{(3)} & = -1 ,\nonumber                  \\
	I_{cc'}^{(1)} & = -4, & I_{cc'}^{(2)} & = 2,  & I_{cc'}^{(3)} & = -1 ,\nonumber                  \\
	I_{bc'}^{(1)} & = 2,  & I_{bc'}^{(2)} & = -1, & I_{bc'}^{(3)} & = 1 . \label{eq:4Intersection16} 
\end{align}
There are 8 SM singlet fields $S_L^i$ and 2
Higgs-like state $H^{\prime}_{u, d}$. 

Let us consider four-point interactions with $b'$ with the following parameters with shifts $l=\frac{k}{2} $ and $\ell=\frac{k}{3} $ taken along the index $k$,
\begin{align}\label{deltas16}
	\delta & =\frac{i}{I_{ab}^{(1)}} +\frac{j}{I_{ca}^{(1)}} +\frac{k}{I_{bc}^{(1)}} +l , \nonumber               \\
	       & = \frac{i}{3}-\frac{j}{3} ,                                                                          \\
	d      & =\frac{\imath}{I_{b b'}^{(1)}} +\frac{\jmath}{I_{bc'}^{(1)}}+\frac{k}{I_{bc}^{(1)}} +\ell ,\nonumber \\
	       & = \frac{\imath}{8}+\frac{\jmath}{2} ,                                                                
\end{align}
the matrix elements $a_{i,j,\imath}$ on the first torus from the four-point functions results in the following classical 4-point contribution to the mass matrix,
\begin{equation}\label{eq:4point16} 
	Z_{4cl} =  \left(
	\begin{array}{ccc}
		F_0 u_1 w_1+F_6 u_7 w_1+F_1 u_6 w_2 & F_5 u_6 w_1+F_0 u_5 w_2 & F_4 u_5 w_1+F_7 u_4 w_2             \\
		F_5 u_6 w_1+F_0 u_5 w_2             & F_4 u_5 w_1+F_7 u_4 w_2 & F_3 u_4 w_1+F_6 u_3 w_2             \\
		F_4 u_5 w_1+F_7 u_4 w_2             & F_3 u_4 w_1+F_6 u_3 w_2 & F_2 u_3 w_1+F_5 u_2 w_2+F_3 u_8 w_2 \\
	\end{array}
	\right)
\end{equation} 
where $u_i, w_j$ are the VEVs and the couplings are defined as,
\begin{align}\label{eq:4couplings16}
	F_{i} & \equiv  \vartheta \left[\begin{array}{c} 
	\epsilon^{(1)}+\frac{i}{8}\\  \phi^{(1)} \end{array} \right]
	(\frac{6 J^{\text{(1)}}}{\alpha '}),\qquad i={0,\dots,|I_{b b'}^{(1)}|-1}.
\end{align}

Since, we have already fitted the up-type quark matrix $|M_{3u}|$ exactly, so its 4-point correction should be zero,
\begin{equation}
	|M_{4u}| = 0 ,
\end{equation}
which is true by setting all up-type VEVs $u_u^i$ and $w_u^i$ to be zero. 
Therefore, we are essentially concerned with fitting down-type quarks in such a way that corresponding corrections for the charged-leptons remain negligible. The desired solution can be readily obtained by setting $\epsilon^{(1)}_{4d}=0$ with the following values of the VEVs,
\begin{equation}\label{eq:UdWd16} 
	\begin{array}{l}
		u^d{}_1 = 0.00002386  \\
		u^d{}_2 = 0           \\
		u^d{}_3 = 0           \\
		u^d{}_4 = -0.00155239 \\
		u^d{}_5 = 0           \\
		u^d{}_6 = 0           \\
		u^d{}_7 = 0           \\
		u^d{}_8 = 0           \\
	\end{array}
	\quad,\quad
	\begin{array}{l}
		w^d{}_1 = 1 \\
		w^d{}_2 = 1 \\
	\end{array}
\end{equation}
The 4-point correction to the down-type quarks' masses is given by,
\begin{equation}\label{M4d_16}
	|M_{4d}|=m_b \left(
	\begin{array}{ccc}
		0.001193 & 0.        & -0.0178   \\
		0.       & -0.0178   & -0.000215 \\
		-0.0178  & -0.000215 & 0.        \\
	\end{array}
	\right)
\end{equation}
which can be added to the matrix obtained from 3-point functions \eqref{eq:Down3_16} as,
\begin{equation}\label{M34d_16}
	|M_{3d}|+|M_{4d}|=m_b \left(
	\begin{array}{ccc}
		0.00141   & -0.000252 & -0.017697 \\
		0.002252  & 0.028     & -0.000215 \\
		-0.017812 & -0.000214 & 1.        \\
	\end{array}
	\right) \sim D_d ~,
\end{equation} 
However, we also need to keep the corrections to charged-leptons' masses to be negligible by setting $\epsilon^{(1)}_{4e}=0$,
\begin{equation}\label{M4e_16}
	|M_{4e}|\sim \left(
	\begin{array}{ccc}
		0.000755  & 0.        & -0.011266 \\
		0.        & -0.011266 & -0.000136 \\
		-0.011266 & -0.000136 & 0.        \\
	\end{array}
	\right)\sim 0.
\end{equation} 
Although the results appears to be an near-exact it should be noted that we have assumed a strictly symmetric CKM matrix \eqref{eq:VCKM} and therefore, the matching is only approximate for a general asymmetric matrix.
\subsection{Model 16-dual}\label{sec:model-16.5}
In Model~\hyperref[model16.5]{16-dual} the three-point Yukawa couplings arise from the triplet intersections from the branes ${a, b, c}$ on the first two-torus ($r=1$) with 6 pairs of Higgs from $\mathcal{N}=2$ subsector.

\begin{figure}[t]
	\centering
	\includegraphics[width=\textwidth]{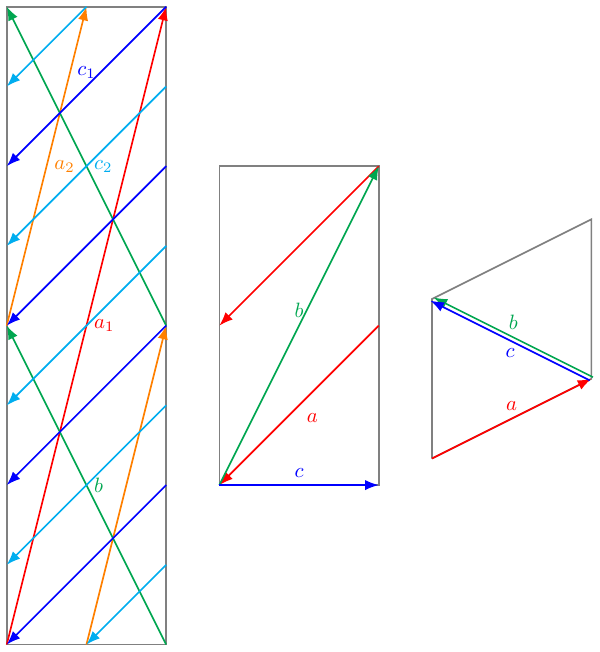}
	\caption{Brane configuration for the three two-tori in Model~\hyperref[model16.5]{16-dual} where the third two-torus is tilted. Fermion mass hierarchies result from the intersections on the first two-torus.}  \label{Fig.16.5}
\end{figure}

Yukawa matrices for the Model~\hyperref[model16.5]{16-dual} are of rank 3 and the three intersections required to form the disk diagrams for the Yukawa couplings all occur on the first torus as shown in figure~\ref{Fig.16.5}. The other two-tori only contribute an overall constant that has no effect in computing the fermion mass ratios. Thus, it is sufficient for our purpose to only focus on the first torus to explain the masses and the mixing in the standard model fermions.

\subsubsection{3-point Yukawa mass-matrices for Model~\hyperref[model16.5]{16-dual}}
From the wrapping numbers listed in table~\ref{model16.5}, the relevant intersection numbers are calculated as,
\begin{align}\label{eq:Intersections.16.5} 
	I_{ab}^{(1)} & =3,  &   
	I_{ab}^{(2)} & =-1, &   
	I_{ab}^{(3)}&=1, \nonumber\\
	I_{bc}^{(1)} & =6,  &   
	I_{bc}^{(2)} & =-1, &   
	I_{bc}^{(3)}&=0, \nonumber\\
	I_{ca}^{(1)} & =-3, &   
	I_{ca}^{(2)} & =-1, &   
	I_{ca}^{(3)}&=-1, 
\end{align}
As the intersection numbers are not coprime, we define the greatest common divisor,
$d^{(1)}=g.c.d.(I_{ab}^{(1)},I_{bc}^{(1)},I_{ca}^{(1)})=3$. Thus, the arguments of the modular theta function as defined in \eqref{eqn:Yinput} can be written as,
\begin{align}
	\delta^{(1)} & = \frac{i^{(1)}}{3} + \frac{j^{(1)}}{-3} + \frac{k^{(1)}}{6} +\frac{1}{18} \left(-6 \epsilon _a^{\text{(1)}}+3 \epsilon _b^{\text{(1)}}-3 \epsilon _c^{\text{(1)}}\right)+ \frac{s^{(1)}}{3},  \label{eq:delta.16.5} \\
	\phi^{(1)}   & = \frac{1}{3} \left(6 \theta _a^{\text{(1)}}-3 \theta _b^{\text{(1)}}+3 \theta _c^{\text{(1)}}\right) =0,                                                                                                            \\
	\kappa^{(1)} & = \frac{6 J^{\text{(1)}}}{\alpha '}, \label{eq:kappa.16.5}                                                                                                                                                           
\end{align}
and recalling \eqref{eq:ijk}, we have $i=\{0,\dots ,2\}$, $j=\{0,\dots ,2\}$ and $k=\{0,\dots ,5\}$ which respectively index the left-handed fermions, the right-handed fermions and the Higgs fields. Clearly, there arise 6 Higgs fields from the $bc$ sector.

The second-last term in the right side of \eqref{eq:delta.16.5} can be used to redefine the shift on the torus as
\begin{align}
	\epsilon^{(1)} & \equiv \frac{1}{18} \left(-6 \epsilon _a^{\text{(1)}}+3 \epsilon _b^{\text{(1)}}-3 \epsilon _c^{\text{(1)}}\right). \label{shifts16.5} 
\end{align}  
The selection rule for the occurrence of a trilinear Yukawa coupling for a given set of indices is given as,
\begin{equation}\label{selection-rule.16.5}
	i^{(1)} + j^{(1)} + k^{(1)} = 0\; \mathrm{ mod }\; 3.
\end{equation}
Then the suitable rank-3 mass-matrix can be determined by choosing the specific value of $s^{(1)}=j$. 
\begin{align}
	\delta^{(1)} & =  \frac{i}{3}+\frac{k}{6} \label{eq:s1.16.5} 
\end{align}
Here, we will ignore other equivalent cases of solutions or cases with rank-1 problem. The mass matrices for up quarks, down quarks and charged leptons have the following general form:
\begin{align}
	Z_{3} &= Z_{3q} \left(
	\begin{array}{ccc}
	T_0 v_1+T_3 v_4 & T_2 v_3+T_5 v_6 & T_1 v_2+T_4 v_5 \\
	T_4 v_3+T_1 v_6 & T_3 v_2+T_0 v_5 & T_2 v_1+T_5 v_4 \\
	T_5 v_2+T_2 v_5 & T_4 v_1+T_1 v_4 & T_0 v_3+T_3 v_6 \\
	\end{array}
	\right),\label{eq:s-i16.5}
\end{align} 
where $v_i = \left\langle H_{i} \right\rangle$ and the three-point coupling functions are given in terms of Jacobi theta function of the third kind as, 

\begin{align}\label{eq:3couplings16.5}
	T_k & \equiv  \vartheta \left[\begin{array}{c} 
	\epsilon^{(1)}+\frac{k}{6}\\  \phi^{(1)} \end{array} \right]
	(\frac{6 J^{\text{(1)}}}{\alpha '}),  \quad   k={0,\cdots,5}.
\end{align}
  
\subsubsection{Fermion masses and mixings from 3-point functions in Model~\hyperref[model16.5]{16-dual}}
In order to accommodate the fermion masses and the quark mixings, we need to fit the up-type quarks mixing matrix \eqref{eq:mixing-quarks}, the down-type quarks matrix \eqref{eq:mass-downquarks} and the masses of the charged-leptons \eqref{eq:mass-chargedleptons}.
 
We set the K\"{a}hler modulus on the first two-torus defined in \eqref{eq:kappa.16.5} as $\kappa^{(1)} = 30 $  and evaluate the couplings functions \eqref{eq:3couplings16.5} by setting geometric brane position parameters as $\epsilon^{(1)}_{u} = 0$, $\epsilon^{(1)}_{d} = 0.0696666$ and $\epsilon^{(1)}_{e} = 0$ which yields a nearest fit for the following VEVs,
\begin{equation}\label{eq:VEVs_16.5} 
	\begin{array}{ll}
		v^u{}_1 = 0.000291336 , & v^d{}_1 = 6.8572\times 10^{-6}   \\
		v^u{}_2 = 0.117993    , & v^d{}_2 = -5.61839\times 10^{-7} \\
		v^u{}_3 = 0.998234    , & v^d{}_3 = 0.0316                 \\
		v^u{}_4 = 0.566674    , & v^d{}_4 = -2.66199\times 10^{-9} \\
		v^u{}_5 = 0.00552729  , & v^d{}_5 = 0.00144728             \\
		v^u{}_6 = 0.0163423   , & v^d{}_6 = -0.0000122672          \\
	\end{array}
\end{equation}
\begin{align} 
	|M_{3u}|&=m_t \left(
	\begin{array}{ccc}
	0.000291  & 0.00122  & 0.008608  \\
	0.00122   & 0.005527 & 0.041338  \\
	0.008608  & 0.041338 & 0.998234  \\
	\end{array}
	\right)\sim M_u ~, \label{eq:Up3_16.5} \\
	|M_{3e}|&=m_\tau \left(
	\begin{array}{ccc}
	0.000217  & 0.       & 0.        \\
	0.        & 0.0458   & 0.        \\
	0.        & 0.       & 1.        \\
	\end{array}
	\right)\sim D_e~,\label{eq:Leptons3_16.5} \\ 
	|M_{3d}|&=m_b \left(
	\begin{array}{ccc}
	0.000217  & 0.002252 & -0.000012 \\
	-0.000252 & 0.0458   & 0.        \\
	0.000103  & 0.       & 1.        \\
	\end{array}
	\right)~.\label{eq:Down3_16.5}
\end{align}
While the masses of up-type quarks and the charged leptons are fitted, in the down-type quarks matrix, only the mass of the bottom quark can be fitted with three-point couplings only. The masses of the charm and the down quarks are not fitted. Notice, that these results are only at the tree-level and there could indeed be other corrections, such as those coming from higher-dimensional operators, which may contribute most greatly to the charm and the down quarks' masses since they are lighter.

\subsubsection{4-point corrections in Model~\hyperref[model16.5]{16-dual}}
 
The four-point couplings in Model~\hyperref[model16.5]{16-dual} in table~\ref{model16.5} can come from considering interactions of ${a, b, c}$ with $b'$ or $c'$ on the first two-torus as can be seen from the following intersection numbers,
\begin{align}
	I_{bb'}^{(1)} & = -4, & I_{bb'}^{(2)} & = 2,  & I_{bb'}^{(3)} & = -1 ,\nonumber                    \\
	I_{cc'}^{(1)} & = 8,  & I_{cc'}^{(2)} & = 0,  & I_{cc'}^{(3)} & = -1 ,\nonumber                    \\
	I_{bc'}^{(1)} & = 2,  & I_{bc'}^{(2)} & = -1, & I_{bc'}^{(3)} & = 1 . \label{eq:4Intersection16.5} 
\end{align}
There are 8 SM singlet fields $S_L^i$ and 2
Higgs-like state $H^{\prime}_{u, d}$. 

Let us consider four-point interactions with $c'$ with the following parameters with shifts $l=-\frac{k}{2} $ and $\ell=-\frac{k}{3} $ taken along the index $k$,
\begin{align}\label{deltas16.5}
	\delta & =\frac{i}{I_{ab}^{(1)}} +\frac{j}{I_{ca}^{(1)}} +\frac{k}{I_{bc}^{(1)}} +l , \nonumber               \\
	       & = \frac{i}{3}-\frac{j}{3} ,                                                                          \\
	d      & =\frac{\imath}{I_{c c'}^{(1)}} +\frac{\jmath}{I_{bc'}^{(1)}}+\frac{k}{I_{bc}^{(1)}} +\ell ,\nonumber \\
	       & = \frac{\imath}{8}+\frac{\jmath}{2} ,                                                                
\end{align}
the matrix elements $a_{i,j,\imath}$ on the first torus from the four-point functions results in the following classical 4-point contribution to the mass matrix,
\begin{equation}\label{eq:4point16.5} 
	Z_{4cl} =  \left(
	\begin{array}{ccc}
		F_0 u_1 w_1+F_6 u_7 w_1+F_1 u_6 w_2 & F_5 u_6 w_1+F_0 u_5 w_2 & F_4 u_5 w_1+F_7 u_4 w_2             \\
		F_5 u_6 w_1+F_0 u_5 w_2             & F_4 u_5 w_1+F_7 u_4 w_2 & F_3 u_4 w_1+F_6 u_3 w_2             \\
		F_4 u_5 w_1+F_7 u_4 w_2             & F_3 u_4 w_1+F_6 u_3 w_2 & F_2 u_3 w_1+F_5 u_2 w_2+F_3 u_8 w_2 \\
	\end{array}
	\right)
\end{equation} 
where $u_i, w_j$ are the VEVs and the couplings are defined as,
\begin{align}\label{eq:4couplings16.5}
	F_{i} & \equiv  \vartheta \left[\begin{array}{c} 
	\epsilon^{(1)}+\frac{i}{8}\\  \phi^{(1)} \end{array} \right]
	(\frac{6 J^{\text{(1)}}}{\alpha '}),\qquad i={0,\dots,|I_{c c'}^{(1)}|-1}.
\end{align}

Since, we have already fitted the up-type quark matrix $|M_{3u}|$ exactly, so its 4-point correction should be zero,
\begin{equation}
	|M_{4u}| = 0 ,
\end{equation}
which is true by setting all up-type VEVs $u_u^i$ and $w_u^i$ to be zero. 
Therefore, we are essentially concerned with fitting down-type quarks in such a way that corresponding corrections for the charged-leptons remain negligible. The desired solution can be readily obtained by setting $\epsilon^{(1)}_{4d}=0$ with the following values of the VEVs,
\begin{equation}\label{eq:UdWd16.5} 
	\begin{array}{l}
		u^d{}_1 = 0.00002386  \\
		u^d{}_2 = 0           \\
		u^d{}_3 = 0           \\
		u^d{}_4 = -0.00155239 \\
		u^d{}_5 = 0           \\
		u^d{}_6 = 0           \\
		u^d{}_7 = 0           \\
		u^d{}_8 = 0           \\
	\end{array}
	\quad,\quad
	\begin{array}{l}
		w^d{}_1 = 1 \\
		w^d{}_2 = 1 \\
	\end{array}
\end{equation}
The 4-point correction to the down-type quarks' masses is given by,
\begin{equation}\label{M4d_16.5}
	|M_{4d}|=m_b \left(
	\begin{array}{ccc}
		0.001193 & 0.      & -0.0178 \\
		0.       & -0.0178 & 0.      \\
		-0.0178  & 0.      & 0.      \\
	\end{array}
	\right)
\end{equation}
which can be added to the matrix obtained from 3-point functions \eqref{eq:Down3_16.5} as,
\begin{equation}\label{M34d_16.5}
	|M_{3d}|+|M_{4d}|=m_b \left(
	\begin{array}{ccc}
		0.00141   & 0.002252 & -0.017812 \\
		-0.000252 & 0.028    & 0.        \\
		-0.017697 & 0.       & 1.        \\
	\end{array}
	\right) \sim D_d ~,
\end{equation} 
However, we also need to keep the corrections to charged-leptons' masses to be negligible by setting $\epsilon^{(1)}_{4e}=0$,
\begin{equation}\label{M4e_16.5}
	|M_{4e}|\sim \left(
	\begin{array}{ccc}
		0.000755  & 0.        & -0.011266 \\
		0.        & -0.011266 & 0.        \\
		-0.011266 & 0.        & 0.        \\
	\end{array}
	\right)\sim 0.
\end{equation} 
Although the results appears to be an near-exact it should be noted that we have assumed a strictly symmetric CKM matrix \eqref{eq:VCKM} and therefore, the matching is only approximate for a general asymmetric matrix.
\section{Models with 9 Higgs from $\mathcal{N}=2$ sector}\label{sec:9Higgs}
\subsection{Model 17}\label{sec:model-17}
In Model~\hyperref[model17]{17} the three-point Yukawa couplings arise from the triplet intersections from the branes ${a, b, c}$ on the second two-torus ($r=2$) with 9 pairs of Higgs from $\mathcal{N}=2$ subsector.

\begin{figure}[t]
	\centering
	\includegraphics[width=\textwidth]{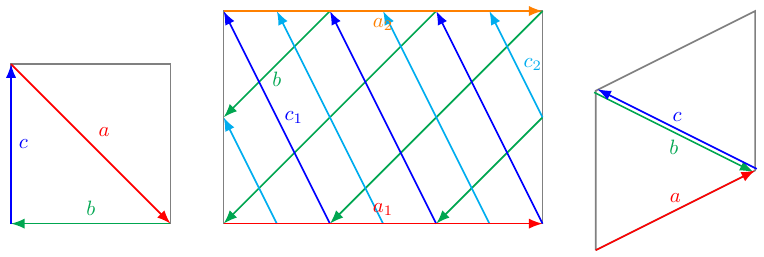}
	\caption{Brane configuration for the three two-tori in Model~\hyperref[model17]{17} where the third two-torus is tilted. Fermion mass hierarchies result from the intersections on the second two-torus.}  \label{Fig.17}
\end{figure}

Yukawa matrices for the Model~\hyperref[model17]{17} are of rank 3 and the three intersections required to form the disk diagrams for the Yukawa couplings all occur on the second torus as shown in figure~\ref{Fig.17}. The other two-tori only contribute an overall constant that has no effect in computing the fermion mass ratios. Thus, it is sufficient for our purpose to only focus on the second torus to explain the masses and the mixing in the standard model fermions.

\subsubsection{3-point Yukawa mass-matrices for Model~\hyperref[model17]{17}}
From the wrapping numbers listed in table~\ref{model17}, the relevant intersection numbers are calculated as,
\begin{align}\label{eq:Intersections.17} 
	I_{ab}^{(1)} & =-1, &   
	I_{ab}^{(2)} & =-3, &   
	I_{ab}^{(3)}&=-1, \nonumber\\
	I_{bc}^{(1)} & =-1, &   
	I_{bc}^{(2)} & =-9, &   
	I_{bc}^{(3)}&=0, \nonumber\\
	I_{ca}^{(1)} & =-1, &   
	I_{ca}^{(2)} & =-3, &   
	I_{ca}^{(3)}&=-1, 
\end{align}
As the intersection numbers are not coprime, we define the greatest common divisor,
$d^{(1)}=g.c.d.(I_{ab}^{(2)},I_{bc}^{(2)},I_{ca}^{(2)})=3$. Thus, the arguments of the modular theta function as defined in \eqref{eqn:Yinput} can be written as,
\begin{align}
	\delta^{(2)} & = \frac{i^{(2)}}{-3} + \frac{j^{(2)}}{-3} + \frac{k^{(2)}}{-9} +\frac{1}{27} \left(9 \epsilon _a^{\text{(2)}}+3 \epsilon _b^{\text{(2)}}+3 \epsilon _c^{\text{(2)}}\right)+ \frac{s^{(2)}}{3},  \label{eq:delta.17} \\
	\phi^{(2)}   & = \frac{1}{3} \left(-9 \theta _a^{\text{(2)}}-3 \theta _b^{\text{(2)}}-3 \theta _c^{\text{(2)}}\right) =0,                                                                                                          \\
	\kappa^{(2)} & = \frac{9 J^{\text{(2)}}}{\alpha '}, \label{eq:kappa.17}                                                                                                                                                            
\end{align}
and recalling \eqref{eq:ijk}, we have $i=\{0,\dots ,2\}$, $j=\{0,\dots ,2\}$ and $k=\{0,\dots ,8\}$ which respectively index the left-handed fermions, the right-handed fermions and the Higgs fields. Clearly, there arise 9 Higgs fields from the $bc$ sector.

The second-last term in the right side of \eqref{eq:delta.17} can be used to redefine the shift on the torus as
\begin{align}
	\epsilon^{(2)} & \equiv \frac{1}{27} \left(9 \epsilon _a^{\text{(2)}}+3 \epsilon _b^{\text{(2)}}+3 \epsilon _c^{\text{(2)}}\right). \label{shifts17} 
\end{align}  
The selection rule for the occurrence of a trilinear Yukawa coupling for a given set of indices is given as,
\begin{equation}\label{selection-rule.17}
	i^{(2)} + j^{(2)} + k^{(2)} = 0\; \mathrm{ mod }\; 3.
\end{equation}
Then the suitable rank-3 mass-matrix can be determined by choosing the specific value of $s^{(2)}=i$. 
\begin{align}
	\delta^{(2)} & =  -\frac{j}{3}-\frac{k}{9} \label{eq:s1.17} 
\end{align}
Here, we will ignore other equivalent cases of solutions or cases with rank-1 problem. The mass matrices for up quarks, down quarks and charged leptons have the following general form:
\begin{align}
	Z_{3} &= Z_{3q} \left(
	\begin{array}{ccc}
	T_0 v_1+T_6 v_4+T_3 v_7 & T_7 v_3+T_4 v_6+T_1 v_9 & T_8 v_2+T_5 v_5+T_2 v_8 \\
	T_4 v_3+T_1 v_6+T_7 v_9 & T_5 v_2+T_2 v_5+T_8 v_8 & T_6 v_1+T_3 v_4+T_0 v_7 \\
	T_2 v_2+T_8 v_5+T_5 v_8 & T_3 v_1+T_0 v_4+T_6 v_7 & T_1 v_3+T_7 v_6+T_4 v_9 \\
	\end{array}
	\right),\label{eq:s-i17}
\end{align} 
where $v_i = \left\langle H_{i} \right\rangle$ and the three-point coupling functions are given in terms of Jacobi theta function of the third kind as, 

\begin{align}\label{eq:3couplings17}
	T_k & \equiv  \vartheta \left[\begin{array}{c} 
	\epsilon^{(2)}+\frac{k}{9}\\  \phi^{(2)} \end{array} \right]
	(\frac{9 J^{\text{(2)}}}{\alpha '}),  \quad   k={0,\cdots,8}.
\end{align}
  
\subsubsection{Fermion masses and mixings from 3-point functions in Model~\hyperref[model17]{17}}
In order to accommodate the fermion masses and the quark mixings, we need to fit the up-type quarks mixing matrix \eqref{eq:mixing-quarks}, the down-type quarks matrix \eqref{eq:mass-downquarks} and the masses of the charged-leptons \eqref{eq:mass-chargedleptons}.
 
We set the K\"{a}hler modulus on the second two-torus defined in \eqref{eq:kappa.17} as $\kappa^{(2)} = 45 $  and evaluate the couplings functions \eqref{eq:3couplings17} by setting geometric brane position parameters as $\epsilon^{(2)}_{u} = 0$, $\epsilon^{(2)}_{d} = 0$ and $\epsilon^{(2)}_{e} = -0.0156645$ which yields a nearest fit for the following VEVs,
\begin{equation}\label{eq:VEVs_17} 
	\begin{array}{ll}
		v^u{}_1 = 0.000291324 , & v^d{}_1 = 0.0000282               \\
		v^u{}_2 = 0.0491357   , & v^d{}_2 = -0.0000170692           \\
		v^u{}_3 = 5.71763     , & v^d{}_3 = 0.114556                \\
		v^u{}_4 = 0.0413384   , & v^d{}_4 = -4.24979\times 10^{-12} \\
		v^u{}_5 = 0.0490413   , & v^d{}_5 = 9.08351\times 10^{-8}   \\
		v^u{}_6 = 0.00711503  , & v^d{}_6 = 3.24411\times 10^{-6}   \\
		v^u{}_7 = 0.0413384   , & v^d{}_7 = -4.24979\times 10^{-12} \\
		v^u{}_8 = 0.0313982   , & v^d{}_8 = 0.00320756              \\
		v^u{}_9 = -0.0234364  , & v^d{}_9 = -0.000609616            \\
	\end{array}
\end{equation}
\begin{align} 
	|M_{3u}|&=m_t \left(
	\begin{array}{ccc}
	0.000291  & 0.00122  & 0.008608          \\
	0.00122   & 0.005527 & 0.041338          \\
	0.008608  & 0.041338 & 0.998234          \\
	\end{array}
	\right)\sim M_u ~, \label{eq:Up3_17} \\
	|M_{3d}|&=m_b \left(
	\begin{array}{ccc}
	0.00141   & 0.       & 0.                \\
	0.        & 0.028    & 0.                \\
	0.        & 0.       & 1.                \\
	\end{array}
	\right)\sim D_d ~,\label{eq:Down3_17} \\ 
	|M_{3e}|&=m_\tau \left(
	\begin{array}{ccc}
	0.000862  & 0.000022 & -1.\times 10^{-6} \\
	-0.004106 & 0.010464 & 0.                \\
	0.000188  & 0.       & 1.                \\
	\end{array}
	\right)~.\label{eq:Leptons3_17}
\end{align}While the masses of up-type quarks and the down-type quarks are fitted, in the charged-lepton matrix, only the mass of the tau can be fitted with three-point couplings only. The masses of the muon and the electron are not fitted. Notice, that these results are only at the tree-level and there could indeed be other corrections, such as those coming from higher-dimensional operators, which may contribute most greatly to the muon and the electron masses since they are lighter.

\subsubsection{4-point corrections in Model~\hyperref[model17]{17}}
 
The four-point couplings in Model~\hyperref[model17]{17} in table~\ref{model17} can come from considering interactions of ${a, b, c}$ with $b'$ or $c'$ on the second two-torus as can be seen from the following intersection numbers,
\begin{align}
	I_{bb'}^{(1)} & = 0, & I_{bb'}^{(2)} & = 12, & I_{bb'}^{(3)} & = -1 ,\nonumber                   \\
	I_{cc'}^{(1)} & = 0, & I_{cc'}^{(2)} & = -6, & I_{cc'}^{(3)} & = -1 ,\nonumber                   \\
	I_{bc'}^{(1)} & = 1, & I_{bc'}^{(2)} & = 3,  & I_{bc'}^{(3)} & = -1 . \label{eq:4Intersection17} 
\end{align}
There are 6 SM singlet fields $S_L^i$ and 3
Higgs-like state $H^{\prime}_{u, d}$. 

Let us consider four-point interactions with $c'$ with the following parameters with shifts $l=\frac{k}{3} $ and $\ell=\frac{k}{3} $ taken along the index $k$,
\begin{align}\label{deltas17}
	\delta & =\frac{i}{I_{ab}^{(2)}} +\frac{j}{I_{ca}^{(2)}} +\frac{k}{I_{bc}^{(2)}} +l , \nonumber               \\
	       & = -\frac{i}{3}-\frac{j}{3} ,                                                                         \\
	d      & =\frac{\imath}{I_{c c'}^{(2)}} +\frac{\jmath}{I_{bc'}^{(2)}}+\frac{k}{I_{bc}^{(2)}} +\ell ,\nonumber \\
	       & = \frac{\jmath}{3}-\frac{\imath}{6} ,                                                                
\end{align}
the matrix elements $a_{i,j,\imath}$ on the second torus from the four-point functions results in the following classical 4-point contribution to the mass matrix,
\begin{equation}\label{eq:4point17} 
	Z_{4cl} =  
	\resizebox{.95\textwidth}{!}{
		$\begin{aligned}
			  & \left(                 
			\begin{array}{c}
			F_0 u_1 w_1+F_3 u_4 w_1+F_0 u_3 w_2+F_3 u_6 w_2+F_3 u_2 w_3+F_0 u_5 w_3\\
			F_4 u_3 w_1+F_1 u_6 w_1+F_1 u_2 w_2+F_4 u_5 w_2+F_4 u_1 w_3+F_1 u_4 w_3\\
			F_5 u_2 w_1+F_2 u_5 w_1+F_2 u_1 w_2+F_5 u_4 w_2+F_2 u_3 w_3+F_5 u_6 w_3\\
			\end{array}\right.\\
			  & \begin{array}{c}       
			F_4 u_3 w_1+F_1 u_6 w_1+F_1 u_2 w_2+F_4 u_5 w_2+F_4 u_1 w_3+F_1 u_4 w_3\\
			F_5 u_2 w_1+F_2 u_5 w_1+F_2 u_1 w_2+F_5 u_4 w_2+F_2 u_3 w_3+F_5 u_6 w_3\\
			F_0 u_1 w_1+F_3 u_4 w_1+F_0 u_3 w_2+F_3 u_6 w_2+F_3 u_2 w_3+F_0 u_5 w_3\\
			\end{array}\\
			  & \left.\begin{array}{c} 
			F_5 u_2 w_1+F_2 u_5 w_1+F_2 u_1 w_2+F_5 u_4 w_2+F_2 u_3 w_3+F_5 u_6 w_3\\
			F_0 u_1 w_1+F_3 u_4 w_1+F_0 u_3 w_2+F_3 u_6 w_2+F_3 u_2 w_3+F_0 u_5 w_3\\
			F_4 u_3 w_1+F_1 u_6 w_1+F_1 u_2 w_2+F_4 u_5 w_2+F_4 u_1 w_3+F_1 u_4 w_3\\
			\end{array}\right),
		\end{aligned}$
	} 
\end{equation} 
where $u_i, w_j$ are the VEVs and the couplings are defined as,
\begin{align}\label{eq:4couplings17}
	F_{i} & \equiv  \vartheta \left[\begin{array}{c} 
	\epsilon^{(2)}+\frac{i}{6}\\  \phi^{(2)} \end{array} \right]
	(\frac{9 J^{\text{(2)}}}{\alpha '}),\qquad i={0,\dots,|I_{c c'}^{(2)}|-1}.
\end{align}

Since, we have already fitted the up-type quark matrix $|M_{3u}|$ exactly, so its 4-point correction should be zero,
\begin{equation}
	|M_{4u}| = 0 ,
\end{equation}
which is true by setting all up-type VEVs $u_u^i$ and $w_u^i$ to be zero. 
Therefore, we are essentially concerned with fitting charged-leptons in such a way that corresponding corrections for the down-type quarks remain negligible.
The desired solution can be readily obtained by setting $\epsilon^{(2)}_{4e}=0$ with the following values of the VEVs,
\begin{equation}\label{eq:UdWd17} 
	\begin{array}{l}
		u^d{}_1 = 3704.69   \\
		u^d{}_2 = 0.0283361 \\
		u^d{}_3 = -3704.69  \\
		u^d{}_4 = 0         \\
		u^d{}_5 = 0         \\
		u^d{}_6 = 0         \\
	\end{array}
	\quad,\quad
	\begin{array}{l}
		w^d{}_1 = 1 \\
		w^d{}_2 = 1 \\
		w^d{}_3 = 0 \\
	\end{array}
\end{equation}
The 4-point correction to the charged-leptons' masses is given by,
\begin{equation}\label{M4e_17}
	|M_{4e}|=m_\tau \left(
	\begin{array}{ccc}
		-0.000645 & 0.        & 0.035336  \\
		0.        & 0.035336  & -0.000645 \\
		0.035336  & -0.000645 & 0.        \\
	\end{array}
	\right)
\end{equation}
which can be added to the matrix obtained from 3-point functions \eqref{eq:Leptons3_17} as,
\begin{equation}\label{M34e_17}
	|M_{3e}|+|M_{4e}|=m_\tau \left(
	\begin{array}{ccc}
		0.000217  & 0.000022  & 0.035335  \\
		-0.004106 & 0.0458    & -0.000645 \\
		0.035524  & -0.000645 & 1.        \\
	\end{array}
	\right) \sim D_e ~,
\end{equation} 
The corrections to down-type quarks' masses can be made negligible by setting $\epsilon^{(1)}_{4d}=0$,
\begin{equation}\label{M4d_17}
	|M_{4d}|\sim \left(
	\begin{array}{ccc}
		-0.001019 & 0.        & 0.05583   \\
		0.        & 0.05583   & -0.001019 \\
		0.05583   & -0.001019 & 0.        \\
	\end{array}
	\right)\sim 0.
\end{equation} Therefore, we can achieve near-exact matching of fermion masses and mixings.
\subsection{Model 17-dual}\label{sec:model-17.5}
In Model~\hyperref[model17.5]{17-dual} the three-point Yukawa couplings arise from the triplet intersections from the branes ${a, b, c}$ on the second two-torus ($r=2$) with 9 pairs of Higgs from $\mathcal{N}=2$ subsector.

\begin{figure}[t]
	\centering
	\includegraphics[width=\textwidth]{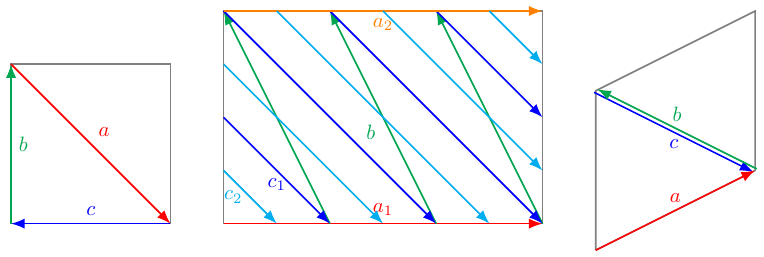}
	\caption{Brane configuration for the three two-tori in Model~\hyperref[model17.5]{17-dual} where the third two-torus is tilted. Fermion mass hierarchies result from the intersections on the second two-torus.}  \label{Fig.17.5}
\end{figure}

Yukawa matrices for the Model~\hyperref[model17.5]{17-dual} are of rank 3 and the three intersections required to form the disk diagrams for the Yukawa couplings all occur on the second torus as shown in figure~\ref{Fig.17.5}. The other two-tori only contribute an overall constant that has no effect in computing the fermion mass ratios. Thus, it is sufficient for our purpose to only focus on the second torus to explain the masses and the mixing in the standard model fermions.

\subsubsection{3-point Yukawa mass-matrices for Model~\hyperref[model17.5]{17-dual}}
From the wrapping numbers listed in table~\ref{model17.5}, the relevant intersection numbers are calculated as,
\begin{align}\label{eq:Intersections.17.5} 
	I_{ab}^{(1)} & =1, &   
	I_{ab}^{(2)} & =3, &   
	I_{ab}^{(3)}&=1, \nonumber\\
	I_{bc}^{(1)} & =1, &   
	I_{bc}^{(2)} & =9, &   
	I_{bc}^{(3)}&=0, \nonumber\\
	I_{ca}^{(1)} & =1, &   
	I_{ca}^{(2)} & =3, &   
	I_{ca}^{(3)}&=1, 
\end{align}
As the intersection numbers are not coprime, we define the greatest common divisor,
$d^{(1)}=g.c.d.(I_{ab}^{(2)},I_{bc}^{(2)},I_{ca}^{(2)})=3$. Thus, the arguments of the modular theta function as defined in \eqref{eqn:Yinput} can be written as,
\begin{align}
	\delta^{(2)} & = \frac{i^{(2)}}{3} + \frac{j^{(2)}}{3} + \frac{k^{(2)}}{9} +\frac{1}{27} \left(9 \epsilon _a^{\text{(2)}}+3 \epsilon _b^{\text{(2)}}+3 \epsilon _c^{\text{(2)}}\right)+ \frac{s^{(2)}}{3},  \label{eq:delta.17.5} \\
	\phi^{(2)}   & = \frac{1}{3} \left(9 \theta _a^{\text{(2)}}+3 \theta _b^{\text{(2)}}+3 \theta _c^{\text{(2)}}\right) =0,                                                                                                          \\
	\kappa^{(2)} & = \frac{9 J^{\text{(2)}}}{\alpha '}, \label{eq:kappa.17.5}                                                                                                                                                         
\end{align}
and recalling \eqref{eq:ijk}, we have $i=\{0,\dots ,2\}$, $j=\{0,\dots ,2\}$ and $k=\{0,\dots ,8\}$ which respectively index the left-handed fermions, the right-handed fermions and the Higgs fields. Clearly, there arise 9 Higgs fields from the $bc$ sector.

The second-last term in the right side of \eqref{eq:delta.17.5} can be used to redefine the shift on the torus as
\begin{align}
	\epsilon^{(2)} & \equiv \frac{1}{27} \left(9 \epsilon _a^{\text{(2)}}+3 \epsilon _b^{\text{(2)}}+3 \epsilon _c^{\text{(2)}}\right). \label{shifts17.5} 
\end{align}  
The selection rule for the occurrence of a trilinear Yukawa coupling for a given set of indices is given as,
\begin{equation}\label{selection-rule.17.5}
	i^{(2)} + j^{(2)} + k^{(2)} = 0\; \mathrm{ mod }\; 3.
\end{equation}
Then the suitable rank-3 mass-matrix can be determined by choosing the specific value of $s^{(2)}=-i$. 
\begin{align}
	\delta^{(2)} & =  \frac{j}{3}+\frac{k}{9} \label{eq:s1.17.5} 
\end{align}
Here, we will ignore other equivalent cases of solutions or cases with rank-1 problem. The mass matrices for up quarks, down quarks and charged leptons have the following general form:
\begin{align}
	Z_{3} &= Z_{3q} \left(
	\begin{array}{ccc}
	T_0 v_1+T_3 v_4+T_6 v_7 & T_2 v_3+T_5 v_6+T_8 v_9 & T_1 v_2+T_4 v_5+T_7 v_8 \\
	T_5 v_3+T_8 v_6+T_2 v_9 & T_4 v_2+T_7 v_5+T_1 v_8 & T_3 v_1+T_6 v_4+T_0 v_7 \\
	T_7 v_2+T_1 v_5+T_4 v_8 & T_6 v_1+T_0 v_4+T_3 v_7 & T_8 v_3+T_2 v_6+T_5 v_9 \\
	\end{array}
	\right),\label{eq:s-i17.5}
\end{align} 
where $v_i = \left\langle H_{i} \right\rangle$ and the three-point coupling functions are given in terms of Jacobi theta function of the third kind as, 

\begin{align}\label{eq:3couplings17.5}
	T_k & \equiv  \vartheta \left[\begin{array}{c} 
	\epsilon^{(2)}+\frac{k}{9}\\  \phi^{(2)} \end{array} \right]
	(\frac{9 J^{\text{(2)}}}{\alpha '}),  \quad   k={0,\cdots,8}.
\end{align}
  
\subsubsection{Fermion masses and mixings from 3-point functions in Model~\hyperref[model17.5]{17-dual}}
In order to accommodate the fermion masses and the quark mixings, we need to fit the up-type quarks mixing matrix \eqref{eq:mixing-quarks}, the down-type quarks matrix \eqref{eq:mass-downquarks} and the masses of the charged-leptons \eqref{eq:mass-chargedleptons}.
 
We set the K\"{a}hler modulus on the second two-torus defined in \eqref{eq:kappa.17.5} as $\kappa^{(2)} = 45 $  and evaluate the couplings functions \eqref{eq:3couplings17.5} by setting geometric brane position parameters as $\epsilon^{(2)}_{u} = 0$, $\epsilon^{(2)}_{d} = 0$ and $\epsilon^{(2)}_{e} = 0.0156645$ which yields a nearest fit for the following VEVs,
\begin{equation}\label{eq:VEVs_17.5} 
	\begin{array}{ll}
		v^u{}_1 = 0.000291324 , & v^d{}_1 = 0.0000282               \\
		v^u{}_2 = 0.0491357   , & v^d{}_2 = -0.0000170692           \\
		v^u{}_3 = 5.71763     , & v^d{}_3 = 0.114556                \\
		v^u{}_4 = 0.0413384   , & v^d{}_4 = -4.24979\times 10^{-12} \\
		v^u{}_5 = 0.0490413   , & v^d{}_5 = 9.08351\times 10^{-8}   \\
		v^u{}_6 = 0.00711503  , & v^d{}_6 = 3.24411\times 10^{-6}   \\
		v^u{}_7 = 0.0413384   , & v^d{}_7 = -4.24979\times 10^{-12} \\
		v^u{}_8 = 0.0313982   , & v^d{}_8 = 0.00320756              \\
		v^u{}_9 = -0.0234364  , & v^d{}_9 = -0.000609616            \\
	\end{array}
\end{equation}
\begin{align} 
	|M_{3u}|&=m_t \left(
	\begin{array}{ccc}
	0.000291  & 0.00122  & 0.008608          \\
	0.00122   & 0.005527 & 0.041338          \\
	0.008608  & 0.041338 & 0.998234          \\
	\end{array}
	\right)\sim M_u ~, \label{eq:Up3_17.5} \\
	|M_{3d}|&=m_b \left(
	\begin{array}{ccc}
	0.00141   & 0.       & 0.                \\
	0.        & 0.028    & 0.                \\
	0.        & 0.       & 1.                \\
	\end{array}
	\right)\sim D_d ~,\label{eq:Down3_17.5} \\ 
	|M_{3e}|&=m_\tau \left(
	\begin{array}{ccc}
	0.000862  & 0.000022 & -1.\times 10^{-6} \\
	-0.004106 & 0.010464 & 0.                \\
	0.000188  & 0.       & 1.                \\
	\end{array}
	\right)~.\label{eq:Leptons3_17.5}
\end{align}While the masses of up-type quarks and the down-type quarks are fitted, in the charged-lepton matrix, only the mass of the tau can be fitted with three-point couplings only. The masses of the muon and the electron are not fitted. Notice, that these results are only at the tree-level and there could indeed be other corrections, such as those coming from higher-dimensional operators, which may contribute most greatly to the muon and the electron masses since they are lighter.

\subsubsection{4-point corrections in Model~\hyperref[model17.5]{17-dual}}
 
The four-point couplings in Model~\hyperref[model17.5]{17-dual} in table~\ref{model17.5} can come from considering interactions of ${a, b, c}$ with $b'$ or $c'$ on the second two-torus as can be seen from the following intersection numbers,
\begin{align}
	I_{bb'}^{(1)} & = 0, & I_{bb'}^{(2)} & = -6, & I_{bb'}^{(3)} & = -1 ,\nonumber                     \\
	I_{cc'}^{(1)} & = 0, & I_{cc'}^{(2)} & = 12, & I_{cc'}^{(3)} & = -1 ,\nonumber                     \\
	I_{bc'}^{(1)} & = 1, & I_{bc'}^{(2)} & = 3,  & I_{bc'}^{(3)} & = -1 . \label{eq:4Intersection17.5} 
\end{align}
There are 6 SM singlet fields $S_L^i$ and 3
Higgs-like state $H^{\prime}_{u, d}$. 

Let us consider four-point interactions with $b'$ with the following parameters with shifts $l=-\frac{k}{3} $ and $\ell=-\frac{k}{3} $ taken along the index $k$,
\begin{align}\label{deltas17.5}
	\delta & =\frac{i}{I_{ab}^{(2)}} +\frac{j}{I_{ca}^{(2)}} +\frac{k}{I_{bc}^{(2)}} +l , \nonumber               \\
	       & = \frac{i}{3}+\frac{j}{3} ,                                                                          \\
	d      & =\frac{\imath}{I_{b b'}^{(2)}} +\frac{\jmath}{I_{bc'}^{(2)}}+\frac{k}{I_{bc}^{(2)}} +\ell ,\nonumber \\
	       & = \frac{\jmath}{3}-\frac{\imath}{6} ,                                                                
\end{align}
the matrix elements $a_{i,j,\imath}$ on the second torus from the four-point functions results in the following classical 4-point contribution to the mass matrix,
\begin{equation}\label{eq:4point17.5} 
	Z_{4cl} =  
	\resizebox{.95\textwidth}{!}{
		$\begin{aligned}
			  & \left(                 
			\begin{array}{c}
			F_0 u_1 w_1+F_3 u_4 w_1+F_0 u_3 w_2+F_3 u_6 w_2+F_3 u_2 w_3+F_0 u_5 w_3\\
			F_4 u_3 w_1+F_1 u_6 w_1+F_1 u_2 w_2+F_4 u_5 w_2+F_4 u_1 w_3+F_1 u_4 w_3\\
			F_5 u_2 w_1+F_2 u_5 w_1+F_2 u_1 w_2+F_5 u_4 w_2+F_2 u_3 w_3+F_5 u_6 w_3\\
			\end{array}\right.\\
			  & \begin{array}{c}       
			F_4 u_3 w_1+F_1 u_6 w_1+F_1 u_2 w_2+F_4 u_5 w_2+F_4 u_1 w_3+F_1 u_4 w_3\\
			F_5 u_2 w_1+F_2 u_5 w_1+F_2 u_1 w_2+F_5 u_4 w_2+F_2 u_3 w_3+F_5 u_6 w_3\\
			F_0 u_1 w_1+F_3 u_4 w_1+F_0 u_3 w_2+F_3 u_6 w_2+F_3 u_2 w_3+F_0 u_5 w_3\\
			\end{array}\\
			  & \left.\begin{array}{c} 
			F_5 u_2 w_1+F_2 u_5 w_1+F_2 u_1 w_2+F_5 u_4 w_2+F_2 u_3 w_3+F_5 u_6 w_3\\
			F_0 u_1 w_1+F_3 u_4 w_1+F_0 u_3 w_2+F_3 u_6 w_2+F_3 u_2 w_3+F_0 u_5 w_3\\
			F_4 u_3 w_1+F_1 u_6 w_1+F_1 u_2 w_2+F_4 u_5 w_2+F_4 u_1 w_3+F_1 u_4 w_3\\
			\end{array}\right),
		\end{aligned}$
	} 
\end{equation} 
where $u_i, w_j$ are the VEVs and the couplings are defined as,
\begin{align}\label{eq:4couplings17.5}
	F_{i} & \equiv  \vartheta \left[\begin{array}{c} 
	\epsilon^{(2)}+\frac{i}{6}\\  \phi^{(2)} \end{array} \right]
	(\frac{9 J^{\text{(2)}}}{\alpha '}),\qquad i={0,\dots,|I_{b b'}^{(2)}|-1}.
\end{align}

Since, we have already fitted the up-type quark matrix $|M_{3u}|$ exactly, so its 4-point correction should be zero,
\begin{equation}
	|M_{4u}| = 0 ,
\end{equation}
which is true by setting all up-type VEVs $u_u^i$ and $w_u^i$ to be zero. 
Therefore, we are essentially concerned with fitting charged-leptons in such a way that corresponding corrections for the down-type quarks remain negligible.
The desired solution can be readily obtained by setting $\epsilon^{(2)}_{4e}=0$ with the following values of the VEVs,
\begin{equation}\label{eq:UdWd17.5} 
	\begin{array}{l}
		u^d{}_1 = 3704.69   \\
		u^d{}_2 = 0.0283361 \\
		u^d{}_3 = -3704.69  \\
		u^d{}_4 = 0         \\
		u^d{}_5 = 0         \\
		u^d{}_6 = 0         \\
	\end{array}
	\quad,\quad
	\begin{array}{l}
		w^d{}_1 = 1 \\
		w^d{}_2 = 1 \\
		w^d{}_3 = 0 \\
	\end{array}
\end{equation}
The 4-point correction to the charged-leptons' masses is given by,
\begin{equation}\label{M4e_17.5}
	|M_{4e}|=m_\tau \left(
	\begin{array}{ccc}
		-0.000645 & 0.        & 0.035336  \\
		0.        & 0.035336  & -0.000645 \\
		0.035336  & -0.000645 & 0.        \\
	\end{array}
	\right)
\end{equation}
which can be added to the matrix obtained from 3-point functions \eqref{eq:Leptons3_17.5} as,
\begin{equation}\label{M34e_17.5}
	|M_{3e}|+|M_{4e}|=m_\tau \left(
	\begin{array}{ccc}
		0.000217  & 0.000022  & 0.035335  \\
		-0.004106 & 0.0458    & -0.000645 \\
		0.035524  & -0.000645 & 1.        \\
	\end{array}
	\right) \sim D_e ~,
\end{equation} 
The corrections to down-type quarks' masses can be made negligible by setting $\epsilon^{(1)}_{4d}=0$,
\begin{equation}\label{M4d_17.5}
	|M_{4d}|\sim \left(
	\begin{array}{ccc}
		-0.001019 & 0.        & 0.05583   \\
		0.        & 0.05583   & -0.001019 \\
		0.05583   & -0.001019 & 0.        \\
	\end{array}
	\right)\sim 0.
\end{equation} Therefore, we can achieve near-exact matching of fermion masses and mixings.
\subsection{Model 18}\label{sec:model-18}
In Model~\hyperref[model18]{18} the three-point Yukawa couplings arise from the triplet intersections from the branes ${a, b, c}$ on the first two-torus ($r=1$) with 9 pairs of Higgs from $\mathcal{N}=2$ subsector.

\begin{figure}[t]
	\centering
	\includegraphics[width=\textwidth]{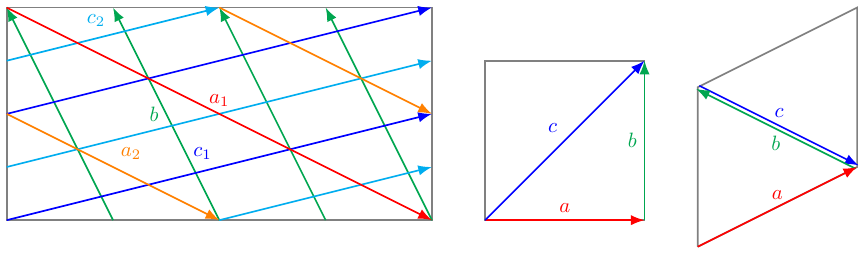}
	\caption{Brane configuration for the three two-tori in Model~\hyperref[model18]{18} where the third two-torus is tilted. Fermion mass hierarchies result from the intersections on the first two-torus.}  \label{Fig.18}
\end{figure}

Yukawa matrices for the Model~\hyperref[model18]{18} are of rank 3 and the three intersections required to form the disk diagrams for the Yukawa couplings all occur on the first torus as shown in figure~\ref{Fig.18}. The other two-tori only contribute an overall constant that has no effect in computing the fermion mass ratios. Thus, it is sufficient for our purpose to only focus on the first torus to explain the masses and the mixing in the standard model fermions.

\subsubsection{3-point Yukawa mass-matrices for Model~\hyperref[model18]{18}}
From the wrapping numbers listed in table~\ref{model18}, the relevant intersection numbers are calculated as,
\begin{align}\label{eq:Intersections.18} 
	I_{ab}^{(1)} & =3,  &   
	I_{ab}^{(2)} & =1,  &   
	I_{ab}^{(3)}&=1, \nonumber\\
	I_{bc}^{(1)} & =-9, &   
	I_{bc}^{(2)} & =-1, &   
	I_{bc}^{(3)}&=0, \nonumber\\
	I_{ca}^{(1)} & =-3, &   
	I_{ca}^{(2)} & =-1, &   
	I_{ca}^{(3)}&=1, 
\end{align}
As the intersection numbers are not coprime, we define the greatest common divisor,
$d^{(1)}=g.c.d.(I_{ab}^{(1)},I_{bc}^{(1)},I_{ca}^{(1)})=3$. Thus, the arguments of the modular theta function as defined in \eqref{eqn:Yinput} can be written as,
\begin{align}
	\delta^{(1)} & = \frac{i^{(1)}}{3} + \frac{j^{(1)}}{-3} + \frac{k^{(1)}}{-9} +\frac{1}{27} \left(-9 \epsilon _a^{\text{(1)}}-3 \epsilon _b^{\text{(1)}}+3 \epsilon _c^{\text{(1)}}\right)+ \frac{s^{(1)}}{3},  \label{eq:delta.18} \\
	\phi^{(1)}   & = \frac{1}{3} \left(-9 \theta _a^{\text{(1)}}-3 \theta _b^{\text{(1)}}+3 \theta _c^{\text{(1)}}\right) =0,                                                                                                          \\
	\kappa^{(1)} & = \frac{9 J^{\text{(1)}}}{\alpha '}, \label{eq:kappa.18}                                                                                                                                                            
\end{align}
and recalling \eqref{eq:ijk}, we have $i=\{0,\dots ,2\}$, $j=\{0,\dots ,2\}$ and $k=\{0,\dots ,8\}$ which respectively index the left-handed fermions, the right-handed fermions and the Higgs fields. Clearly, there arise 9 Higgs fields from the $bc$ sector.

The second-last term in the right side of \eqref{eq:delta.18} can be used to redefine the shift on the torus as
\begin{align}
	\epsilon^{(1)} & \equiv \frac{1}{27} \left(-9 \epsilon _a^{\text{(1)}}-3 \epsilon _b^{\text{(1)}}+3 \epsilon _c^{\text{(1)}}\right). \label{shifts18} 
\end{align}  
The selection rule for the occurrence of a trilinear Yukawa coupling for a given set of indices is given as,
\begin{equation}\label{selection-rule.18}
	i^{(1)} + j^{(1)} + k^{(1)} = 0\; \mathrm{ mod }\; 3.
\end{equation}
Then the suitable rank-3 mass-matrix can be determined by choosing the specific value of $s^{(1)}=j$. 
\begin{align}
	\delta^{(1)} & =  \frac{i}{3}-\frac{k}{9} \label{eq:s1.18} 
\end{align}
Here, we will ignore other equivalent cases of solutions or cases with rank-1 problem. The mass matrices for up quarks, down quarks and charged leptons have the following general form:
\begin{align}
	Z_{3} &= Z_{3q} \left(
	\begin{array}{ccc}
	T_0 v_1+T_6 v_4+T_3 v_7 & T_7 v_3+T_4 v_6+T_1 v_9 & T_8 v_2+T_5 v_5+T_2 v_8 \\
	T_1 v_3+T_7 v_6+T_4 v_9 & T_2 v_2+T_8 v_5+T_5 v_8 & T_3 v_1+T_0 v_4+T_6 v_7 \\
	T_5 v_2+T_2 v_5+T_8 v_8 & T_6 v_1+T_3 v_4+T_0 v_7 & T_4 v_3+T_1 v_6+T_7 v_9 \\
	\end{array}
	\right),\label{eq:s-i18}
\end{align} 
where $v_i = \left\langle H_{i} \right\rangle$ and the three-point coupling functions are given in terms of Jacobi theta function of the third kind as, 

\begin{align}\label{eq:3couplings18}
	T_k & \equiv  \vartheta \left[\begin{array}{c} 
	\epsilon^{(1)}+\frac{k}{9}\\  \phi^{(1)} \end{array} \right]
	(\frac{9 J^{\text{(1)}}}{\alpha '}),  \quad   k={0,\cdots,8}.
\end{align}
  
\subsubsection{Fermion masses and mixings from 3-point functions in Model~\hyperref[model18]{18}}
In order to accommodate the fermion masses and the quark mixings, we need to fit the up-type quarks mixing matrix \eqref{eq:mixing-quarks}, the down-type quarks matrix \eqref{eq:mass-downquarks} and the masses of the charged-leptons \eqref{eq:mass-chargedleptons}.
 
We set the K\"{a}hler modulus on the first two-torus defined in \eqref{eq:kappa.18} as $\kappa^{(1)} = 45 $  and evaluate the couplings functions \eqref{eq:3couplings18} by setting geometric brane position parameters as $\epsilon^{(1)}_{u} = 0$, $\epsilon^{(1)}_{d} = 0$ and $\epsilon^{(1)}_{e} = -0.0156645$ which yields a nearest fit for the following VEVs,
\begin{equation}\label{eq:VEVs_18} 
	\begin{array}{ll}
		v^u{}_1 = 0.000291324 , & v^d{}_1 = 0.0000282               \\
		v^u{}_2 = 0.0490413   , & v^d{}_2 = 9.08351\times 10^{-8}   \\
		v^u{}_3 = -0.0234364  , & v^d{}_3 = -0.000609616            \\
		v^u{}_4 = 0.0413384   , & v^d{}_4 = -4.24979\times 10^{-12} \\
		v^u{}_5 = 0.0313982   , & v^d{}_5 = 0.00320756              \\
		v^u{}_6 = 5.71763     , & v^d{}_6 = 0.114556                \\
		v^u{}_7 = 0.0413384   , & v^d{}_7 = -4.24979\times 10^{-12} \\
		v^u{}_8 = 0.0491357   , & v^d{}_8 = -0.0000170692           \\
		v^u{}_9 = 0.00711503  , & v^d{}_9 = 3.24411\times 10^{-6}   \\
	\end{array}
\end{equation}
\begin{align} 
	|M_{3u}|&=m_t \left(
	\begin{array}{ccc}
	0.000291          & 0.00122   & 0.008608 \\
	0.00122           & 0.005527  & 0.041338 \\
	0.008608          & 0.041338  & 0.998234 \\
	\end{array}
	\right)\sim M_u ~, \label{eq:Up3_18} \\
	|M_{3d}|&=m_b \left(
	\begin{array}{ccc}
	0.00141           & 0.        & 0.       \\
	0.                & 0.028     & 0.       \\
	0.                & 0.        & 1.       \\
	\end{array}
	\right)\sim D_d ~,\label{eq:Down3_18} \\ 
	|M_{3e}|&=m_\tau \left(
	\begin{array}{ccc}
	0.000862          & -0.004106 & 0.000188 \\
	0.000022          & 0.010464  & 0.       \\
	-1.\times 10^{-6} & 0.        & 1.       \\
	\end{array}
	\right)~.\label{eq:Leptons3_18}
\end{align}While the masses of up-type quarks and the down-type quarks are fitted, in the charged-lepton matrix, only the mass of the tau can be fitted with three-point couplings only. The masses of the muon and the electron are not fitted. Notice, that these results are only at the tree-level and there could indeed be other corrections, such as those coming from higher-dimensional operators, which may contribute most greatly to the muon and the electron masses since they are lighter.

\subsubsection{4-point corrections in Model~\hyperref[model18]{18}}
 
The four-point couplings in Model~\hyperref[model18]{18} in table~\ref{model18} can come from considering interactions of ${a, b, c}$ with $b'$ or $c'$ on the first two-torus as can be seen from the following intersection numbers,
\begin{align}
	I_{bb'}^{(1)} & = -8, & I_{bb'}^{(2)} & = 0,  & I_{bb'}^{(3)} & = -1 ,\nonumber                   \\
	I_{cc'}^{(1)} & = 4,  & I_{cc'}^{(2)} & = 2,  & I_{cc'}^{(3)} & = -1 ,\nonumber                   \\
	I_{bc'}^{(1)} & = -7, & I_{bc'}^{(2)} & = -1, & I_{bc'}^{(3)} & = -1 . \label{eq:4Intersection18} 
\end{align}
There are 8 SM singlet fields $S_L^i$ and 7
Higgs-like state $H^{\prime}_{u, d}$. 

Let us consider four-point interactions with $b'$ with the following parameters with shifts $l=\frac{k}{3} $ and $\ell=\frac{k}{9} $ taken along the index $k$,
\begin{align}\label{deltas18}
	\delta & =\frac{i}{I_{ab}^{(1)}} +\frac{j}{I_{ca}^{(1)}} +\frac{k}{I_{bc}^{(1)}} +l , \nonumber               \\
	       & = \frac{i}{3}-\frac{j}{3} ,                                                                          \\
	d      & =\frac{\imath}{I_{b b'}^{(1)}} +\frac{\jmath}{I_{bc'}^{(1)}}+\frac{k}{I_{bc}^{(1)}} +\ell ,\nonumber \\
	       & = -\frac{\imath}{8}-\frac{\jmath}{7} ,                                                               
\end{align}
the matrix elements $a_{i,j,\imath}$ on the first torus from the four-point functions results in the following classical 4-point contribution to the mass matrix,
\begin{equation}\label{eq:4point18} 
	Z_{4cl} =  
	\resizebox{.95\textwidth}{!}{
		$\begin{aligned}
			  & \left(                 
			\begin{array}{c}
			F_0 u_1 w_1+F_{35} u_4 w_1+F_{14} u_7 w_1+F_{34} u_3 w_2+F_{13} u_6 w_2+F_{33} u_2 w_3+F_{12} u_5 w_3+F_{47} u_8 w_3+F_{32} u_1 w_4+F_{11} u_4 w_4+F_{46} u_7 w_4+F_{10} u_3 w_5+F_{45} u_6 w_5+F_9 u_2 w_6+F_{44} u_5 w_6+F_{23} u_8 w_6+F_8 u_1 w_7+F_{43} u_4 w_7+F_{22} u_7 w_7\\
			F_{42} u_3 w_1+F_{21} u_6 w_1+F_{41} u_2 w_2+F_{20} u_5 w_2+F_{55} u_8 w_2+F_{40} u_1 w_3+F_{19} u_4 w_3+F_{54} u_7 w_3+F_{18} u_3 w_4+F_{53} u_6 w_4+F_{17} u_2 w_5+F_{52} u_5 w_5+F_{31} u_8 w_5+F_{16} u_1 w_6+F_{51} u_4 w_6+F_{30} u_7 w_6+F_{50} u_3 w_7+F_{29} u_6 w_7\\
			F_{49} u_2 w_1+F_{28} u_5 w_1+F_7 u_8 w_1+F_{48} u_1 w_2+F_{27} u_4 w_2+F_6 u_7 w_2+F_{26} u_3 w_3+F_5 u_6 w_3+F_{25} u_2 w_4+F_4 u_5 w_4+F_{39} u_8 w_4+F_{24} u_1 w_5+F_3 u_4 w_5+F_{38} u_7 w_5+F_2 u_3 w_6+F_{37} u_6 w_6+F_1 u_2 w_7+F_{36} u_5 w_7+F_{15} u_8 w_7\\
			\end{array}\right.\\
			  & \begin{array}{c}       
			F_{42} u_3 w_1+F_{21} u_6 w_1+F_{41} u_2 w_2+F_{20} u_5 w_2+F_{55} u_8 w_2+F_{40} u_1 w_3+F_{19} u_4 w_3+F_{54} u_7 w_3+F_{18} u_3 w_4+F_{53} u_6 w_4+F_{17} u_2 w_5+F_{52} u_5 w_5+F_{31} u_8 w_5+F_{16} u_1 w_6+F_{51} u_4 w_6+F_{30} u_7 w_6+F_{50} u_3 w_7+F_{29} u_6 w_7\\
			F_{49} u_2 w_1+F_{28} u_5 w_1+F_7 u_8 w_1+F_{48} u_1 w_2+F_{27} u_4 w_2+F_6 u_7 w_2+F_{26} u_3 w_3+F_5 u_6 w_3+F_{25} u_2 w_4+F_4 u_5 w_4+F_{39} u_8 w_4+F_{24} u_1 w_5+F_3 u_4 w_5+F_{38} u_7 w_5+F_2 u_3 w_6+F_{37} u_6 w_6+F_1 u_2 w_7+F_{36} u_5 w_7+F_{15} u_8 w_7\\
			F_0 u_1 w_1+F_{35} u_4 w_1+F_{14} u_7 w_1+F_{34} u_3 w_2+F_{13} u_6 w_2+F_{33} u_2 w_3+F_{12} u_5 w_3+F_{47} u_8 w_3+F_{32} u_1 w_4+F_{11} u_4 w_4+F_{46} u_7 w_4+F_{10} u_3 w_5+F_{45} u_6 w_5+F_9 u_2 w_6+F_{44} u_5 w_6+F_{23} u_8 w_6+F_8 u_1 w_7+F_{43} u_4 w_7+F_{22} u_7 w_7\\
			\end{array}\\
			  & \left.\begin{array}{c} 
			F_{49} u_2 w_1+F_{28} u_5 w_1+F_7 u_8 w_1+F_{48} u_1 w_2+F_{27} u_4 w_2+F_6 u_7 w_2+F_{26} u_3 w_3+F_5 u_6 w_3+F_{25} u_2 w_4+F_4 u_5 w_4+F_{39} u_8 w_4+F_{24} u_1 w_5+F_3 u_4 w_5+F_{38} u_7 w_5+F_2 u_3 w_6+F_{37} u_6 w_6+F_1 u_2 w_7+F_{36} u_5 w_7+F_{15} u_8 w_7\\
			F_0 u_1 w_1+F_{35} u_4 w_1+F_{14} u_7 w_1+F_{34} u_3 w_2+F_{13} u_6 w_2+F_{33} u_2 w_3+F_{12} u_5 w_3+F_{47} u_8 w_3+F_{32} u_1 w_4+F_{11} u_4 w_4+F_{46} u_7 w_4+F_{10} u_3 w_5+F_{45} u_6 w_5+F_9 u_2 w_6+F_{44} u_5 w_6+F_{23} u_8 w_6+F_8 u_1 w_7+F_{43} u_4 w_7+F_{22} u_7 w_7\\
			F_{42} u_3 w_1+F_{21} u_6 w_1+F_{41} u_2 w_2+F_{20} u_5 w_2+F_{55} u_8 w_2+F_{40} u_1 w_3+F_{19} u_4 w_3+F_{54} u_7 w_3+F_{18} u_3 w_4+F_{53} u_6 w_4+F_{17} u_2 w_5+F_{52} u_5 w_5+F_{31} u_8 w_5+F_{16} u_1 w_6+F_{51} u_4 w_6+F_{30} u_7 w_6+F_{50} u_3 w_7+F_{29} u_6 w_7\\
			\end{array}\right),
		\end{aligned}$
	} 
\end{equation} 
where $u_i, w_j$ are the VEVs and the couplings are defined as,
\begin{align}\label{eq:4couplings18}
	F_{7i} & \equiv  \vartheta \left[\begin{array}{c} 
	\epsilon^{(1)}+\frac{i}{8}\\  \phi^{(1)} \end{array} \right]
	(\frac{9 J^{\text{(1)}}}{\alpha '}),\qquad i={0,\dots,|I_{b b'}^{(1)}|-1}.
\end{align}

Since, we have already fitted the up-type quark matrix $|M_{3u}|$ exactly, so its 4-point correction should be zero,
\begin{equation}
	|M_{4u}| = 0 ,
\end{equation}
which is true by setting all up-type VEVs $u_u^i$ and $w_u^i$ to be zero. 
Therefore, we are essentially concerned with fitting charged-leptons in such a way that corresponding corrections for the down-type quarks remain negligible.
The desired solution can be readily obtained by setting $\epsilon^{(1)}_{4e}=0$ with the following values of the VEVs,
\begin{equation}\label{eq:UdWd18} 
	\begin{array}{l}
		u^d{}_1 = -0.000870512  \\
		u^d{}_2 = 0.00862204    \\
		u^d{}_3 = -0.0000102392 \\
		u^d{}_4 = 0             \\
		u^d{}_5 = 0             \\
		u^d{}_6 = 0             \\
		u^d{}_7 = 0             \\
		u^d{}_8 = 0             \\
	\end{array}
	\quad,\quad
	\begin{array}{l}
		w^d{}_1 = 1 \\
		w^d{}_2 = 1 \\
		w^d{}_3 = 0 \\
		w^d{}_4 = 1 \\
		w^d{}_5 = 1 \\
		w^d{}_6 = 1 \\
		w^d{}_7 = 1 \\
	\end{array}
\end{equation}
The 4-point correction to the charged-leptons' masses is given by,
\begin{equation}\label{M4e_18}
	|M_{4e}|=m_\tau \left(
	\begin{array}{ccc}
		-0.000645 & 0.        & 0.035336  \\
		0.        & 0.035336  & -0.000645 \\
		0.035336  & -0.000645 & 0.        \\
	\end{array}
	\right)
\end{equation}
which can be added to the matrix obtained from 3-point functions \eqref{eq:Leptons3_18} as,
\begin{equation}\label{M34e_18}
	|M_{3e}|+|M_{4e}|=m_\tau \left(
	\begin{array}{ccc}
		0.000217 & -0.004106 & 0.035524  \\
		0.000022 & 0.0458    & -0.000645 \\
		0.035335 & -0.000645 & 1.        \\
	\end{array}
	\right) \sim D_e ~,
\end{equation} 
The corrections to down-type quarks' masses can be made negligible by setting $\epsilon^{(1)}_{4d}=0$,
\begin{equation}\label{M4d_18}
	|M_{4d}|\sim \left(
	\begin{array}{ccc}
		-0.001019 & 0.        & 0.05583   \\
		0.        & 0.05583   & -0.001019 \\
		0.05583   & -0.001019 & 0.        \\
	\end{array}
	\right)\sim 0.
\end{equation} Therefore, we can achieve near-exact matching of fermion masses and mixings.
\subsection{Model 18-dual}\label{sec:model-18.5}
In Model~\hyperref[model18.5]{18-dual} the three-point Yukawa couplings arise from the triplet intersections from the branes ${a, b, c}$ on the first two-torus ($r=1$) with 9 pairs of Higgs from $\mathcal{N}=2$ subsector.

\begin{figure}[t]
	\centering
	\includegraphics[width=\textwidth]{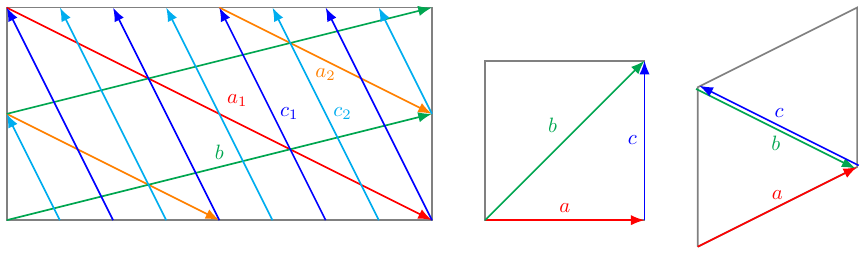}
	\caption{Brane configuration for the three two-tori in Model~\hyperref[model18.5]{18-dual} where the third two-torus is tilted. Fermion mass hierarchies result from the intersections on the first two-torus.}  \label{Fig.18.5}
\end{figure}

Yukawa matrices for the Model~\hyperref[model18.5]{18-dual} are of rank 3 and the three intersections required to form the disk diagrams for the Yukawa couplings all occur on the first torus as shown in figure~\ref{Fig.18.5}. The other two-tori only contribute an overall constant that has no effect in computing the fermion mass ratios. Thus, it is sufficient for our purpose to only focus on the first torus to explain the masses and the mixing in the standard model fermions.

\subsubsection{3-point Yukawa mass-matrices for Model~\hyperref[model18.5]{18-dual}}
From the wrapping numbers listed in table~\ref{model18.5}, the relevant intersection numbers are calculated as,
\begin{align}\label{eq:Intersections.18.5} 
	I_{ab}^{(1)} & =3,  &   
	I_{ab}^{(2)} & =1,  &   
	I_{ab}^{(3)}&=-1, \nonumber\\
	I_{bc}^{(1)} & =9,  &   
	I_{bc}^{(2)} & =1,  &   
	I_{bc}^{(3)}&=0, \nonumber\\
	I_{ca}^{(1)} & =-3, &   
	I_{ca}^{(2)} & =-1, &   
	I_{ca}^{(3)}&=-1, 
\end{align}
As the intersection numbers are not coprime, we define the greatest common divisor,
$d^{(1)}=g.c.d.(I_{ab}^{(1)},I_{bc}^{(1)},I_{ca}^{(1)})=3$. Thus, the arguments of the modular theta function as defined in \eqref{eqn:Yinput} can be written as,
\begin{align}
	\delta^{(1)} & = \frac{i^{(1)}}{3} + \frac{j^{(1)}}{-3} + \frac{k^{(1)}}{9} +\frac{1}{27} \left(-9 \epsilon _a^{\text{(1)}}+3 \epsilon _b^{\text{(1)}}-3 \epsilon _c^{\text{(1)}}\right)+ \frac{s^{(1)}}{3},  \label{eq:delta.18.5} \\
	\phi^{(1)}   & = \frac{1}{3} \left(9 \theta _a^{\text{(1)}}-3 \theta _b^{\text{(1)}}+3 \theta _c^{\text{(1)}}\right) =0,                                                                                                            \\
	\kappa^{(1)} & = \frac{9 J^{\text{(1)}}}{\alpha '}, \label{eq:kappa.18.5}                                                                                                                                                           
\end{align}
and recalling \eqref{eq:ijk}, we have $i=\{0,\dots ,2\}$, $j=\{0,\dots ,2\}$ and $k=\{0,\dots ,8\}$ which respectively index the left-handed fermions, the right-handed fermions and the Higgs fields. Clearly, there arise 9 Higgs fields from the $bc$ sector.

The second-last term in the right side of \eqref{eq:delta.18.5} can be used to redefine the shift on the torus as
\begin{align}
	\epsilon^{(1)} & \equiv \frac{1}{27} \left(-9 \epsilon _a^{\text{(1)}}+3 \epsilon _b^{\text{(1)}}-3 \epsilon _c^{\text{(1)}}\right). \label{shifts18.5} 
\end{align}  
The selection rule for the occurrence of a trilinear Yukawa coupling for a given set of indices is given as,
\begin{equation}\label{selection-rule.18.5}
	i^{(1)} + j^{(1)} + k^{(1)} = 0\; \mathrm{ mod }\; 3.
\end{equation}
Then the suitable rank-3 mass-matrix can be determined by choosing the specific value of $s^{(1)}=-i$. 
\begin{align}
	\delta^{(1)} & =  \frac{k}{9}-\frac{j}{3} \label{eq:s1.18.5} 
\end{align}
Here, we will ignore other equivalent cases of solutions or cases with rank-1 problem. The mass matrices for up quarks, down quarks and charged leptons have the following general form:
\begin{align}
	Z_{3} &= Z_{3q} \left(
	\begin{array}{ccc}
	T_0 v_1+T_3 v_4+T_6 v_7 & T_2 v_3+T_5 v_6+T_8 v_9 & T_1 v_2+T_4 v_5+T_7 v_8 \\
	T_8 v_3+T_2 v_6+T_5 v_9 & T_7 v_2+T_1 v_5+T_4 v_8 & T_6 v_1+T_0 v_4+T_3 v_7 \\
	T_4 v_2+T_7 v_5+T_1 v_8 & T_3 v_1+T_6 v_4+T_0 v_7 & T_5 v_3+T_8 v_6+T_2 v_9 \\
	\end{array}
	\right),\label{eq:s-i18.5}
\end{align} 
where $v_i = \left\langle H_{i} \right\rangle$ and the three-point coupling functions are given in terms of Jacobi theta function of the third kind as, 

\begin{align}\label{eq:3couplings18.5}
	T_k & \equiv  \vartheta \left[\begin{array}{c} 
	\epsilon^{(1)}+\frac{k}{9}\\  \phi^{(1)} \end{array} \right]
	(\frac{9 J^{\text{(1)}}}{\alpha '}),  \quad   k={0,\cdots,8}.
\end{align}
  
\subsubsection{Fermion masses and mixings from 3-point functions in Model~\hyperref[model18.5]{18-dual}}
In order to accommodate the fermion masses and the quark mixings, we need to fit the up-type quarks mixing matrix \eqref{eq:mixing-quarks}, the down-type quarks matrix \eqref{eq:mass-downquarks} and the masses of the charged-leptons \eqref{eq:mass-chargedleptons}.
 
We set the K\"{a}hler modulus on the first two-torus defined in \eqref{eq:kappa.18.5} as $\kappa^{(1)} = 45 $  and evaluate the couplings functions \eqref{eq:3couplings18.5} by setting geometric brane position parameters as $\epsilon^{(1)}_{u} = 0$, $\epsilon^{(1)}_{d} = 0$ and $\epsilon^{(1)}_{e} = 0.0156645$ which yields a nearest fit for the following VEVs,
\begin{equation}\label{eq:VEVs_18.5} 
	\begin{array}{ll}
		v^u{}_1 = 0.000291324 , & v^d{}_1 = 0.0000282               \\
		v^u{}_2 = 0.0490413   , & v^d{}_2 = 9.08351\times 10^{-8}   \\
		v^u{}_3 = -0.0234364  , & v^d{}_3 = -0.000609616            \\
		v^u{}_4 = 0.0413384   , & v^d{}_4 = -4.24979\times 10^{-12} \\
		v^u{}_5 = 0.0313982   , & v^d{}_5 = 0.00320756              \\
		v^u{}_6 = 5.71763     , & v^d{}_6 = 0.114556                \\
		v^u{}_7 = 0.0413384   , & v^d{}_7 = -4.24979\times 10^{-12} \\
		v^u{}_8 = 0.0491357   , & v^d{}_8 = -0.0000170692           \\
		v^u{}_9 = 0.00711503  , & v^d{}_9 = 3.24411\times 10^{-6}   \\
	\end{array}
\end{equation}
\begin{align} 
	|M_{3u}|&=m_t \left(
	\begin{array}{ccc}
	0.000291          & 0.00122   & 0.008608 \\
	0.00122           & 0.005527  & 0.041338 \\
	0.008608          & 0.041338  & 0.998234 \\
	\end{array}
	\right)\sim M_u ~, \label{eq:Up3_18.5} \\
	|M_{3d}|&=m_b \left(
	\begin{array}{ccc}
	0.00141           & 0.        & 0.       \\
	0.                & 0.028     & 0.       \\
	0.                & 0.        & 1.       \\
	\end{array}
	\right)\sim D_d ~,\label{eq:Down3_18.5} \\ 
	|M_{3e}|&=m_\tau \left(
	\begin{array}{ccc}
	0.000862          & -0.004106 & 0.000188 \\
	0.000022          & 0.010464  & 0.       \\
	-1.\times 10^{-6} & 0.        & 1.       \\
	\end{array}
	\right)~.\label{eq:Leptons3_18.5}
\end{align}While the masses of up-type quarks and the down-type quarks are fitted, in the charged-lepton matrix, only the mass of the tau can be fitted with three-point couplings only. The masses of the muon and the electron are not fitted. Notice, that these results are only at the tree-level and there could indeed be other corrections, such as those coming from higher-dimensional operators, which may contribute most greatly to the muon and the electron masses since they are lighter.

\subsubsection{4-point corrections in Model~\hyperref[model18.5]{18-dual}}
 
The four-point couplings in Model~\hyperref[model18.5]{18-dual} in table~\ref{model18.5} can come from considering interactions of ${a, b, c}$ with $b'$ or $c'$ on the first two-torus as can be seen from the following intersection numbers,
\begin{align}
	I_{bb'}^{(1)} & = 4,  & I_{bb'}^{(2)} & = 2,  & I_{bb'}^{(3)} & = -1 ,\nonumber                     \\
	I_{cc'}^{(1)} & = -8, & I_{cc'}^{(2)} & = 0,  & I_{cc'}^{(3)} & = -1 ,\nonumber                     \\
	I_{bc'}^{(1)} & = -7, & I_{bc'}^{(2)} & = -1, & I_{bc'}^{(3)} & = -1 . \label{eq:4Intersection18.5} 
\end{align}
There are 8 SM singlet fields $S_L^i$ and 7
Higgs-like state $H^{\prime}_{u, d}$. 

Let us consider four-point interactions with $c'$ with the following parameters with shifts $l=-\frac{k}{3} $ and $\ell=-\frac{k}{9} $ taken along the index $k$,
\begin{align}\label{deltas18.5}
	\delta & =\frac{i}{I_{ab}^{(1)}} +\frac{j}{I_{ca}^{(1)}} +\frac{k}{I_{bc}^{(1)}} +l , \nonumber               \\
	       & = \frac{i}{3}-\frac{j}{3} ,                                                                          \\
	d      & =\frac{\imath}{I_{c c'}^{(1)}} +\frac{\jmath}{I_{bc'}^{(1)}}+\frac{k}{I_{bc}^{(1)}} +\ell ,\nonumber \\
	       & = -\frac{\imath}{8}-\frac{\jmath}{7} ,                                                               
\end{align}
the matrix elements $a_{i,j,\imath}$ on the first torus from the four-point functions results in the following classical 4-point contribution to the mass matrix,
\begin{equation}\label{eq:4point18.5} 
	Z_{4cl} =  
	\resizebox{.95\textwidth}{!}{
		$\begin{aligned}
			  & \left(                 
			\begin{array}{c}
			F_0 u_1 w_1+F_{35} u_4 w_1+F_{14} u_7 w_1+F_{34} u_3 w_2+F_{13} u_6 w_2+F_{33} u_2 w_3+F_{12} u_5 w_3+F_{47} u_8 w_3+F_{32} u_1 w_4+F_{11} u_4 w_4+F_{46} u_7 w_4+F_{10} u_3 w_5+F_{45} u_6 w_5+F_9 u_2 w_6+F_{44} u_5 w_6+F_{23} u_8 w_6+F_8 u_1 w_7+F_{43} u_4 w_7+F_{22} u_7 w_7\\
			F_{42} u_3 w_1+F_{21} u_6 w_1+F_{41} u_2 w_2+F_{20} u_5 w_2+F_{55} u_8 w_2+F_{40} u_1 w_3+F_{19} u_4 w_3+F_{54} u_7 w_3+F_{18} u_3 w_4+F_{53} u_6 w_4+F_{17} u_2 w_5+F_{52} u_5 w_5+F_{31} u_8 w_5+F_{16} u_1 w_6+F_{51} u_4 w_6+F_{30} u_7 w_6+F_{50} u_3 w_7+F_{29} u_6 w_7\\
			F_{49} u_2 w_1+F_{28} u_5 w_1+F_7 u_8 w_1+F_{48} u_1 w_2+F_{27} u_4 w_2+F_6 u_7 w_2+F_{26} u_3 w_3+F_5 u_6 w_3+F_{25} u_2 w_4+F_4 u_5 w_4+F_{39} u_8 w_4+F_{24} u_1 w_5+F_3 u_4 w_5+F_{38} u_7 w_5+F_2 u_3 w_6+F_{37} u_6 w_6+F_1 u_2 w_7+F_{36} u_5 w_7+F_{15} u_8 w_7\\
			\end{array}\right.\\
			  & \begin{array}{c}       
			F_{42} u_3 w_1+F_{21} u_6 w_1+F_{41} u_2 w_2+F_{20} u_5 w_2+F_{55} u_8 w_2+F_{40} u_1 w_3+F_{19} u_4 w_3+F_{54} u_7 w_3+F_{18} u_3 w_4+F_{53} u_6 w_4+F_{17} u_2 w_5+F_{52} u_5 w_5+F_{31} u_8 w_5+F_{16} u_1 w_6+F_{51} u_4 w_6+F_{30} u_7 w_6+F_{50} u_3 w_7+F_{29} u_6 w_7\\
			F_{49} u_2 w_1+F_{28} u_5 w_1+F_7 u_8 w_1+F_{48} u_1 w_2+F_{27} u_4 w_2+F_6 u_7 w_2+F_{26} u_3 w_3+F_5 u_6 w_3+F_{25} u_2 w_4+F_4 u_5 w_4+F_{39} u_8 w_4+F_{24} u_1 w_5+F_3 u_4 w_5+F_{38} u_7 w_5+F_2 u_3 w_6+F_{37} u_6 w_6+F_1 u_2 w_7+F_{36} u_5 w_7+F_{15} u_8 w_7\\
			F_0 u_1 w_1+F_{35} u_4 w_1+F_{14} u_7 w_1+F_{34} u_3 w_2+F_{13} u_6 w_2+F_{33} u_2 w_3+F_{12} u_5 w_3+F_{47} u_8 w_3+F_{32} u_1 w_4+F_{11} u_4 w_4+F_{46} u_7 w_4+F_{10} u_3 w_5+F_{45} u_6 w_5+F_9 u_2 w_6+F_{44} u_5 w_6+F_{23} u_8 w_6+F_8 u_1 w_7+F_{43} u_4 w_7+F_{22} u_7 w_7\\
			\end{array}\\
			  & \left.\begin{array}{c} 
			F_{49} u_2 w_1+F_{28} u_5 w_1+F_7 u_8 w_1+F_{48} u_1 w_2+F_{27} u_4 w_2+F_6 u_7 w_2+F_{26} u_3 w_3+F_5 u_6 w_3+F_{25} u_2 w_4+F_4 u_5 w_4+F_{39} u_8 w_4+F_{24} u_1 w_5+F_3 u_4 w_5+F_{38} u_7 w_5+F_2 u_3 w_6+F_{37} u_6 w_6+F_1 u_2 w_7+F_{36} u_5 w_7+F_{15} u_8 w_7\\
			F_0 u_1 w_1+F_{35} u_4 w_1+F_{14} u_7 w_1+F_{34} u_3 w_2+F_{13} u_6 w_2+F_{33} u_2 w_3+F_{12} u_5 w_3+F_{47} u_8 w_3+F_{32} u_1 w_4+F_{11} u_4 w_4+F_{46} u_7 w_4+F_{10} u_3 w_5+F_{45} u_6 w_5+F_9 u_2 w_6+F_{44} u_5 w_6+F_{23} u_8 w_6+F_8 u_1 w_7+F_{43} u_4 w_7+F_{22} u_7 w_7\\
			F_{42} u_3 w_1+F_{21} u_6 w_1+F_{41} u_2 w_2+F_{20} u_5 w_2+F_{55} u_8 w_2+F_{40} u_1 w_3+F_{19} u_4 w_3+F_{54} u_7 w_3+F_{18} u_3 w_4+F_{53} u_6 w_4+F_{17} u_2 w_5+F_{52} u_5 w_5+F_{31} u_8 w_5+F_{16} u_1 w_6+F_{51} u_4 w_6+F_{30} u_7 w_6+F_{50} u_3 w_7+F_{29} u_6 w_7\\
			\end{array}\right),
		\end{aligned}$
	} 
\end{equation} 
where $u_i, w_j$ are the VEVs and the couplings are defined as,
\begin{align}\label{eq:4couplings18.5}
	F_{7i} & \equiv  \vartheta \left[\begin{array}{c} 
	\epsilon^{(1)}+\frac{i}{8}\\  \phi^{(1)} \end{array} \right]
	(\frac{9 J^{\text{(1)}}}{\alpha '}),\qquad i={0,\dots,|I_{c c'}^{(1)}|-1}.
\end{align}

Since, we have already fitted the up-type quark matrix $|M_{3u}|$ exactly, so its 4-point correction should be zero,
\begin{equation}
	|M_{4u}| = 0 ,
\end{equation}
which is true by setting all up-type VEVs $u_u^i$ and $w_u^i$ to be zero. 
Therefore, we are essentially concerned with fitting charged-leptons in such a way that corresponding corrections for the down-type quarks remain negligible.
The desired solution can be readily obtained by setting $\epsilon^{(1)}_{4e}=0$ with the following values of the VEVs,
\begin{equation}\label{eq:UdWd18.5} 
	\begin{array}{l}
		u^d{}_1 = -0.000117911           \\
		u^d{}_2 = 0.00100719             \\
		u^d{}_3 = -1.17467\times 10^{-6} \\
		u^d{}_4 = 0                      \\
		u^d{}_5 = 0                      \\
		u^d{}_6 = 0                      \\
		u^d{}_7 = 0                      \\
		u^d{}_8 = 0                      \\
	\end{array}
	\quad,\quad
	\begin{array}{l}
		w^d{}_1 = 1 \\
		w^d{}_2 = 1 \\
		w^d{}_3 = 0 \\
		w^d{}_4 = 1 \\
		w^d{}_5 = 1 \\
		w^d{}_6 = 1 \\
		w^d{}_7 = 1 \\
	\end{array}
\end{equation}
The 4-point correction to the charged-leptons' masses is given by,
\begin{equation}\label{M4e_18.5}
	|M_{4e}|=m_\tau \left(
	\begin{array}{ccc}
		-0.000645 & 0.        & 0.035336  \\
		0.        & 0.035336  & -0.000645 \\
		0.035336  & -0.000645 & 0.        \\
	\end{array}
	\right)
\end{equation}
which can be added to the matrix obtained from 3-point functions \eqref{eq:Leptons3_18.5} as,
\begin{equation}\label{M34e_18.5}
	|M_{3e}|+|M_{4e}|=m_\tau \left(
	\begin{array}{ccc}
		0.000217 & -0.004106 & 0.035524  \\
		0.000022 & 0.0458    & -0.000645 \\
		0.035335 & -0.000645 & 1.        \\
	\end{array}
	\right) \sim D_e ~,
\end{equation} 
The corrections to down-type quarks' masses can be made negligible by setting $\epsilon^{(1)}_{4d}=0$,
\begin{equation}\label{M4d_18.5}
	|M_{4d}|\sim \left(
	\begin{array}{ccc}
		-0.001019 & 0.        & 0.05583   \\
		0.        & 0.05583   & -0.001019 \\
		0.05583   & -0.001019 & 0.        \\
	\end{array}
	\right)\sim 0.
\end{equation} Therefore, we can achieve near-exact matching of fermion masses and mixings.
\subsection{Model 19}\label{sec:model-19}
In Model~\hyperref[model19]{19} the three-point Yukawa couplings arise from the triplet intersections from the branes ${a, b, c}$ on the first two-torus ($r=1$) with 9 pairs of Higgs from $\mathcal{N}=2$ subsector.

\begin{figure}[t]
	\centering
	\includegraphics[width=\textwidth]{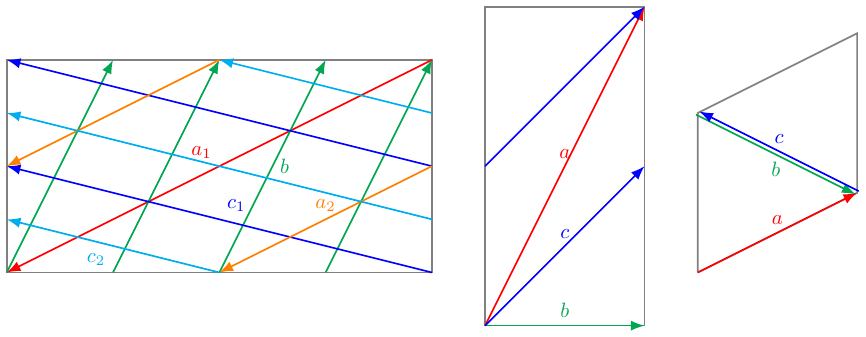}
	\caption{Brane configuration for the three two-tori in Model~\hyperref[model19]{19} where the third two-torus is tilted. Fermion mass hierarchies result from the intersections on the first two-torus.}  \label{Fig.19}
\end{figure}

Yukawa matrices for the Model~\hyperref[model19]{19} are of rank 3 and the three intersections required to form the disk diagrams for the Yukawa couplings all occur on the first torus as shown in figure~\ref{Fig.19}. The other two-tori only contribute an overall constant that has no effect in computing the fermion mass ratios. Thus, it is sufficient for our purpose to only focus on the first torus to explain the masses and the mixing in the standard model fermions.

\subsubsection{3-point Yukawa mass-matrices for Model~\hyperref[model19]{19}}
From the wrapping numbers listed in table~\ref{model19}, the relevant intersection numbers are calculated as,
\begin{align}\label{eq:Intersections.19} 
	I_{ab}^{(1)} & =-3, &   
	I_{ab}^{(2)} & =-1, &   
	I_{ab}^{(3)}&=-1, \nonumber\\
	I_{bc}^{(1)} & =9,  &   
	I_{bc}^{(2)} & =1,  &   
	I_{bc}^{(3)}&=0, \nonumber\\
	I_{ca}^{(1)} & =3,  &   
	I_{ca}^{(2)} & =1,  &   
	I_{ca}^{(3)}&=-1, 
\end{align}
As the intersection numbers are not coprime, we define the greatest common divisor,
$d^{(1)}=g.c.d.(I_{ab}^{(1)},I_{bc}^{(1)},I_{ca}^{(1)})=3$. Thus, the arguments of the modular theta function as defined in \eqref{eqn:Yinput} can be written as,
\begin{align}
	\delta^{(1)} & = \frac{i^{(1)}}{-3} + \frac{j^{(1)}}{3} + \frac{k^{(1)}}{9} +\frac{1}{27} \left(-9 \epsilon _a^{\text{(1)}}-3 \epsilon _b^{\text{(1)}}+3 \epsilon _c^{\text{(1)}}\right)+ \frac{s^{(1)}}{3},  \label{eq:delta.19} \\
	\phi^{(1)}   & = \frac{1}{3} \left(9 \theta _a^{\text{(1)}}+3 \theta _b^{\text{(1)}}-3 \theta _c^{\text{(1)}}\right) =0,                                                                                                          \\
	\kappa^{(1)} & = \frac{9 J^{\text{(1)}}}{\alpha '}, \label{eq:kappa.19}                                                                                                                                                           
\end{align}
and recalling \eqref{eq:ijk}, we have $i=\{0,\dots ,2\}$, $j=\{0,\dots ,2\}$ and $k=\{0,\dots ,8\}$ which respectively index the left-handed fermions, the right-handed fermions and the Higgs fields. Clearly, there arise 9 Higgs fields from the $bc$ sector.

The second-last term in the right side of \eqref{eq:delta.19} can be used to redefine the shift on the torus as
\begin{align}
	\epsilon^{(1)} & \equiv \frac{1}{27} \left(-9 \epsilon _a^{\text{(1)}}-3 \epsilon _b^{\text{(1)}}+3 \epsilon _c^{\text{(1)}}\right). \label{shifts19} 
\end{align}  
The selection rule for the occurrence of a trilinear Yukawa coupling for a given set of indices is given as,
\begin{equation}\label{selection-rule.19}
	i^{(1)} + j^{(1)} + k^{(1)} = 0\; \mathrm{ mod }\; 3.
\end{equation}
Then the suitable rank-3 mass-matrix can be determined by choosing the specific value of $s^{(1)}=-j$. 
\begin{align}
	\delta^{(1)} & =  \frac{k}{9}-\frac{i}{3} \label{eq:s1.19} 
\end{align}
Here, we will ignore other equivalent cases of solutions or cases with rank-1 problem. The mass matrices for up quarks, down quarks and charged leptons have the following general form:
\begin{align}
	Z_{3} &= Z_{3q} \left(
	\begin{array}{ccc}
	T_0 v_1+T_3 v_4+T_6 v_7 & T_2 v_3+T_5 v_6+T_8 v_9 & T_1 v_2+T_4 v_5+T_7 v_8 \\
	T_8 v_3+T_2 v_6+T_5 v_9 & T_7 v_2+T_1 v_5+T_4 v_8 & T_6 v_1+T_0 v_4+T_3 v_7 \\
	T_4 v_2+T_7 v_5+T_1 v_8 & T_3 v_1+T_6 v_4+T_0 v_7 & T_5 v_3+T_8 v_6+T_2 v_9 \\
	\end{array}
	\right),\label{eq:s-i19}
\end{align} 
where $v_i = \left\langle H_{i} \right\rangle$ and the three-point coupling functions are given in terms of Jacobi theta function of the third kind as, 

\begin{align}\label{eq:3couplings19}
	T_k & \equiv  \vartheta \left[\begin{array}{c} 
	\epsilon^{(1)}+\frac{k}{9}\\  \phi^{(1)} \end{array} \right]
	(\frac{9 J^{\text{(1)}}}{\alpha '}),  \quad   k={0,\cdots,8}.
\end{align}
  
\subsubsection{Fermion masses and mixings from 3-point functions in Model~\hyperref[model19]{19}}
In order to accommodate the fermion masses and the quark mixings, we need to fit the up-type quarks mixing matrix \eqref{eq:mixing-quarks}, the down-type quarks matrix \eqref{eq:mass-downquarks} and the masses of the charged-leptons \eqref{eq:mass-chargedleptons}.
 
We set the K\"{a}hler modulus on the first two-torus defined in \eqref{eq:kappa.19} as $\kappa^{(1)} = 45 $  and evaluate the couplings functions \eqref{eq:3couplings19} by setting geometric brane position parameters as $\epsilon^{(1)}_{u} = 0$, $\epsilon^{(1)}_{d} = 0$ and $\epsilon^{(1)}_{e} = 0.0156645$ which yields a nearest fit for the following VEVs,
\begin{equation}\label{eq:VEVs_19} 
	\begin{array}{ll}
		v^u{}_1 = 0.000291324 , & v^d{}_1 = 0.0000282               \\
		v^u{}_2 = 0.0490413   , & v^d{}_2 = 9.08351\times 10^{-8}   \\
		v^u{}_3 = -0.0234364  , & v^d{}_3 = -0.000609616            \\
		v^u{}_4 = 0.0413384   , & v^d{}_4 = -4.24979\times 10^{-12} \\
		v^u{}_5 = 0.0313982   , & v^d{}_5 = 0.00320756              \\
		v^u{}_6 = 5.71763     , & v^d{}_6 = 0.114556                \\
		v^u{}_7 = 0.0413384   , & v^d{}_7 = -4.24979\times 10^{-12} \\
		v^u{}_8 = 0.0491357   , & v^d{}_8 = -0.0000170692           \\
		v^u{}_9 = 0.00711503  , & v^d{}_9 = 3.24411\times 10^{-6}   \\
	\end{array}
\end{equation}
\begin{align} 
	|M_{3u}|&=m_t \left(
	\begin{array}{ccc}
	0.000291          & 0.00122   & 0.008608 \\
	0.00122           & 0.005527  & 0.041338 \\
	0.008608          & 0.041338  & 0.998234 \\
	\end{array}
	\right)\sim M_u ~, \label{eq:Up3_19} \\
	|M_{3d}|&=m_b \left(
	\begin{array}{ccc}
	0.00141           & 0.        & 0.       \\
	0.                & 0.028     & 0.       \\
	0.                & 0.        & 1.       \\
	\end{array}
	\right)\sim D_d ~,\label{eq:Down3_19} \\ 
	|M_{3e}|&=m_\tau \left(
	\begin{array}{ccc}
	0.000862          & -0.004106 & 0.000188 \\
	0.000022          & 0.010464  & 0.       \\
	-1.\times 10^{-6} & 0.        & 1.       \\
	\end{array}
	\right)~.\label{eq:Leptons3_19}
\end{align}While the masses of up-type quarks and the down-type quarks are fitted, in the charged-lepton matrix, only the mass of the tau can be fitted with three-point couplings only. The masses of the muon and the electron are not fitted. Notice, that these results are only at the tree-level and there could indeed be other corrections, such as those coming from higher-dimensional operators, which may contribute most greatly to the muon and the electron masses since they are lighter.

\subsubsection{4-point corrections in Model~\hyperref[model19]{19}}
 
The four-point couplings in Model~\hyperref[model19]{19} in table~\ref{model19} can come from considering interactions of ${a, b, c}$ with $b'$ or $c'$ on the first two-torus as can be seen from the following intersection numbers,
\begin{align}
	I_{bb'}^{(1)} & = 8,  & I_{bb'}^{(2)} & = 0,  & I_{bb'}^{(3)} & = -1 ,\nonumber                   \\
	I_{cc'}^{(1)} & = -4, & I_{cc'}^{(2)} & = 4,  & I_{cc'}^{(3)} & = -1 ,\nonumber                   \\
	I_{bc'}^{(1)} & = 7,  & I_{bc'}^{(2)} & = -1, & I_{bc'}^{(3)} & = -1 . \label{eq:4Intersection19} 
\end{align}
There are 8 SM singlet fields $S_L^i$ and 7
Higgs-like state $H^{\prime}_{u, d}$. 

Let us consider four-point interactions with $b'$ with the following parameters with shifts $l=-\frac{k}{3} $ and $\ell=-\frac{k}{9} $ taken along the index $k$,
\begin{align}\label{deltas19}
	\delta & =\frac{i}{I_{ab}^{(1)}} +\frac{j}{I_{ca}^{(1)}} +\frac{k}{I_{bc}^{(1)}} +l , \nonumber               \\
	       & = \frac{j}{3}-\frac{i}{3} ,                                                                          \\
	d      & =\frac{\imath}{I_{b b'}^{(1)}} +\frac{\jmath}{I_{bc'}^{(1)}}+\frac{k}{I_{bc}^{(1)}} +\ell ,\nonumber \\
	       & = \frac{\imath}{8}+\frac{\jmath}{7} ,                                                                
\end{align}
the matrix elements $a_{i,j,\imath}$ on the first torus from the four-point functions results in the following classical 4-point contribution to the mass matrix,
\begin{equation}\label{eq:4point19} 
	Z_{4cl} =  
	\resizebox{.95\textwidth}{!}{
		$\begin{aligned}
			  & \left(                 
			\begin{array}{c}
			F_0 u_1 w_1+F_{21} u_4 w_1+F_{42} u_7 w_1+F_{22} u_3 w_2+F_{43} u_6 w_2+F_{23} u_2 w_3+F_{44} u_5 w_3+F_9 u_8 w_3+F_{24} u_1 w_4+F_{45} u_4 w_4+F_{10} u_7 w_4+F_{46} u_3 w_5+F_{11} u_6 w_5+F_{47} u_2 w_6+F_{12} u_5 w_6+F_{33} u_8 w_6+F_{48} u_1 w_7+F_{13} u_4 w_7+F_{34} u_7 w_7\\
			F_{14} u_3 w_1+F_{35} u_6 w_1+F_{15} u_2 w_2+F_{36} u_5 w_2+F_1 u_8 w_2+F_{16} u_1 w_3+F_{37} u_4 w_3+F_2 u_7 w_3+F_{38} u_3 w_4+F_3 u_6 w_4+F_{39} u_2 w_5+F_4 u_5 w_5+F_{25} u_8 w_5+F_{40} u_1 w_6+F_5 u_4 w_6+F_{26} u_7 w_6+F_6 u_3 w_7+F_{27} u_6 w_7\\
			F_7 u_2 w_1+F_{28} u_5 w_1+F_{49} u_8 w_1+F_8 u_1 w_2+F_{29} u_4 w_2+F_{50} u_7 w_2+F_{30} u_3 w_3+F_{51} u_6 w_3+F_{31} u_2 w_4+F_{52} u_5 w_4+F_{17} u_8 w_4+F_{32} u_1 w_5+F_{53} u_4 w_5+F_{18} u_7 w_5+F_{54} u_3 w_6+F_{19} u_6 w_6+F_{55} u_2 w_7+F_{20} u_5 w_7+F_{41} u_8 w_7\\
			\end{array}\right.\\
			  & \begin{array}{c}       
			F_{14} u_3 w_1+F_{35} u_6 w_1+F_{15} u_2 w_2+F_{36} u_5 w_2+F_1 u_8 w_2+F_{16} u_1 w_3+F_{37} u_4 w_3+F_2 u_7 w_3+F_{38} u_3 w_4+F_3 u_6 w_4+F_{39} u_2 w_5+F_4 u_5 w_5+F_{25} u_8 w_5+F_{40} u_1 w_6+F_5 u_4 w_6+F_{26} u_7 w_6+F_6 u_3 w_7+F_{27} u_6 w_7\\
			F_7 u_2 w_1+F_{28} u_5 w_1+F_{49} u_8 w_1+F_8 u_1 w_2+F_{29} u_4 w_2+F_{50} u_7 w_2+F_{30} u_3 w_3+F_{51} u_6 w_3+F_{31} u_2 w_4+F_{52} u_5 w_4+F_{17} u_8 w_4+F_{32} u_1 w_5+F_{53} u_4 w_5+F_{18} u_7 w_5+F_{54} u_3 w_6+F_{19} u_6 w_6+F_{55} u_2 w_7+F_{20} u_5 w_7+F_{41} u_8 w_7\\
			F_0 u_1 w_1+F_{21} u_4 w_1+F_{42} u_7 w_1+F_{22} u_3 w_2+F_{43} u_6 w_2+F_{23} u_2 w_3+F_{44} u_5 w_3+F_9 u_8 w_3+F_{24} u_1 w_4+F_{45} u_4 w_4+F_{10} u_7 w_4+F_{46} u_3 w_5+F_{11} u_6 w_5+F_{47} u_2 w_6+F_{12} u_5 w_6+F_{33} u_8 w_6+F_{48} u_1 w_7+F_{13} u_4 w_7+F_{34} u_7 w_7\\
			\end{array}\\
			  & \left.\begin{array}{c} 
			F_7 u_2 w_1+F_{28} u_5 w_1+F_{49} u_8 w_1+F_8 u_1 w_2+F_{29} u_4 w_2+F_{50} u_7 w_2+F_{30} u_3 w_3+F_{51} u_6 w_3+F_{31} u_2 w_4+F_{52} u_5 w_4+F_{17} u_8 w_4+F_{32} u_1 w_5+F_{53} u_4 w_5+F_{18} u_7 w_5+F_{54} u_3 w_6+F_{19} u_6 w_6+F_{55} u_2 w_7+F_{20} u_5 w_7+F_{41} u_8 w_7\\
			F_0 u_1 w_1+F_{21} u_4 w_1+F_{42} u_7 w_1+F_{22} u_3 w_2+F_{43} u_6 w_2+F_{23} u_2 w_3+F_{44} u_5 w_3+F_9 u_8 w_3+F_{24} u_1 w_4+F_{45} u_4 w_4+F_{10} u_7 w_4+F_{46} u_3 w_5+F_{11} u_6 w_5+F_{47} u_2 w_6+F_{12} u_5 w_6+F_{33} u_8 w_6+F_{48} u_1 w_7+F_{13} u_4 w_7+F_{34} u_7 w_7\\
			F_{14} u_3 w_1+F_{35} u_6 w_1+F_{15} u_2 w_2+F_{36} u_5 w_2+F_1 u_8 w_2+F_{16} u_1 w_3+F_{37} u_4 w_3+F_2 u_7 w_3+F_{38} u_3 w_4+F_3 u_6 w_4+F_{39} u_2 w_5+F_4 u_5 w_5+F_{25} u_8 w_5+F_{40} u_1 w_6+F_5 u_4 w_6+F_{26} u_7 w_6+F_6 u_3 w_7+F_{27} u_6 w_7\\
			\end{array}\right),
		\end{aligned}$
	} 
\end{equation} 
where $u_i, w_j$ are the VEVs and the couplings are defined as,
\begin{align}\label{eq:4couplings19}
	F_{7i} & \equiv  \vartheta \left[\begin{array}{c} 
	\epsilon^{(1)}+\frac{i}{8}\\  \phi^{(1)} \end{array} \right]
	(\frac{9 J^{\text{(1)}}}{\alpha '}),\qquad i={0,\dots,|I_{b b'}^{(1)}|-1}.
\end{align}

Since, we have already fitted the up-type quark matrix $|M_{3u}|$ exactly, so its 4-point correction should be zero,
\begin{equation}
	|M_{4u}| = 0 ,
\end{equation}
which is true by setting all up-type VEVs $u_u^i$ and $w_u^i$ to be zero. 
Therefore, we are essentially concerned with fitting charged-leptons in such a way that corresponding corrections for the down-type quarks remain negligible.
The desired solution can be readily obtained by setting $\epsilon^{(1)}_{4e}=0$ with the following values of the VEVs,
\begin{equation}\label{eq:UdWd19} 
	\begin{array}{l}
		u^d{}_1 = -0.0000211192          \\
		u^d{}_2 = 0.00509483             \\
		u^d{}_3 = -7.40909\times 10^{-7} \\
		u^d{}_4 = 0                      \\
		u^d{}_5 = 0                      \\
		u^d{}_6 = 0                      \\
		u^d{}_7 = 0                      \\
		u^d{}_8 = 0                      \\
	\end{array}
	\quad,\quad
	\begin{array}{l}
		w^d{}_1 = 1 \\
		w^d{}_2 = 1 \\
		w^d{}_3 = 0 \\
		w^d{}_4 = 1 \\
		w^d{}_5 = 1 \\
		w^d{}_6 = 1 \\
		w^d{}_7 = 1 \\
	\end{array}
\end{equation}
The 4-point correction to the charged-leptons' masses is given by,
\begin{equation}\label{M4e_19}
	|M_{4e}|=m_\tau \left(
	\begin{array}{ccc}
		-0.000645 & 0.        & 0.035336  \\
		0.        & 0.035336  & -0.000645 \\
		0.035336  & -0.000645 & 0.        \\
	\end{array}
	\right)
\end{equation}
which can be added to the matrix obtained from 3-point functions \eqref{eq:Leptons3_19} as,
\begin{equation}\label{M34e_19}
	|M_{3e}|+|M_{4e}|=m_\tau \left(
	\begin{array}{ccc}
		0.000217 & -0.004106 & 0.035524  \\
		0.000022 & 0.0458    & -0.000645 \\
		0.035335 & -0.000645 & 1.        \\
	\end{array}
	\right) \sim D_e ~,
\end{equation} 
The corrections to down-type quarks' masses can be made negligible by setting $\epsilon^{(1)}_{4d}=0$,
\begin{equation}\label{M4d_19}
	|M_{4d}|\sim \left(
	\begin{array}{ccc}
		-0.001019 & 0.        & 0.05583   \\
		0.        & 0.05583   & -0.001019 \\
		0.05583   & -0.001019 & 0.        \\
	\end{array}
	\right)\sim 0.
\end{equation} Therefore, we can achieve near-exact matching of fermion masses and mixings.
\subsection{Model 19-dual}\label{sec:model-19.5}
In Model~\hyperref[model19.5]{19-dual} the three-point Yukawa couplings arise from the triplet intersections from the branes ${a, b, c}$ on the first two-torus ($r=1$) with 9 pairs of Higgs from $\mathcal{N}=2$ subsector.

\begin{figure}[t]
	\centering
	\includegraphics[width=\textwidth]{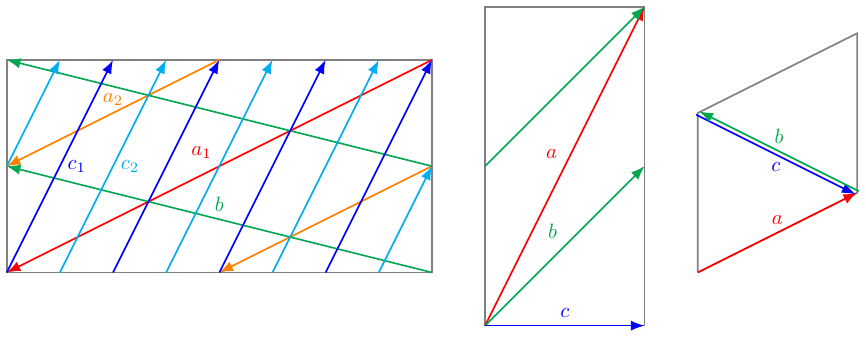}
	\caption{Brane configuration for the three two-tori in Model~\hyperref[model19.5]{19-dual} where the third two-torus is tilted. Fermion mass hierarchies result from the intersections on the first two-torus.}  \label{Fig.19.5}
\end{figure}

Yukawa matrices for the Model~\hyperref[model19.5]{19-dual} are of rank 3 and the three intersections required to form the disk diagrams for the Yukawa couplings all occur on the first torus as shown in figure~\ref{Fig.19.5}. The other two-tori only contribute an overall constant that has no effect in computing the fermion mass ratios. Thus, it is sufficient for our purpose to only focus on the first torus to explain the masses and the mixing in the standard model fermions.

\subsubsection{3-point Yukawa mass-matrices for Model~\hyperref[model19.5]{19-dual}}
From the wrapping numbers listed in table~\ref{model19.5}, the relevant intersection numbers are calculated as,
\begin{align}\label{eq:Intersections.19.5} 
	I_{ab}^{(1)} & =-3, &   
	I_{ab}^{(2)} & =-1, &   
	I_{ab}^{(3)}&=1, \nonumber\\
	I_{bc}^{(1)} & =-9, &   
	I_{bc}^{(2)} & =-1, &   
	I_{bc}^{(3)}&=0, \nonumber\\
	I_{ca}^{(1)} & =3,  &   
	I_{ca}^{(2)} & =1,  &   
	I_{ca}^{(3)}&=1, 
\end{align}
As the intersection numbers are not coprime, we define the greatest common divisor,
$d^{(1)}=g.c.d.(I_{ab}^{(1)},I_{bc}^{(1)},I_{ca}^{(1)})=3$. Thus, the arguments of the modular theta function as defined in \eqref{eqn:Yinput} can be written as,
\begin{align}
	\delta^{(1)} & = \frac{i^{(1)}}{-3} + \frac{j^{(1)}}{3} + \frac{k^{(1)}}{-9} +\frac{1}{27} \left(-9 \epsilon _a^{\text{(1)}}+3 \epsilon _b^{\text{(1)}}-3 \epsilon _c^{\text{(1)}}\right)+ \frac{s^{(1)}}{3},  \label{eq:delta.19.5} \\
	\phi^{(1)}   & = \frac{1}{3} \left(-9 \theta _a^{\text{(1)}}+3 \theta _b^{\text{(1)}}-3 \theta _c^{\text{(1)}}\right) =0,                                                                                                            \\
	\kappa^{(1)} & = \frac{9 J^{\text{(1)}}}{\alpha '}, \label{eq:kappa.19.5}                                                                                                                                                            
\end{align}
and recalling \eqref{eq:ijk}, we have $i=\{0,\dots ,2\}$, $j=\{0,\dots ,2\}$ and $k=\{0,\dots ,8\}$ which respectively index the left-handed fermions, the right-handed fermions and the Higgs fields. Clearly, there arise 9 Higgs fields from the $bc$ sector.

The second-last term in the right side of \eqref{eq:delta.19.5} can be used to redefine the shift on the torus as
\begin{align}
	\epsilon^{(1)} & \equiv \frac{1}{27} \left(-9 \epsilon _a^{\text{(1)}}+3 \epsilon _b^{\text{(1)}}-3 \epsilon _c^{\text{(1)}}\right). \label{shifts19.5} 
\end{align}  
The selection rule for the occurrence of a trilinear Yukawa coupling for a given set of indices is given as,
\begin{equation}\label{selection-rule.19.5}
	i^{(1)} + j^{(1)} + k^{(1)} = 0\; \mathrm{ mod }\; 3.
\end{equation}
Then the suitable rank-3 mass-matrix can be determined by choosing the specific value of $s^{(1)}=i$. 
\begin{align}
	\delta^{(1)} & =  \frac{j}{3}-\frac{k}{9} \label{eq:s1.19.5} 
\end{align}
Here, we will ignore other equivalent cases of solutions or cases with rank-1 problem. The mass matrices for up quarks, down quarks and charged leptons have the following general form:
\begin{align}
	Z_{3} &= Z_{3q} \left(
	\begin{array}{ccc}
	T_0 v_1+T_6 v_4+T_3 v_7 & T_7 v_3+T_4 v_6+T_1 v_9 & T_8 v_2+T_5 v_5+T_2 v_8 \\
	T_1 v_3+T_7 v_6+T_4 v_9 & T_2 v_2+T_8 v_5+T_5 v_8 & T_3 v_1+T_0 v_4+T_6 v_7 \\
	T_5 v_2+T_2 v_5+T_8 v_8 & T_6 v_1+T_3 v_4+T_0 v_7 & T_4 v_3+T_1 v_6+T_7 v_9 \\
	\end{array}
	\right),\label{eq:s-i19.5}
\end{align} 
where $v_i = \left\langle H_{i} \right\rangle$ and the three-point coupling functions are given in terms of Jacobi theta function of the third kind as, 

\begin{align}\label{eq:3couplings19.5}
	T_k & \equiv  \vartheta \left[\begin{array}{c} 
	\epsilon^{(1)}+\frac{k}{9}\\  \phi^{(1)} \end{array} \right]
	(\frac{9 J^{\text{(1)}}}{\alpha '}),  \quad   k={0,\cdots,8}.
\end{align}
  
\subsubsection{Fermion masses and mixings from 3-point functions in Model~\hyperref[model19.5]{19-dual}}
In order to accommodate the fermion masses and the quark mixings, we need to fit the up-type quarks mixing matrix \eqref{eq:mixing-quarks}, the down-type quarks matrix \eqref{eq:mass-downquarks} and the masses of the charged-leptons \eqref{eq:mass-chargedleptons}.
 
We set the K\"{a}hler modulus on the first two-torus defined in \eqref{eq:kappa.19.5} as $\kappa^{(1)} = 45 $  and evaluate the couplings functions \eqref{eq:3couplings19.5} by setting geometric brane position parameters as $\epsilon^{(1)}_{u} = 0$, $\epsilon^{(1)}_{d} = 0$ and $\epsilon^{(1)}_{e} = -0.0156645$ which yields a nearest fit for the following VEVs,
\begin{equation}\label{eq:VEVs_19.5} 
	\begin{array}{ll}
		v^u{}_1 = 0.000291324 , & v^d{}_1 = 0.0000282               \\
		v^u{}_2 = 0.0490413   , & v^d{}_2 = 9.08351\times 10^{-8}   \\
		v^u{}_3 = -0.0234364  , & v^d{}_3 = -0.000609616            \\
		v^u{}_4 = 0.0413384   , & v^d{}_4 = -4.24979\times 10^{-12} \\
		v^u{}_5 = 0.0313982   , & v^d{}_5 = 0.00320756              \\
		v^u{}_6 = 5.71763     , & v^d{}_6 = 0.114556                \\
		v^u{}_7 = 0.0413384   , & v^d{}_7 = -4.24979\times 10^{-12} \\
		v^u{}_8 = 0.0491357   , & v^d{}_8 = -0.0000170692           \\
		v^u{}_9 = 0.00711503  , & v^d{}_9 = 3.24411\times 10^{-6}   \\
	\end{array}
\end{equation}
\begin{align} 
	|M_{3u}|&=m_t \left(
	\begin{array}{ccc}
	0.000291          & 0.00122   & 0.008608 \\
	0.00122           & 0.005527  & 0.041338 \\
	0.008608          & 0.041338  & 0.998234 \\
	\end{array}
	\right)\sim M_u ~, \label{eq:Up3_19.5} \\
	|M_{3d}|&=m_b \left(
	\begin{array}{ccc}
	0.00141           & 0.        & 0.       \\
	0.                & 0.028     & 0.       \\
	0.                & 0.        & 1.       \\
	\end{array}
	\right)\sim D_d ~,\label{eq:Down3_19.5} \\ 
	|M_{3e}|&=m_\tau \left(
	\begin{array}{ccc}
	0.000862          & -0.004106 & 0.000188 \\
	0.000022          & 0.010464  & 0.       \\
	-1.\times 10^{-6} & 0.        & 1.       \\
	\end{array}
	\right)~.\label{eq:Leptons3_19.5}
\end{align}While the masses of up-type quarks and the down-type quarks are fitted, in the charged-lepton matrix, only the mass of the tau can be fitted with three-point couplings only. The masses of the muon and the electron are not fitted. Notice, that these results are only at the tree-level and there could indeed be other corrections, such as those coming from higher-dimensional operators, which may contribute most greatly to the muon and the electron masses since they are lighter.

\subsubsection{4-point corrections in Model~\hyperref[model19.5]{19-dual}}
 
The four-point couplings in Model~\hyperref[model19.5]{19-dual} in table~\ref{model19.5} can come from considering interactions of ${a, b, c}$ with $b'$ or $c'$ on the first two-torus as can be seen from the following intersection numbers,
\begin{align}
	I_{bb'}^{(1)} & = -4, & I_{bb'}^{(2)} & = 4,  & I_{bb'}^{(3)} & = -1 ,\nonumber                     \\
	I_{cc'}^{(1)} & = 8,  & I_{cc'}^{(2)} & = 0,  & I_{cc'}^{(3)} & = -1 ,\nonumber                     \\
	I_{bc'}^{(1)} & = 7,  & I_{bc'}^{(2)} & = -1, & I_{bc'}^{(3)} & = -1 . \label{eq:4Intersection19.5} 
\end{align}
There are 8 SM singlet fields $S_L^i$ and 7
Higgs-like state $H^{\prime}_{u, d}$. 

Let us consider four-point interactions with $c'$ with the following parameters with shifts $l=\frac{k}{3} $ and $\ell=\frac{k}{9} $ taken along the index $k$,
\begin{align}\label{deltas19.5}
	\delta & =\frac{i}{I_{ab}^{(1)}} +\frac{j}{I_{ca}^{(1)}} +\frac{k}{I_{bc}^{(1)}} +l , \nonumber               \\
	       & = \frac{j}{3}-\frac{i}{3} ,                                                                          \\
	d      & =\frac{\imath}{I_{c c'}^{(1)}} +\frac{\jmath}{I_{bc'}^{(1)}}+\frac{k}{I_{bc}^{(1)}} +\ell ,\nonumber \\
	       & = \frac{\imath}{8}+\frac{\jmath}{7} ,                                                                
\end{align}
the matrix elements $a_{i,j,\imath}$ on the first torus from the four-point functions results in the following classical 4-point contribution to the mass matrix,
\begin{equation}\label{eq:4point19.5} 
	Z_{4cl} =  
	\resizebox{.95\textwidth}{!}{
		$\begin{aligned}
			  & \left(                 
			\begin{array}{c}
			F_0 u_1 w_1+F_{21} u_4 w_1+F_{42} u_7 w_1+F_{22} u_3 w_2+F_{43} u_6 w_2+F_{23} u_2 w_3+F_{44} u_5 w_3+F_9 u_8 w_3+F_{24} u_1 w_4+F_{45} u_4 w_4+F_{10} u_7 w_4+F_{46} u_3 w_5+F_{11} u_6 w_5+F_{47} u_2 w_6+F_{12} u_5 w_6+F_{33} u_8 w_6+F_{48} u_1 w_7+F_{13} u_4 w_7+F_{34} u_7 w_7\\
			F_{14} u_3 w_1+F_{35} u_6 w_1+F_{15} u_2 w_2+F_{36} u_5 w_2+F_1 u_8 w_2+F_{16} u_1 w_3+F_{37} u_4 w_3+F_2 u_7 w_3+F_{38} u_3 w_4+F_3 u_6 w_4+F_{39} u_2 w_5+F_4 u_5 w_5+F_{25} u_8 w_5+F_{40} u_1 w_6+F_5 u_4 w_6+F_{26} u_7 w_6+F_6 u_3 w_7+F_{27} u_6 w_7\\
			F_7 u_2 w_1+F_{28} u_5 w_1+F_{49} u_8 w_1+F_8 u_1 w_2+F_{29} u_4 w_2+F_{50} u_7 w_2+F_{30} u_3 w_3+F_{51} u_6 w_3+F_{31} u_2 w_4+F_{52} u_5 w_4+F_{17} u_8 w_4+F_{32} u_1 w_5+F_{53} u_4 w_5+F_{18} u_7 w_5+F_{54} u_3 w_6+F_{19} u_6 w_6+F_{55} u_2 w_7+F_{20} u_5 w_7+F_{41} u_8 w_7\\
			\end{array}\right.\\
			  & \begin{array}{c}       
			F_{14} u_3 w_1+F_{35} u_6 w_1+F_{15} u_2 w_2+F_{36} u_5 w_2+F_1 u_8 w_2+F_{16} u_1 w_3+F_{37} u_4 w_3+F_2 u_7 w_3+F_{38} u_3 w_4+F_3 u_6 w_4+F_{39} u_2 w_5+F_4 u_5 w_5+F_{25} u_8 w_5+F_{40} u_1 w_6+F_5 u_4 w_6+F_{26} u_7 w_6+F_6 u_3 w_7+F_{27} u_6 w_7\\
			F_7 u_2 w_1+F_{28} u_5 w_1+F_{49} u_8 w_1+F_8 u_1 w_2+F_{29} u_4 w_2+F_{50} u_7 w_2+F_{30} u_3 w_3+F_{51} u_6 w_3+F_{31} u_2 w_4+F_{52} u_5 w_4+F_{17} u_8 w_4+F_{32} u_1 w_5+F_{53} u_4 w_5+F_{18} u_7 w_5+F_{54} u_3 w_6+F_{19} u_6 w_6+F_{55} u_2 w_7+F_{20} u_5 w_7+F_{41} u_8 w_7\\
			F_0 u_1 w_1+F_{21} u_4 w_1+F_{42} u_7 w_1+F_{22} u_3 w_2+F_{43} u_6 w_2+F_{23} u_2 w_3+F_{44} u_5 w_3+F_9 u_8 w_3+F_{24} u_1 w_4+F_{45} u_4 w_4+F_{10} u_7 w_4+F_{46} u_3 w_5+F_{11} u_6 w_5+F_{47} u_2 w_6+F_{12} u_5 w_6+F_{33} u_8 w_6+F_{48} u_1 w_7+F_{13} u_4 w_7+F_{34} u_7 w_7\\
			\end{array}\\
			  & \left.\begin{array}{c} 
			F_7 u_2 w_1+F_{28} u_5 w_1+F_{49} u_8 w_1+F_8 u_1 w_2+F_{29} u_4 w_2+F_{50} u_7 w_2+F_{30} u_3 w_3+F_{51} u_6 w_3+F_{31} u_2 w_4+F_{52} u_5 w_4+F_{17} u_8 w_4+F_{32} u_1 w_5+F_{53} u_4 w_5+F_{18} u_7 w_5+F_{54} u_3 w_6+F_{19} u_6 w_6+F_{55} u_2 w_7+F_{20} u_5 w_7+F_{41} u_8 w_7\\
			F_0 u_1 w_1+F_{21} u_4 w_1+F_{42} u_7 w_1+F_{22} u_3 w_2+F_{43} u_6 w_2+F_{23} u_2 w_3+F_{44} u_5 w_3+F_9 u_8 w_3+F_{24} u_1 w_4+F_{45} u_4 w_4+F_{10} u_7 w_4+F_{46} u_3 w_5+F_{11} u_6 w_5+F_{47} u_2 w_6+F_{12} u_5 w_6+F_{33} u_8 w_6+F_{48} u_1 w_7+F_{13} u_4 w_7+F_{34} u_7 w_7\\
			F_{14} u_3 w_1+F_{35} u_6 w_1+F_{15} u_2 w_2+F_{36} u_5 w_2+F_1 u_8 w_2+F_{16} u_1 w_3+F_{37} u_4 w_3+F_2 u_7 w_3+F_{38} u_3 w_4+F_3 u_6 w_4+F_{39} u_2 w_5+F_4 u_5 w_5+F_{25} u_8 w_5+F_{40} u_1 w_6+F_5 u_4 w_6+F_{26} u_7 w_6+F_6 u_3 w_7+F_{27} u_6 w_7\\
			\end{array}\right),
		\end{aligned}$
	} 
\end{equation} 
where $u_i, w_j$ are the VEVs and the couplings are defined as,
\begin{align}\label{eq:4couplings19.5}
	F_{7i} & \equiv  \vartheta \left[\begin{array}{c} 
	\epsilon^{(1)}+\frac{i}{8}\\  \phi^{(1)} \end{array} \right]
	(\frac{9 J^{\text{(1)}}}{\alpha '}),\qquad i={0,\dots,|I_{c c'}^{(1)}|-1}.
\end{align}

Since, we have already fitted the up-type quark matrix $|M_{3u}|$ exactly, so its 4-point correction should be zero,
\begin{equation}
	|M_{4u}| = 0 ,
\end{equation}
which is true by setting all up-type VEVs $u_u^i$ and $w_u^i$ to be zero. 
Therefore, we are essentially concerned with fitting charged-leptons in such a way that corresponding corrections for the down-type quarks remain negligible.
The desired solution can be readily obtained by setting $\epsilon^{(1)}_{4e}=0$ with the following values of the VEVs,
\begin{equation}\label{eq:UdWd19.5} 
	\begin{array}{l}
		u^d{}_1 = -0.0000211192          \\
		u^d{}_2 = 0.00509483             \\
		u^d{}_3 = -7.40909\times 10^{-7} \\
		u^d{}_4 = 0                      \\
		u^d{}_5 = 0                      \\
		u^d{}_6 = 0                      \\
		u^d{}_7 = 0                      \\
		u^d{}_8 = 0                      \\
	\end{array}
	\quad,\quad
	\begin{array}{l}
		w^d{}_1 = 1 \\
		w^d{}_2 = 1 \\
		w^d{}_3 = 0 \\
		w^d{}_4 = 1 \\
		w^d{}_5 = 1 \\
		w^d{}_6 = 1 \\
		w^d{}_7 = 1 \\
	\end{array}
\end{equation}
The 4-point correction to the charged-leptons' masses is given by,
\begin{equation}\label{M4e_19.5}
	|M_{4e}|=m_\tau \left(
	\begin{array}{ccc}
		-0.000645 & 0.        & 0.035336  \\
		0.        & 0.035336  & -0.000645 \\
		0.035336  & -0.000645 & 0.        \\
	\end{array}
	\right)
\end{equation}
which can be added to the matrix obtained from 3-point functions \eqref{eq:Leptons3_19.5} as,
\begin{equation}\label{M34e_19.5}
	|M_{3e}|+|M_{4e}|=m_\tau \left(
	\begin{array}{ccc}
		0.000217 & -0.004106 & 0.035524  \\
		0.000022 & 0.0458    & -0.000645 \\
		0.035335 & -0.000645 & 1.        \\
	\end{array}
	\right) \sim D_e ~,
\end{equation} 
The corrections to down-type quarks' masses can be made negligible by setting $\epsilon^{(1)}_{4d}=0$,
\begin{equation}\label{M4d_19.5}
	|M_{4d}|\sim \left(
	\begin{array}{ccc}
		-0.001019 & 0.        & 0.05583   \\
		0.        & 0.05583   & -0.001019 \\
		0.05583   & -0.001019 & 0.        \\
	\end{array}
	\right)\sim 0.
\end{equation} Therefore, we can achieve near-exact matching of fermion masses and mixings.
\subsection{Model 20}\label{sec:model-20}
In Model~\hyperref[model20]{20} the three-point Yukawa couplings arise from the triplet intersections from the branes ${a, b, c}$ on the second two-torus ($r=2$) with 9 pairs of Higgs from $\mathcal{N}=2$ subsector.

\begin{figure}[t]
	\centering
	\includegraphics[width=\textwidth]{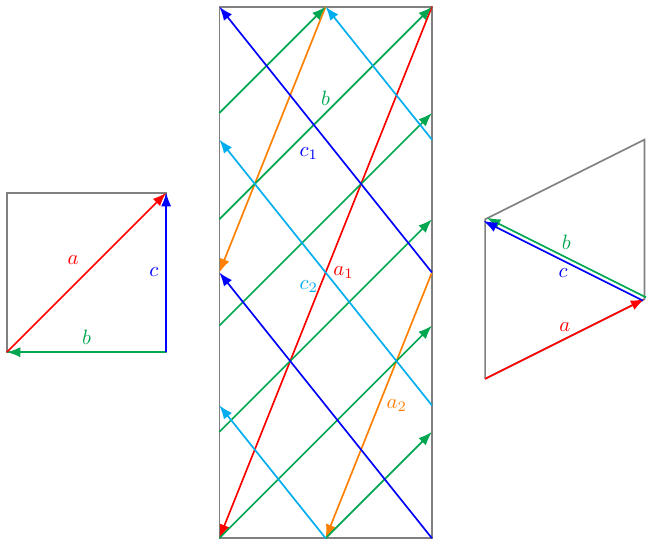}
	\caption{Brane configuration for the three two-tori in Model~\hyperref[model20]{20} where the third two-torus is tilted. Fermion mass hierarchies result from the intersections on the second two-torus.}  \label{Fig.20}
\end{figure}

Yukawa matrices for the Model~\hyperref[model20]{20} are of rank 3 and the three intersections required to form the disk diagrams for the Yukawa couplings all occur on the second torus as shown in figure~\ref{Fig.20}. The other two-tori only contribute an overall constant that has no effect in computing the fermion mass ratios. Thus, it is sufficient for our purpose to only focus on the second torus to explain the masses and the mixing in the standard model fermions.

\subsubsection{3-point Yukawa mass-matrices for Model~\hyperref[model20]{20}}
From the wrapping numbers listed in table~\ref{model20}, the relevant intersection numbers are calculated as,
\begin{align}\label{eq:Intersections.20} 
	I_{ab}^{(1)} & =1,  &   
	I_{ab}^{(2)} & =3,  &   
	I_{ab}^{(3)}&=1, \nonumber\\
	I_{bc}^{(1)} & =-1, &   
	I_{bc}^{(2)} & =9,  &   
	I_{bc}^{(3)}&=0, \nonumber\\
	I_{ca}^{(1)} & =-1, &   
	I_{ca}^{(2)} & =3,  &   
	I_{ca}^{(3)}&=-1, 
\end{align}
As the intersection numbers are not coprime, we define the greatest common divisor,
$d^{(1)}=g.c.d.(I_{ab}^{(2)},I_{bc}^{(2)},I_{ca}^{(2)})=3$. Thus, the arguments of the modular theta function as defined in \eqref{eqn:Yinput} can be written as,
\begin{align}
	\delta^{(2)} & = \frac{i^{(2)}}{3} + \frac{j^{(2)}}{3} + \frac{k^{(2)}}{9} +\frac{1}{27} \left(9 \epsilon _a^{\text{(2)}}+3 \epsilon _b^{\text{(2)}}+3 \epsilon _c^{\text{(2)}}\right)+ \frac{s^{(2)}}{3},  \label{eq:delta.20} \\
	\phi^{(2)}   & = \frac{1}{3} \left(9 \theta _a^{\text{(2)}}+3 \theta _b^{\text{(2)}}+3 \theta _c^{\text{(2)}}\right) =0,                                                                                                        \\
	\kappa^{(2)} & = \frac{9 J^{\text{(2)}}}{\alpha '}, \label{eq:kappa.20}                                                                                                                                                         
\end{align}
and recalling \eqref{eq:ijk}, we have $i=\{0,\dots ,2\}$, $j=\{0,\dots ,2\}$ and $k=\{0,\dots ,8\}$ which respectively index the left-handed fermions, the right-handed fermions and the Higgs fields. Clearly, there arise 9 Higgs fields from the $bc$ sector.

The second-last term in the right side of \eqref{eq:delta.20} can be used to redefine the shift on the torus as
\begin{align}
	\epsilon^{(2)} & \equiv \frac{1}{27} \left(9 \epsilon _a^{\text{(2)}}+3 \epsilon _b^{\text{(2)}}+3 \epsilon _c^{\text{(2)}}\right). \label{shifts20} 
\end{align}  
The selection rule for the occurrence of a trilinear Yukawa coupling for a given set of indices is given as,
\begin{equation}\label{selection-rule.20}
	i^{(2)} + j^{(2)} + k^{(2)} = 0\; \mathrm{ mod }\; 3.
\end{equation}
Then the suitable rank-3 mass-matrix can be determined by choosing the specific value of $s^{(2)}=-i$. 
\begin{align}
	\delta^{(2)} & =  \frac{j}{3}+\frac{k}{9} \label{eq:s1.20} 
\end{align}
Here, we will ignore other equivalent cases of solutions or cases with rank-1 problem. The mass matrices for up quarks, down quarks and charged leptons have the following general form:
\begin{align}
	Z_{3} &= Z_{3q} \left(
	\begin{array}{ccc}
	T_0 v_1+T_3 v_4+T_6 v_7 & T_2 v_3+T_5 v_6+T_8 v_9 & T_1 v_2+T_4 v_5+T_7 v_8 \\
	T_5 v_3+T_8 v_6+T_2 v_9 & T_4 v_2+T_7 v_5+T_1 v_8 & T_3 v_1+T_6 v_4+T_0 v_7 \\
	T_7 v_2+T_1 v_5+T_4 v_8 & T_6 v_1+T_0 v_4+T_3 v_7 & T_8 v_3+T_2 v_6+T_5 v_9 \\
	\end{array}
	\right),\label{eq:s-i20}
\end{align} 
where $v_i = \left\langle H_{i} \right\rangle$ and the three-point coupling functions are given in terms of Jacobi theta function of the third kind as, 

\begin{align}\label{eq:3couplings20}
	T_k & \equiv  \vartheta \left[\begin{array}{c} 
	\epsilon^{(2)}+\frac{k}{9}\\  \phi^{(2)} \end{array} \right]
	(\frac{9 J^{\text{(2)}}}{\alpha '}),  \quad   k={0,\cdots,8}.
\end{align}
  
\subsubsection{Fermion masses and mixings from 3-point functions in Model~\hyperref[model20]{20}}
In order to accommodate the fermion masses and the quark mixings, we need to fit the up-type quarks mixing matrix \eqref{eq:mixing-quarks}, the down-type quarks matrix \eqref{eq:mass-downquarks} and the masses of the charged-leptons \eqref{eq:mass-chargedleptons}.
 
We set the K\"{a}hler modulus on the second two-torus defined in \eqref{eq:kappa.20} as $\kappa^{(2)} = 45 $  and evaluate the couplings functions \eqref{eq:3couplings20} by setting geometric brane position parameters as $\epsilon^{(2)}_{u} = 0$, $\epsilon^{(2)}_{d} = 0$ and $\epsilon^{(2)}_{e} = 0.0156645$ which yields a nearest fit for the following VEVs,
\begin{equation}\label{eq:VEVs_20} 
	\begin{array}{ll}
		v^u{}_1 = 0.000291324 , & v^d{}_1 = 0.0000282               \\
		v^u{}_2 = 0.0491357   , & v^d{}_2 = -0.0000170692           \\
		v^u{}_3 = 5.71763     , & v^d{}_3 = 0.114556                \\
		v^u{}_4 = 0.0413384   , & v^d{}_4 = -4.24979\times 10^{-12} \\
		v^u{}_5 = 0.0490413   , & v^d{}_5 = 9.08351\times 10^{-8}   \\
		v^u{}_6 = 0.00711503  , & v^d{}_6 = 3.24411\times 10^{-6}   \\
		v^u{}_7 = 0.0413384   , & v^d{}_7 = -4.24979\times 10^{-12} \\
		v^u{}_8 = 0.0313982   , & v^d{}_8 = 0.00320756              \\
		v^u{}_9 = -0.0234364  , & v^d{}_9 = -0.000609616            \\
	\end{array}
\end{equation}
\begin{align} 
	|M_{3u}|&=m_t \left(
	\begin{array}{ccc}
	0.000291  & 0.00122  & 0.008608          \\
	0.00122   & 0.005527 & 0.041338          \\
	0.008608  & 0.041338 & 0.998234          \\
	\end{array}
	\right)\sim M_u ~, \label{eq:Up3_20} \\
	|M_{3d}|&=m_b \left(
	\begin{array}{ccc}
	0.00141   & 0.       & 0.                \\
	0.        & 0.028    & 0.                \\
	0.        & 0.       & 1.                \\
	\end{array}
	\right)\sim D_d ~,\label{eq:Down3_20} \\ 
	|M_{3e}|&=m_\tau \left(
	\begin{array}{ccc}
	0.000862  & 0.000022 & -1.\times 10^{-6} \\
	-0.004106 & 0.010464 & 0.                \\
	0.000188  & 0.       & 1.                \\
	\end{array}
	\right)~.\label{eq:Leptons3_20}
\end{align}While the masses of up-type quarks and the down-type quarks are fitted, in the charged-lepton matrix, only the mass of the tau can be fitted with three-point couplings only. The masses of the muon and the electron are not fitted. Notice, that these results are only at the tree-level and there could indeed be other corrections, such as those coming from higher-dimensional operators, which may contribute most greatly to the muon and the electron masses since they are lighter.

\subsubsection{4-point corrections in Model~\hyperref[model20]{20}}
 
The four-point couplings in Model~\hyperref[model20]{20} in table~\ref{model20} can come from considering interactions of ${a, b, c}$ with $b'$ or $c'$ on the second two-torus as can be seen from the following intersection numbers,
\begin{align}
	I_{bb'}^{(1)} & = 0, & I_{bb'}^{(2)} & = 20, & I_{bb'}^{(3)} & = -1 ,\nonumber                  \\
	I_{cc'}^{(1)} & = 0, & I_{cc'}^{(2)} & = -4, & I_{cc'}^{(3)} & = -1 ,\nonumber                  \\
	I_{bc'}^{(1)} & = 1, & I_{bc'}^{(2)} & = -1, & I_{bc'}^{(3)} & = 1 . \label{eq:4Intersection20} 
\end{align}
There are 4 SM singlet fields $S_L^i$ and 1
Higgs-like state $H^{\prime}_{u, d}$. 

Let us consider four-point interactions with $c'$ with the following parameters with shifts $l=-\frac{k}{3} $ and $\ell=-\frac{k}{9} $ taken along the index $k$,
\begin{align}\label{deltas20}
	\delta & =\frac{i}{I_{ab}^{(2)}} +\frac{j}{I_{ca}^{(2)}} +\frac{k}{I_{bc}^{(2)}} +l , \nonumber               \\
	       & = \frac{i}{3}+\frac{j}{3} ,                                                                          \\
	d      & =\frac{\imath}{I_{c c'}^{(2)}} +\frac{\jmath}{I_{bc'}^{(2)}}+\frac{k}{I_{bc}^{(2)}} +\ell ,\nonumber \\
	       & = -\frac{\imath}{4}-\jmath ,                                                                         
\end{align}
the matrix elements $a_{i,j,\imath}$ on the second torus from the four-point functions results in the following classical 4-point contribution to the mass matrix,
\begin{equation}\label{eq:4point20} 
	Z_{4cl} =  \left(
	\begin{array}{ccc}
		F_0 u_1 w_1+F_1 u_4 w_1 & F_2 u_3 w_1             & F_3 u_2 w_1             \\
		F_2 u_3 w_1             & F_3 u_2 w_1             & F_0 u_1 w_1+F_1 u_4 w_1 \\
		F_3 u_2 w_1             & F_0 u_1 w_1+F_1 u_4 w_1 & F_2 u_3 w_1             \\
	\end{array}
	\right)
\end{equation} 
where $u_i, w_j$ are the VEVs and the couplings are defined as,
\begin{align}\label{eq:4couplings20}
	F_{i} & \equiv  \vartheta \left[\begin{array}{c} 
	\epsilon^{(2)}+\frac{i}{4}\\  \phi^{(2)} \end{array} \right]
	(\frac{9 J^{\text{(2)}}}{\alpha '}),\qquad i={0,\dots,|I_{c c'}^{(2)}|-1}.
\end{align}

Since, we have already fitted the up-type quark matrix $|M_{3u}|$ exactly, so its 4-point correction should be zero,
\begin{equation}
	|M_{4u}| = 0 ,
\end{equation}
which is true by setting all up-type VEVs $u_u^i$ and $w_u^i$ to be zero. 
Therefore, we are essentially concerned with fitting charged-leptons in such a way that corresponding corrections for the down-type quarks remain negligible.
The desired solution can be readily obtained by setting $\epsilon^{(2)}_{4e}=0$ with the following values of the VEVs,
\begin{equation}\label{eq:UdWd20} 
	\begin{array}{l}
		u^d{}_1 = -0.0000203813 \\
		u^d{}_2 = 0.00111661    \\
		u^d{}_3 = 0             \\
		u^d{}_4 = 0             \\
	\end{array}
	\quad,\quad
	\begin{array}{l}
		w^d{}_1 = 1 \\
	\end{array}
\end{equation}
The 4-point correction to the charged-leptons' masses is given by,
\begin{equation}\label{M4e_20}
	|M_{4e}|=m_\tau \left(
	\begin{array}{ccc}
		-0.000645 & 0.        & 0.035336  \\
		0.        & 0.035336  & -0.000645 \\
		0.035336  & -0.000645 & 0.        \\
	\end{array}
	\right)
\end{equation}
which can be added to the matrix obtained from 3-point functions \eqref{eq:Leptons3_20} as,
\begin{equation}\label{M34e_20}
	|M_{3e}|+|M_{4e}|=m_\tau \left(
	\begin{array}{ccc}
		0.000217  & 0.000022  & 0.035335  \\
		-0.004106 & 0.0458    & -0.000645 \\
		0.035524  & -0.000645 & 1.        \\
	\end{array}
	\right) \sim D_e ~,
\end{equation} 
The corrections to down-type quarks' masses can be made negligible by setting $\epsilon^{(1)}_{4d}=0$,
\begin{equation}\label{M4d_20}
	|M_{4d}|\sim \left(
	\begin{array}{ccc}
		-0.001019 & 0.        & 0.05583   \\
		0.        & 0.05583   & -0.001019 \\
		0.05583   & -0.001019 & 0.        \\
	\end{array}
	\right)\sim 0.
\end{equation} Therefore, we can achieve near-exact matching of fermion masses and mixings.
\subsection{Model 20-dual}\label{sec:model-20.5}
In Model~\hyperref[model20.5]{20-dual} the three-point Yukawa couplings arise from the triplet intersections from the branes ${a, b, c}$ on the second two-torus ($r=2$) with 9 pairs of Higgs from $\mathcal{N}=2$ subsector.

\begin{figure}[t]
	\centering
	\includegraphics[width=\textwidth]{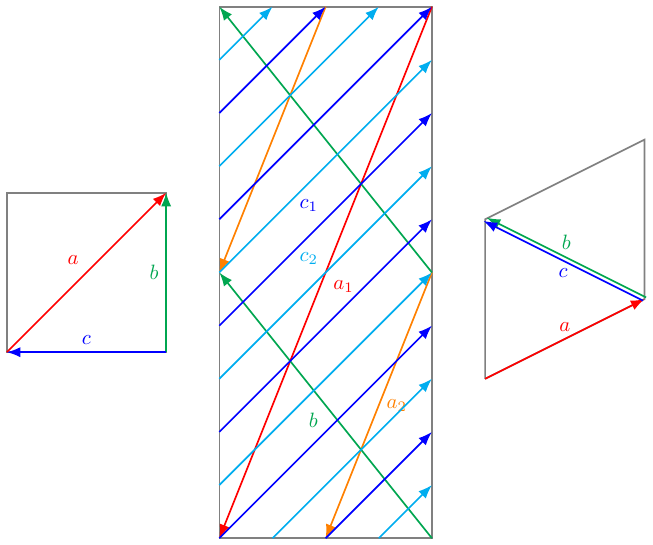}
	\caption{Brane configuration for the three two-tori in Model~\hyperref[model20.5]{20-dual} where the third two-torus is tilted. Fermion mass hierarchies result from the intersections on the second two-torus.}  \label{Fig.20.5}
\end{figure}

Yukawa matrices for the Model~\hyperref[model20.5]{20-dual} are of rank 3 and the three intersections required to form the disk diagrams for the Yukawa couplings all occur on the second torus as shown in figure~\ref{Fig.20.5}. The other two-tori only contribute an overall constant that has no effect in computing the fermion mass ratios. Thus, it is sufficient for our purpose to only focus on the second torus to explain the masses and the mixing in the standard model fermions.

\subsubsection{3-point Yukawa mass-matrices for Model~\hyperref[model20.5]{20-dual}}
From the wrapping numbers listed in table~\ref{model20.5}, the relevant intersection numbers are calculated as,
\begin{align}\label{eq:Intersections.20.5} 
	I_{ab}^{(1)} & =1,  &   
	I_{ab}^{(2)} & =-3, &   
	I_{ab}^{(3)}&=1, \nonumber\\
	I_{bc}^{(1)} & =1,  &   
	I_{bc}^{(2)} & =-9, &   
	I_{bc}^{(3)}&=0, \nonumber\\
	I_{ca}^{(1)} & =-1, &   
	I_{ca}^{(2)} & =-3, &   
	I_{ca}^{(3)}&=-1, 
\end{align}
As the intersection numbers are not coprime, we define the greatest common divisor,
$d^{(1)}=g.c.d.(I_{ab}^{(2)},I_{bc}^{(2)},I_{ca}^{(2)})=3$. Thus, the arguments of the modular theta function as defined in \eqref{eqn:Yinput} can be written as,
\begin{align}
	\delta^{(2)} & = \frac{i^{(2)}}{-3} + \frac{j^{(2)}}{-3} + \frac{k^{(2)}}{-9} +\frac{1}{27} \left(9 \epsilon _a^{\text{(2)}}+3 \epsilon _b^{\text{(2)}}+3 \epsilon _c^{\text{(2)}}\right)+ \frac{s^{(2)}}{3},  \label{eq:delta.20.5} \\
	\phi^{(2)}   & = \frac{1}{3} \left(-9 \theta _a^{\text{(2)}}-3 \theta _b^{\text{(2)}}-3 \theta _c^{\text{(2)}}\right) =0,                                                                                                            \\
	\kappa^{(2)} & = \frac{9 J^{\text{(2)}}}{\alpha '}, \label{eq:kappa.20.5}                                                                                                                                                            
\end{align}
and recalling \eqref{eq:ijk}, we have $i=\{0,\dots ,2\}$, $j=\{0,\dots ,2\}$ and $k=\{0,\dots ,8\}$ which respectively index the left-handed fermions, the right-handed fermions and the Higgs fields. Clearly, there arise 9 Higgs fields from the $bc$ sector.

The second-last term in the right side of \eqref{eq:delta.20.5} can be used to redefine the shift on the torus as
\begin{align}
	\epsilon^{(2)} & \equiv \frac{1}{27} \left(9 \epsilon _a^{\text{(2)}}+3 \epsilon _b^{\text{(2)}}+3 \epsilon _c^{\text{(2)}}\right). \label{shifts20.5} 
\end{align}  
The selection rule for the occurrence of a trilinear Yukawa coupling for a given set of indices is given as,
\begin{equation}\label{selection-rule.20.5}
	i^{(2)} + j^{(2)} + k^{(2)} = 0\; \mathrm{ mod }\; 3.
\end{equation}
Then the suitable rank-3 mass-matrix can be determined by choosing the specific value of $s^{(2)}=i$. 
\begin{align}
	\delta^{(2)} & =  -\frac{j}{3}-\frac{k}{9} \label{eq:s1.20.5} 
\end{align}
Here, we will ignore other equivalent cases of solutions or cases with rank-1 problem. The mass matrices for up quarks, down quarks and charged leptons have the following general form:
\begin{align}
	Z_{3} &= Z_{3q} \left(
	\begin{array}{ccc}
	T_0 v_1+T_6 v_4+T_3 v_7 & T_7 v_3+T_4 v_6+T_1 v_9 & T_8 v_2+T_5 v_5+T_2 v_8 \\
	T_4 v_3+T_1 v_6+T_7 v_9 & T_5 v_2+T_2 v_5+T_8 v_8 & T_6 v_1+T_3 v_4+T_0 v_7 \\
	T_2 v_2+T_8 v_5+T_5 v_8 & T_3 v_1+T_0 v_4+T_6 v_7 & T_1 v_3+T_7 v_6+T_4 v_9 \\
	\end{array}
	\right),\label{eq:s-i20.5}
\end{align} 
where $v_i = \left\langle H_{i} \right\rangle$ and the three-point coupling functions are given in terms of Jacobi theta function of the third kind as, 

\begin{align}\label{eq:3couplings20.5}
	T_k & \equiv  \vartheta \left[\begin{array}{c} 
	\epsilon^{(2)}+\frac{k}{9}\\  \phi^{(2)} \end{array} \right]
	(\frac{9 J^{\text{(2)}}}{\alpha '}),  \quad   k={0,\cdots,8}.
\end{align}
  
\subsubsection{Fermion masses and mixings from 3-point functions in Model~\hyperref[model20.5]{20-dual}}
In order to accommodate the fermion masses and the quark mixings, we need to fit the up-type quarks mixing matrix \eqref{eq:mixing-quarks}, the down-type quarks matrix \eqref{eq:mass-downquarks} and the masses of the charged-leptons \eqref{eq:mass-chargedleptons}.
 
We set the K\"{a}hler modulus on the second two-torus defined in \eqref{eq:kappa.20.5} as $\kappa^{(2)} = 45 $  and evaluate the couplings functions \eqref{eq:3couplings20.5} by setting geometric brane position parameters as $\epsilon^{(2)}_{u} = 0$, $\epsilon^{(2)}_{d} = 0$ and $\epsilon^{(2)}_{e} = -0.0156645$ which yields a nearest fit for the following VEVs,
\begin{equation}\label{eq:VEVs_20.5} 
	\begin{array}{ll}
		v^u{}_1 = 0.000291324 , & v^d{}_1 = 0.0000282               \\
		v^u{}_2 = 0.0491357   , & v^d{}_2 = -0.0000170692           \\
		v^u{}_3 = 5.71763     , & v^d{}_3 = 0.114556                \\
		v^u{}_4 = 0.0413384   , & v^d{}_4 = -4.24979\times 10^{-12} \\
		v^u{}_5 = 0.0490413   , & v^d{}_5 = 9.08351\times 10^{-8}   \\
		v^u{}_6 = 0.00711503  , & v^d{}_6 = 3.24411\times 10^{-6}   \\
		v^u{}_7 = 0.0413384   , & v^d{}_7 = -4.24979\times 10^{-12} \\
		v^u{}_8 = 0.0313982   , & v^d{}_8 = 0.00320756              \\
		v^u{}_9 = -0.0234364  , & v^d{}_9 = -0.000609616            \\
	\end{array}
\end{equation}
\begin{align} 
	|M_{3u}|&=m_t \left(
	\begin{array}{ccc}
	0.000291  & 0.00122  & 0.008608          \\
	0.00122   & 0.005527 & 0.041338          \\
	0.008608  & 0.041338 & 0.998234          \\
	\end{array}
	\right)\sim M_u ~, \label{eq:Up3_20.5} \\
	|M_{3d}|&=m_b \left(
	\begin{array}{ccc}
	0.00141   & 0.       & 0.                \\
	0.        & 0.028    & 0.                \\
	0.        & 0.       & 1.                \\
	\end{array}
	\right)\sim D_d ~,\label{eq:Down3_20.5} \\ 
	|M_{3e}|&=m_\tau \left(
	\begin{array}{ccc}
	0.000862  & 0.000022 & -1.\times 10^{-6} \\
	-0.004106 & 0.010464 & 0.                \\
	0.000188  & 0.       & 1.                \\
	\end{array}
	\right)~.\label{eq:Leptons3_20.5}
\end{align}While the masses of up-type quarks and the down-type quarks are fitted, in the charged-lepton matrix, only the mass of the tau can be fitted with three-point couplings only. The masses of the muon and the electron are not fitted. Notice, that these results are only at the tree-level and there could indeed be other corrections, such as those coming from higher-dimensional operators, which may contribute most greatly to the muon and the electron masses since they are lighter.

\subsubsection{4-point corrections in Model~\hyperref[model20.5]{20-dual}}
 
The four-point couplings in Model~\hyperref[model20.5]{20-dual} in table~\ref{model20.5} can come from considering interactions of ${a, b, c}$ with $b'$ or $c'$ on the second two-torus as can be seen from the following intersection numbers,
\begin{align}
	I_{bb'}^{(1)} & = 0, & I_{bb'}^{(2)} & = -4, & I_{bb'}^{(3)} & = -1 ,\nonumber                    \\
	I_{cc'}^{(1)} & = 0, & I_{cc'}^{(2)} & = 20, & I_{cc'}^{(3)} & = -1 ,\nonumber                    \\
	I_{bc'}^{(1)} & = 1, & I_{bc'}^{(2)} & = -1, & I_{bc'}^{(3)} & = 1 . \label{eq:4Intersection20.5} 
\end{align}
There are 4 SM singlet fields $S_L^i$ and 1
Higgs-like state $H^{\prime}_{u, d}$. 

Let us consider four-point interactions with $b'$ with the following parameters with shifts $l=\frac{k}{3} $ and $\ell=\frac{k}{9} $ taken along the index $k$,
\begin{align}\label{deltas20.5}
	\delta & =\frac{i}{I_{ab}^{(2)}} +\frac{j}{I_{ca}^{(2)}} +\frac{k}{I_{bc}^{(2)}} +l , \nonumber               \\
	       & = -\frac{i}{3}-\frac{j}{3} ,                                                                         \\
	d      & =\frac{\imath}{I_{b b'}^{(2)}} +\frac{\jmath}{I_{bc'}^{(2)}}+\frac{k}{I_{bc}^{(2)}} +\ell ,\nonumber \\
	       & = -\frac{\imath}{4}-\jmath ,                                                                         
\end{align}
the matrix elements $a_{i,j,\imath}$ on the second torus from the four-point functions results in the following classical 4-point contribution to the mass matrix,
\begin{equation}\label{eq:4point20.5} 
	Z_{4cl} =  \left(
	\begin{array}{ccc}
		F_0 u_1 w_1+F_1 u_4 w_1 & F_2 u_3 w_1             & F_3 u_2 w_1             \\
		F_2 u_3 w_1             & F_3 u_2 w_1             & F_0 u_1 w_1+F_1 u_4 w_1 \\
		F_3 u_2 w_1             & F_0 u_1 w_1+F_1 u_4 w_1 & F_2 u_3 w_1             \\
	\end{array}
	\right)
\end{equation} 
where $u_i, w_j$ are the VEVs and the couplings are defined as,
\begin{align}\label{eq:4couplings20.5}
	F_{i} & \equiv  \vartheta \left[\begin{array}{c} 
	\epsilon^{(2)}+\frac{i}{4}\\  \phi^{(2)} \end{array} \right]
	(\frac{9 J^{\text{(2)}}}{\alpha '}),\qquad i={0,\dots,|I_{b b'}^{(2)}|-1}.
\end{align}

Since, we have already fitted the up-type quark matrix $|M_{3u}|$ exactly, so its 4-point correction should be zero,
\begin{equation}
	|M_{4u}| = 0 ,
\end{equation}
which is true by setting all up-type VEVs $u_u^i$ and $w_u^i$ to be zero. 
Therefore, we are essentially concerned with fitting charged-leptons in such a way that corresponding corrections for the down-type quarks remain negligible.
The desired solution can be readily obtained by setting $\epsilon^{(2)}_{4e}=0$ with the following values of the VEVs,
\begin{equation}\label{eq:UdWd20.5} 
	\begin{array}{l}
		u^d{}_1 = -0.0000203813 \\
		u^d{}_2 = 7.6773        \\
		u^d{}_3 = 0             \\
		u^d{}_4 = 0             \\
	\end{array}
	\quad,\quad
	\begin{array}{l}
		w^d{}_1 = 1 \\
	\end{array}
\end{equation}
The 4-point correction to the charged-leptons' masses is given by,
\begin{equation}\label{M4e_20.5}
	|M_{4e}|=m_\tau \left(
	\begin{array}{ccc}
		-0.000645 & 0.        & 0.035336  \\
		0.        & 0.035336  & -0.000645 \\
		0.035336  & -0.000645 & 0.        \\
	\end{array}
	\right)
\end{equation}
which can be added to the matrix obtained from 3-point functions \eqref{eq:Leptons3_20.5} as,
\begin{equation}\label{M34e_20.5}
	|M_{3e}|+|M_{4e}|=m_\tau \left(
	\begin{array}{ccc}
		0.000217  & 0.000022  & 0.035335  \\
		-0.004106 & 0.0458    & -0.000645 \\
		0.035524  & -0.000645 & 1.        \\
	\end{array}
	\right) \sim D_e ~,
\end{equation} 
The corrections to down-type quarks' masses can be made negligible by setting $\epsilon^{(1)}_{4d}=0$,
\begin{equation}\label{M4d_20.5}
	|M_{4d}|\sim \left(
	\begin{array}{ccc}
		-0.001019 & 0.        & 0.05583   \\
		0.        & 0.05583   & -0.001019 \\
		0.05583   & -0.001019 & 0.        \\
	\end{array}
	\right)\sim 0.
\end{equation} Therefore, we can achieve near-exact matching of fermion masses and mixings.
\section{Models with 12 Higgs from $\mathcal{N}=2$ sector}\label{sec:12Higgs}
\subsection{Model 21}\label{sec:model-21}
In Model~\hyperref[model21]{21} the three-point Yukawa couplings arise from the triplet intersections from the branes ${a, b, c}$ on the second two-torus ($r=2$) with 12 pairs of Higgs from $\mathcal{N}=2$ subsector.

\begin{figure}[t]
	\centering
	\includegraphics[width=\textwidth]{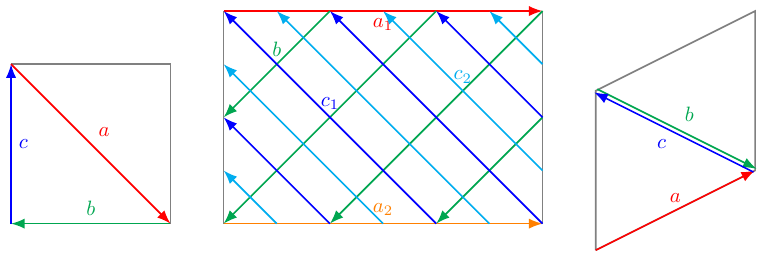}
	\caption{Brane configuration for the three two-tori in Model~\hyperref[model21]{21} where the third two-torus is tilted. Fermion mass hierarchies result from the intersections on the second two-torus.}  \label{Fig.21}
\end{figure}

Yukawa matrices for the Model~\hyperref[model21]{21} are of rank 3 and the three intersections required to form the disk diagrams for the Yukawa couplings all occur on the second torus as shown in figure~\ref{Fig.21}. The other two-tori only contribute an overall constant that has no effect in computing the fermion mass ratios. Thus, it is sufficient for our purpose to only focus on the second torus to explain the masses and the mixing in the standard model fermions.

\subsubsection{3-point Yukawa mass-matrices for Model~\hyperref[model21]{21}}
From the wrapping numbers listed in table~\ref{model21}, the relevant intersection numbers are calculated as,
\begin{align}\label{eq:Intersections.21} 
	I_{ab}^{(1)} & =-1,  &   
	I_{ab}^{(2)} & =-3,  &   
	I_{ab}^{(3)}&=-1, \nonumber\\
	I_{bc}^{(1)} & =-1,  &   
	I_{bc}^{(2)} & =-12, &   
	I_{bc}^{(3)}&=0, \nonumber\\
	I_{ca}^{(1)} & =-1,  &   
	I_{ca}^{(2)} & =-3,  &   
	I_{ca}^{(3)}&=-1, 
\end{align}
As the intersection numbers are not coprime, we define the greatest common divisor,
$d^{(1)}=g.c.d.(I_{ab}^{(2)},I_{bc}^{(2)},I_{ca}^{(2)})=3$. Thus, the arguments of the modular theta function as defined in \eqref{eqn:Yinput} can be written as,
\begin{align}
	\delta^{(2)} & = \frac{i^{(2)}}{-3} + \frac{j^{(2)}}{-3} + \frac{k^{(2)}}{-12} +\frac{1}{36} \left(12 \epsilon _a^{\text{(2)}}+3 \epsilon _b^{\text{(2)}}+3 \epsilon _c^{\text{(2)}}\right)+ \frac{s^{(2)}}{3},  \label{eq:delta.21} \\
	\phi^{(2)}   & = \frac{1}{3} \left(-12 \theta _a^{\text{(2)}}-3 \theta _b^{\text{(2)}}-3 \theta _c^{\text{(2)}}\right) =0,                                                                                                           \\
	\kappa^{(2)} & = \frac{12 J^{\text{(2)}}}{\alpha '}, \label{eq:kappa.21}                                                                                                                                                             
\end{align}
and recalling \eqref{eq:ijk}, we have $i=\{0,\dots ,2\}$, $j=\{0,\dots ,2\}$ and $k=\{0,\dots ,11\}$ which respectively index the left-handed fermions, the right-handed fermions and the Higgs fields. Clearly, there arise 12 Higgs fields from the $bc$ sector.

The second-last term in the right side of \eqref{eq:delta.21} can be used to redefine the shift on the torus as
\begin{align}
	\epsilon^{(2)} & \equiv \frac{1}{36} \left(12 \epsilon _a^{\text{(2)}}+3 \epsilon _b^{\text{(2)}}+3 \epsilon _c^{\text{(2)}}\right). \label{shifts21} 
\end{align}  
The selection rule for the occurrence of a trilinear Yukawa coupling for a given set of indices is given as,
\begin{equation}\label{selection-rule.21}
	i^{(2)} + j^{(2)} + k^{(2)} = 0\; \mathrm{ mod }\; 3.
\end{equation}
Then the suitable rank-3 mass-matrix can be determined by choosing the specific value of $s^{(2)}=j-i$. 
\begin{align}
	\delta^{(2)} & =  -\frac{2 i}{3}-\frac{k}{12} \label{eq:s1.21} 
\end{align}
Here, we will ignore other equivalent cases of solutions or cases with rank-1 problem. The mass matrices for up quarks, down quarks and charged leptons have the following general form:
\begin{align}
	Z_{3} &= Z_{3q} \left(
	\begin{array}{ccc}
	T_0 v_1+T_9 v_4+T_6 v_7+T_3 v_{10}    & T_{10} v_3+T_7 v_6+T_4 v_9+T_1 v_{12} & T_{11} v_2+T_8 v_5+T_5 v_8+T_2 v_{11} \\
	T_2 v_3+T_{11} v_6+T_8 v_9+T_5 v_{12} & T_3 v_2+T_0 v_5+T_9 v_8+T_6 v_{11}    & T_4 v_1+T_1 v_4+T_{10} v_7+T_7 v_{10} \\
	T_7 v_2+T_4 v_5+T_1 v_8+T_{10} v_{11} & T_8 v_1+T_5 v_4+T_2 v_7+T_{11} v_{10} & T_6 v_3+T_3 v_6+T_0 v_9+T_9 v_{12}    \\
	\end{array}
	\right),\label{eq:s-i21}
\end{align} 
where $v_i = \left\langle H_{i} \right\rangle$ and the three-point coupling functions are given in terms of Jacobi theta function of the third kind as, 

\begin{align}\label{eq:3couplings21}
	T_k & \equiv  \vartheta \left[\begin{array}{c} 
	\epsilon^{(2)}+\frac{k}{12}\\  \phi^{(2)} \end{array} \right]
	(\frac{12 J^{\text{(2)}}}{\alpha '}),  \quad   k={0,\cdots,11}.
\end{align}
  
\subsubsection{Fermion masses and mixings from 3-point functions in Model~\hyperref[model21]{21}}
In order to accommodate the fermion masses and the quark mixings, we need to fit the up-type quarks mixing matrix \eqref{eq:mixing-quarks}, the down-type quarks matrix \eqref{eq:mass-downquarks} and the masses of the charged-leptons \eqref{eq:mass-chargedleptons}.
 
We set the K\"{a}hler modulus on the second two-torus defined in \eqref{eq:kappa.21} as $\kappa^{(2)} = 64 $  and evaluate the couplings functions \eqref{eq:3couplings21} by setting geometric brane position parameters as $\epsilon^{(2)}_{u} = 0$, $\epsilon^{(2)}_{d} = 0$ and $\epsilon^{(2)}_{e} = \frac{1}{2}$ which yields an \emph{exact} fitting for the following VEVs,
\begin{equation}\label{eq:VEVs_21} 
	\begin{array}{ll}
		v^u{}_1 = 0.000291336               , & v^d{}_1 = 0.0000282                 \\
		v^u{}_2 = -0.0000838211             , & v^d{}_2 = -0.0000219475             \\
		v^u{}_3 = 1.32337                   , & v^d{}_3 = 0.0316                    \\
		v^u{}_4 = -4.42684\times 10^{-6}    , & v^d{}_4 = -1.03987\times 10^{-7}    \\
		v^u{}_5 = 0.00552729                , & v^d{}_5 = 0.00056                   \\
		v^u{}_6 = -0.0151378                , & v^d{}_6 = -0.000479202              \\
		v^u{}_7 = 11.0134                   , & v^d{}_7 = 6.8572\times 10^{-6}      \\
		v^u{}_8 = -0.0000838211             , & v^d{}_8 = -0.0000219475             \\
		v^u{}_9 = 0.998234                  , & v^d{}_9 = 0.02                      \\
		v^u{}_{10} = -4.42684\times 10^{-6} , & v^d{}_{10} = -1.03987\times 10^{-7} \\
		v^u{}_{11} = 2.29873                , & v^d{}_{11} = 0.00144728             \\
		v^u{}_{12} = -0.0151378             , & v^d{}_{12} = -0.000479202           \\
	\end{array}
\end{equation}
\begin{align} 
	|M_{3u}|&=m_t \left(
	\begin{array}{ccc}
	0.000291  & 0.00122          & 0.008608         \\
	0.00122   & 0.005527         & 0.041338         \\
	0.008608  & 0.041338         & 0.998234         \\
	\end{array}
	\right)\sim M_u ~, \label{eq:Up3_21} \\
	|M_{3n}|&=m_\nu \left(
	\begin{array}{ccc}
	11.0134   & 0.               & 0.               \\
	0.        & 2.29873          & 0.               \\
	0.        & 0.               & 1.32337          \\
	\end{array}
	\right)~,\label{eq:nu3_21} \\|M_{3d}|&=m_b \left(
	\begin{array}{ccc}
	0.00141   & 0.               & 0.               \\
	0.        & 0.028            & 0.               \\
	0.        & 0.               & 1.               \\
	\end{array}
	\right)\sim D_d ~,\label{eq:Down3_21} \\ 
	|M_{3e}|&=m_\tau \left(
	\begin{array}{ccc}
	0.000217  & -0.001378        & -0.000105        \\
	-0.001378 & 0.0458           & 3.\times 10^{-6} \\
	-0.000105 & 3.\times 10^{-6} & 1.               \\
	\end{array}
	\right)\sim D_e~.\label{eq:Leptons3_21}
\end{align} Since the intersection number $I_{bc'}^{(2)}=0$, there are no further corrections from the four-point functions.
\subsection{Model 22}\label{sec:model-22}
In Model~\hyperref[model22]{22} the three-point Yukawa couplings arise from the triplet intersections from the branes ${a, b, c}$ on the first two-torus ($r=1$) with 12 pairs of Higgs from $\mathcal{N}=2$ subsector.

\begin{figure}[t]
	\centering
	\includegraphics[width=\textwidth]{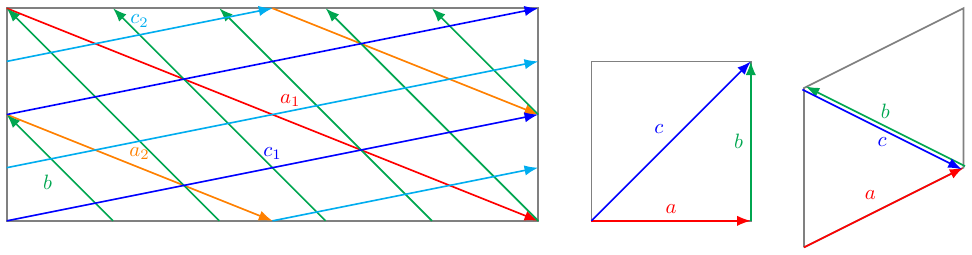}
	\caption{Brane configuration for the three two-tori in Model~\hyperref[model22]{22} where the third two-torus is tilted. Fermion mass hierarchies result from the intersections on the first two-torus.}  \label{Fig.22}
\end{figure}

Yukawa matrices for the Model~\hyperref[model22]{22} are of rank 3 and the three intersections required to form the disk diagrams for the Yukawa couplings all occur on the first torus as shown in figure~\ref{Fig.22}. The other two-tori only contribute an overall constant that has no effect in computing the fermion mass ratios. Thus, it is sufficient for our purpose to only focus on the first torus to explain the masses and the mixing in the standard model fermions.

\subsubsection{3-point Yukawa mass-matrices for Model~\hyperref[model22]{22}}
From the wrapping numbers listed in table~\ref{model22}, the relevant intersection numbers are calculated as,
\begin{align}\label{eq:Intersections.22} 
	I_{ab}^{(1)} & =3,   &   
	I_{ab}^{(2)} & =1,   &   
	I_{ab}^{(3)}&=1, \nonumber\\
	I_{bc}^{(1)} & =-12, &   
	I_{bc}^{(2)} & =-1,  &   
	I_{bc}^{(3)}&=0, \nonumber\\
	I_{ca}^{(1)} & =-3,  &   
	I_{ca}^{(2)} & =-1,  &   
	I_{ca}^{(3)}&=1, 
\end{align}
As the intersection numbers are not coprime, we define the greatest common divisor,
$d^{(1)}=g.c.d.(I_{ab}^{(1)},I_{bc}^{(1)},I_{ca}^{(1)})=3$. Thus, the arguments of the modular theta function as defined in \eqref{eqn:Yinput} can be written as,
\begin{align}
	\delta^{(1)} & = \frac{i^{(1)}}{3} + \frac{j^{(1)}}{-3} + \frac{k^{(1)}}{-12} +\frac{1}{36} \left(-12 \epsilon _a^{\text{(1)}}-3 \epsilon _b^{\text{(1)}}+3 \epsilon _c^{\text{(1)}}\right)+ \frac{s^{(1)}}{3},  \label{eq:delta.22} \\
	\phi^{(1)}   & = \frac{1}{3} \left(-12 \theta _a^{\text{(1)}}-3 \theta _b^{\text{(1)}}+3 \theta _c^{\text{(1)}}\right) =0,                                                                                                           \\
	\kappa^{(1)} & = \frac{12 J^{\text{(1)}}}{\alpha '}, \label{eq:kappa.22}                                                                                                                                                             
\end{align}
and recalling \eqref{eq:ijk}, we have $i=\{0,\dots ,2\}$, $j=\{0,\dots ,2\}$ and $k=\{0,\dots ,11\}$ which respectively index the left-handed fermions, the right-handed fermions and the Higgs fields. Clearly, there arise 12 Higgs fields from the $bc$ sector.

The second-last term in the right side of \eqref{eq:delta.22} can be used to redefine the shift on the torus as
\begin{align}
	\epsilon^{(1)} & \equiv \frac{1}{36} \left(-12 \epsilon _a^{\text{(1)}}-3 \epsilon _b^{\text{(1)}}+3 \epsilon _c^{\text{(1)}}\right). \label{shifts22} 
\end{align}  
The selection rule for the occurrence of a trilinear Yukawa coupling for a given set of indices is given as,
\begin{equation}\label{selection-rule.22}
	i^{(1)} + j^{(1)} + k^{(1)} = 0\; \mathrm{ mod }\; 3.
\end{equation}
Then the suitable rank-3 mass-matrix can be determined by choosing the specific value of $s^{(1)}=j$. 
\begin{align}
	\delta^{(1)} & =  \frac{i}{3}-\frac{k}{12} \label{eq:s1.22} 
\end{align}
Here, we will ignore other equivalent cases of solutions or cases with rank-1 problem. The mass matrices for up quarks, down quarks and charged leptons have the following general form:
\begin{align}
	Z_{3} &= Z_{3q} \left(
	\begin{array}{ccc}
	T_0 v_1+T_9 v_4+T_6 v_7+T_3 v_{10}    & T_{10} v_3+T_7 v_6+T_4 v_9+T_1 v_{12} & T_{11} v_2+T_8 v_5+T_5 v_8+T_2 v_{11} \\
	T_2 v_3+T_{11} v_6+T_8 v_9+T_5 v_{12} & T_3 v_2+T_0 v_5+T_9 v_8+T_6 v_{11}    & T_4 v_1+T_1 v_4+T_{10} v_7+T_7 v_{10} \\
	T_7 v_2+T_4 v_5+T_1 v_8+T_{10} v_{11} & T_8 v_1+T_5 v_4+T_2 v_7+T_{11} v_{10} & T_6 v_3+T_3 v_6+T_0 v_9+T_9 v_{12}    \\
	\end{array}
	\right),\label{eq:s-i22}
\end{align} 
where $v_i = \left\langle H_{i} \right\rangle$ and the three-point coupling functions are given in terms of Jacobi theta function of the third kind as, 

\begin{align}\label{eq:3couplings22}
	T_k & \equiv  \vartheta \left[\begin{array}{c} 
	\epsilon^{(1)}+\frac{k}{12}\\  \phi^{(1)} \end{array} \right]
	(\frac{12 J^{\text{(1)}}}{\alpha '}),  \quad   k={0,\cdots,11}.
\end{align}
  
\subsubsection{Fermion masses and mixings from 3-point functions in Model~\hyperref[model22]{22}}
In order to accommodate the fermion masses and the quark mixings, we need to fit the up-type quarks mixing matrix \eqref{eq:mixing-quarks}, the down-type quarks matrix \eqref{eq:mass-downquarks} and the masses of the charged-leptons \eqref{eq:mass-chargedleptons}.
 
We set the K\"{a}hler modulus on the first two-torus defined in \eqref{eq:kappa.22} as $\kappa^{(1)} = 64 $  and evaluate the couplings functions \eqref{eq:3couplings22} by setting geometric brane position parameters as $\epsilon^{(1)}_{u} = 0$, $\epsilon^{(1)}_{d} = 0$ and $\epsilon^{(1)}_{e} = \frac{1}{2}$ which yields an \emph{exact} fitting for the following VEVs,
\begin{equation}\label{eq:VEVs_22} 
	\begin{array}{ll}
		v^u{}_1 = 0.000291336               , & v^d{}_1 = 0.0000282001     \\
		v^u{}_2 = -0.0000838211             , & v^d{}_2 = -0.000163191     \\
		v^u{}_3 = 1.32337                   , & v^d{}_3 = 0.0173018        \\
		v^u{}_4 = -4.42684\times 10^{-6}    , & v^d{}_4 = -0.0000173322    \\
		v^u{}_5 = 0.00552729                , & v^d{}_5 = 0.000560001      \\
		v^u{}_6 = -0.0151378                , & v^d{}_6 = -0.000262376     \\
		v^u{}_7 = 11.0134                   , & v^d{}_7 = 0.00114294       \\
		v^u{}_8 = -0.0000838211             , & v^d{}_8 = -0.000163191     \\
		v^u{}_9 = 0.998234                  , & v^d{}_9 = 0.02             \\
		v^u{}_{10} = -4.42684\times 10^{-6} , & v^d{}_{10} = -0.0000173322 \\
		v^u{}_{11} = 2.29873                , & v^d{}_{11} = 0.0107613     \\
		v^u{}_{12} = -0.0151378             , & v^d{}_{12} = -0.000262376  \\
	\end{array}
\end{equation}
\begin{align} 
	|M_{3u}|&=m_t \left(
	\begin{array}{ccc}
	0.000291  & 0.00122   & 0.008608  \\
	0.00122   & 0.005527  & 0.041338  \\
	0.008608  & 0.041338  & 0.998234  \\
	\end{array}
	\right)\sim M_u ~, \label{eq:Up3_22} \\
	|M_{3n}|&=m_\nu \left(
	\begin{array}{ccc}
	11.0134   & 0.        & 0.        \\
	0.        & 2.29873   & 0.        \\
	0.        & 0.        & 1.32337   \\
	\end{array}
	\right)~,\label{eq:nu3_22} \\|M_{3d}|&=m_b \left(
	\begin{array}{ccc}
	0.00141   & 0.        & 0.        \\
	0.        & 0.028     & 0.        \\
	0.        & 0.        & 1.        \\
	\end{array}
	\right)\sim D_d ~,\label{eq:Down3_22} \\ 
	|M_{3e}|&=m_\tau \left(
	\begin{array}{ccc}
	0.036169  & 0.00032   & -0.001212 \\
	0.00032   & 0.340547  & -0.000132 \\
	-0.001212 & -0.000132 & 0.547527  \\
	\end{array}
	\right)\sim D_e~.\label{eq:Leptons3_22}
\end{align} Notice, that these results are only at the tree-level and there could indeed be other corrections, such as those coming from higher-dimensional operators, which may contribute most greatly to the charm and the down quarks' masses since they are lighter. In order to preserve the exact-fitting of up-quarks mixed form matrix and the neutrino-masses already achieved from 3-point functions, we can switch off the corrections from the 4-point functions by taking all up-type VEVs to be zero. However, the down-type VEVS from 4-point functions will be needed to explain the leptons' mixing.

\subsubsection{4-point corrections in Model~\hyperref[model22]{22}}
 
The four-point couplings in Model~\hyperref[model22]{22} in table~\ref{model22} can come from considering interactions of ${a, b, c}$ with $b'$ or $c'$ on the first two-torus as can be seen from the following intersection numbers,
\begin{align}
	I_{bb'}^{(1)} & = -20, & I_{bb'}^{(2)} & = 0,  & I_{bb'}^{(3)} & = -1 ,\nonumber                   \\
	I_{cc'}^{(1)} & = 4,   & I_{cc'}^{(2)} & = 2,  & I_{cc'}^{(3)} & = -1 ,\nonumber                   \\
	I_{bc'}^{(1)} & = -8,  & I_{bc'}^{(2)} & = -1, & I_{bc'}^{(3)} & = -1 . \label{eq:4Intersection22} 
\end{align}
There are 20 SM singlet fields $S_L^i$ and 8
Higgs-like state $H^{\prime}_{u, d}$. 

Let us consider four-point interactions with $b'$ with the following parameters with shifts $l=\frac{k}{4} $ and $\ell=\frac{k}{3} $ taken along the index $k$,
\begin{align}\label{deltas22}
	\delta & =\frac{i}{I_{ab}^{(1)}} +\frac{j}{I_{ca}^{(1)}} +\frac{k}{I_{bc}^{(1)}} +l , \nonumber               \\
	       & = \frac{i}{3}-\frac{j}{3} ,                                                                          \\
	d      & =\frac{\imath}{I_{b b'}^{(1)}} +\frac{\jmath}{I_{bc'}^{(1)}}+\frac{k}{I_{bc}^{(1)}} +\ell ,\nonumber \\
	       & = -\frac{\imath}{20}-\frac{\jmath}{8} ,                                                              
\end{align}
the matrix elements $a_{i,j,\imath}$ on the first torus from the four-point functions results in the following classical 4-point contribution to the mass matrix,
\begin{equation}\label{eq:4point22} 
	Z_{4cl} =  
	\resizebox{.95\textwidth}{!}{
		$\begin{aligned}
			  & \left(                 
			\begin{array}{c}
			F_0 u_1 w_1+F_{16} u_{13} w_1+F_{13} u_{12} w_2+F_{10} u_{11} w_3+F_7 u_{10} w_4+F_4 u_9 w_5+F_1 u_8 w_6+F_{17} u_{20} w_6+F_{38} u_7 w_7+F_{14} u_{19} w_7+F_{35} u_6 w_8+F_{11} u_{18} w_8\\
			F_{18} u_{12} w_1+F_{15} u_{11} w_2+F_{12} u_{10} w_3+F_9 u_9 w_4+F_6 u_8 w_5+F_{22} u_{20} w_5+F_3 u_7 w_6+F_{19} u_{19} w_6+F_0 u_6 w_7+F_{16} u_{18} w_7+F_{37} u_5 w_8+F_{13} u_{17} w_8\\
			F_{20} u_{11} w_1+F_{17} u_{10} w_2+F_{14} u_9 w_3+F_{11} u_8 w_4+F_{27} u_{20} w_4+F_8 u_7 w_5+F_{24} u_{19} w_5+F_5 u_6 w_6+F_{21} u_{18} w_6+F_2 u_5 w_7+F_{18} u_{17} w_7+F_{39} u_4 w_8+F_{15} u_{16} w_8\\
			\end{array}\right.\\
			  & \begin{array}{c}       
			F_{18} u_{12} w_1+F_{15} u_{11} w_2+F_{12} u_{10} w_3+F_9 u_9 w_4+F_6 u_8 w_5+F_{22} u_{20} w_5+F_3 u_7 w_6+F_{19} u_{19} w_6+F_0 u_6 w_7+F_{16} u_{18} w_7+F_{37} u_5 w_8+F_{13} u_{17} w_8\\
			F_{20} u_{11} w_1+F_{17} u_{10} w_2+F_{14} u_9 w_3+F_{11} u_8 w_4+F_{27} u_{20} w_4+F_8 u_7 w_5+F_{24} u_{19} w_5+F_5 u_6 w_6+F_{21} u_{18} w_6+F_2 u_5 w_7+F_{18} u_{17} w_7+F_{39} u_4 w_8+F_{15} u_{16} w_8\\
			F_{22} u_{10} w_1+F_{19} u_9 w_2+F_{16} u_8 w_3+F_{32} u_{20} w_3+F_{13} u_7 w_4+F_{29} u_{19} w_4+F_{10} u_6 w_5+F_{26} u_{18} w_5+F_7 u_5 w_6+F_{23} u_{17} w_6+F_4 u_4 w_7+F_{20} u_{16} w_7+F_1 u_3 w_8+F_{17} u_{15} w_8\\
			\end{array}\\
			  & \left.\begin{array}{c} 
			F_{20} u_{11} w_1+F_{17} u_{10} w_2+F_{14} u_9 w_3+F_{11} u_8 w_4+F_{27} u_{20} w_4+F_8 u_7 w_5+F_{24} u_{19} w_5+F_5 u_6 w_6+F_{21} u_{18} w_6+F_2 u_5 w_7+F_{18} u_{17} w_7+F_{39} u_4 w_8+F_{15} u_{16} w_8\\
			F_{22} u_{10} w_1+F_{19} u_9 w_2+F_{16} u_8 w_3+F_{32} u_{20} w_3+F_{13} u_7 w_4+F_{29} u_{19} w_4+F_{10} u_6 w_5+F_{26} u_{18} w_5+F_7 u_5 w_6+F_{23} u_{17} w_6+F_4 u_4 w_7+F_{20} u_{16} w_7+F_1 u_3 w_8+F_{17} u_{15} w_8\\
			F_{24} u_9 w_1+F_{21} u_8 w_2+F_{37} u_{20} w_2+F_{18} u_7 w_3+F_{34} u_{19} w_3+F_{15} u_6 w_4+F_{31} u_{18} w_4+F_{12} u_5 w_5+F_{28} u_{17} w_5+F_9 u_4 w_6+F_{25} u_{16} w_6+F_6 u_3 w_7+F_{22} u_{15} w_7+F_3 u_2 w_8+F_{19} u_{14} w_8\\
			\end{array}\right),
		\end{aligned}$
	} 
\end{equation} 
where $u_i, w_j$ are the VEVs and the couplings are defined as,
\begin{align}\label{eq:4couplings22}
	F_{8i} & \equiv  \vartheta \left[\begin{array}{c} 
	\epsilon^{(1)}+\frac{i}{20}\\  \phi^{(1)} \end{array} \right]
	(\frac{12 J^{\text{(1)}}}{\alpha '}),\qquad i={0,\dots,|I_{b b'}^{(1)}|-1}.
\end{align}

Since, we have already fitted the up-type quark matrix $|M_{3u}|$ exactly, so its 4-point correction should be zero,
\begin{equation}
	|M_{4u}| = 0 ,
\end{equation}
which is true by setting all up-type VEVs $u_u^i$ and $w_u^i$ to be zero. 
Therefore, we are essentially concerned with fitting charged-leptons in such a way that corresponding corrections for the down-type quarks remain negligible.
The desired solution can be readily obtained by setting $\epsilon^{(1)}_{4e}=0$ with the following values of the VEVs,
\begin{equation}\label{eq:UdWd22} 
	\begin{array}{l}
		u^d{}_1 = 0          \\
		u^d{}_2 = 0          \\
		u^d{}_3 = 0.0158255  \\
		u^d{}_4 = 0.00615937 \\
		u^d{}_5 = 0.0254406  \\
		u^d{}_6 = 0          \\
		u^d{}_7 = 0          \\
		u^d{}_8 = 0          \\
		u^d{}_9 = 0          \\
		u^d{}_{10} = 0       \\
		u^d{}_{11} = 0       \\
		u^d{}_{12} = 0       \\
		u^d{}_{13} = 0       \\
		u^d{}_{14} = 0       \\
		u^d{}_{15} = 0       \\
		u^d{}_{16} = 0       \\
		u^d{}_{17} = 0       \\
		u^d{}_{18} = 0       \\
		u^d{}_{19} = 0       \\
		u^d{}_{20} = 0       \\
	\end{array}
	\quad,\quad
	\begin{array}{l}
		w^d{}_1 = 0 \\
		w^d{}_2 = 0 \\
		w^d{}_3 = 0 \\
		w^d{}_4 = 0 \\
		w^d{}_5 = 0 \\
		w^d{}_6 = 0 \\
		w^d{}_7 = 0 \\
		w^d{}_8 = 1 \\
	\end{array}
\end{equation}
The 4-point correction to the charged-leptons' masses is given by,
\begin{equation}\label{M4e_22}
	|M_{4e}|=m_\tau \left(
	\begin{array}{ccc}
		0.       & 0.107805 & 0.11791  \\
		0.107805 & 0.11791  & 0.500807 \\
		0.11791  & 0.500807 & 0.       \\
	\end{array}
	\right)
\end{equation}
which can be added to the matrix obtained from 3-point functions \eqref{eq:Leptons3_22} as,
\begin{equation}\label{M34e_22}
	|M_{3e}|+|M_{4e}|=m_\tau \left(
	\begin{array}{ccc}
		0.036169 & 0.108125 & 0.116698 \\
		0.108125 & 0.458457 & 0.500675 \\
		0.116698 & 0.500675 & 0.547527 \\
	\end{array}
	\right) \sim D_e ~,
\end{equation} 
The corrections to down-type quarks' masses can be made negligible by setting $\epsilon^{(1)}_{4d}=\frac{1}{2}$,
\begin{equation}\label{M4d_22}
	|M_{4d}|\sim \left(
	\begin{array}{ccc}
		0. & 0. & 0. \\
		0. & 0. & 0. \\
		0. & 0. & 0. \\
	\end{array}
	\right)\sim 0.
\end{equation} Therefore, we can achieve near-exact matching of fermion masses and mixings.
\subsection{Model 22-dual}\label{sec:model-22.5}
In Model~\hyperref[model22.5]{22-dual} the three-point Yukawa couplings arise from the triplet intersections from the branes ${a, b, c}$ on the first two-torus ($r=1$) with 12 pairs of Higgs from $\mathcal{N}=2$ subsector.

\begin{figure}[t]
	\centering
	\includegraphics[width=\textwidth]{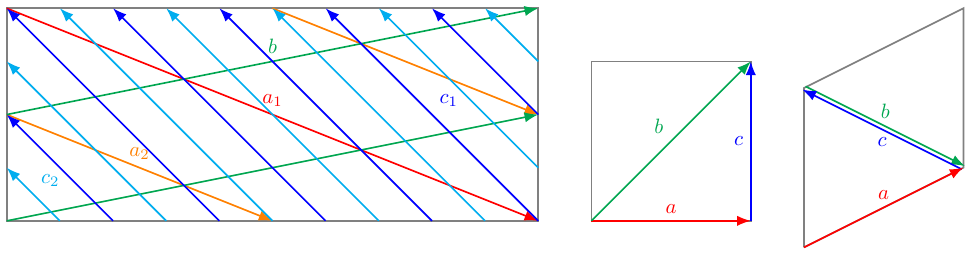}
	\caption{Brane configuration for the three two-tori in Model~\hyperref[model22.5]{22-dual} where the third two-torus is tilted. Fermion mass hierarchies result from the intersections on the first two-torus.}  \label{Fig.22.5}
\end{figure}

Yukawa matrices for the Model~\hyperref[model22.5]{22-dual} are of rank 3 and the three intersections required to form the disk diagrams for the Yukawa couplings all occur on the first torus as shown in figure~\ref{Fig.22.5}. The other two-tori only contribute an overall constant that has no effect in computing the fermion mass ratios. Thus, it is sufficient for our purpose to only focus on the first torus to explain the masses and the mixing in the standard model fermions.

\subsubsection{3-point Yukawa mass-matrices for Model~\hyperref[model22.5]{22-dual}}
From the wrapping numbers listed in table~\ref{model22.5}, the relevant intersection numbers are calculated as,
\begin{align}\label{eq:Intersections.22.5} 
	I_{ab}^{(1)} & =3,  &   
	I_{ab}^{(2)} & =1,  &   
	I_{ab}^{(3)}&=-1, \nonumber\\
	I_{bc}^{(1)} & =12, &   
	I_{bc}^{(2)} & =1,  &   
	I_{bc}^{(3)}&=0, \nonumber\\
	I_{ca}^{(1)} & =-3, &   
	I_{ca}^{(2)} & =-1, &   
	I_{ca}^{(3)}&=-1, 
\end{align}
As the intersection numbers are not coprime, we define the greatest common divisor,
$d^{(1)}=g.c.d.(I_{ab}^{(1)},I_{bc}^{(1)},I_{ca}^{(1)})=3$. Thus, the arguments of the modular theta function as defined in \eqref{eqn:Yinput} can be written as,
\begin{align}
	\delta^{(1)} & = \frac{i^{(1)}}{3} + \frac{j^{(1)}}{-3} + \frac{k^{(1)}}{12} +\frac{1}{36} \left(-12 \epsilon _a^{\text{(1)}}+3 \epsilon _b^{\text{(1)}}-3 \epsilon _c^{\text{(1)}}\right)+ \frac{s^{(1)}}{3},  \label{eq:delta.22.5} \\
	\phi^{(1)}   & = \frac{1}{3} \left(12 \theta _a^{\text{(1)}}-3 \theta _b^{\text{(1)}}+3 \theta _c^{\text{(1)}}\right) =0,                                                                                                             \\
	\kappa^{(1)} & = \frac{12 J^{\text{(1)}}}{\alpha '}, \label{eq:kappa.22.5}                                                                                                                                                            
\end{align}
and recalling \eqref{eq:ijk}, we have $i=\{0,\dots ,2\}$, $j=\{0,\dots ,2\}$ and $k=\{0,\dots ,11\}$ which respectively index the left-handed fermions, the right-handed fermions and the Higgs fields. Clearly, there arise 12 Higgs fields from the $bc$ sector.

The second-last term in the right side of \eqref{eq:delta.22.5} can be used to redefine the shift on the torus as
\begin{align}
	\epsilon^{(1)} & \equiv \frac{1}{36} \left(-12 \epsilon _a^{\text{(1)}}+3 \epsilon _b^{\text{(1)}}-3 \epsilon _c^{\text{(1)}}\right). \label{shifts22.5} 
\end{align}  
The selection rule for the occurrence of a trilinear Yukawa coupling for a given set of indices is given as,
\begin{equation}\label{selection-rule.22.5}
	i^{(1)} + j^{(1)} + k^{(1)} = 0\; \mathrm{ mod }\; 3.
\end{equation}
Then the suitable rank-3 mass-matrix can be determined by choosing the specific value of $s^{(1)}=-i$. 
\begin{align}
	\delta^{(1)} & =  \frac{k}{12}-\frac{j}{3} \label{eq:s1.22.5} 
\end{align}
Here, we will ignore other equivalent cases of solutions or cases with rank-1 problem. The mass matrices for up quarks, down quarks and charged leptons have the following general form:
\begin{align}
	Z_{3} &= Z_{3q} \left(
	\begin{array}{ccc}
	T_0 v_1+T_3 v_4+T_6 v_7+T_9 v_{10}    & T_2 v_3+T_5 v_6+T_8 v_9+T_{11} v_{12} & T_1 v_2+T_4 v_5+T_7 v_8+T_{10} v_{11} \\
	T_{10} v_3+T_1 v_6+T_4 v_9+T_7 v_{12} & T_9 v_2+T_0 v_5+T_3 v_8+T_6 v_{11}    & T_8 v_1+T_{11} v_4+T_2 v_7+T_5 v_{10} \\
	T_5 v_2+T_8 v_5+T_{11} v_8+T_2 v_{11} & T_4 v_1+T_7 v_4+T_{10} v_7+T_1 v_{10} & T_6 v_3+T_9 v_6+T_0 v_9+T_3 v_{12}    \\
	\end{array}
	\right),\label{eq:s-i22.5}
\end{align} 
where $v_i = \left\langle H_{i} \right\rangle$ and the three-point coupling functions are given in terms of Jacobi theta function of the third kind as, 

\begin{align}\label{eq:3couplings22.5}
	T_k & \equiv  \vartheta \left[\begin{array}{c} 
	\epsilon^{(1)}+\frac{k}{12}\\  \phi^{(1)} \end{array} \right]
	(\frac{12 J^{\text{(1)}}}{\alpha '}),  \quad   k={0,\cdots,11}.
\end{align}
  
\subsubsection{Fermion masses and mixings from 3-point functions in Model~\hyperref[model22.5]{22-dual}}
In order to accommodate the fermion masses and the quark mixings, we need to fit the up-type quarks mixing matrix \eqref{eq:mixing-quarks}, the down-type quarks matrix \eqref{eq:mass-downquarks} and the masses of the charged-leptons \eqref{eq:mass-chargedleptons}.
 
We set the K\"{a}hler modulus on the first two-torus defined in \eqref{eq:kappa.22.5} as $\kappa^{(1)} = 64 $  and evaluate the couplings functions \eqref{eq:3couplings22.5} by setting geometric brane position parameters as $\epsilon^{(1)}_{u} = 0$, $\epsilon^{(1)}_{d} = 0$ and $\epsilon^{(1)}_{e} = \frac{1}{2}$ which yields an \emph{exact} fitting for the following VEVs,
\begin{equation}\label{eq:VEVs_22.5} 
	\begin{array}{ll}
		v^u{}_1 = 0.000291336               , & v^d{}_1 = 0.0000282001     \\
		v^u{}_2 = -0.0000838211             , & v^d{}_2 = -0.000163191     \\
		v^u{}_3 = 1.32337                   , & v^d{}_3 = 0.0173018        \\
		v^u{}_4 = -4.42684\times 10^{-6}    , & v^d{}_4 = -0.0000173322    \\
		v^u{}_5 = 0.00552729                , & v^d{}_5 = 0.000560001      \\
		v^u{}_6 = -0.0151378                , & v^d{}_6 = -0.000262376     \\
		v^u{}_7 = 11.0134                   , & v^d{}_7 = 0.00114294       \\
		v^u{}_8 = -0.0000838211             , & v^d{}_8 = -0.000163191     \\
		v^u{}_9 = 0.998234                  , & v^d{}_9 = 0.02             \\
		v^u{}_{10} = -4.42684\times 10^{-6} , & v^d{}_{10} = -0.0000173322 \\
		v^u{}_{11} = 2.29873                , & v^d{}_{11} = 0.0107613     \\
		v^u{}_{12} = -0.0151378             , & v^d{}_{12} = -0.000262376  \\
	\end{array}
\end{equation}
\begin{align} 
	|M_{3u}|&=m_t \left(
	\begin{array}{ccc}
	0.000291  & 0.00122   & 0.008608  \\
	0.00122   & 0.005527  & 0.041338  \\
	0.008608  & 0.041338  & 0.998234  \\
	\end{array}
	\right)\sim M_u ~, \label{eq:Up3_22.5} \\
	|M_{3n}|&=m_\nu \left(
	\begin{array}{ccc}
	11.0134   & 0.        & 0.        \\
	0.        & 2.29873   & 0.        \\
	0.        & 0.        & 1.32337   \\
	\end{array}
	\right)~,\label{eq:nu3_22.5} \\|M_{3d}|&=m_b \left(
	\begin{array}{ccc}
	0.00141   & 0.        & 0.        \\
	0.        & 0.028     & 0.        \\
	0.        & 0.        & 1.        \\
	\end{array}
	\right)\sim D_d ~,\label{eq:Down3_22.5} \\ 
	|M_{3e}|&=m_\tau \left(
	\begin{array}{ccc}
	0.036169  & 0.00032   & -0.001212 \\
	0.00032   & 0.340547  & -0.000132 \\
	-0.001212 & -0.000132 & 0.547527  \\
	\end{array}
	\right)\sim D_e~.\label{eq:Leptons3_22.5}
\end{align} Notice, that these results are only at the tree-level and there could indeed be other corrections, such as those coming from higher-dimensional operators, which may contribute most greatly to the charm and the down quarks' masses since they are lighter. In order to preserve the exact-fitting of up-quarks mixed form matrix and the neutrino-masses already achieved from 3-point functions, we can switch off the corrections from the 4-point functions by taking all up-type VEVs to be zero. However, the down-type VEVS from 4-point functions will be needed to explain the leptons' mixing.

\subsubsection{4-point corrections in Model~\hyperref[model22.5]{22-dual}}
 
The four-point couplings in Model~\hyperref[model22.5]{22-dual} in table~\ref{model22.5} can come from considering interactions of ${a, b, c}$ with $b'$ or $c'$ on the first two-torus as can be seen from the following intersection numbers,
\begin{align}
	I_{bb'}^{(1)} & = 4,   & I_{bb'}^{(2)} & = 2,  & I_{bb'}^{(3)} & = -1 ,\nonumber                     \\
	I_{cc'}^{(1)} & = -20, & I_{cc'}^{(2)} & = 0,  & I_{cc'}^{(3)} & = -1 ,\nonumber                     \\
	I_{bc'}^{(1)} & = -8,  & I_{bc'}^{(2)} & = -1, & I_{bc'}^{(3)} & = -1 . \label{eq:4Intersection22.5} 
\end{align}
There are 20 SM singlet fields $S_L^i$ and 8
Higgs-like state $H^{\prime}_{u, d}$. 

Let us consider four-point interactions with $c'$ with the following parameters with shifts $l=-\frac{k}{4} $ and $\ell=-\frac{k}{3} $ taken along the index $k$,
\begin{align}\label{deltas22.5}
	\delta & =\frac{i}{I_{ab}^{(1)}} +\frac{j}{I_{ca}^{(1)}} +\frac{k}{I_{bc}^{(1)}} +l , \nonumber               \\
	       & = \frac{i}{3}-\frac{j}{3} ,                                                                          \\
	d      & =\frac{\imath}{I_{c c'}^{(1)}} +\frac{\jmath}{I_{bc'}^{(1)}}+\frac{k}{I_{bc}^{(1)}} +\ell ,\nonumber \\
	       & = -\frac{\imath}{20}-\frac{\jmath}{8} ,                                                              
\end{align}
the matrix elements $a_{i,j,\imath}$ on the first torus from the four-point functions results in the following classical 4-point contribution to the mass matrix,
\begin{equation}\label{eq:4point22.5} 
	Z_{4cl} =  
	\resizebox{.95\textwidth}{!}{
		$\begin{aligned}
			  & \left(                 
			\begin{array}{c}
			F_0 u_1 w_1+F_{16} u_{13} w_1+F_{13} u_{12} w_2+F_{10} u_{11} w_3+F_7 u_{10} w_4+F_4 u_9 w_5+F_1 u_8 w_6+F_{17} u_{20} w_6+F_{38} u_7 w_7+F_{14} u_{19} w_7+F_{35} u_6 w_8+F_{11} u_{18} w_8\\
			F_{18} u_{12} w_1+F_{15} u_{11} w_2+F_{12} u_{10} w_3+F_9 u_9 w_4+F_6 u_8 w_5+F_{22} u_{20} w_5+F_3 u_7 w_6+F_{19} u_{19} w_6+F_0 u_6 w_7+F_{16} u_{18} w_7+F_{37} u_5 w_8+F_{13} u_{17} w_8\\
			F_{20} u_{11} w_1+F_{17} u_{10} w_2+F_{14} u_9 w_3+F_{11} u_8 w_4+F_{27} u_{20} w_4+F_8 u_7 w_5+F_{24} u_{19} w_5+F_5 u_6 w_6+F_{21} u_{18} w_6+F_2 u_5 w_7+F_{18} u_{17} w_7+F_{39} u_4 w_8+F_{15} u_{16} w_8\\
			\end{array}\right.\\
			  & \begin{array}{c}       
			F_{18} u_{12} w_1+F_{15} u_{11} w_2+F_{12} u_{10} w_3+F_9 u_9 w_4+F_6 u_8 w_5+F_{22} u_{20} w_5+F_3 u_7 w_6+F_{19} u_{19} w_6+F_0 u_6 w_7+F_{16} u_{18} w_7+F_{37} u_5 w_8+F_{13} u_{17} w_8\\
			F_{20} u_{11} w_1+F_{17} u_{10} w_2+F_{14} u_9 w_3+F_{11} u_8 w_4+F_{27} u_{20} w_4+F_8 u_7 w_5+F_{24} u_{19} w_5+F_5 u_6 w_6+F_{21} u_{18} w_6+F_2 u_5 w_7+F_{18} u_{17} w_7+F_{39} u_4 w_8+F_{15} u_{16} w_8\\
			F_{22} u_{10} w_1+F_{19} u_9 w_2+F_{16} u_8 w_3+F_{32} u_{20} w_3+F_{13} u_7 w_4+F_{29} u_{19} w_4+F_{10} u_6 w_5+F_{26} u_{18} w_5+F_7 u_5 w_6+F_{23} u_{17} w_6+F_4 u_4 w_7+F_{20} u_{16} w_7+F_1 u_3 w_8+F_{17} u_{15} w_8\\
			\end{array}\\
			  & \left.\begin{array}{c} 
			F_{20} u_{11} w_1+F_{17} u_{10} w_2+F_{14} u_9 w_3+F_{11} u_8 w_4+F_{27} u_{20} w_4+F_8 u_7 w_5+F_{24} u_{19} w_5+F_5 u_6 w_6+F_{21} u_{18} w_6+F_2 u_5 w_7+F_{18} u_{17} w_7+F_{39} u_4 w_8+F_{15} u_{16} w_8\\
			F_{22} u_{10} w_1+F_{19} u_9 w_2+F_{16} u_8 w_3+F_{32} u_{20} w_3+F_{13} u_7 w_4+F_{29} u_{19} w_4+F_{10} u_6 w_5+F_{26} u_{18} w_5+F_7 u_5 w_6+F_{23} u_{17} w_6+F_4 u_4 w_7+F_{20} u_{16} w_7+F_1 u_3 w_8+F_{17} u_{15} w_8\\
			F_{24} u_9 w_1+F_{21} u_8 w_2+F_{37} u_{20} w_2+F_{18} u_7 w_3+F_{34} u_{19} w_3+F_{15} u_6 w_4+F_{31} u_{18} w_4+F_{12} u_5 w_5+F_{28} u_{17} w_5+F_9 u_4 w_6+F_{25} u_{16} w_6+F_6 u_3 w_7+F_{22} u_{15} w_7+F_3 u_2 w_8+F_{19} u_{14} w_8\\
			\end{array}\right),
		\end{aligned}$
	} 
\end{equation} 
where $u_i, w_j$ are the VEVs and the couplings are defined as,
\begin{align}\label{eq:4couplings22.5}
	F_{8i} & \equiv  \vartheta \left[\begin{array}{c} 
	\epsilon^{(1)}+\frac{i}{20}\\  \phi^{(1)} \end{array} \right]
	(\frac{12 J^{\text{(1)}}}{\alpha '}),\qquad i={0,\dots,|I_{c c'}^{(1)}|-1}.
\end{align}

Since, we have already fitted the up-type quark matrix $|M_{3u}|$ exactly, so its 4-point correction should be zero,
\begin{equation}
	|M_{4u}| = 0 ,
\end{equation}
which is true by setting all up-type VEVs $u_u^i$ and $w_u^i$ to be zero. 
Therefore, we are essentially concerned with fitting charged-leptons in such a way that corresponding corrections for the down-type quarks remain negligible.
The desired solution can be readily obtained by setting $\epsilon^{(1)}_{4e}=0$ with the following values of the VEVs,
\begin{equation}\label{eq:UdWd22.5} 
	\begin{array}{l}
		u^d{}_1 = 0          \\
		u^d{}_2 = 0          \\
		u^d{}_3 = 0.0158255  \\
		u^d{}_4 = 0.00615937 \\
		u^d{}_5 = 0.0254406  \\
		u^d{}_6 = 0          \\
		u^d{}_7 = 0          \\
		u^d{}_8 = 0          \\
		u^d{}_9 = 0          \\
		u^d{}_{10} = 0       \\
		u^d{}_{11} = 0       \\
		u^d{}_{12} = 0       \\
		u^d{}_{13} = 0       \\
		u^d{}_{14} = 0       \\
		u^d{}_{15} = 0       \\
		u^d{}_{16} = 0       \\
		u^d{}_{17} = 0       \\
		u^d{}_{18} = 0       \\
		u^d{}_{19} = 0       \\
		u^d{}_{20} = 0       \\
	\end{array}
	\quad,\quad
	\begin{array}{l}
		w^d{}_1 = 0 \\
		w^d{}_2 = 0 \\
		w^d{}_3 = 0 \\
		w^d{}_4 = 0 \\
		w^d{}_5 = 0 \\
		w^d{}_6 = 0 \\
		w^d{}_7 = 0 \\
		w^d{}_8 = 1 \\
	\end{array}
\end{equation}
The 4-point correction to the charged-leptons' masses is given by,
\begin{equation}\label{M4e_22.5}
	|M_{4e}|=m_\tau \left(
	\begin{array}{ccc}
		0.       & 0.107805 & 0.11791  \\
		0.107805 & 0.11791  & 0.500807 \\
		0.11791  & 0.500807 & 0.       \\
	\end{array}
	\right)
\end{equation}
which can be added to the matrix obtained from 3-point functions \eqref{eq:Leptons3_22.5} as,
\begin{equation}\label{M34e_22.5}
	|M_{3e}|+|M_{4e}|=m_\tau \left(
	\begin{array}{ccc}
		0.036169 & 0.108125 & 0.116698 \\
		0.108125 & 0.458457 & 0.500675 \\
		0.116698 & 0.500675 & 0.547527 \\
	\end{array}
	\right) \sim D_e ~,
\end{equation} 
The corrections to down-type quarks' masses can be made negligible by setting $\epsilon^{(1)}_{4d}=\frac{1}{2}$,
\begin{equation}\label{M4d_22.5}
	|M_{4d}|\sim \left(
	\begin{array}{ccc}
		0. & 0. & 0. \\
		0. & 0. & 0. \\
		0. & 0. & 0. \\
	\end{array}
	\right)\sim 0.
\end{equation} Therefore, we can achieve near-exact matching of fermion masses and mixings. 

\subsection{Prediction of Dirac-Neutrino Masses from 12-Higgs models}
It is clear from \eqref{eq:nu3_21}, \eqref{eq:nu3_22} and \eqref{eq:nu3_22.5} that the mass-eigenvalues of the Dirac-neutrinos are set by the VEVs $(v^u_7,~ v^u_{11},~ v^u_3)$ with coupling $T^\nu_6=1$ upto an overall scale $m_\nu$ that is to be fixed by experimental constraints. It is also evident that the Dirac-neutrinos are predicted to be in normal ordering. $T^\nu_6=1$ is a nice feature as it avoids extra fine-tuning, given that the neutrinos are already several orders of magnitude lighter than their quark and lepton counterparts. 

The experimental constraints \eqref{eq:constraints} for the normal ordering ($m_3\gg m_2>m_1$) are naturally satisfied by setting $m_\nu = 4.57761$~meV, 
\begin{mdframed}\setlength\abovedisplayskip{0pt} 
	\begin{align} 
		\Rightarrow (m_3,~ m_2,~ m_1) & = (50.4,~10.5,~6.1)\pm 0.1~\mathrm{meV},\nonumber                         \\
		\Delta m_{21}^2               & = 74.0~\mathrm{meV}^2, \nonumber                                          \\
		\Delta m_{31}^2               & = +2505~\mathrm{meV}^2, \quad \mathrm{(NO)} \nonumber                     \\
		m_1 + m_2 + m_3               & = 67.0~\mathrm{meV}, \quad \mathrm{(Dirac~with~NO)}. \label{eq:neutrinos} 
	\end{align}  
\end{mdframed}
The prediction of Dirac-neutrino-masses is robust, as the ratios of neutrino-masses are essentially determined by the up-quarks matrix \eqref{eq:mass-upquarks} that serves as an input into the up-quarks mixing matrix \eqref{eq:mixing-quarks} given that the CKM matrix is now known with high precision. Although the uncertainties in \eqref{eq:mass-upquarks} can be significant since the unification-scale is not known precisely, however the experimental constraints \eqref{eq:constraints} can mitigate these uncertainties. Consequently, the uncertainties in \eqref{eq:mass-upquarks} translate into the uncertainty in the K\"{a}hler modulus $\kappa^{(1)} = 64 \pm 2 $, while the overall uncertainty in neutrino-masses remains within $\pm 0.1$~meV.  

Recent evidence from the swampland program \cite{Vafa:2005ui}, particularly from the non-SUSY AdS instability conjecture \cite{Ooguri:2016pdq} and the light fermion conjecture \cite{Gonzalo:2021fma, Gonzalo:2021zsp}, building on the earlier work of \cite{Arkani-Hamed:2007ryu, Arnold:2010qz} suggests that without additional chiral fermions with tiny masses, neutrinos must be of Dirac-type together with a bound on the lightest neutrino mass given by the cosmological constant scale as, $m_{\nu}^{\rm lightest} \lesssim \Lambda^{1/4}$ \cite{Ibanez:2017kvh, Hamada:2017yji, Gonzalo:2018tpb, Castellano:2023qhp, Harada:2022ijm} \footnote{A stronger bound of $m_1 < 4.1~\mathrm{meV}$ (NO) was found for 2D compactification on a torus in \cite{Hamada:2017yji, Ibanez:2017kvh}. However unlike in 3D, there may be several choices of compactifications in 2D e.g. torus $\T^2$, orbifolds $\T^2/\Z_N$, 2D spheres $S^2$ with or without fluxes \cite{Arnold:2010qz, Gonzalo:2018tpb} which thus renders the 2D bounds less predictive. Note that the bounds are sensitive to the value of $\Lambda$, here taken to be $2.6 \times 10^{-47}$ GeV${^4}$, having large uncertainties owing to the impreciseness in the determination of Hubble constant.} 
\begin{align}\label{eq:Swampland-bounds}
	m_1 < 7.7~\mathrm{meV}~ \mathrm{(NO)}, \qquad \sum_{i=1}^3 m_i & = 60 \textnormal{--} 70~\mathrm{meV}~ \mathrm{(NO)}. 
\end{align} 
The 3D Casimir energy of the SM compactified on a circle receives a positive contribution from the lightest neutrino, which is necessary to avoid unstable non-supersymmetric AdS vacua. This constraint is only satisfied for Dirac neutrinos, which carry 4 degrees of freedom, unlike Majorana neutrinos, which only have 2 and cannot compensate for the 4 bosonic degrees of freedom from the photon and the graviton.  
 
Comparing the swampland bounds \eqref{eq:Swampland-bounds} and our results of Dirac-neutrino masses \eqref{eq:neutrinos}, our universe avoids AdS vacua in 3D as the mass of the lightest neutrino turns out to be less than the threshold value \eqref{eq:Swampland-bounds} and the sum of the masses of three Dirac-neutrino also falls within the range given by the multiple point criticality principle which requires the 3D dS vacuum to be close to the flat vacuum \cite{Hamada:2017yji}. In analogy with the transition between ice and water being of first order the slush exists at $0^\circ$C. Conversely, if the temperature happens to be suspiciously close to zero, it is because of the existence of such as a slush \cite{Froggatt:1995rt}. Since the AdS and the dS vacua are separated by infinite distance in the moduli-space \cite{Lust:2019zwm}, any transition between them is of first-order, making the multiple point criticality principle applicable to our universe. 

Note that the masses of Dirac-neutrinos are derived by three-point functions whereas the leptons' mixing need four-point functions which are suppressed by the string-scale $M_{\rm S}$ \eqref{eq:WY4}. These higher-dimensional operators may link neutrino-mixings with the dark-dimension scenario \cite{Montero:2022prj} motivated by the emergent strings conjecture \cite{Lee:2019wij}. Dark dimension relates dark matter (5D gravitons), dark energy ($\Lambda$) and axion decay constant ($f_a\lesssim \hat{M}_5$) with the scale of lightest-neutrino ($m_1$) \cite{Gonzalo:2022jac, Gendler:2024gdo}. Taking $m_1=6.1$~meV in the following relations, 
\begin{align}\label{eq:Darkdim}
	\hat{M}_5 = m_1^{1/3}M_{\rm pl}^{2/3}, \qquad l_5 = m_{1}^{-1} = \lambda\,\Lambda^{-1/4} , 
\end{align}
the species-scale in 5D is set at $\hat{M}_5 =9.7\times 10^8$~GeV, resulting in the size of the dark-dimension to be 32~$\mu$m and the thickness of the brane to be $2.0\times 10^{-23}$~cm. Note that there can be large uncertainties in $\Lambda$ that can affect the value of the coupling $\lambda$ but the value of $\hat{M}_5$ is precise due to the preciseness in the value of $m_1$. Accordingly, deviation from the gravitational inverse-square law are predicted below 32~$\mu$m. Experimentally, no deviations have been detected above $38.6$~$\mu$m at 2$\sigma$ \cite{Lee:2020zjt}, however, it is hoped to be probed in near-future. Furthermore, the dark dimension scale can also be related to the scale of supersymmetry breaking \cite{Anchordoqui:2023oqm} which will be discussed in \cite{Sabir:2024jsx}.

\section{Conclusion}\label{sec:conclusion}

\begin{table}[ht]
	\begin{threeparttable}
		\centering
		\begin{tabular}{|l|c|c|c|c|c|c|}
			\hline
			\textbf{ }                          & \textbf{Up-}     & \textbf{Down-}   & \textbf{Charged-} & \textbf{CKM-}    & \textbf{Dirac-}\dag & \textbf{PMNS-}   \\
			\textbf{Model}                      & \textbf{Quarks'} & \textbf{Quarks'} & \textbf{leptons'} & \textbf{mixings} & \textbf{Neutrino}   & \textbf{mixings} \\
			\textbf{ }                          & \textbf{masses}  & \textbf{masses}  & \textbf{masses}   &                  & \textbf{masses}     &                  \\
			\hline
			\hyperref[model13]{13}\tnote{\textcolor{red}{*}}              & Yes              & Yes              & Approx.           & No               & No                  & No               \\
			\hline
			\hyperref[model14]{14}\tnote{\ddag} & Yes              & Approx.          & Yes               & Approx.          & No                  & No               \\
			\hyperref[model15]{15}              & Yes              & Yes              & Yes               & Approx.          & No                  & No               \\
			\hyperref[model15.5]{15-dual}       & Yes              & Yes              & Yes               & Approx.          & No                  & No               \\
			\hyperref[model16]{16}              & Yes              & Yes              & Yes               & Approx.          & No                  & No               \\
			\hyperref[model16.5]{16-dual}       & Yes              & Yes              & Yes               & Approx.          & No                  & No               \\
			\hline
			\hyperref[model17]{17}              & Yes              & Yes              & Yes               & Yes              & No                  & No               \\
			\hyperref[model17.5]{17-dual}       & Yes              & Yes              & Yes               & Yes              & No                  & No               \\
			\hyperref[model18]{18}              & Yes              & Yes              & Yes               & Yes              & No                  & No               \\
			\hyperref[model18.5]{18-dual}       & Yes              & Yes              & Yes               & Yes              & No                  & No               \\
			\hyperref[model19]{19}              & Yes              & Yes              & Yes               & Yes              & No                  & No               \\
			\hyperref[model19.5]{19-dual}       & Yes              & Yes              & Yes               & Yes              & No                  & No               \\
			\hyperref[model20]{20}              & Yes              & Yes              & Yes               & Yes              & No                  & No               \\
			\hyperref[model20.5]{20-dual}       & Yes              & Yes              & Yes               & Yes              & No                  & No               \\
			\hline
			\hyperref[model21]{21}\tnote{\ddag} & Yes              & Yes              & Yes               & Yes              & Yes                 & No               \\
			\hyperref[model22]{22}              & Yes              & Yes              & Yes               & Yes              & Yes                 & Yes              \\
			\hyperref[model22.5]{22-dual}       & Yes              & Yes              & Yes               & Yes              & Yes                 & Yes              \\
			\hline
		\end{tabular}
		\begin{tablenotes}
			\item[\dag] Majorana masses can be obtained for all models via type-I seesaw mechanism.
            \item[\textcolor{red}{*}] Only bulk-Higgs are considered because $\mathcal{N}=2$ sector Higgs are not available.
			\item[\ddag] Only three-point functions are considered because viable four-point-couplings are not available.
		\end{tablenotes}
	\end{threeparttable}
\caption{Summary of possible 3-point and the 4-point Yukawa interactions to match the fermions masses and mixings in the landscape of three-family supersymmetric Pati-Salam models from intersecting D6-Branes on a type IIA $\T^6/(\Z_2\times \Z_2)$ orientifold.}
		\label{tab:summary}
\end{table}

We have systematically analyzed all viable models in the complete landscape of three-family supersymmetric Pati-Salam intersecting D6-brane models on a $\T^6/(\Z_2\times \Z_2)$ orientifold in type IIA string theory. We have presented the detailed particle spectra of all 33 models and have calculated the three-point and four-point couplings to accommodate standard model fermion masses and mixings. It is found that only 17 models contain viable Yukawa textures such that viable models split into four classes. The first class consists of a single model with 3 bulk Higgs fields while the remaining three classes contain Higgs fields from $\mathcal{N}=2$ sector such that there are five models with 6 Higgs, eight models with 9 Higgs and three models with 12 Higgs field. Table \ref{tab:summary} summarizes the main findings from all viable models. 

Remarkably, the class of models with 12 Higgs pairs naturally predicts the Dirac-neutrino masses in normal ordering consistent with both the experimental constraints as well as the bounds from the swampland program. With twelve Higgs from the $\mathcal{N}=2$ sector, the models precisely accommodates all SM fermion masses and mixings along with Dirac neutrino masses in NO $(50.4,~10.5,~6.1) \pm 0.1$~meV consistent with both experimental constraints as well as the swampland bounds on the mass of lightest neutrino and the sum of neutrinos according to the non-supersymmetric AdS instability conjecture and the multiple point criticality principle respectively. Yukawa couplings from the $\mathcal{N}=2$ sector also evade the dangerous infinite distance limits, thereby avoiding the decompactification of extra-dimensions. This constitutes the first precise prediction of Dirac neutrino masses from a consistent string theory setup. An experimental confirmation of the heaviest neutrino-mass at $\sim 50$~meV will thus validate this class of models. 

Yukawa couplings depend on the geometric position of the stacks of D6-branes and the K\"{a}hler moduli. A mechanism to stabilize these open string moduli is needed to eliminate the non-chiral open string states associated to the brane positions and the Wilson lines in the low energy spectrum. This may be accomplished in the case of type II compactifications on $\T^6/(\Z_2\times \Z'_2)$ background possessing rigid cycles \cite{Blumenhagen:2005tn}. We will report on the progress in this direction in near future \cite{Mansha:2025gvr}.

\FloatBarrier

\acknowledgments{TL is supported in part by the National Key Research and Development Program of China Grant No. 2020YFC2201504, by the Projects No. 11875062, No. 11947302, No. 12047503, and No. 12275333 supported by the National Natural Science Foundation of China, by the Key Research Program of the Chinese Academy of Sciences, Grant No. XDPB15, by the Scientific Instrument Developing Project of the Chinese Academy of Sciences, Grant No. YJKYYQ20190049, and by the International Partnership Program of Chinese Academy of Sciences for Grand Challenges, Grant No. 112311KYSB20210012. AM is supported by the Guangdong Basic and Applied Basic Research Foundation (Grant No. 2021B1515130007), Shenzhen Natural Science Fund (the Stable Support Plan Program 20220810130956001). Z.-W. Wang is supported in part by the hundred talented program at University of Electronic Science and Technology of China and by the National Natural Science Foundation of China (Grant No. 12475105).}

\appendix 

\section{Independent Supersymmetric Pati-Salam Models}\label{appA} 

Following is the complete list of the $33$ independent three-family ${\cal N}=1$ supersymmetric Pati-Salam models with distinct allowed gauge coupling relations arising from intersecting D6-branes on a type IIA $\T^6/(\Z_2\times \Z_2)$ orientifold \cite{He:2021gug}. We have used type-I T-duality to list models in such a way that the three-point Yukawa interactions arise from the $(a,b,c)$ triplet. As a result some of the $a$, $b$ and $c$ branes are replaced by their corresponding orientifold images $a'$, $b'$ and $c'$. We have sorted the models with the maximum number of viable Higgs on any of the three two-tori with viable Yukawa mass-matrix, followed by the occurrence of highest wrapping numbers, followed by the occurrence of other wrapping number in the ascending order. 

For 11 out of 33 models, interchanging the $b$ and $c$ stacks results in dual models with distinct gauge coupling relations as can be discerned from the captions of the tables provided that the constraint, $I_{ac'} = 0$, is satisfied after the $b\leftrightarrow c$ exchange. Two models viz. \hyperref[model14]{14} and \hyperref[model21]{21} are self-dual due to exact or partial gauge coupling unification while eleven other models do not have duals because the constraint, $I_{ac'} = 0$, is not satisfied after the $b\leftrightarrow c$ exchange. The wrapping numbers $(n^3,l^3)$ in the third two-torus can be converted to $(n^3,m^3)$ using \eqref{basis-l-m} as $m^3 = \frac{l^3-n^3}{2}$.
In the case where dual models exists, e.g. model \hyperref[model1]{1} and \hyperref[model1.5]{1-dual}, the main model is chosen to contain the highest wrapping number in stack $b$, accordingly in the dual model, the highest wrapping number is found in stack $c$.


\begin{table}[ht]\footnotesize\centering 
$
\begin{array}{|c|c|c|c|c|c|c|c|c|c|c|}
\hline\multicolumn{2}{|c|}{\text{Model~\hyperref[spec1]{1}}} & \multicolumn{9}{c|}{ {\U(4)_C\times \U(2)_L \times \U(2)_R\times \USp(4)_2\times \USp(4)_4} }\\
\hline\rm{stack} & N & (n^1,l^1)\times(n^2,l^2)\times(n^3,l^3) & n_{\yng(2)}& n_{\yng(1,1)_{}} & b & b' & c & c' & 2 &  4\\
\hline
a & 8 & (1, -1)\times (1, 0)\times (1, 1) & 0 & 0  & 3 & 0 & -3 & 0 & -1 &  1\\
b & 4 &  (0, 1)\times (-1, 3)\times (-1, 1) & 2 & -2  & \text{-} & \text{-} & 0 & -4 & 1 &  0\\
c & 4 &  (-2, -1)\times (-1, -1)\times (1, -1) & 2 & 6  & \text{-} & \text{-} & \text{-} & \text{-} & 1 &  -2\\
\hline
 &   &  & \multicolumn{8}{c|}{ x_A = \frac{1}{3}x_B = x_C = \frac{1}{3}x_D }\\
2 & 4 &  (1, 0)\times (0, -1)\times (0, 2) & \multicolumn{8}{c|}{ ~\beta^g_2=-2,~\beta^g_4=-2, }\\
 &   &   & \multicolumn{8}{c|}{ \chi_1=1 , ~ \chi_2=\frac{1}{3} , ~ \chi_3=2 }\\
4 & 4  &   (0, -1)\times (0, 1)\times (2, 0) & \multicolumn{8}{c|}{}\\
\hline  
\end{array}$
\caption{D6-brane configurations and intersection numbers of Model~\hyperref[spec1]{1}, and its MSSM gauge coupling relation is $g_a^2=g_b^2=\frac{5}{3}g_c^2=\frac{25}{19}\frac{5 g_Y^2}{3}=4 \sqrt{\frac{2}{3}} \, \pi \,  e^{\phi _4}$.}
\label{model1}
\end{table}


\begin{table}[ht]\footnotesize\centering 
$\begin{array}{|c|c|c|c|c|c|c|c|c|c|c|}
\hline\multicolumn{2}{|c|}{\text{Model~\hyperref[spec1.5]{1-dual}}} & \multicolumn{9}{c|}{ {\U(4)_C\times \U(2)_L \times \U(2)_R\times \USp(4)_2\times \USp(4)_4} }\\
\hline\rm{stack} & N & (n^1,l^1)\times(n^2,l^2)\times(n^3,l^3) & n_{\yng(2)}& n_{\yng(1,1)_{}} & b & b' & c & c' & 2 &  4\\
\hline
a & 8 & (1, -1)\times (1, 0)\times (1, 1) & 0 & 0  & -3 & 0 & 3 & 0 & -1 &  1\\
b & 4 &  (-2, -1)\times (-1, -1)\times (1, -1) & 2 & 6  & \text{-} & \text{-} & 0 & -4 & 1 &  -2\\
c & 4 &  (0, 1)\times (-1, 3)\times (-1, 1) & 2 & -2  & \text{-} & \text{-} & \text{-} & \text{-} & 1 &  0\\
\hline
 &   &  & \multicolumn{8}{c|}{ x_A = \frac{1}{3}x_B = x_C = \frac{1}{3}x_D }\\
2 & 4 &  (1, 0)\times (0, -1)\times (0, 2) & \multicolumn{8}{c|}{ ~\beta^g_2=-2,~\beta^g_4=-2, }\\
 &   &   & \multicolumn{8}{c|}{ \chi_1=1 , ~ \chi_2=\frac{1}{3} , ~ \chi_3=2 }\\
4 & 4  &   (0, -1)\times (0, 1)\times (2, 0) & \multicolumn{8}{c|}{}\\
\hline  
\end{array}$
\caption{D6-brane configurations and intersection numbers of Model~\hyperref[spec1.5]{1-dual}, and its MSSM gauge coupling relation is $g_a^2=\frac{5}{3}g_b^2=g_c^2=\frac{5 g_Y^2}{3}=4 \sqrt{\frac{2}{3}} \, \pi \,  e^{\phi _4}$.}
\label{model1.5}
\end{table}


\begin{table}[ht]\footnotesize\centering 
$\begin{array}{|c|c|c|c|c|c|c|c|c|c|c|c|c|}
\hline\multicolumn{2}{|c|}{\text{Model~\hyperref[spec2]{2}}} & \multicolumn{11}{c|}{ {\U(4)_C\times \U(2)_L \times \U(2)_R\times \USp(2)_1\times \USp(2)_2\times \USp(2)_3\times \USp(2)_4} }\\
\hline\rm{stack} & N & (n^1,l^1)\times(n^2,l^2)\times(n^3,l^3) & n_{\yng(2)}& n_{\yng(1,1)_{}} & b & b' & c & c' & 1 &  2 &  3 &  4\\
\hline
a & 8 & (1, -1)\times (1, 0)\times (1, 1) & 0 & 0  & -2 & -1 & 3 & 0 & 0 &  -1 &  0 &  1\\
b & 4 &  (-1, 0)\times (-1, -1)\times (1, -3) & 2 & -2  & \text{-} & \text{-} & -4 & -4 & 0 &  0 &  1 &  -3\\
c & 4 &  (0, 1)\times (-1, 3)\times (-1, 1) & 2 & -2  & \text{-} & \text{-} & \text{-} & \text{-} & -3 &  1 &  0 &  0\\
\hline
1 & 2  &  (1, 0)\times (1, 0)\times (2, 0) & \multicolumn{10}{c|}{ x_A = \frac{1}{3}x_B = \frac{1}{9}x_C = \frac{1}{3}x_D }\\
2 & 2 &  (1, 0)\times (0, -1)\times (0, 2) & \multicolumn{10}{c|}{ \beta^g_1=-3,~\beta^g_2=-3,~\beta^g_3=-5,~\beta^g_4=-1, }\\
3 & 2  &  (0, -1)\times (1, 0)\times (0, 2)  & \multicolumn{10}{c|}{ \chi_1=\frac{1}{3} , ~ \chi_2=1 , ~ \chi_3=\frac{2}{3} }\\
4 & 2  &   (0, -1)\times (0, 1)\times (2, 0) & \multicolumn{10}{c|}{}\\
\hline  
\end{array}$
\caption{D6-brane configurations and intersection numbers of Model~\hyperref[spec2]{2}, and its MSSM gauge coupling relation is $g_a^2=\frac{9}{5}g_b^2=g_c^2=\frac{5 g_Y^2}{3}=\frac{12}{5} \sqrt{2} \, \pi \,  e^{\phi _4}$.}
\label{model2}
\end{table}


\begin{table}[ht]\footnotesize\centering 
$\begin{array}{|c|c|c|c|c|c|c|c|c|c|}
\hline\multicolumn{2}{|c|}{\text{Model~\hyperref[spec3]{3}}} & \multicolumn{8}{c|}{ {\U(4)_C\times \U(2)_L \times \U(2)_R\times \USp(4)_4} }\\
\hline\rm{stack} & N & (n^1,l^1)\times(n^2,l^2)\times(n^3,l^3) & n_{\yng(2)}& n_{\yng(1,1)_{}} & b & b' & c & c' & 4\\
\hline
a & 8 & (-2, 1)\times (-1, 0)\times (1, 1) & -1 & 1  & -3 & 0 & 3 & 0 & 2\\
b & 4 &  (1, 0)\times (1, 3)\times (1, -1) & -2 & 2  & \text{-} & \text{-} & 0 & 4 & -1\\
c & 4 &  (-1, 2)\times (-1, 1)\times (-1, 1) & 2 & 6  & \text{-} & \text{-} & \text{-} & \text{-} & 1\\
\hline
 &   &  & \multicolumn{7}{c|}{ x_A = \frac{4}{3}x_B = 8x_C = \frac{8}{3}x_D }\\
  &   &   & \multicolumn{7}{c|}{ ~\beta^g_4=0, }\\
 &   &   & \multicolumn{7}{c|}{ \chi_1=4 , ~ \chi_2=\frac{2}{3} , ~ \chi_3=4 }\\
4 & 4  &   (0, -1)\times (0, 1)\times (2, 0) & \multicolumn{7}{c|}{}\\
\hline  
\end{array}$
\caption{D6-brane configurations and intersection numbers of Model~\hyperref[spec3]{3}, and its MSSM gauge coupling relation is $g_a^2=\frac{1}{2}g_b^2=\frac{13}{6}g_c^2=\frac{65}{44}\frac{5 g_Y^2}{3}=\frac{16}{5} \sqrt{\frac{2}{3}} \, \pi \,  e^{\phi _4}$.}
\label{model3}
\end{table}


\begin{table}[ht]\footnotesize\centering 
$\begin{array}{|c|c|c|c|c|c|c|c|c|c|}
\hline\multicolumn{2}{|c|}{\text{Model~\hyperref[spec3.5]{3-dual}}} & \multicolumn{8}{c|}{ {\U(4)_C\times \U(2)_L \times \U(2)_R\times \USp(4)_4} }\\
\hline\rm{stack} & N & (n^1,l^1)\times(n^2,l^2)\times(n^3,l^3) & n_{\yng(2)}& n_{\yng(1,1)_{}} & b & b' & c & c' & 4\\
\hline
a & 8 & (-2, 1)\times (-1, 0)\times (1, 1) & -1 & 1  & 3 & 0 & -3 & 0 & 2\\
b & 4 &  (-1, 2)\times (-1, 1)\times (-1, 1) & 2 & 6  & \text{-} & \text{-} & 0 & 4 & 1\\
c & 4 &  (1, 0)\times (1, 3)\times (1, -1) & -2 & 2  & \text{-} & \text{-} & \text{-} & \text{-} & -1\\
\hline
 &   &  & \multicolumn{7}{c|}{ x_A = \frac{4}{3}x_B = 8x_C = \frac{8}{3}x_D }\\
  &   &   & \multicolumn{7}{c|}{ ~\beta^g_4=0, }\\
 &   &   & \multicolumn{7}{c|}{ \chi_1=4 , ~ \chi_2=\frac{2}{3} , ~ \chi_3=4 }\\
4 & 4  &   (0, -1)\times (0, 1)\times (2, 0) & \multicolumn{7}{c|}{}\\
\hline  
\end{array}$
\caption{D6-brane configurations and intersection numbers of Model~\hyperref[spec3.5]{3-dual}, and its MSSM gauge coupling relation is $g_a^2=\frac{13}{6}g_b^2=\frac{1}{2}g_c^2=\frac{5}{8}\frac{5 g_Y^2}{3}=\frac{16}{5} \sqrt{\frac{2}{3}} \, \pi \,  e^{\phi _4}$.}
\label{model3.5}
\end{table}


\begin{table}[ht]\footnotesize\centering 
$\begin{array}{|c|c|c|c|c|c|c|c|c|c|}
\hline\multicolumn{2}{|c|}{\text{Model~\hyperref[spec4]{4}}} & \multicolumn{8}{c|}{ {\U(4)_C\times \U(2)_L \times \U(2)_R\times \USp(4)_4} }\\
\hline\rm{stack} & N & (n^1,l^1)\times(n^2,l^2)\times(n^3,l^3) & n_{\yng(2)}& n_{\yng(1,1)_{}} & b & b' & c & c' & 4\\
\hline
a & 8 & (2, -1)\times (1, 0)\times (1, 1) & -1 & 1  & -2 & -1 & 3 & 0 & 2\\
b & 4 &  (1, 0)\times (1, 1)\times (1, -3) & 2 & -2  & \text{-} & \text{-} & -4 & 0 & -3\\
c & 4 &  (-1, 2)\times (-1, 1)\times (-1, 1) & 2 & 6  & \text{-} & \text{-} & \text{-} & \text{-} & 1\\
\hline
 &   &  & \multicolumn{7}{c|}{ x_A = 2x_B = \frac{4}{3}x_C = 4x_D }\\
  &   &   & \multicolumn{7}{c|}{ ~\beta^g_4=2, }\\
 &   &   & \multicolumn{7}{c|}{ \chi_1=2 \sqrt{\frac{2}{3}} , ~ \chi_2=\sqrt{6} , ~ \chi_3=2 \sqrt{\frac{2}{3}} }\\
4 & 4  &   (0, -1)\times (0, 1)\times (2, 0) & \multicolumn{7}{c|}{}\\
\hline  
\end{array}$
\caption{D6-brane configurations and intersection numbers of Model~\hyperref[spec4]{4}, and its MSSM gauge coupling relation is $g_a^2=\frac{21}{10}g_b^2=\frac{7}{2}g_c^2=\frac{7}{4}\frac{5 g_Y^2}{3}=\frac{8}{5} 6^{3/4} \, \pi \,  e^{\phi _4}$.}
\label{model4}
\end{table}


\begin{table}[ht]\footnotesize\centering 
$\begin{array}{|c|c|c|c|c|c|c|c|c|c|c|c|}
\hline\multicolumn{2}{|c|}{\text{Model~\hyperref[spec5]{5}}} & \multicolumn{10}{c|}{ {\U(4)_C\times \U(2)_L \times \U(2)_R\times \USp(2)_1\times \USp(2)_2\times \USp(2)_4} }\\
\hline\rm{stack} & N & (n^1,l^1)\times(n^2,l^2)\times(n^3,l^3) & n_{\yng(2)}& n_{\yng(1,1)_{}} & b & b' & c & c' & 1 &  2 &  4\\
\hline
a & 8 & (1, -1)\times (1, 0)\times (1, 1) & 0 & 0  & -2 & -1 & 3 & 0 & 0 &  -1 &  1\\
b & 4 &  (-1, 0)\times (-1, -1)\times (1, -3) & 2 & -2  & \text{-} & \text{-} & -5 & -2 & 0 &  0 &  -3\\
c & 4 &  (0, 1)\times (-2, 3)\times (-1, 1) & 1 & -1  & \text{-} & \text{-} & \text{-} & \text{-} & -3 &  2 &  0\\
\hline
1 & 2  &  (1, 0)\times (1, 0)\times (2, 0) & \multicolumn{9}{c|}{ x_A = \frac{2}{3}x_B = \frac{2}{9}x_C = \frac{2}{3}x_D }\\
2 & 2 &  (1, 0)\times (0, -1)\times (0, 2) & \multicolumn{9}{c|}{ \beta^g_1=-3,~\beta^g_2=-2,~\beta^g_4=-1, }\\
 &   &   & \multicolumn{9}{c|}{ \chi_1=\frac{\sqrt{2}}{3} , ~ \chi_2=\sqrt{2} , ~ \chi_3=\frac{2 \sqrt{2}}{3} }\\
4 & 2  &   (0, -1)\times (0, 1)\times (2, 0) & \multicolumn{9}{c|}{}\\
\hline  
\end{array}$
\caption{D6-brane configurations and intersection numbers of Model~\hyperref[spec5]{5}, and its MSSM gauge coupling relation is $g_a^2=\frac{27}{11}g_b^2=2g_c^2=\frac{10}{7}\frac{5 g_Y^2}{3}=\frac{48}{11} \sqrt[4]{2} \, \pi \,  e^{\phi _4}$.}
\label{model5}
\end{table}


\begin{table}[ht]\footnotesize\centering 
$\begin{array}{|c|c|c|c|c|c|c|c|c|c|c|c|}
\hline\multicolumn{2}{|c|}{\text{Model~\hyperref[spec6]{6}}} & \multicolumn{10}{c|}{ {\U(4)_C\times \U(2)_L \times \U(2)_R\times \USp(2)_2\times \USp(2)_3\times \USp(2)_4} }\\
\hline\rm{stack} & N & (n^1,l^1)\times(n^2,l^2)\times(n^3,l^3) & n_{\yng(2)}& n_{\yng(1,1)_{}} & b & b' & c & c' & 2 &  3 &  4\\
\hline
a & 8 & (1, -1)\times (1, 0)\times (1, 1) & 0 & 0  & -2 & -1 & 3 & 0 & -1 &  0 &  1\\
b & 4 &  (-1, 0)\times (-2, -1)\times (1, -3) & 5 & -5  & \text{-} & \text{-} & -7 & -10 & 0 &  1 &  -6\\
c & 4 &  (0, 1)\times (-1, 3)\times (-1, 1) & 2 & -2  & \text{-} & \text{-} & \text{-} & \text{-} & 1 &  0 &  0\\
\hline
 &   &  & \multicolumn{9}{c|}{ x_A = \frac{1}{3}x_B = \frac{1}{18}x_C = \frac{1}{3}x_D }\\
2 & 2 &  (1, 0)\times (0, -1)\times (0, 2) & \multicolumn{9}{c|}{ ~\beta^g_2=-3,~\beta^g_3=-5,~\beta^g_4=2, }\\
3 & 2  &  (0, -1)\times (1, 0)\times (0, 2)  & \multicolumn{9}{c|}{ \chi_1=\frac{1}{3 \sqrt{2}} , ~ \chi_2=\sqrt{2} , ~ \chi_3=\frac{\sqrt{2}}{3} }\\
4 & 2  &   (0, -1)\times (0, 1)\times (2, 0) & \multicolumn{9}{c|}{}\\
\hline  
\end{array}$
\caption{D6-brane configurations and intersection numbers of Model~\hyperref[spec6]{6}, and its MSSM gauge coupling relation is $g_a^2=\frac{54}{19}g_b^2=g_c^2=\frac{5 g_Y^2}{3}=\frac{48}{19} \sqrt[4]{2} \, \pi \,  e^{\phi _4}$.}
\label{model6}
\end{table}


\begin{table}[ht]\footnotesize\centering 
$\begin{array}{|c|c|c|c|c|c|c|c|c|c|c|}
\hline\multicolumn{2}{|c|}{\text{Model~\hyperref[spec7]{7}}} & \multicolumn{9}{c|}{ {\U(4)_C\times \U(2)_L \times \U(2)_R\times \USp(2)_2\times \USp(2)_4} }\\
\hline\rm{stack} & N & (n^1,l^1)\times(n^2,l^2)\times(n^3,l^3) & n_{\yng(2)}& n_{\yng(1,1)_{}} & b & b' & c & c' & 2 &  4\\
\hline
a & 8 & (1, -1)\times (1, 0)\times (1, 1) & 0 & 0  & -2 & -1 & 3 & 0 & -1 &  1\\
b & 4 &  (-1, 0)\times (-2, -1)\times (1, -3) & 5 & -5  & \text{-} & \text{-} & -8 & -8 & 0 &  -6\\
c & 4 &  (0, 1)\times (-2, 3)\times (-1, 1) & 1 & -1  & \text{-} & \text{-} & \text{-} & \text{-} & 2 &  0\\
\hline
 &   &  & \multicolumn{8}{c|}{ x_A = \frac{2}{3}x_B = \frac{1}{9}x_C = \frac{2}{3}x_D }\\
2 & 2 &  (1, 0)\times (0, -1)\times (0, 2) & \multicolumn{8}{c|}{ ~\beta^g_2=-2,~\beta^g_4=2, }\\
 &   &   & \multicolumn{8}{c|}{ \chi_1=\frac{1}{3} , ~ \chi_2=2 , ~ \chi_3=\frac{2}{3} }\\
4 & 2  &   (0, -1)\times (0, 1)\times (2, 0) & \multicolumn{8}{c|}{}\\
\hline  
\end{array}$
\caption{D6-brane configurations and intersection numbers of Model~\hyperref[spec7]{7}, and its MSSM gauge coupling relation is $g_a^2=\frac{18}{5}g_b^2=2g_c^2=\frac{10}{7}\frac{5 g_Y^2}{3}=\frac{24 \, \pi \,  e^{\phi _4}}{5}$.}
\label{model7}
\end{table}


\begin{table}[ht]\footnotesize\centering 
$\begin{array}{|c|c|c|c|c|c|c|c|c|c|c|}
\hline\multicolumn{2}{|c|}{\text{Model~\hyperref[spec8]{8}}} & \multicolumn{9}{c|}{ {\U(4)_C\times \U(2)_L \times \U(2)_R\times \USp(2)_1\times \USp(4)_3} }\\
\hline\rm{stack} & N & (n^1,l^1)\times(n^2,l^2)\times(n^3,l^3) & n_{\yng(2)}& n_{\yng(1,1)_{}} & b & b' & c & c' & 1 &  3\\
\hline
a & 8 & (1, -1)\times (-1, 0)\times (-1, -1) & 0 & 0  & -2 & -1 & 3 & 0 & 0 &  0\\
b & 4 &  (1, 0)\times (1, 1)\times (1, -3) & 2 & -2  & \text{-} & \text{-} & -4 & -8 & 0 &  1\\
c & 4 &  (-1, 4)\times (0, 1)\times (-1, 1) & 3 & -3  & \text{-} & \text{-} & \text{-} & \text{-} & -4 &  1\\
\hline
1 & 2  &  (1, 0)\times (1, 0)\times (2, 0) & \multicolumn{8}{c|}{ x_A = \frac{3}{4}x_B = \frac{1}{4}x_C = \frac{3}{4}x_D }\\
  &   &   & \multicolumn{8}{c|}{ \beta^g_1=-2,~\beta^g_3=-4, }\\
3 & 4  &  (0, -1)\times (1, 0)\times (0, 2)  & \multicolumn{8}{c|}{ \chi_1=\frac{1}{2} , ~ \chi_2=\frac{3}{2} , ~ \chi_3=1 }\\
 &   &    & \multicolumn{8}{c|}{}\\
\hline  
\end{array}$
\caption{D6-brane configurations and intersection numbers of Model~\hyperref[spec8]{8}, and its MSSM gauge coupling relation is $g_a^2=\frac{13}{5}g_b^2=3g_c^2=\frac{5}{3}\frac{5 g_Y^2}{3}=\frac{16}{5} \sqrt{3} \, \pi \,  e^{\phi _4}$.}
\label{model8}
\end{table}


\begin{table}[ht]\footnotesize\centering 
$\begin{array}{|c|c|c|c|c|c|c|c|c|c|c|}
\hline\multicolumn{2}{|c|}{\text{Model~\hyperref[spec9]{9}}} & \multicolumn{9}{c|}{ {\U(4)_C\times \U(2)_L \times \U(2)_R\times \USp(2)_1\times \USp(4)_3} }\\
\hline\rm{stack} & N & (n^1,l^1)\times(n^2,l^2)\times(n^3,l^3) & n_{\yng(2)}& n_{\yng(1,1)_{}} & b & b' & c & c' & 1 &  3\\
\hline
a & 8 & (-1, 1)\times (-1, 0)\times (1, 1) & 0 & 0  & 3 & 0 & -3 & 0 & 0 &  0\\
b & 4 &  (-1, 4)\times (0, 1)\times (-1, 1) & 3 & -3  & \text{-} & \text{-} & 0 & -4 & -4 &  1\\
c & 4 &  (1, 0)\times (1, 3)\times (1, -1) & -2 & 2  & \text{-} & \text{-} & \text{-} & \text{-} & 0 &  3\\
\hline
1 & 2  &  (1, 0)\times (1, 0)\times (2, 0) & \multicolumn{8}{c|}{ x_A = \frac{1}{12}x_B = \frac{1}{4}x_C = \frac{1}{12}x_D }\\
  &   &   & \multicolumn{8}{c|}{ \beta^g_1=-2,~\beta^g_3=-2, }\\
3 & 4  &  (0, -1)\times (1, 0)\times (0, 2)  & \multicolumn{8}{c|}{ \chi_1=\frac{1}{2} , ~ \chi_2=\frac{1}{6} , ~ \chi_3=1 }\\
 &   &    & \multicolumn{8}{c|}{}\\
\hline  
\end{array}$
\caption{D6-brane configurations and intersection numbers of Model~\hyperref[spec9]{9}, and its MSSM gauge coupling relation is $g_a^2=\frac{1}{3}g_b^2=g_c^2=\frac{5 g_Y^2}{3}=\frac{16 \, \pi \,  e^{\phi _4}}{5 \sqrt{3}}$.}
\label{model9}
\end{table}


\begin{table}[ht]\footnotesize\centering 
$\begin{array}{|c|c|c|c|c|c|c|c|c|c|c|}
\hline\multicolumn{2}{|c|}{\text{Model~\hyperref[spec9.5]{9-dual}}} & \multicolumn{9}{c|}{ {\U(4)_C\times \U(2)_L \times \U(2)_R\times \USp(2)_1\times \USp(4)_3} }\\
\hline\rm{stack} & N & (n^1,l^1)\times(n^2,l^2)\times(n^3,l^3) & n_{\yng(2)}& n_{\yng(1,1)_{}} & b & b' & c & c' & 1 &  3\\
\hline
a & 8 & (-1, 1)\times (-1, 0)\times (1, 1) & 0 & 0  & -3 & 0 & 3 & 0 & 0 &  0\\
b & 4 &  (1, 0)\times (1, 3)\times (1, -1) & -2 & 2  & \text{-} & \text{-} & 0 & -4 & 0 &  3\\
c & 4 &  (-1, 4)\times (0, 1)\times (-1, 1) & 3 & -3  & \text{-} & \text{-} & \text{-} & \text{-} & -4 &  1\\
\hline
1 & 2  &  (1, 0)\times (1, 0)\times (2, 0) & \multicolumn{8}{c|}{ x_A = \frac{1}{12}x_B = \frac{1}{4}x_C = \frac{1}{12}x_D }\\
  &   &   & \multicolumn{8}{c|}{ \beta^g_1=-2,~\beta^g_3=-2, }\\
3 & 4  &  (0, -1)\times (1, 0)\times (0, 2)  & \multicolumn{8}{c|}{ \chi_1=\frac{1}{2} , ~ \chi_2=\frac{1}{6} , ~ \chi_3=1 }\\
 &   &    & \multicolumn{8}{c|}{}\\
\hline  
\end{array}$
\caption{D6-brane configurations and intersection numbers of Model~\hyperref[spec9.5]{9-dual}, and its MSSM gauge coupling relation is $g_a^2=g_b^2=\frac{1}{3}g_c^2=\frac{5}{11}\frac{5 g_Y^2}{3}=\frac{16 \, \pi \,  e^{\phi _4}}{5 \sqrt{3}}$.}
\label{model9.5}
\end{table}


\begin{table}[ht]\footnotesize\centering 
$\begin{array}{|c|c|c|c|c|c|c|c|c|c|}
\hline\multicolumn{2}{|c|}{\text{Model~\hyperref[spec10]{10}}} & \multicolumn{8}{c|}{ {\U(4)_C\times \U(2)_L \times \U(2)_R\times \USp(4)_3} }\\
\hline\rm{stack} & N & (n^1,l^1)\times(n^2,l^2)\times(n^3,l^3) & n_{\yng(2)}& n_{\yng(1,1)_{}} & b & b' & c & c' & 3\\
\hline
a & 8 & (1, -1)\times (-1, 0)\times (-1, -1) & 0 & 0  & -2 & -1 & 3 & 0 & 0\\
b & 4 &  (1, 0)\times (2, 1)\times (1, -3) & 5 & -5  & \text{-} & \text{-} & -8 & -16 & 1\\
c & 4 &  (-1, 4)\times (0, 1)\times (-1, 1) & 3 & -3  & \text{-} & \text{-} & \text{-} & \text{-} & 1\\
\hline
 &   &  & \multicolumn{7}{c|}{ x_A = \frac{3}{2}x_B = \frac{1}{4}x_C = \frac{3}{2}x_D }\\
  &   &   & \multicolumn{7}{c|}{ ~\beta^g_3=-4, }\\
3 & 4  &  (0, -1)\times (1, 0)\times (0, 2)  & \multicolumn{7}{c|}{ \chi_1=\frac{1}{2} , ~ \chi_2=3 , ~ \chi_3=1 }\\
 &   &    & \multicolumn{7}{c|}{}\\
\hline  
\end{array}$
\caption{D6-brane configurations and intersection numbers of Model~\hyperref[spec10]{10}, and its MSSM gauge coupling relation is $g_a^2=\frac{26}{5}g_b^2=6g_c^2=2\frac{5 g_Y^2}{3}=\frac{16}{5} \sqrt{6} \, \pi \,  e^{\phi _4}$.}
\label{model10}
\end{table}


\begin{table}[ht]\footnotesize\centering 
$\begin{array}{|c|c|c|c|c|c|c|c|c|c|}
\hline\multicolumn{2}{|c|}{\text{Model~\hyperref[spec11]{11}}} & \multicolumn{8}{c|}{ {\U(4)_C\times \U(2)_L \times \U(2)_R\times \USp(4)_3} }\\
\hline\rm{stack} & N & (n^1,l^1)\times(n^2,l^2)\times(n^3,l^3) & n_{\yng(2)}& n_{\yng(1,1)_{}} & b & b' & c & c' & 3\\
\hline
a & 8 & (1, -1)\times (-1, 0)\times (-1, -1) & 0 & 0  & 3 & 0 & -3 & 0 & 0\\
b & 4 &  (-1, 4)\times (0, 1)\times (-1, 1) & 3 & -3  & \text{-} & \text{-} & 0 & -8 & 1\\
c & 4 &  (1, 0)\times (2, 3)\times (1, -1) & -1 & 1  & \text{-} & \text{-} & \text{-} & \text{-} & 3\\
\hline
 &   &  & \multicolumn{7}{c|}{ x_A = \frac{1}{6}x_B = \frac{1}{4}x_C = \frac{1}{6}x_D }\\
  &   &   & \multicolumn{7}{c|}{ ~\beta^g_3=-2, }\\
3 & 4  &  (0, -1)\times (1, 0)\times (0, 2)  & \multicolumn{7}{c|}{ \chi_1=\frac{1}{2} , ~ \chi_2=\frac{1}{3} , ~ \chi_3=1 }\\
 &   &    & \multicolumn{7}{c|}{}\\
\hline  
\end{array}$
\caption{D6-brane configurations and intersection numbers of Model~\hyperref[spec11]{11}, and its MSSM gauge coupling relation is $g_a^2=\frac{2}{3}g_b^2=2g_c^2=\frac{10}{7}\frac{5 g_Y^2}{3}=\frac{16}{5} \sqrt{\frac{2}{3}} \, \pi \,  e^{\phi _4}$.}
\label{model11}
\end{table}


\begin{table}[ht]\footnotesize\centering 
$\begin{array}{|c|c|c|c|c|c|c|c|c|c|}
\hline\multicolumn{2}{|c|}{\text{Model~\hyperref[spec11.5]{11-dual}}} & \multicolumn{8}{c|}{ {\U(4)_C\times \U(2)_L \times \U(2)_R\times \USp(4)_3} }\\
\hline\rm{stack} & N & (n^1,l^1)\times(n^2,l^2)\times(n^3,l^3) & n_{\yng(2)}& n_{\yng(1,1)_{}} & b & b' & c & c' & 3\\
\hline
a & 8 & (1, -1)\times (-1, 0)\times (-1, -1) & 0 & 0  & -3 & 0 & 3 & 0 & 0\\
b & 4 &  (1, 0)\times (2, 3)\times (1, -1) & -1 & 1  & \text{-} & \text{-} & 0 & -8 & 3\\
c & 4 &  (-1, 4)\times (0, 1)\times (-1, 1) & 3 & -3  & \text{-} & \text{-} & \text{-} & \text{-} & 1\\
\hline
 &   &  & \multicolumn{7}{c|}{ x_A = \frac{1}{6}x_B = \frac{1}{4}x_C = \frac{1}{6}x_D }\\
  &   &   & \multicolumn{7}{c|}{ ~\beta^g_3=-2, }\\
3 & 4  &  (0, -1)\times (1, 0)\times (0, 2)  & \multicolumn{7}{c|}{ \chi_1=\frac{1}{2} , ~ \chi_2=\frac{1}{3} , ~ \chi_3=1 }\\
 &   &    & \multicolumn{7}{c|}{}\\
\hline  
\end{array}$
\caption{D6-brane configurations and intersection numbers of Model~\hyperref[spec11.5]{11-dual}, and its MSSM gauge coupling relation is $g_a^2=2g_b^2=\frac{2}{3}g_c^2=\frac{10}{13}\frac{5 g_Y^2}{3}=\frac{16}{5} \sqrt{\frac{2}{3}} \, \pi \,  e^{\phi _4}$.}
\label{model11.5}
\end{table}


\begin{table}[ht]\footnotesize\centering 
$\begin{array}{|c|c|c|c|c|c|c|c|c|c|}
\hline\multicolumn{2}{|c|}{\text{Model~\hyperref[spec12]{12}}} & \multicolumn{8}{c|}{ {\U(4)_C\times \U(2)_L \times \U(2)_R\times \USp(2)_3} }\\
\hline\rm{stack} & N & (n^1,l^1)\times(n^2,l^2)\times(n^3,l^3) & n_{\yng(2)}& n_{\yng(1,1)_{}} & b & b' & c & c' & 3\\
\hline
a & 8 & (1, 1)\times (1, 0)\times (1, -1) & 0 & 0  & 3 & -6 & 3 & 0 & 0\\
b & 4 &  (-2, -1)\times (0, -1)\times (-5, -1) & 9 & -9  & \text{-} & \text{-} & -8 & 0 & -10\\
c & 4 &  (-2, 1)\times (-1, 1)\times (1, 1) & -2 & -6  & \text{-} & \text{-} & \text{-} & \text{-} & -2\\
\hline
 &   &  & \multicolumn{7}{c|}{ x_A = \frac{5}{6}x_B = 10x_C = \frac{5}{6}x_D }\\
  &   &   & \multicolumn{7}{c|}{ ~\beta^g_3=6, }\\
3 & 2  &  (0, -1)\times (1, 0)\times (0, 2)  & \multicolumn{7}{c|}{ \chi_1=\sqrt{10} , ~ \chi_2=\frac{\sqrt{\frac{5}{2}}}{6} , ~ \chi_3=2 \sqrt{10} }\\
 &   &    & \multicolumn{7}{c|}{}\\
\hline  
\end{array}$
\caption{D6-brane configurations and intersection numbers of Model~\hyperref[spec12]{12}, and its MSSM gauge coupling relation is $g_a^2=\frac{35}{66}g_b^2=\frac{7}{6}g_c^2=\frac{35}{32}\frac{5 g_Y^2}{3}=\frac{8 \sqrt[4]{2} 5^{3/4} \, \pi \,  e^{\phi _4}}{11 \sqrt{3}}$.}
\label{model12}
\end{table}


\begin{table}[ht]\footnotesize\centering 
$\begin{array}{|c|c|c|c|c|c|c|c|c|c|c|}
\hline\multicolumn{2}{|c|}{\text{Model~\hyperref[spec13]{13}}} & \multicolumn{9}{c|}{ {\U(4)_C\times \U(2)_L \times \U(2)_R\times \USp(2)_1\times \USp(2)_4} }\\
\hline\rm{stack} & N & (n^1,l^1)\times(n^2,l^2)\times(n^3,l^3) & n_{\yng(2)}& n_{\yng(1,1)_{}} & b & b' & c & c' & 1 &  4\\
\hline
a & 8 & (-1, -1)\times (1, 1)\times (1, 1) & 0 & -4  & 6 & -3 & -3 & 0 & 1 &  -1\\
b & 4 &  (-1, 2)\times (-1, 0)\times (5, 1) & 9 & -9  & \text{-} & \text{-} & -9 & -10 & 0 &  1\\
c & 4 &  (-2, 1)\times (0, -1)\times (1, -1) & -1 & 1  & \text{-} & \text{-} & \text{-} & \text{-} & -1 &  0\\
\hline
1 & 2  &  (1, 0)\times (1, 0)\times (2, 0) & \multicolumn{8}{c|}{ x_A = 22x_B = 2x_C = \frac{11}{5}x_D }\\
  &   &   & \multicolumn{8}{c|}{ \beta^g_1=-3,~\beta^g_4=-3, }\\
 &   &   & \multicolumn{8}{c|}{ \chi_1=\frac{1}{\sqrt{5}} , ~ \chi_2=\frac{11}{\sqrt{5}} , ~ \chi_3=4 \sqrt{5} }\\
4 & 2  &   (0, -1)\times (0, 1)\times (2, 0) & \multicolumn{8}{c|}{}\\
\hline  
\end{array}$
\caption{D6-brane configurations and intersection numbers of Model~\hyperref[spec13]{13}, and its MSSM gauge coupling relation is $g_a^2=\frac{5}{14}g_b^2=\frac{11}{6}g_c^2=\frac{11}{8}\frac{5 g_Y^2}{3}=\frac{8}{63} 5^{3/4} \sqrt{11} \, \pi \,  e^{\phi _4}$.}
\label{model13}
\end{table}


\begin{table}[ht]\footnotesize\centering 
$\begin{array}{|c|c|c|c|c|c|c|c|c|c|c|c|c|}
\hline\multicolumn{2}{|c|}{\text{Model~\hyperref[spec14]{14}}} & \multicolumn{11}{c|}{ {\U(4)_C\times \U(2)_L \times \U(2)_R\times \USp(2)_1\times \USp(2)_2\times \USp(2)_3\times \USp(2)_4} }\\
\hline\rm{stack} & N & (n^1,l^1)\times(n^2,l^2)\times(n^3,l^3) & n_{\yng(2)}& n_{\yng(1,1)_{}} & b & b' & c & c' & 1 &  2 &  3 &  4\\
\hline
a & 8 & (1, -1)\times (1, 0)\times (1, 1) & 0 & 0  & -3 & 0 & 3 & 0 & 0 &  -1 &  0 &  1\\
b & 4 &  (-1, 0)\times (-1, -3)\times (1, -1) & -2 & 2  & \text{-} & \text{-} & 0 & 0 & 0 &  0 &  3 &  -1\\
c & 4 &  (0, 1)\times (-1, 3)\times (-1, 1) & 2 & -2  & \text{-} & \text{-} & \text{-} & \text{-} & -3 &  1 &  0 &  0\\
\hline
1 & 2  &  (1, 0)\times (1, 0)\times (2, 0) & \multicolumn{10}{c|}{ x_A = \frac{1}{3}x_B = x_C = \frac{1}{3}x_D }\\
2 & 2 &  (1, 0)\times (0, -1)\times (0, 2) & \multicolumn{10}{c|}{ \beta^g_1=-3,~\beta^g_2=-3,~\beta^g_3=-3,~\beta^g_4=-3, }\\
3 & 2  &  (0, -1)\times (1, 0)\times (0, 2)  & \multicolumn{10}{c|}{ \chi_1=1 , ~ \chi_2=\frac{1}{3} , ~ \chi_3=2 }\\
4 & 2  &   (0, -1)\times (0, 1)\times (2, 0) & \multicolumn{10}{c|}{}\\
\hline  
\end{array}$
\caption{D6-brane configurations and intersection numbers of Model~\hyperref[spec14]{14}, and its MSSM gauge coupling relation is $g_a^2=g_b^2=g_c^2=\frac{5 g_Y^2}{3}=4 \sqrt{\frac{2}{3}} \, \pi \,  e^{\phi _4}$.}
\label{model14}
\end{table}


\begin{table}[ht]\footnotesize\centering 
$\begin{array}{|c|c|c|c|c|c|c|c|c|c|c|c|}
\hline\multicolumn{2}{|c|}{\text{Model~\hyperref[spec15]{15}}} & \multicolumn{10}{c|}{ {\U(4)_C\times \U(2)_L \times \U(2)_R\times \USp(4)_1\times \USp(2)_3\times \USp(2)_4} }\\
\hline\rm{stack} & N & (n^1,l^1)\times(n^2,l^2)\times(n^3,l^3) & n_{\yng(2)}& n_{\yng(1,1)_{}} & b & b' & c & c' & 1 &  3 &  4\\
\hline
a & 8 & (-1, 1)\times (1, -1)\times (1, -1) & 0 & 4  & -3 & 0 & 3 & 0 & -1 &  1 &  1\\
b & 4 &  (-4, 1)\times (-1, 0)\times (1, 1) & -3 & 3  & \text{-} & \text{-} & 0 & 2 & 0 &  0 &  4\\
c & 4 &  (-2, -1)\times (0, 1)\times (1, 1) & 1 & -1  & \text{-} & \text{-} & \text{-} & \text{-} & 1 &  -2 &  0\\
\hline
1 & 4  &  (1, 0)\times (1, 0)\times (2, 0) & \multicolumn{9}{c|}{ x_A = \frac{5}{2}x_B = 2x_C = 10x_D }\\
  &   &   & \multicolumn{9}{c|}{ \beta^g_1=-3,~\beta^g_3=-2,~\beta^g_4=0, }\\
3 & 2  &  (0, -1)\times (1, 0)\times (0, 2)  & \multicolumn{9}{c|}{ \chi_1=2 \sqrt{2} , ~ \chi_2=\frac{5}{\sqrt{2}} , ~ \chi_3=\sqrt{2} }\\
4 & 2  &   (0, -1)\times (0, 1)\times (2, 0) & \multicolumn{9}{c|}{}\\
\hline  
\end{array}$
\caption{D6-brane configurations and intersection numbers of Model~\hyperref[spec15]{15}, and its MSSM gauge coupling relation is $g_a^2=\frac{4}{9}g_b^2=\frac{10}{9}g_c^2=\frac{50}{47}\frac{5 g_Y^2}{3}=\frac{16}{27} 2^{3/4} \sqrt{5} \, \pi \,  e^{\phi _4}$.}
\label{model15}
\end{table}


\begin{table}[ht]\footnotesize\centering 
$\begin{array}{|c|c|c|c|c|c|c|c|c|c|c|c|}
\hline\multicolumn{2}{|c|}{\text{Model~\hyperref[spec15.5]{15-dual}}} & \multicolumn{10}{c|}{ {\U(4)_C\times \U(2)_L \times \U(2)_R\times \USp(4)_1\times \USp(2)_2\times \USp(2)_4} }\\
\hline\rm{stack} & N & (n^1,l^1)\times(n^2,l^2)\times(n^3,l^3) & n_{\yng(2)}& n_{\yng(1,1)_{}} & b & b' & c & c' & 1 &  2 &  4\\
\hline
a & 8 & (1, 1)\times (-1, -1)\times (1, 1) & 0 & -4  & -3 & 0 & 3 & 0 & 1 &  -1 &  -1\\
b & 4 &  (0, 1)\times (-2, 1)\times (-1, 1) & -1 & 1  & \text{-} & \text{-} & 0 & -2 & -1 &  2 &  0\\
c & 4 &  (-1, 0)\times (4, 1)\times (-1, 1) & 3 & -3  & \text{-} & \text{-} & \text{-} & \text{-} & 0 &  0 &  -4\\
\hline
1 & 4  &  (1, 0)\times (1, 0)\times (2, 0) & \multicolumn{9}{c|}{ x_A = 2x_B = \frac{5}{2}x_C = 10x_D }\\
2 & 2 &  (1, 0)\times (0, -1)\times (0, 2) & \multicolumn{9}{c|}{ \beta^g_1=-3,~\beta^g_2=-2,~\beta^g_4=0, }\\
 &   &   & \multicolumn{9}{c|}{ \chi_1=\frac{5}{\sqrt{2}} , ~ \chi_2=2 \sqrt{2} , ~ \chi_3=\sqrt{2} }\\
4 & 2  &   (0, -1)\times (0, 1)\times (2, 0) & \multicolumn{9}{c|}{}\\
\hline  
\end{array}$
\caption{D6-brane configurations and intersection numbers of Model~\hyperref[spec15.5]{15-dual}, and its MSSM gauge coupling relation is $g_a^2=\frac{10}{9}g_b^2=\frac{4}{9}g_c^2=\frac{4}{7}\frac{5 g_Y^2}{3}=\frac{16}{27} 2^{3/4} \sqrt{5} \, \pi \,  e^{\phi _4}$.}
\label{model15.5}
\end{table}


\begin{table}[ht]\footnotesize\centering 
$\begin{array}{|c|c|c|c|c|c|c|c|c|c|c|}
\hline\multicolumn{2}{|c|}{\text{Model~\hyperref[spec16]{16}}} & \multicolumn{9}{c|}{ {\U(4)_C\times \U(2)_L \times \U(2)_R\times \USp(4)_1\times \USp(2)_4} }\\
\hline\rm{stack} & N & (n^1,l^1)\times(n^2,l^2)\times(n^3,l^3) & n_{\yng(2)}& n_{\yng(1,1)_{}} & b & b' & c & c' & 1 &  4\\
\hline
a & 8 & (1, 1)\times (-2, -1)\times (1, 1) & 0 & -8  & 3 & 0 & -3 & 0 & 1 &  -2\\
b & 4 &  (-4, -1)\times (1, 0)\times (-1, 1) & 3 & -3  & \text{-} & \text{-} & 0 & -2 & 0 &  -4\\
c & 4 &  (-2, 1)\times (1, 1)\times (-1, 1) & -2 & -6  & \text{-} & \text{-} & \text{-} & \text{-} & -1 &  -2\\
\hline
1 & 4  &  (1, 0)\times (1, 0)\times (2, 0) & \multicolumn{8}{c|}{ x_A = \frac{13}{2}x_B = \frac{13}{8}x_C = 26x_D }\\
  &   &   & \multicolumn{8}{c|}{ \beta^g_1=-3,~\beta^g_4=4, }\\
 &   &   & \multicolumn{8}{c|}{ \chi_1=\sqrt{\frac{13}{2}} , ~ \chi_2=2 \sqrt{26} , ~ \chi_3=\frac{\sqrt{\frac{13}{2}}}{2} }\\
4 & 2  &   (0, -1)\times (0, 1)\times (2, 0) & \multicolumn{8}{c|}{}\\
\hline  
\end{array}$
\caption{D6-brane configurations and intersection numbers of Model~\hyperref[spec16]{16}, and its MSSM gauge coupling relation is $g_a^2=\frac{1}{6}g_b^2=\frac{7}{6}g_c^2=\frac{35}{32}\frac{5 g_Y^2}{3}=\frac{16}{135} 26^{3/4} \, \pi \,  e^{\phi _4}$.}
\label{model16}
\end{table}


\begin{table}[ht]\footnotesize\centering 
$\begin{array}{|c|c|c|c|c|c|c|c|c|c|c|}
\hline\multicolumn{2}{|c|}{\text{Model~\hyperref[spec16.5]{16-dual}}} & \multicolumn{9}{c|}{ {\U(4)_C\times \U(2)_L \times \U(2)_R\times \USp(4)_1\times \USp(2)_4} }\\
\hline\rm{stack} & N & (n^1,l^1)\times(n^2,l^2)\times(n^3,l^3) & n_{\yng(2)}& n_{\yng(1,1)_{}} & b & b' & c & c' & 1 &  4\\
\hline
a & 8 & (1, 1)\times (-2, -1)\times (1, 1) & 0 & -8  & -3 & 0 & 3 & 0 & 1 &  -2\\
b & 4 &  (-2, 1)\times (1, 1)\times (-1, 1) & -2 & -6  & \text{-} & \text{-} & 0 & -2 & -1 &  -2\\
c & 4 &  (-4, -1)\times (1, 0)\times (-1, 1) & 3 & -3  & \text{-} & \text{-} & \text{-} & \text{-} & 0 &  -4\\
\hline
1 & 4  &  (1, 0)\times (1, 0)\times (2, 0) & \multicolumn{8}{c|}{ x_A = \frac{13}{2}x_B = \frac{13}{8}x_C = 26x_D }\\
  &   &   & \multicolumn{8}{c|}{ \beta^g_1=-3,~\beta^g_4=4, }\\
 &   &   & \multicolumn{8}{c|}{ \chi_1=\sqrt{\frac{13}{2}} , ~ \chi_2=2 \sqrt{26} , ~ \chi_3=\frac{\sqrt{\frac{13}{2}}}{2} }\\
4 & 2  &   (0, -1)\times (0, 1)\times (2, 0) & \multicolumn{8}{c|}{}\\
\hline  
\end{array}$
\caption{D6-brane configurations and intersection numbers of Model~\hyperref[spec16.5]{16-dual}, and its MSSM gauge coupling relation is $g_a^2=\frac{7}{6}g_b^2=\frac{1}{6}g_c^2=\frac{1}{4}\frac{5 g_Y^2}{3}=\frac{16}{135} 26^{3/4} \, \pi \,  e^{\phi _4}$.}
\label{model16.5}
\end{table}


\begin{table}[ht]\footnotesize\centering 
$\begin{array}{|c|c|c|c|c|c|c|c|c|c|c|c|}
\hline\multicolumn{2}{|c|}{\text{Model~\hyperref[spec17]{17}}} & \multicolumn{10}{c|}{ {\U(4)_C\times \U(2)_L \times \U(2)_R\times \USp(2)_2\times \USp(2)_3\times \USp(2)_4} }\\
\hline\rm{stack} & N & (n^1,l^1)\times(n^2,l^2)\times(n^3,l^3) & n_{\yng(2)}& n_{\yng(1,1)_{}} & b & b' & c & c' & 2 &  3 &  4\\
\hline
a & 8 & (1, -1)\times (1, 0)\times (1, 1) & 0 & 0  & -3 & 0 & 3 & 0 & -1 &  0 &  1\\
b & 4 &  (-1, 0)\times (-2, -3)\times (1, -1) & -1 & 1  & \text{-} & \text{-} & 0 & -3 & 0 &  3 &  -2\\
c & 4 &  (0, 1)\times (-1, 3)\times (-1, 1) & 2 & -2  & \text{-} & \text{-} & \text{-} & \text{-} & 1 &  0 &  0\\
\hline
 &   &  & \multicolumn{9}{c|}{ x_A = \frac{1}{3}x_B = \frac{1}{2}x_C = \frac{1}{3}x_D }\\
2 & 2 &  (1, 0)\times (0, -1)\times (0, 2) & \multicolumn{9}{c|}{ ~\beta^g_2=-3,~\beta^g_3=-3,~\beta^g_4=-2, }\\
3 & 2  &  (0, -1)\times (1, 0)\times (0, 2)  & \multicolumn{9}{c|}{ \chi_1=\frac{1}{\sqrt{2}} , ~ \chi_2=\frac{\sqrt{2}}{3} , ~ \chi_3=\sqrt{2} }\\
4 & 2  &   (0, -1)\times (0, 1)\times (2, 0) & \multicolumn{9}{c|}{}\\
\hline  
\end{array}$
\caption{D6-brane configurations and intersection numbers of Model~\hyperref[spec17]{17}, and its MSSM gauge coupling relation is $g_a^2=2g_b^2=g_c^2=\frac{5 g_Y^2}{3}=\frac{16 \sqrt[4]{2} \, \pi \,  e^{\phi _4}}{3 \sqrt{3}}$.}
\label{model17}
\end{table}


\begin{table}[ht]\footnotesize\centering 
$\begin{array}{|c|c|c|c|c|c|c|c|c|c|c|c|}
\hline\multicolumn{2}{|c|}{\text{Model~\hyperref[spec17.5]{17-dual}}} & \multicolumn{10}{c|}{ {\U(4)_C\times \U(2)_L \times \U(2)_R\times \USp(2)_2\times \USp(2)_3\times \USp(2)_4} }\\
\hline\rm{stack} & N & (n^1,l^1)\times(n^2,l^2)\times(n^3,l^3) & n_{\yng(2)}& n_{\yng(1,1)_{}} & b & b' & c & c' & 2 &  3 &  4\\
\hline
a & 8 & (1, -1)\times (1, 0)\times (1, 1) & 0 & 0  & 3 & 0 & -3 & 0 & -1 &  0 &  1\\
b & 4 &  (0, 1)\times (-1, 3)\times (-1, 1) & 2 & -2  & \text{-} & \text{-} & 0 & -3 & 1 &  0 &  0\\
c & 4 &  (-1, 0)\times (-2, -3)\times (1, -1) & -1 & 1  & \text{-} & \text{-} & \text{-} & \text{-} & 0 &  3 &  -2\\
\hline
 &   &  & \multicolumn{9}{c|}{ x_A = \frac{1}{3}x_B = \frac{1}{2}x_C = \frac{1}{3}x_D }\\
2 & 2 &  (1, 0)\times (0, -1)\times (0, 2) & \multicolumn{9}{c|}{ ~\beta^g_2=-3,~\beta^g_3=-3,~\beta^g_4=-2, }\\
3 & 2  &  (0, -1)\times (1, 0)\times (0, 2)  & \multicolumn{9}{c|}{ \chi_1=\frac{1}{\sqrt{2}} , ~ \chi_2=\frac{\sqrt{2}}{3} , ~ \chi_3=\sqrt{2} }\\
4 & 2  &   (0, -1)\times (0, 1)\times (2, 0) & \multicolumn{9}{c|}{}\\
\hline  
\end{array}$
\caption{D6-brane configurations and intersection numbers of Model~\hyperref[spec17.5]{17-dual}, and its MSSM gauge coupling relation is $g_a^2=g_b^2=2g_c^2=\frac{10}{7}\frac{5 g_Y^2}{3}=\frac{16 \sqrt[4]{2} \, \pi \,  e^{\phi _4}}{3 \sqrt{3}}$.}
\label{model17.5}
\end{table}


\begin{table}[ht]\footnotesize\centering 
$\begin{array}{|c|c|c|c|c|c|c|c|c|c|c|c|}
\hline\multicolumn{2}{|c|}{\text{Model~\hyperref[spec18]{18}}} & \multicolumn{10}{c|}{ {\U(4)_C\times \U(2)_L \times \U(2)_R\times \USp(2)_2\times \USp(2)_3\times \USp(2)_4} }\\
\hline\rm{stack} & N & (n^1,l^1)\times(n^2,l^2)\times(n^3,l^3) & n_{\yng(2)}& n_{\yng(1,1)_{}} & b & b' & c & c' & 2 &  3 &  4\\
\hline
a & 8 & (1, -1)\times (1, 0)\times (1, 1) & 0 & 0  & 3 & 0 & -3 & 0 & -1 &  0 &  1\\
b & 4 &  (-1, 4)\times (0, 1)\times (-1, 1) & 3 & -3  & \text{-} & \text{-} & 0 & -7 & 0 &  1 &  0\\
c & 4 &  (2, 1)\times (1, 1)\times (1, -1) & 2 & 6  & \text{-} & \text{-} & \text{-} & \text{-} & 1 &  2 &  -2\\
\hline
 &   &  & \multicolumn{9}{c|}{ x_A = \frac{1}{9}x_B = \frac{1}{4}x_C = \frac{1}{9}x_D }\\
2 & 2 &  (1, 0)\times (0, -1)\times (0, 2) & \multicolumn{9}{c|}{ ~\beta^g_2=-3,~\beta^g_3=-3,~\beta^g_4=-2, }\\
3 & 2  &  (0, -1)\times (1, 0)\times (0, 2)  & \multicolumn{9}{c|}{ \chi_1=\frac{1}{2} , ~ \chi_2=\frac{2}{9} , ~ \chi_3=1 }\\
4 & 2  &   (0, -1)\times (0, 1)\times (2, 0) & \multicolumn{9}{c|}{}\\
\hline  
\end{array}$
\caption{D6-brane configurations and intersection numbers of Model~\hyperref[spec18]{18}, and its MSSM gauge coupling relation is $g_a^2=\frac{4}{9}g_b^2=\frac{17}{9}g_c^2=\frac{85}{61}\frac{5 g_Y^2}{3}=\frac{32 \, \pi \,  e^{\phi _4}}{15}$.}
\label{model18}
\end{table}


\begin{table}[ht]\footnotesize\centering 
$\begin{array}{|c|c|c|c|c|c|c|c|c|c|c|c|}
\hline\multicolumn{2}{|c|}{\text{Model~\hyperref[spec18.5]{18-dual}}} & \multicolumn{10}{c|}{ {\U(4)_C\times \U(2)_L \times \U(2)_R\times \USp(2)_2\times \USp(2)_3\times \USp(2)_4} }\\
\hline\rm{stack} & N & (n^1,l^1)\times(n^2,l^2)\times(n^3,l^3) & n_{\yng(2)}& n_{\yng(1,1)_{}} & b & b' & c & c' & 2 &  3 &  4\\
\hline
a & 8 & (1, -1)\times (1, 0)\times (1, 1) & 0 & 0  & -3 & 0 & 3 & 0 & -1 &  0 &  1\\
b & 4 &  (2, 1)\times (1, 1)\times (1, -1) & 2 & 6  & \text{-} & \text{-} & 0 & -7 & 1 &  2 &  -2\\
c & 4 &  (-1, 4)\times (0, 1)\times (-1, 1) & 3 & -3  & \text{-} & \text{-} & \text{-} & \text{-} & 0 &  1 &  0\\
\hline
 &   &  & \multicolumn{9}{c|}{ x_A = \frac{1}{9}x_B = \frac{1}{4}x_C = \frac{1}{9}x_D }\\
2 & 2 &  (1, 0)\times (0, -1)\times (0, 2) & \multicolumn{9}{c|}{ ~\beta^g_2=-3,~\beta^g_3=-3,~\beta^g_4=-2, }\\
3 & 2  &  (0, -1)\times (1, 0)\times (0, 2)  & \multicolumn{9}{c|}{ \chi_1=\frac{1}{2} , ~ \chi_2=\frac{2}{9} , ~ \chi_3=1 }\\
4 & 2  &   (0, -1)\times (0, 1)\times (2, 0) & \multicolumn{9}{c|}{}\\
\hline  
\end{array}$
\caption{D6-brane configurations and intersection numbers of Model~\hyperref[spec18.5]{18-dual}, and its MSSM gauge coupling relation is $g_a^2=\frac{17}{9}g_b^2=\frac{4}{9}g_c^2=\frac{4}{7}\frac{5 g_Y^2}{3}=\frac{32 \, \pi \,  e^{\phi _4}}{15}$.}
\label{model18.5}
\end{table}


\begin{table}[ht]\footnotesize\centering 
$\begin{array}{|c|c|c|c|c|c|c|c|c|c|c|}
\hline\multicolumn{2}{|c|}{\text{Model~\hyperref[spec19]{19}}} & \multicolumn{9}{c|}{ {\U(4)_C\times \U(2)_L \times \U(2)_R\times \USp(2)_1\times \USp(2)_4} }\\
\hline\rm{stack} & N & (n^1,l^1)\times(n^2,l^2)\times(n^3,l^3) & n_{\yng(2)}& n_{\yng(1,1)_{}} & b & b' & c & c' & 1 &  4\\
\hline
a & 8 & (-1, -1)\times (1, 1)\times (1, 1) & 0 & -4  & -3 & 0 & 3 & 0 & 1 &  -1\\
b & 4 &  (1, 4)\times (1, 0)\times (1, -1) & -3 & 3  & \text{-} & \text{-} & 0 & 7 & 0 &  -1\\
c & 4 &  (-2, 1)\times (2, 1)\times (-1, 1) & -3 & -13  & \text{-} & \text{-} & \text{-} & \text{-} & -1 &  -4\\
\hline
1 & 2  &  (1, 0)\times (1, 0)\times (2, 0) & \multicolumn{8}{c|}{ x_A = 28x_B = \frac{28}{23}x_C = 7x_D }\\
  &   &   & \multicolumn{8}{c|}{ \beta^g_1=-3,~\beta^g_4=1, }\\
 &   &   & \multicolumn{8}{c|}{ \chi_1=\sqrt{\frac{7}{23}} , ~ \chi_2=\sqrt{161} , ~ \chi_3=8 \sqrt{\frac{7}{23}} }\\
4 & 2  &   (0, -1)\times (0, 1)\times (2, 0) & \multicolumn{8}{c|}{}\\
\hline  
\end{array}$
\caption{D6-brane configurations and intersection numbers of Model~\hyperref[spec19]{19}, and its MSSM gauge coupling relation is $g_a^2=\frac{1}{6}g_b^2=\frac{11}{6}g_c^2=\frac{11}{8}\frac{5 g_Y^2}{3}=\frac{8}{405} \sqrt{2} 161^{3/4} \, \pi \,  e^{\phi _4}$.}
\label{model19}
\end{table}


\begin{table}[ht]\footnotesize\centering 
$\begin{array}{|c|c|c|c|c|c|c|c|c|c|c|}
\hline\multicolumn{2}{|c|}{\text{Model~\hyperref[spec19.5]{19-dual}}} & \multicolumn{9}{c|}{ {\U(4)_C\times \U(2)_L \times \U(2)_R\times \USp(2)_1\times \USp(2)_4} }\\
\hline\rm{stack} & N & (n^1,l^1)\times(n^2,l^2)\times(n^3,l^3) & n_{\yng(2)}& n_{\yng(1,1)_{}} & b & b' & c & c' & 1 &  4\\
\hline
a & 8 & (-1, -1)\times (1, 1)\times (1, 1) & 0 & -4  & 3 & 0 & -3 & 0 & 1 &  -1\\
b & 4 &  (-2, 1)\times (2, 1)\times (-1, 1) & -3 & -13  & \text{-} & \text{-} & 0 & 7 & -1 &  -4\\
c & 4 &  (1, 4)\times (1, 0)\times (1, -1) & -3 & 3  & \text{-} & \text{-} & \text{-} & \text{-} & 0 &  -1\\
\hline
1 & 2  &  (1, 0)\times (1, 0)\times (2, 0) & \multicolumn{8}{c|}{ x_A = 28x_B = \frac{28}{23}x_C = 7x_D }\\
  &   &   & \multicolumn{8}{c|}{ \beta^g_1=-3,~\beta^g_4=1, }\\
 &   &   & \multicolumn{8}{c|}{ \chi_1=\sqrt{\frac{7}{23}} , ~ \chi_2=\sqrt{161} , ~ \chi_3=8 \sqrt{\frac{7}{23}} }\\
4 & 2  &   (0, -1)\times (0, 1)\times (2, 0) & \multicolumn{8}{c|}{}\\
\hline  
\end{array}$
\caption{D6-brane configurations and intersection numbers of Model~\hyperref[spec19.5]{19-dual}, and its MSSM gauge coupling relation is $g_a^2=\frac{11}{6}g_b^2=\frac{1}{6}g_c^2=\frac{1}{4}\frac{5 g_Y^2}{3}=\frac{8}{405} \sqrt{2} 161^{3/4} \, \pi \,  e^{\phi _4}$.}
\label{model19.5}
\end{table}


\begin{table}[ht]\footnotesize\centering 
$\begin{array}{|c|c|c|c|c|c|c|c|c|c|c|}
\hline\multicolumn{2}{|c|}{\text{Model~\hyperref[spec20]{20}}} & \multicolumn{9}{c|}{ {\U(4)_C\times \U(2)_L \times \U(2)_R\times \USp(2)_1\times \USp(2)_4} }\\
\hline\rm{stack} & N & (n^1,l^1)\times(n^2,l^2)\times(n^3,l^3) & n_{\yng(2)}& n_{\yng(1,1)_{}} & b & b' & c & c' & 1 &  4\\
\hline
a & 8 & (1, 1)\times (-1, -1)\times (1, 1) & 0 & -4  & 3 & 0 & -3 & 0 & 1 &  -1\\
b & 4 &  (-1, 0)\times (5, 2)\times (-1, 1) & 3 & -3  & \text{-} & \text{-} & 0 & -1 & 0 &  -5\\
c & 4 &  (0, 1)\times (-2, 1)\times (-1, 1) & -1 & 1  & \text{-} & \text{-} & \text{-} & \text{-} & -1 &  0\\
\hline
1 & 2  &  (1, 0)\times (1, 0)\times (2, 0) & \multicolumn{8}{c|}{ x_A = 2x_B = \frac{14}{5}x_C = 7x_D }\\
  &   &   & \multicolumn{8}{c|}{ \beta^g_1=-3,~\beta^g_4=1, }\\
 &   &   & \multicolumn{8}{c|}{ \chi_1=\frac{7}{\sqrt{5}} , ~ \chi_2=\sqrt{5} , ~ \chi_3=\frac{4}{\sqrt{5}} }\\
4 & 2  &   (0, -1)\times (0, 1)\times (2, 0) & \multicolumn{8}{c|}{}\\
\hline  
\end{array}$
\caption{D6-brane configurations and intersection numbers of Model~\hyperref[spec20]{20}, and its MSSM gauge coupling relation is $g_a^2=\frac{5}{6}g_b^2=\frac{7}{6}g_c^2=\frac{35}{32}\frac{5 g_Y^2}{3}=\frac{8}{27} 5^{3/4} \sqrt{7} \, \pi \,  e^{\phi _4}$.}
\label{model20}
\end{table}


\begin{table}[ht]\footnotesize\centering 
$\begin{array}{|c|c|c|c|c|c|c|c|c|c|c|}
\hline\multicolumn{2}{|c|}{\text{Model~\hyperref[spec20.5]{20-dual}}} & \multicolumn{9}{c|}{ {\U(4)_C\times \U(2)_L \times \U(2)_R\times \USp(2)_1\times \USp(2)_4} }\\
\hline\rm{stack} & N & (n^1,l^1)\times(n^2,l^2)\times(n^3,l^3) & n_{\yng(2)}& n_{\yng(1,1)_{}} & b & b' & c & c' & 1 &  4\\
\hline
a & 8 & (1, 1)\times (-1, -1)\times (1, 1) & 0 & -4  & -3 & 0 & 3 & 0 & 1 &  -1\\
b & 4 &  (0, 1)\times (-2, 1)\times (-1, 1) & -1 & 1  & \text{-} & \text{-} & 0 & -1 & -1 &  0\\
c & 4 &  (-1, 0)\times (5, 2)\times (-1, 1) & 3 & -3  & \text{-} & \text{-} & \text{-} & \text{-} & 0 &  -5\\
\hline
1 & 2  &  (1, 0)\times (1, 0)\times (2, 0) & \multicolumn{8}{c|}{ x_A = 2x_B = \frac{14}{5}x_C = 7x_D }\\
  &   &   & \multicolumn{8}{c|}{ \beta^g_1=-3,~\beta^g_4=1, }\\
 &   &   & \multicolumn{8}{c|}{ \chi_1=\frac{7}{\sqrt{5}} , ~ \chi_2=\sqrt{5} , ~ \chi_3=\frac{4}{\sqrt{5}} }\\
4 & 2  &   (0, -1)\times (0, 1)\times (2, 0) & \multicolumn{8}{c|}{}\\
\hline  
\end{array}$
\caption{D6-brane configurations and intersection numbers of Model~\hyperref[spec20.5]{20-dual}, and its MSSM gauge coupling relation is $g_a^2=\frac{7}{6}g_b^2=\frac{5}{6}g_c^2=\frac{25}{28}\frac{5 g_Y^2}{3}=\frac{8}{27} 5^{3/4} \sqrt{7} \, \pi \,  e^{\phi _4}$.}
\label{model20.5}
\end{table}


\begin{table}[ht]\footnotesize\centering 
$\begin{array}{|c|c|c|c|c|c|c|c|c|c|c|}
\hline\multicolumn{2}{|c|}{\text{Model~\hyperref[spec21]{21}}} & \multicolumn{9}{c|}{ {\U(4)_C\times \U(2)_L \times \U(2)_R\times \USp(2)_2\times \USp(2)_4} }\\
\hline\rm{stack} & N & (n^1,l^1)\times(n^2,l^2)\times(n^3,l^3) & n_{\yng(2)}& n_{\yng(1,1)_{}} & b & b' & c & c' & 2 &  4\\
\hline
a & 8 & (1, -1)\times (1, 0)\times (1, 1) & 0 & 0  & -3 & 0 & 3 & 0 & -1 &  1\\
b & 4 &  (-1, 0)\times (-2, -3)\times (1, -1) & -1 & 1  & \text{-} & \text{-} & 0 & 0 & 0 &  -2\\
c & 4 &  (0, 1)\times (-2, 3)\times (-1, 1) & 1 & -1  & \text{-} & \text{-} & \text{-} & \text{-} & 2 &  0\\
\hline
 &   &  & \multicolumn{8}{c|}{ x_A = \frac{2}{3}x_B = x_C = \frac{2}{3}x_D }\\
2 & 2 &  (1, 0)\times (0, -1)\times (0, 2) & \multicolumn{8}{c|}{ ~\beta^g_2=-2,~\beta^g_4=-2, }\\
 &   &   & \multicolumn{8}{c|}{ \chi_1=1 , ~ \chi_2=\frac{2}{3} , ~ \chi_3=2 }\\
4 & 2  &   (0, -1)\times (0, 1)\times (2, 0) & \multicolumn{8}{c|}{}\\
\hline  
\end{array}$
\caption{D6-brane configurations and intersection numbers of Model~\hyperref[spec21]{21}, and its MSSM gauge coupling relation is $g_a^2=2g_b^2=2g_c^2=\frac{10}{7}\frac{5 g_Y^2}{3}=\frac{8 \, \pi \,  e^{\phi _4}}{\sqrt{3}}$.}
\label{model21}
\end{table}


\begin{table}[ht]\footnotesize\centering 
$\begin{array}{|c|c|c|c|c|c|c|c|c|c|}
\hline\multicolumn{2}{|c|}{\text{Model~\hyperref[spec22]{22}}} & \multicolumn{8}{c|}{ {\U(4)_C\times \U(2)_L \times \U(2)_R\times \USp(2)_3} }\\
\hline\rm{stack} & N & (n^1,l^1)\times(n^2,l^2)\times(n^3,l^3) & n_{\yng(2)}& n_{\yng(1,1)_{}} & b & b' & c & c' & 3\\
\hline
a & 8 & (1, -1)\times (1, 0)\times (1, 1) & 0 & 0  & 3 & 0 & -3 & 0 & 0\\
b & 4 &  (-2, 5)\times (0, 1)\times (-1, 1) & 3 & -3  & \text{-} & \text{-} & 0 & -8 & 2\\
c & 4 &  (2, 1)\times (1, 1)\times (1, -1) & 2 & 6  & \text{-} & \text{-} & \text{-} & \text{-} & 2\\
\hline
 &   &  & \multicolumn{7}{c|}{ x_A = \frac{1}{6}x_B = \frac{2}{5}x_C = \frac{1}{6}x_D }\\
  &   &   & \multicolumn{7}{c|}{ ~\beta^g_3=-2, }\\
3 & 2  &  (0, -1)\times (1, 0)\times (0, 2)  & \multicolumn{7}{c|}{ \chi_1=\sqrt{\frac{2}{5}} , ~ \chi_2=\frac{\sqrt{\frac{5}{2}}}{6} , ~ \chi_3=2 \sqrt{\frac{2}{5}} }\\
 &   &    & \multicolumn{7}{c|}{}\\
\hline  
\end{array}$
\caption{D6-brane configurations and intersection numbers of Model~\hyperref[spec22]{22}, and its MSSM gauge coupling relation is $g_a^2=\frac{5}{6}g_b^2=\frac{11}{6}g_c^2=\frac{11}{8}\frac{5 g_Y^2}{3}=\frac{8 \sqrt[4]{2} 5^{3/4} \, \pi \,  e^{\phi _4}}{7 \sqrt{3}}$.}
\label{model22}
\end{table}


\begin{table}[ht]\footnotesize\centering 
$\begin{array}{|c|c|c|c|c|c|c|c|c|c|}
\hline\multicolumn{2}{|c|}{\text{Model~\hyperref[spec22.5]{22-dual}}} & \multicolumn{8}{c|}{ {\U(4)_C\times \U(2)_L \times \U(2)_R\times \USp(2)_3} }\\
\hline\rm{stack} & N & (n^1,l^1)\times(n^2,l^2)\times(n^3,l^3) & n_{\yng(2)}& n_{\yng(1,1)_{}} & b & b' & c & c' & 3\\
\hline
a & 8 & (1, -1)\times (1, 0)\times (1, 1) & 0 & 0  & -3 & 0 & 3 & 0 & 0\\
b & 4 &  (2, 1)\times (1, 1)\times (1, -1) & 2 & 6  & \text{-} & \text{-} & 0 & -8 & 2\\
c & 4 &  (-2, 5)\times (0, 1)\times (-1, 1) & 3 & -3  & \text{-} & \text{-} & \text{-} & \text{-} & 2\\
\hline
 &   &  & \multicolumn{7}{c|}{ x_A = \frac{1}{6}x_B = \frac{2}{5}x_C = \frac{1}{6}x_D }\\
  &   &   & \multicolumn{7}{c|}{ ~\beta^g_3=-2, }\\
3 & 2  &  (0, -1)\times (1, 0)\times (0, 2)  & \multicolumn{7}{c|}{ \chi_1=\sqrt{\frac{2}{5}} , ~ \chi_2=\frac{\sqrt{\frac{5}{2}}}{6} , ~ \chi_3=2 \sqrt{\frac{2}{5}} }\\
 &   &    & \multicolumn{7}{c|}{}\\
\hline  
\end{array}$
\caption{D6-brane configurations and intersection numbers of Model~\hyperref[spec22.5]{22-dual}, and its MSSM gauge coupling relation is $g_a^2=\frac{11}{6}g_b^2=\frac{5}{6}g_c^2=\frac{25}{28}\frac{5 g_Y^2}{3}=\frac{8 \sqrt[4]{2} 5^{3/4} \, \pi \,  e^{\phi _4}}{7 \sqrt{3}}$.}
\label{model22.5}
\end{table}
 
\FloatBarrier
\section{Particle spectra}\label{appB}  
\begin{table}[th]
  \footnotesize\renewcommand{\arraystretch}{1.3}
  \label{spec1}
  \begin{center}
    $
\begin{array}{|c|D{.}{\times}{13}|c|c|c|c|c|c|}
        \hline
        \text{Model~\hyperref[model1]{1}} & \mc{\text{Quantum Number}}     & Q_{4 C} & Q_{2 L} & Q_{2 R} & Q_{\text{em}}       & B-L              & \text{Field} \\
        \hline   
ab                          & 3.(4,\bar{2},1,1,1)                           & 1  & -1 & 0  & -\frac{1}{3},\frac{2}{3},-1,0 & \frac{1}{3},-1 & Q_L,L_L\\
ac                          & 3.(\bar{4},1,2,1,1)                           & -1 & 0  & 1  & \frac{1}{3},-\frac{2}{3},1,0  & -\frac{1}{3},1 & Q_R,L_R\\
b_{\overline{\yng(1,1)_{}}} & 2.(1,\bar{1}_{\overline{\yng(1,1)_{}}},1,1,1) & 0  & -1 & 0  & 0                             & 0              & S_L    \\
c_{\yng(1,1)_{}}            & 6.(1,1,1_{\yng(1,1)_{}},1,1)                  & 0  & 0  & 1  & 0                             & 0              & S_R    \\
b_{\yng(2)_{}}              & 2.(1,3_{\yng(2)_{}},1,1,1)                    & 0  & 2  & 0  & 0,\pm 1                       & 0              & T_L    \\
c_{\yng(2)_{}}              & 2.(1,1,3_{\yng(2)_{}},1,1)                    & 0  & 0  & 2  & 0,\pm 1                       & 0              & T_R    \\
a2                          & 1.(\bar{4},1,1,1,4)                           & -1 & 0  & 0  & -\frac{1}{6},\frac{1}{2}      & -\frac{1}{3},1 & X_C    \\
\hline
      \end{array}$
  \end{center}
  \caption{The chiral and vector-like superfields, and their quantum numbers under the gauge symmetry ${\SU(4)_C\times \SU(2)_L \times \SU(2)_R\times \USp(4)_2\times \USp(4)_4}$.}
\end{table}


\begin{table}[th]
  \footnotesize\renewcommand{\arraystretch}{1.3}
  \label{spec1.5}
  \begin{center}
    $\begin{array}{|c|D{.}{\times}{13}|c|c|c|c|c|c|}
        \hline
        \text{Model~\hyperref[model1.5]{1-dual}} & \mc{\text{Quantum Number}}     & Q_{4 C} & Q_{2 L} & Q_{2 R} & Q_{\text{em}}       & B-L              & \text{Field} \\
        \hline   
ab                          & 3.(\bar{4},2,1,1,1)                           & -1 & 1  & 0  & -\frac{1}{3},\frac{2}{3},-1,0 & \frac{1}{3},-1 & Q_L,L_L\\
ac                          & 3.(4,1,\bar{2},1,1)                           & 1  & 0  & -1 & \frac{1}{3},-\frac{2}{3},1,0  & -\frac{1}{3},1 & Q_R,L_R\\
b_{\yng(1,1)_{}}            & 6.(1,1_{\yng(1,1)_{}},1,1,1)                  & 0  & 1  & 0  & 0                             & 0              & S_L    \\
c_{\overline{\yng(1,1)_{}}} & 2.(1,1,\bar{1}_{\overline{\yng(1,1)_{}}},1,1) & 0  & 0  & -1 & 0                             & 0              & S_R    \\
b_{\yng(2)_{}}              & 2.(1,3_{\yng(2)_{}},1,1,1)                    & 0  & 2  & 0  & 0,\pm 1                       & 0              & T_L    \\
c_{\yng(2)_{}}              & 2.(1,1,3_{\yng(2)_{}},1,1)                    & 0  & 0  & 2  & 0,\pm 1                       & 0              & T_R    \\
a2                          & 1.(\bar{4},1,1,1,4)                           & -1 & 0  & 0  & -\frac{1}{6},\frac{1}{2}      & -\frac{1}{3},1 & X_C    \\
\hline
      \end{array}$
  \end{center}
  \caption{The chiral and vector-like superfields, and their quantum numbers under the gauge symmetry ${\SU(4)_C\times \SU(2)_L \times \SU(2)_R\times \USp(4)_2\times \USp(4)_4}$.}
\end{table}


\begin{table}[ht]
  \footnotesize\renewcommand{\arraystretch}{1.3}
  \label{spec2}
  \begin{center}
    $\begin{array}{|c|D{.}{\times}{17}|c|c|c|c|c|c|}
        \hline
        \text{Model~\hyperref[model2]{2}} & \mc{\text{Quantum Number}}     & Q_{4 C} & Q_{2 L} & Q_{2 R} & Q_{\text{em}}       & B-L              & \text{Field} \\
        \hline   
ab                          & 2.(\bar{4},2,1,1,1,1,1)                           & -1 & 1  & 0  & -\frac{1}{3},\frac{2}{3},-1,0 & \frac{1}{3},-1 & Q_L,L_L\\
a b'                        & 1.(\bar{4},\bar{2},1,1,1,1,1)                     & -1 & -1 & 0  & -\frac{1}{3},\frac{2}{3},-1,0 & \frac{1}{3},-1 & Q_L,L_L\\
ac                          & 3.(4,1,\bar{2},1,1,1,1)                           & 1  & 0  & -1 & \frac{1}{3},-\frac{2}{3},1,0  & -\frac{1}{3},1 & Q_R,L_R\\
b_{\overline{\yng(1,1)_{}}} & 2.(1,\bar{1}_{\overline{\yng(1,1)_{}}},1,1,1,1,1) & 0  & -1 & 0  & 0                             & 0              & S_L    \\
c_{\overline{\yng(1,1)_{}}} & 2.(1,1,\bar{1}_{\overline{\yng(1,1)_{}}},1,1,1,1) & 0  & 0  & -1 & 0                             & 0              & S_R    \\
b_{\yng(2)_{}}              & 2.(1,3_{\yng(2)_{}},1,1,1,1,1)                    & 0  & 2  & 0  & 0,\pm 1                       & 0              & T_L    \\
c_{\yng(2)_{}}              & 2.(1,1,3_{\yng(2)_{}},1,1,1,1)                    & 0  & 0  & 2  & 0,\pm 1                       & 0              & T_R    \\
\hline
      \end{array}$
  \end{center}
  \caption{The chiral and vector-like superfields, and their quantum numbers under the gauge symmetry ${\SU(4)_C\times \SU(2)_L \times \SU(2)_R\times \USp(2)_1\times \USp(2)_2\times \USp(2)_3\times \USp(2)_4}$.}
\end{table}


\begin{table}[ht]
  \footnotesize\renewcommand{\arraystretch}{1.3}
  \label{spec3}
  \begin{center}
    $\begin{array}{|c|D{.}{\times}{12}|c|c|c|c|c|c|}
        \hline
        \text{Model~\hyperref[model3]{3}} & \mc{\text{Quantum Number}}     & Q_{4 C} & Q_{2 L} & Q_{2 R} & Q_{\text{em}}       & B-L              & \text{Field} \\
        \hline   
ab                        & 3.(\bar{4},2,1,1)                               & -1 & 1  & 0  & -\frac{1}{3},\frac{2}{3},-1,0 & \frac{1}{3},-1 & Q_L,L_L\\
ac                        & 3.(4,1,\bar{2},1)                               & 1  & 0  & -1 & \frac{1}{3},-\frac{2}{3},1,0  & -\frac{1}{3},1 & Q_R,L_R\\
a_{\yng(1,1)_{}}          & 1.(6_{\yng(1,1)_{}},1,1,1)                      & 1  & 0  & 0  & \frac{1}{3},-1                & \frac{2}{3},-2 & S_C    \\
b_{\yng(1,1)_{}}          & 2.(1,1_{\yng(1,1)_{}},1,1)                      & 0  & 1  & 0  & 0                             & 0              & S_L    \\
c_{\yng(1,1)_{}}          & 6.(1,1,1_{\yng(1,1)_{}},1)                      & 0  & 0  & 1  & 0                             & 0              & S_R    \\
a_{\overline{\yng(2)_{}}} & 1.(\overline{10}_{\overline{\yng(2)_{}}},1,1,1) & -2 & 0  & 0  & \frac{1}{3},-\frac{1}{3},-1   & \frac{2}{3},-2 & T_C    \\
b_{\overline{\yng(2)_{}}} & 2.(1,\bar{3}_{\overline{\yng(2)_{}}},1,1)       & 0  & -2 & 0  & 0,\pm 1                       & 0              & T_L    \\
\hline
      \end{array}$
  \end{center}
  \caption{The chiral and vector-like superfields, and their quantum numbers under the gauge symmetry ${\SU(4)_C\times \SU(2)_L \times \SU(2)_R\times \USp(4)_4}$.}
\end{table}


\begin{table}[ht]
  \footnotesize\renewcommand{\arraystretch}{1.3}
  \label{spec3.5}
  \begin{center}
    $\begin{array}{|c|D{.}{\times}{12}|c|c|c|c|c|c|}
        \hline
        \text{Model~\hyperref[model3.5]{3-dual}} & \mc{\text{Quantum Number}}     & Q_{4 C} & Q_{2 L} & Q_{2 R} & Q_{\text{em}}       & B-L              & \text{Field} \\
        \hline   
ab                        & 3.(4,\bar{2},1,1)                               & 1  & -1 & 0  & -\frac{1}{3},\frac{2}{3},-1,0 & \frac{1}{3},-1 & Q_L,L_L\\
ac                        & 3.(\bar{4},1,2,1)                               & -1 & 0  & 1  & \frac{1}{3},-\frac{2}{3},1,0  & -\frac{1}{3},1 & Q_R,L_R\\
a_{\yng(1,1)_{}}          & 1.(6_{\yng(1,1)_{}},1,1,1)                      & 1  & 0  & 0  & \frac{1}{3},-1                & \frac{2}{3},-2 & S_C    \\
b_{\yng(1,1)_{}}          & 6.(1,1_{\yng(1,1)_{}},1,1)                      & 0  & 1  & 0  & 0                             & 0              & S_L    \\
c_{\yng(1,1)_{}}          & 2.(1,1,1_{\yng(1,1)_{}},1)                      & 0  & 0  & 1  & 0                             & 0              & S_R    \\
a_{\overline{\yng(2)_{}}} & 1.(\overline{10}_{\overline{\yng(2)_{}}},1,1,1) & -2 & 0  & 0  & \frac{1}{3},-\frac{1}{3},-1   & \frac{2}{3},-2 & T_C    \\
b_{\yng(2)_{}}            & 2.(1,3_{\yng(2)_{}},1,1)                        & 0  & 2  & 0  & 0,\pm 1                       & 0              & T_L    \\
\hline
      \end{array}$
  \end{center}
  \caption{The chiral and vector-like superfields, and their quantum numbers under the gauge symmetry ${\SU(4)_C\times \SU(2)_L \times \SU(2)_R\times \USp(4)_4}$.}
\end{table}


\begin{table}[ht]
  \footnotesize\renewcommand{\arraystretch}{1.3}
  \label{spec4}
  \begin{center}
    $\begin{array}{|c|D{.}{\times}{12}|c|c|c|c|c|c|}
        \hline
        \text{Model~\hyperref[model4]{4}} & \mc{\text{Quantum Number}}     & Q_{4 C} & Q_{2 L} & Q_{2 R} & Q_{\text{em}}       & B-L              & \text{Field} \\
        \hline   
ab                          & 2.(\bar{4},2,1,1)                               & -1 & 1  & 0  & -\frac{1}{3},\frac{2}{3},-1,0 & \frac{1}{3},-1 & Q_L,L_L\\
a b'                        & 1.(\bar{4},\bar{2},1,1)                         & -1 & -1 & 0  & -\frac{1}{3},\frac{2}{3},-1,0 & \frac{1}{3},-1 & Q_L,L_L\\
ac                          & 3.(4,1,\bar{2},1)                               & 1  & 0  & -1 & \frac{1}{3},-\frac{2}{3},1,0  & -\frac{1}{3},1 & Q_R,L_R\\
a_{\yng(1,1)_{}}            & 1.(6_{\yng(1,1)_{}},1,1,1)                      & 1  & 0  & 0  & \frac{1}{3},-1                & \frac{2}{3},-2 & S_C    \\
b_{\overline{\yng(1,1)_{}}} & 2.(1,\bar{1}_{\overline{\yng(1,1)_{}}},1,1)     & 0  & -1 & 0  & 0                             & 0              & S_L    \\
c_{\yng(1,1)_{}}            & 6.(1,1,1_{\yng(1,1)_{}},1)                      & 0  & 0  & 1  & 0                             & 0              & S_R    \\
a_{\overline{\yng(2)_{}}}   & 1.(\overline{10}_{\overline{\yng(2)_{}}},1,1,1) & -2 & 0  & 0  & \frac{1}{3},-\frac{1}{3},-1   & \frac{2}{3},-2 & T_C    \\
\hline
      \end{array}$
  \end{center}
  \caption{The chiral and vector-like superfields, and their quantum numbers under the gauge symmetry ${\SU(4)_C\times \SU(2)_L \times \SU(2)_R\times \USp(4)_4}$.}
\end{table}


\begin{table}[ht]
  \footnotesize\renewcommand{\arraystretch}{1.3}
  \label{spec5}
  \begin{center}
    $\begin{array}{|c|D{.}{\times}{15}|c|c|c|c|c|c|}
        \hline
        \text{Model~\hyperref[model5]{5}} & \mc{\text{Quantum Number}}     & Q_{4 C} & Q_{2 L} & Q_{2 R} & Q_{\text{em}}       & B-L              & \text{Field} \\
        \hline   
ab                          & 2.(\bar{4},2,1,1,1,1)                           & -1 & 1  & 0  & -\frac{1}{3},\frac{2}{3},-1,0 & \frac{1}{3},-1 & Q_L,L_L\\
a b'                        & 1.(\bar{4},\bar{2},1,1,1,1)                     & -1 & -1 & 0  & -\frac{1}{3},\frac{2}{3},-1,0 & \frac{1}{3},-1 & Q_L,L_L\\
ac                          & 3.(4,1,\bar{2},1,1,1)                           & 1  & 0  & -1 & \frac{1}{3},-\frac{2}{3},1,0  & -\frac{1}{3},1 & Q_R,L_R\\
b_{\overline{\yng(1,1)_{}}} & 2.(1,\bar{1}_{\overline{\yng(1,1)_{}}},1,1,1,1) & 0  & -1 & 0  & 0                             & 0              & S_L    \\
c_{\overline{\yng(1,1)_{}}} & 1.(1,1,\bar{1}_{\overline{\yng(1,1)_{}}},1,1,1) & 0  & 0  & -1 & 0                             & 0              & S_R    \\
b_{\yng(2)_{}}              & 2.(1,3_{\yng(2)_{}},1,1,1,1)                    & 0  & 2  & 0  & 0,\pm 1                       & 0              & T_L    \\
c_{\yng(2)_{}}              & 1.(1,1,3_{\yng(2)_{}},1,1,1)                    & 0  & 0  & 2  & 0,\pm 1                       & 0              & T_R    \\
\hline
      \end{array}$
  \end{center}
  \caption{The chiral and vector-like superfields, and their quantum numbers under the gauge symmetry ${\SU(4)_C\times \SU(2)_L \times \SU(2)_R\times \USp(2)_1\times \USp(2)_2\times \USp(2)_4}$.}
\end{table}


\begin{table}[ht]
  \footnotesize\renewcommand{\arraystretch}{1.3}
  \label{spec6}
  \begin{center}
    $\begin{array}{|c|D{.}{\times}{15}|c|c|c|c|c|c|}
        \hline
        \text{Model~\hyperref[model6]{6}} & \mc{\text{Quantum Number}}     & Q_{4 C} & Q_{2 L} & Q_{2 R} & Q_{\text{em}}       & B-L              & \text{Field} \\
        \hline   
ab                          & 2.(\bar{4},2,1,1,1,1)                           & -1 & 1  & 0  & -\frac{1}{3},\frac{2}{3},-1,0 & \frac{1}{3},-1 & Q_L,L_L\\
a b'                        & 1.(\bar{4},\bar{2},1,1,1,1)                     & -1 & -1 & 0  & -\frac{1}{3},\frac{2}{3},-1,0 & \frac{1}{3},-1 & Q_L,L_L\\
ac                          & 3.(4,1,\bar{2},1,1,1)                           & 1  & 0  & -1 & \frac{1}{3},-\frac{2}{3},1,0  & -\frac{1}{3},1 & Q_R,L_R\\
b_{\overline{\yng(1,1)_{}}} & 5.(1,\bar{1}_{\overline{\yng(1,1)_{}}},1,1,1,1) & 0  & -1 & 0  & 0                             & 0              & S_L    \\
c_{\overline{\yng(1,1)_{}}} & 2.(1,1,\bar{1}_{\overline{\yng(1,1)_{}}},1,1,1) & 0  & 0  & -1 & 0                             & 0              & S_R    \\
b_{\yng(2)_{}}              & 5.(1,3_{\yng(2)_{}},1,1,1,1)                    & 0  & 2  & 0  & 0,\pm 1                       & 0              & T_L    \\
c_{\yng(2)_{}}              & 2.(1,1,3_{\yng(2)_{}},1,1,1)                    & 0  & 0  & 2  & 0,\pm 1                       & 0              & T_R    \\
\hline
      \end{array}$
  \end{center}
  \caption{The chiral and vector-like superfields, and their quantum numbers under the gauge symmetry ${\SU(4)_C\times \SU(2)_L \times \SU(2)_R\times \USp(2)_2\times \USp(2)_3\times \USp(2)_4}$.}
\end{table}


\begin{table}[ht]
  \footnotesize\renewcommand{\arraystretch}{1.3}
  \label{spec7}
  \begin{center}
    $\begin{array}{|c|D{.}{\times}{13}|c|c|c|c|c|c|}
        \hline
        \text{Model~\hyperref[model7]{7}} & \mc{\text{Quantum Number}}     & Q_{4 C} & Q_{2 L} & Q_{2 R} & Q_{\text{em}}       & B-L              & \text{Field} \\
        \hline   
ab                          & 2.(\bar{4},2,1,1,1)                           & -1 & 1  & 0  & -\frac{1}{3},\frac{2}{3},-1,0 & \frac{1}{3},-1 & Q_L,L_L\\
a b'                        & 1.(\bar{4},\bar{2},1,1,1)                     & -1 & -1 & 0  & -\frac{1}{3},\frac{2}{3},-1,0 & \frac{1}{3},-1 & Q_L,L_L\\
ac                          & 3.(4,1,\bar{2},1,1)                           & 1  & 0  & -1 & \frac{1}{3},-\frac{2}{3},1,0  & -\frac{1}{3},1 & Q_R,L_R\\
b_{\overline{\yng(1,1)_{}}} & 5.(1,\bar{1}_{\overline{\yng(1,1)_{}}},1,1,1) & 0  & -1 & 0  & 0                             & 0              & S_L    \\
c_{\overline{\yng(1,1)_{}}} & 1.(1,1,\bar{1}_{\overline{\yng(1,1)_{}}},1,1) & 0  & 0  & -1 & 0                             & 0              & S_R    \\
b_{\yng(2)_{}}              & 5.(1,3_{\yng(2)_{}},1,1,1)                    & 0  & 2  & 0  & 0,\pm 1                       & 0              & T_L    \\
c_{\yng(2)_{}}              & 1.(1,1,3_{\yng(2)_{}},1,1)                    & 0  & 0  & 2  & 0,\pm 1                       & 0              & T_R    \\
\hline
      \end{array}$
  \end{center}
  \caption{The chiral and vector-like superfields, and their quantum numbers under the gauge symmetry ${\SU(4)_C\times \SU(2)_L \times \SU(2)_R\times \USp(2)_2\times \USp(2)_4}$.}
\end{table}


\begin{table}[ht]
  \footnotesize\renewcommand{\arraystretch}{1.3}
  \label{spec8}
  \begin{center}
    $\begin{array}{|c|D{.}{\times}{13}|c|c|c|c|c|c|}
        \hline
        \text{Model~\hyperref[model8]{8}} & \mc{\text{Quantum Number}}     & Q_{4 C} & Q_{2 L} & Q_{2 R} & Q_{\text{em}}       & B-L              & \text{Field} \\
        \hline   
ab                          & 2.(\bar{4},2,1,1,1)                           & -1 & 1  & 0  & -\frac{1}{3},\frac{2}{3},-1,0 & \frac{1}{3},-1 & Q_L,L_L\\
a b'                        & 1.(\bar{4},\bar{2},1,1,1)                     & -1 & -1 & 0  & -\frac{1}{3},\frac{2}{3},-1,0 & \frac{1}{3},-1 & Q_L,L_L\\
ac                          & 3.(4,1,\bar{2},1,1)                           & 1  & 0  & -1 & \frac{1}{3},-\frac{2}{3},1,0  & -\frac{1}{3},1 & Q_R,L_R\\
b_{\overline{\yng(1,1)_{}}} & 2.(1,\bar{1}_{\overline{\yng(1,1)_{}}},1,1,1) & 0  & -1 & 0  & 0                             & 0              & S_L    \\
c_{\overline{\yng(1,1)_{}}} & 3.(1,1,\bar{1}_{\overline{\yng(1,1)_{}}},1,1) & 0  & 0  & -1 & 0                             & 0              & S_R    \\
b_{\yng(2)_{}}              & 2.(1,3_{\yng(2)_{}},1,1,1)                    & 0  & 2  & 0  & 0,\pm 1                       & 0              & T_L    \\
c_{\yng(2)_{}}              & 3.(1,1,3_{\yng(2)_{}},1,1)                    & 0  & 0  & 2  & 0,\pm 1                       & 0              & T_R    \\
\hline
      \end{array}$
  \end{center}
  \caption{The chiral and vector-like superfields, and their quantum numbers under the gauge symmetry ${\SU(4)_C\times \SU(2)_L \times \SU(2)_R\times \USp(2)_1\times \USp(4)_3}$.}
\end{table}


\begin{table}[ht]
  \footnotesize\renewcommand{\arraystretch}{1.3}
  \label{spec9}
  \begin{center}
    $\begin{array}{|c|D{.}{\times}{13}|c|c|c|c|c|c|}
        \hline
        \text{Model~\hyperref[model9]{9}} & \mc{\text{Quantum Number}}     & Q_{4 C} & Q_{2 L} & Q_{2 R} & Q_{\text{em}}       & B-L              & \text{Field} \\
        \hline   
ab                          & 3.(4,\bar{2},1,1,1)                           & 1  & -1 & 0  & -\frac{1}{3},\frac{2}{3},-1,0 & \frac{1}{3},-1 & Q_L,L_L\\
ac                          & 3.(\bar{4},1,2,1,1)                           & -1 & 0  & 1  & \frac{1}{3},-\frac{2}{3},1,0  & -\frac{1}{3},1 & Q_R,L_R\\
b_{\overline{\yng(1,1)_{}}} & 3.(1,\bar{1}_{\overline{\yng(1,1)_{}}},1,1,1) & 0  & -1 & 0  & 0                             & 0              & S_L    \\
c_{\yng(1,1)_{}}            & 2.(1,1,1_{\yng(1,1)_{}},1,1)                  & 0  & 0  & 1  & 0                             & 0              & S_R    \\
b_{\yng(2)_{}}              & 3.(1,3_{\yng(2)_{}},1,1,1)                    & 0  & 2  & 0  & 0,\pm 1                       & 0              & T_L    \\
c_{\overline{\yng(2)_{}}}   & 2.(1,1,\bar{3}_{\overline{\yng(2)_{}}},1,1)   & 0  & 0  & -2 & 0,\pm 1                       & 0              & T_R    \\
b1                          & 4.(1,\bar{2},1,2,1)                           & 0  & -1 & 0  & \pm \frac{1}{2}               & 0              & X_L    \\
b3                          & 1.(1,2,1,1,1)                                 & 0  & 1  & 0  & \pm \frac{1}{2}               & 0              & X_L    \\
c3                          & 3.(1,1,2,1,1)                                 & 0  & 0  & 1  & \pm \frac{1}{2}               & 0              & X_R    \\
b c'                        & 4.(1,\bar{2},\bar{2},1,1)                     & 0  & -1 & -1 & -1,0,0,1                      & 0              & H      \\ \hline
      \end{array}$
  \end{center}
  \caption{The chiral and vector-like superfields, and their quantum numbers under the gauge symmetry ${\SU(4)_C\times \SU(2)_L \times \SU(2)_R\times \USp(2)_1\times \USp(4)_3}$.}
\end{table}


\begin{table}[ht]
  \footnotesize\renewcommand{\arraystretch}{1.3}
  \label{spec9.5}
  \begin{center}
    $%
$
  \end{center}
  \caption{The chiral and vector-like superfields, and their quantum numbers under the gauge symmetry ${\SU(4)_C\times \SU(2)_L \times \SU(2)_R\times \USp(4)_3}$.}
\end{table}


\begin{table}[ht]
  \footnotesize\renewcommand{\arraystretch}{1.3}
  \label{spec11}
  \begin{center}
    $%
$
  \end{center}
  \caption{The chiral and vector-like superfields, and their quantum numbers under the gauge symmetry ${\SU(4)_C\times \SU(2)_L \times \SU(2)_R\times \USp(4)_3}$.}
\end{table}


\begin{table}[ht]
  \footnotesize\renewcommand{\arraystretch}{1.3}
  \label{spec11.5}
  \begin{center}
    $%
$
  \end{center}
  \caption{The chiral and vector-like superfields, and their quantum numbers under the gauge symmetry ${\SU(4)_C\times \SU(2)_L \times \SU(2)_R\times \USp(2)_1\times \USp(2)_2\times \USp(2)_3\times \USp(2)_4}$.}
\end{table}


\begin{table}[ht]
  \footnotesize\renewcommand{\arraystretch}{1.3}
  \label{spec15}
  \begin{center}
    $\begin{array}{|c|D{.}{\times}{15}|c|c|c|c|c|c|}
        \hline
        \text{Model~\hyperref[model15]{15}} & \mc{\text{Quantum Number}}     & Q_{4 C} & Q_{2 L} & Q_{2 R} & Q_{\text{em}}       & B-L              & \text{Field} \\
        \hline   
ab                          & 3.(\bar{4},2,1,1,1,1)                           & -1 & 1  & 0  & -\frac{1}{3},\frac{2}{3},-1,0 & \frac{1}{3},-1 & Q_L,L_L\\
ac                          & 3.(4,1,\bar{2},1,1,1)                           & 1  & 0  & -1 & \frac{1}{3},-\frac{2}{3},1,0  & -\frac{1}{3},1 & Q_R,L_R\\
a_{\yng(1,1)_{}}            & 4.(6_{\yng(1,1)_{}},1,1,1,1,1)                  & 1  & 0  & 0  & \frac{1}{3},-1                & \frac{2}{3},-2 & S_C    \\
b_{\yng(1,1)_{}}            & 3.(1,1_{\yng(1,1)_{}},1,1,1,1)                  & 0  & 1  & 0  & 0                             & 0              & S_L    \\
c_{\overline{\yng(1,1)_{}}} & 1.(1,1,\bar{1}_{\overline{\yng(1,1)_{}}},1,1,1) & 0  & 0  & -1 & 0                             & 0              & S_R    \\
b_{\overline{\yng(2)_{}}}   & 3.(1,\bar{3}_{\overline{\yng(2)_{}}},1,1,1,1)   & 0  & -2 & 0  & 0,\pm 1                       & 0              & T_L    \\
c_{\yng(2)_{}}              & 1.(1,1,3_{\yng(2)_{}},1,1,1)                    & 0  & 0  & 2  & 0,\pm 1                       & 0              & T_R    \\
\hline  
      \end{array}$
  \end{center}
  \caption{The chiral and vector-like superfields, and their quantum numbers under the gauge symmetry ${\SU(4)_C\times \SU(2)_L \times \SU(2)_R\times \USp(4)_1\times \USp(2)_3\times \USp(2)_4}$.}
\end{table}


\begin{table}[ht]
  \footnotesize\renewcommand{\arraystretch}{1.3}
  \label{spec15.5}
  \begin{center}
    $%
$
  \end{center}
  \caption{The chiral and vector-like superfields, and their quantum numbers under the gauge symmetry ${\SU(4)_C\times \SU(2)_L \times \SU(2)_R\times \USp(4)_1\times \USp(2)_2\times \USp(2)_4}$.}
\end{table}


\begin{table}[ht]
  \footnotesize\renewcommand{\arraystretch}{1.3}
  \label{spec16}
  \begin{center}
    $%
$
  \end{center}
  \caption{The chiral and vector-like superfields, and their quantum numbers under the gauge symmetry ${\SU(4)_C\times \SU(2)_L \times \SU(2)_R\times \USp(4)_1\times \USp(2)_4}$.}
\end{table}


\begin{table}[ht]
  \footnotesize\renewcommand{\arraystretch}{1.3}
  \label{spec16.5}
  \begin{center}
    $%
$
  \end{center}
  \caption{The chiral and vector-like superfields, and their quantum numbers under the gauge symmetry ${\SU(4)_C\times \SU(2)_L \times \SU(2)_R\times \USp(2)_2\times \USp(2)_3\times \USp(2)_4}$.}
\end{table}


\begin{table}[ht]
  \footnotesize\renewcommand{\arraystretch}{1.3}
  \label{spec17.5}
  \begin{center}
    $%
$
  \end{center}
  \caption{The chiral and vector-like superfields, and their quantum numbers under the gauge symmetry ${\SU(4)_C\times \SU(2)_L \times \SU(2)_R\times \USp(2)_2\times \USp(2)_3\times \USp(2)_4}$.}
\end{table}


\begin{table}[ht]
  \footnotesize\renewcommand{\arraystretch}{1.3}
  \label{spec18.5}
  \begin{center}
    $%
$
  \end{center}
  \caption{The chiral and vector-like superfields, and their quantum numbers under the gauge symmetry ${\SU(4)_C\times \SU(2)_L \times \SU(2)_R\times \USp(2)_1\times \USp(2)_4}$.}
\end{table}


\begin{table}[ht]
  \footnotesize\renewcommand{\arraystretch}{1.3}
  \label{spec19.5}
  \begin{center}
    $%
$
  \end{center}
  \caption{The chiral and vector-like superfields, and their quantum numbers under the gauge symmetry ${\SU(4)_C\times \SU(2)_L \times \SU(2)_R\times \USp(2)_3}$.}
\end{table}


\begin{table}[ht]
  \footnotesize\renewcommand{\arraystretch}{1.3}
  \label{spec22.5}
  \begin{center}
    $%
$
  \end{center}
  \caption{The chiral and vector-like superfields, and their quantum numbers under the gauge symmetry ${\SU(4)_C\times \SU(2)_L \times \SU(2)_R\times \USp(2)_3}$.}
\end{table}

\FloatBarrier
  
\bibliographystyle{JHEP}

\providecommand{\href}[2]{#2}\begingroup\raggedright\endgroup
 																										 
\end{document}